\documentclass[11pt,urlcolor=blue, linkcolor=blue]{article} 
\usepackage[utf8]{inputenc}
\usepackage[colorlinks,linkcolor=blue,citecolor=blue]{hyperref}
\usepackage{latexsym, amssymb, amsmath, amsthm, mathrsfs, bbm,}
\usepackage[all, knot]{xy}
\xyoption{arc}
\usepackage{tikz}
\usepackage{tikz-cd}                     
\usepackage{amsmath,amscd}
\usepackage{sseq}
\usepackage{color}

\def \To{\longrightarrow}
\def \Hom{\operatorname{Hom}}
\def \Tor{\operatorname{Tor}}

\def \H{\operatorname{H}}

\def \F{\mathbb{F}}
\def \C{\mathbb{C}}

\def \Z{\mathbb{Z}}
\def \Q{\mathbb{Q}}
\def \Pin{\textbf{Pin}}
\def \A{\mathcal{A}}
\def \RP{\mathbb{RP}}

\def\Ext{\operatorname{Ext}}
\numberwithin{equation}{section}

\def\doubleunderline#1{\underline{\underline{#1}}}

\newtheorem{theorem}{Theorem}

\newtheorem{definition}[theorem]{Definition}

\newcommand{\eqn}[1]{eqn.~(\ref{#1})}

\usepackage{changepage}

\newcommand{\bea}{\begin{eqnarray}}
\newcommand{\eea}{\end{eqnarray}}

\usepackage[paperheight=279.4mm]{geometry}
\usepackage{anysize}\marginsize{20mm}{20mm}{20mm}{20mm}

\usetikzlibrary{shadings}
\usetikzlibrary{arrows.meta}
\usetikzlibrary{shapes.arrows}
\usepackage{wasysym}

\DeclareMathOperator{\PD}{PD}
\DeclareMathOperator{\ABK}{ABK}
\DeclareMathOperator{\Arf}{Arf}
\DeclareMathOperator{\fSPT}{fSPT}
\DeclareMathOperator{\Bord}{Bord}
\DeclareMathOperator{\Vect}{Vect}
\newcommand{\VectZtwo}{\Vect_{\C}^{\Z_2}}
\newcommand{\Bordspin}[1]{\Bord_{#1}^{Spin}}

\newcommand{\Bordspinbdry}[1]{\Bord_{#1}^{Spin,\partial}}
\newcommand{\dd}{n}
\usepackage{tabu}

\begin{document}

\begin{titlepage}
\begin{center}

{\bf\LARGE{
Fermionic Finite-Group Gauge Theories\\[6mm]
and Interacting Symmetric/Crystalline Orders\\[8mm]
via Cobordisms
}}

\vskip0.5cm 
\Large{Meng Guo$^{1,2,3}$,  Kantaro Ohmori$^{4}$, Pavel Putrov$^{4,5}$,  \\[2mm] 
Zheyan Wan$^{6}$, Juven Wang$^{4,7}$} 

\vskip.2cm
{\small{\textit{$^1${Department of Mathematics, Harvard University, Cambridge, MA 02138, USA}\\}}
}

 \vskip.2cm
{\small{\textit{$^2${Perimeter Institute for Theoretical Physics, Waterloo, ON, N2L 2Y5, Canada}\\}}
}

 \vskip.2cm
{\small{\textit{$^3${Department of Mathematics, University of Toronto, Toronto, ON M5S 2E4, Canada}\\}}
}

\vskip.2cm
 {\small{\textit{$^4$School of Natural Sciences, Institute for Advanced Study, Princeton, NJ 08540, USA}\\}}

\vskip.2cm
{\small{\textit{$^5$
{ICTP, Trieste 34151, Italy}\\}}
}
 
\vskip.2cm
 {\small{\textit{$^6${School of Mathematical Sciences, USTC, Hefei 230026, China}\\}}
}
 
 \vskip.2cm
 {\small{\textit{$^7${Center of Mathematical Sciences and Applications, Harvard University,  Cambridge, MA, USA} \\}}
}

\end{center}

\begin{abstract}

We formulate a family of spin Topological Quantum Filed Theories (spin-TQFTs) as fermionic generalization 
of bosonic Dijkgraaf-Witten TQFTs. They are obtained by gauging $G$-equivariant invertible spin-TQFTs, or, in physics language, 
gauging the interacting fermionic Symmetry Protected Topological states (SPTs) with a finite group $G$ symmetry. We use the fact that the latter are classified by Pontryagin duals to spin-bordism groups of the classifying space $BG$. 
{We also consider unoriented analogues, that is $G$-equivariant invertible pin$^\pm$-TQFTs (fermionic time-reversal-SPTs) and their gauging.} 
We compute these groups for various examples of abelian $G$ using Adams spectral sequence and describe all corresponding TQFTs via certain bordism invariants in dimensions 3, 4, and other.
This gives explicit formulas for the partition functions of spin-TQFTs on closed manifolds with possible extended operators inserted. The results also provide explicit classification of 't Hooft anomalies of fermionic QFTs with finite abelian group symmetries in one dimension lower.
We construct new anomalous boundary deconfined spin-TQFTs (surface fermionic topological orders).
We explore SPT and SET (symmetry enriched topologically ordered) states, and crystalline SPTs protected by space-group (e.g.\ translation $\mathbb{Z}$) 
or point-group (e.g.\ reflection, inversion or rotation $C_m$) symmetries, 
via the layer-stacking 
construction. 
\\[4mm]

\end{abstract}
\end{titlepage}

\tableofcontents   

\section{Introduction and summary}
Topological quantum field theories (TQFTs) play an important role both in physics and mathematics. In physics, they can be considered as the simplest examples of quantum field theories. Unlike the mathematically poor-defined
``physical'' QFTs (in particular, the gauge theories in the  Standard model, that describe 
interactions between the quarks, leptons and gauge mediator bosons), 
TQFTs instead can be mathematically well-defined.  
Moreover, TQFTs not only play the role of toy models, but can also be used to describe topological phases of condensed matter system. Another application of TQFTs is systematic description of anomalies of general (i.e.\ not necessarily topological) QFTs in one less dimension. 

In mathematics, the TQFTs provide a natural framework for topological invariants of manifolds as well as invariants of embedded submanifolds up to ambient isotopy (i.e.\ links in 3-manifolds). The TQFT structure allows calculation of invariants on complicated manifolds via surgery.

In this work, we are interested in study of spin-TQFTs, that is TQFTs that provide invariants of manifolds that depend not just on their topology but also a choice of spin structure. From physics point of view this generalization is very natural, since the physical system often contain fermions (in particular, the matter in the real world is composed of fermions). Therefore the spin structure on the space-time manifold is needed to describe how spinors transform under parallel transport.

From mathematics point of view, the spin-structure is also quite natural, since it provides a lift of $SO(n)$-principle bundle of orthonormal frames in the tangent bundle to a principle bundle with a simply-connected structure group $Spin(n)$. Such extra structure often allows construction of more refined invariants.

The structure of the article is the following. In Section \ref{sec:spin-TQFT}, we review the notion of spin-TQFT and describe a way to produce a family of spin-TQFTs labeled by elements of a spin-bordism groups of a classifying space of a finite group. In Section \ref{sec:spin-invariants}, we review some invariants of spin-manifolds that will be useful for construction of spin-bordism invariants that appear in expressions for spin-TQFT partition function on a closed manifold. In Section \ref{sec:computation}, we compute of spin-bordism groups of classifying spaces of some simple abelian groups. In Section \ref{sec:links}, we consider the invariants of (surface) links produced by such spin-TQFTs. In Section \ref{sec:spin-twisted-bordism}, we calculate bordism groups of manifolds with $Spin\times_{\Z_2} \Z_{2m}$ structure, extending some of the results previously appeared in the literature. In Section \ref{sec:TR-Pin-TQFT}, we briefly consider unoriented analogues of spin-TQFTs, that is as the pin$^\pm$-TQFTs.
In Section \ref{sec:SbTQFT} , we construct some symmetry-preserving spin-TQFTs living on the boundary of one-dimension-higher bulk SPTs, explicitly in a TQFT language.
In Section \ref{sec:crystalline}, we introduce the crystalline-SPTs that correspond to bordism invariants of manifolds with $G$-bundles, where 
the symmetry $G$ involves a spatial translation $\Z$-symmetry, and then relate them to various previously constructed SPTs with finite abelian symmetries.
In Section \ref{sec:cond-mat}, we comment on interpretation of the results in topological quantum matter. 

For notational convention, we abbreviate the $n$-dimensional spacetime as $n$d. However, in certain cases, we may also use the notation from 
condensed matter community denoting $(n'+1)$D as 
an $n'+1$-dimensional space$+$time, where $n=n'+1$.
We also use a shorthand notation fSPTs for the fermionic SPTs.
In Section \ref{sec:cond-mat}, other than fSPTs protected by  internal symmetries (denoted as a finite group $G$-symmetry in this article),
we will also explore crystalline-fSPTs, which means that the
SPTs is also protected by space group  (such as the translational symmetry) or point group symmetries (e.g.\ the rotational symmetry).
We write the fermion parity symmetry as $\Z_2^f$, where $\Z_2^f$ is always implicitly included for any fSPTs. It can also be understood as the center of $Spin(n)$ symmetry that extends $SO(n)$ Euclidean spacetime symmetry.
When we omit writing $\Z_2^f$ but only quote $G$-fSPTs, 
it means fSPTs with $G$ and $\Z_2^f$-symmetry, more precisely it means an invertible spin TQFT with $G \times \Z_2^f$-symmetry.

\section{Spin-TQFTs and fermionic gauge theories}
\label{sec:spin-TQFT}
In this section, for completeness, we briefly review the notions of spin-TQFT and its equivariant version. 

An ordinary $n$-dimensional TQFT can be defined as a symmetric monoidal functor $Z$ from the $n$-dimensional bordism category $\mathrm{Bord}_n$ to the category of complex vector spaces $\text{Vect}_\mathbb{C}$ \cite{atiyah1988topological}. The objects of  $\mathrm{Bord}_n$ are oriented smooth\footnote{There are generalization without theses conditions.} closed $(n-1)$-manifolds. The morphisms are $n$-dimensional bordisms between them modulo diffeomorphisms. A bordism from an $(n-1)$-manifold $M_1^{n-1}$ to an $(n-1)$-manifold $M_2^{n-1}$ is an oriented smooth $n$-manifold $M^n$ with an isomorphism $\partial M^n \cong \bar{M_1}^{n-1} \sqcup {M_2}^{n-1}$ where bar denotes change of orientation. The fact that the functor is symmetric monoidal means that both categories are treated as symmetric monoidal categories and the functor respects this structure. The tensor product product structure in the bordism category is given by disjoint union, which is symmetric. The role of the unit is played by the empty manifold $\emptyset$. This gives a symmetric monoidal structure on $\mathrm{Bord}^n$. On $\mathrm{Vect}_\C$ the symmetric monoidal structure is given by the ordinary tensor product with unit being $\C$. The value of the functor $Z$ on a closed $n$-manifold $M^n$ is then a linear map $Z(M^n):\C\rightarrow \C$, which can be identified with an element of $\C$ itself. The complex number $Z(M^n)$ is often referred to as \textit{partition function}. Physically an objects of the bordism category, an $(n-1)$-dimensional manifolds $M^{n-1}$, is a spatial manifolds on which the theory is quantized and the corresponding vector space, that is the value of the functor $Z(M^{n-1})$, is the Hilbert space of the quantum theory\footnote{In physics literature such spaces are often denoted by a different symbol, e.g.\ $\mathcal{H}(M^{n-1})$, while $Z$ is only used for partition function, that is the value of the functor $Z$ on a closed $n$-manifold.}.

 This definition has various natural generalizations which involve a choice of some additional structures on manifolds (provided a structure on an $n$-manifold induces a structure on its $(n-1)$-dimensional boundary). In particular, one can consider manifolds with spin structure, that is a lift of $SO(n)$ orthonormal frame bundle to $Spin(n)$ principle bundle, with respect to the extension $\Z_2\rightarrow Spin(n) \rightarrow SO(n)$. A spin structure on a manifold induces a natural spin structure on its boundary (see e.g.\ \cite{kirby1990pin}), therefore one can define the corresponding bordism category of spin manifolds $\mathrm{Bord}_n^{Spin}$. The vector spaces associated to $(n-1)$-dimensional manifolds can be equipped with $\Z_2$-grading: $Z(M^{n-1})=Z^0(M^{n-1})\oplus Z^1(M^{n-1})$. Physically the even and odd parts of vector spaces are bosonic and fermionic states respectively in the Hilbert space $Z(M^{n-1})$.  One can therefore give the following definition (cf.\ \cite{blanchet1992invariants,blanchet1996topological,beliakova2014spin}): 
\begin{definition}
 An $n$-dimensional spin-TQFT is symmetric monoidal functor
 \begin{equation}
   Z:\; \mathrm{Bord}_{n}^{Spin} \rightarrow \mathrm{Vect}_{\C}^{\Z_2}
 \end{equation} 
where $\mathrm{Bord}_{n}^{Spin}$ is the category of $n$-dimensional spin-bordisms and $\mathrm{Vect}_{\C}^{\Z_2}$ is the category of $\Z_2$-graded vector spaces\footnote{The morphisms are grading preserving linear maps.}, satisfying
\begin{equation}
 Z(M^{n-1}\times S^1_{\pm}) =  \dim Z^{0}(M^{n-1})\pm\dim Z^{1}(M^{n-1})
 \label{spin-statistics}
\end{equation} 
for any closed $(n-1)$-dimensional spin-manifold $M^{n-1}$, where $S^1_{\pm}$ is a circle with even/odd-spin structure.
\end{definition}
The last condition can be understood as topological spin-statistics constraint. The notion can be also generalized to non-orientable manifolds and bordisms between them. The analog of the spin-structure is pin$^\pm$ structure, the lift of $O(n)$ bundle with respect to one of the two non-trivial central extensions $\Z_2\rightarrow Pin^\pm(n) \rightarrow O(n)$. The extensions can be distinguished by the following commutative diagram
\begin{equation}
\begin{tikzcd}
\Z_2 \ar[r] \ar[d,equal] & Pin^\pm(n) \ar[r] & O(n) \\
\Z_2 \ar[r]  & Pin^\pm(1) \ar[r] \ar[u, hook] & O(1) \ar[u,hook]
\end{tikzcd}
\label{pin-groups}
\end{equation} 
where $O(1)\cong \Z_2$ and $Pin^+(1)\cong \Z_2\times \Z_2$, $Pin^-(1)\cong \Z_4$. 
In this work, we firstly focus on orientable case, and later discuss non-orientable pin$^{\pm}$ cases in Section \ref{sec:TR-Pin-TQFT}.

Another natural generalization involves a choice of (isomorphism class of) principal $G$-bundle over manifolds, where $G$ is topological group. Equivalently, one can consider maps to its classifying space\footnote{Which can be defined as a connected topological space (unique up to homotopy) satisfying $\pi_1(BG)=1$, $\pi_{i}(BG)=0,\;i>1$} $BG$ up to homotopy. One can consider the corresponding bordism category $\mathrm{Bord}_n(BG)$ where objects are pairs $(M^{n-1},f:M^{n-1}\rightarrow BG)$. The morphisms between $(M^{n-1}_1,f_1)$ and $(M^{n-1}_2,f_2)$ are pairs $(M^n,g:M^n\rightarrow BG)$ such that\footnote{As before, the bordism data contains a choice of the isomorphism between $M^{n-1}_2$ and the boundary, however we will usually not indicate it explicitly.} $\partial M^n\cong\bar{M}^{n-1}_1 \sqcup M^{n-1}_2$, $g|_{\bar{M}^{n-1}_1}=f_1$, $g|_{\bar{M}^{n-1}_2}=f_2$. Then $G$-\textit{equivariant} TQFT can be defined as a symmetric monoidal functor $\mathrm{Bord}_n(BG) \rightarrow \text{Vect}_{\mathbb{C}}$. Physically the group $G$ has meaning of global symmetry group of the theory.\footnote{In the sense that one can couple the theory to a background $G$ gauge field. There is no condition that $G$ acts faithfully on the operators of the theory.}

One can also, of course, combine the notions above and consider $G$-equivariant spin-TQFT which is a symmetric monoidal functor $\mathrm{Bord}_n^{Spin}(BG) \rightarrow \text{Vect}_{\mathbb{C}}^{\Z_2}$. Even more generally, one can consider the bordism category for any extension $H_n\rightarrow SO(n)$ (or $O(n)$ in non-orientable case) and the corresponding bordism category $\text{Bord}_n^{H_n}$ where manifolds are equipped with $H_n$ tangential structures, that is principle $H_n$-bundles that are lifts of $SO(n)$ orthonormal tangent frame bundles. Then one can define TQFT with symmetry $H_n$ as a functor from $\text{Bord}_n^{H_n}$ to the category of complex vector spaces (see \cite{FH} for details). The case of $G$-equivariant spin-TQFT is then the particular case with $H_n=Spin(n)\times G$.

In what follows we will also use the notion of an \textit{invertible} TQFT. This is a TQFT such that the value of the functor on any $(n-1)$-manifold is a one dimensional vector space (that is, in general non-canonically, isomorphic to $\C$) and the value on any bordism is an invertible homomorphism. Invertible TQFTs form an abelian group, while all TQFTs only form a monoid. The product $Z_1\cdot Z_2$ of TQFT functors $Z_i$ is defined by $(Z_1\cdot Z_2)(M)=Z_1(M)\otimes Z_2(M)$. Physically taking the product of two quantum field theories means stacking them together without interaction.

\subsection{Gauging}
\label{sec:gauging}

In this work, we are interested in obtaining non-trivial spin-TQFTs by \textit{gauging} $G$-equivariant invertible spin-TQFTs with finite symmetry group $G$. The theories obtained this way can be understood as a direct generalization of Dijkgraaf-Witten TQFTs \cite{Dijkgraaf:1989pz} to the case of spin-TQFTs. From physics point of view an invertible $G$-equivariant TQFT can be understood as a classical field theory, where the partition function is given by the exponentiated action which depends on the background $G$ gauge field, while the corresponding gauged TQFT is the quantum theory where the gauge field is dynamical. Formally, the gauging procedure can be understood as the following map
\begin{equation}
\begin{array}{ccc}
\left\{\begin{array}{c}
          \text{invertible $G$-equivariant}\\
          \text{spin-TQFTs}
        \end{array}
\right\} 
&
\longrightarrow
&
\left\{ \text{spin-TQFTs}
\right\} 
\\
\\
Z & \longmapsto & Z_\text{gauged}
\end{array}
\label{gauging-map}
\end{equation} 
such that the values of functors on closed $n$-manifolds (i.e.\ partition function) are related as follows:
\begin{equation}
 Z_\text{gauged}(M^n) = \sum_{[f] \in [M^n,BG]} \frac{1}{|\mathrm{Aut}(f)|}\,
 Z(M^n,f)
 \label{gauged-TQFT-closed}
\end{equation} 
where the sum is performed over $[M^n,BG]$, homotopy classes of maps from $M^n$ to $BG$, and $\mathrm{Aut}(f)$ denotes the automorphism group of a principle $G$-bundle over $M^n$ corresponding to the map $f:M^n\rightarrow BG$. Equivalently, $\mathrm{Aut}(f)$ can be understood as the fundamental group of the corresponding connected component of $f$ in the space of maps $\mathrm{Maps}(M^n,BG)$. 
In the case when $G$ is abelian, there is an isomorphism $[M^n,BG]\cong H^1(M^3,G)$ which is the property of the Eilenberg-MacLane space $K(G,1)\equiv BG$. 
More explicitly, the function $f:M^n\rightarrow BG$ corresponds to an element $f^*(e)\in H^1(M^n,G)$ where $e$ is a generator of $H^1(BG,G)$. In this paper, however, we will use the same symbol (e.g.\ $f$ above) for both a function $M^n\rightarrow BG$ and the corresponding element in $H^1(M^n,BG)$. In the abelian case $\mathrm{Aut}(f)=G$ for any $f$.  

The relation (\ref{gauged-TQFT-closed}) can be extended to the full functors analogously to Dijkgraaf-Witten theories \cite{Freed:1991bn,LurieLectures}. In particular, the values of the functor on the objects are related as follows:
\begin{equation}
 Z_{\text{gauged}}(M^{n-1})=\bigoplus_{[f]\in C_{M^{n-1}}}Z(M^{n-1},f)
 \label{gauged-hilbert-space}
\end{equation} 
where $C_{M^{n-1}}$ is the following subset of the set of homotopy classes of maps $M^{n-1}\rightarrow BG$:
\begin{equation}
  C_{M^{n-1}}:=\left\{
   [f]\;\Big|
   \begin{array}{c}
\;Z(M^{n-1}\times S^1,g)\in \C \text{ is the same} \\
\text{for all } g, \text{ s.t.\ } g|_{M^{n-1}\times \text{pt}} \sim f
   \end{array}
  \right\} \subset [M^{n-1},BG].
\end{equation}

This subset of $[M^{n-1},BG]$ has the following meaning. From functoriality of $Z$ it follows that $Z(M^{n-1},f)$ forms a (one dimensional) representation of $\mathrm{Aut}(f)$. The representation is realized as follows. As was mentioned above, an element $[h]\in \mathrm{Aut}(f)$ correspond to a closed path in $\mathrm{Maps}(M^{n-1},BG)$ starting and ending at $f$, considered up to homotopy. Each such path gives a function $h:[0,1]\rightarrow \mathrm{Maps}(M^{n-1},BG)$, or, equivalently, $h':[0,1]\times M^{n-1} \rightarrow BG$, such that $h'|_{0}=h'|_{1}=f$. Then 
\begin{equation}
 Z([0,1]\times M^{n-1},h'): Z(M^{n-1},f)\rightarrow Z(M^{n-1},f)
\end{equation} 
provides the action of $h\in \mathrm{Aut}(f)$ on $Z(M^{n-1},f)$. Then $C_{M^{n-1}}$ appearing in (\ref{gauged-hilbert-space}) can be also defined as the subset of homotopy classes of functions  $f:M^{n-1}\rightarrow BG$ such that $Z(M^{n-1},f)$ is a trivial representation of $\mathrm{Aut}(f)$. Physically this can be interpreted as the Gauss law constraint. One can easily see that the condition (\ref{spin-statistics}) is automatically satisfied. In particular, 
\begin{multline}
 Z_\text{gauged}(M^{n-1}\times S^1_\pm) = \sum_{[g] \in [M^{n-1}\times S^1,BG]} \frac{1}{|\mathrm{Aut}(g)|}\,
 Z(M^{n-1}\times S^1_\pm,fg)
 = \\
 \sum_{[f] \in [M^{n-1},BG]} \frac{1}{|\mathrm{Aut}(f)|} \sum_{[h] \in \text{Aut}(f)}
 Z(M^{n-1}\times S^1_\pm,h')=
 \sum_{[f] \in C_{M^{n-1}}} 
 Z(M^{n-1}\times S^1_\pm,\text{pr}^*f)
\end{multline} 
where the map $h'$ is related to the element $h\in \mathrm{Aut}(f)$ as above and $\text{pr}^*$ is the pullback with respect to the projection map $\text{pr}:\,M^{n-1}\times S^1\rightarrow M^{n-1}$. In the last equality we used the fact that $Z(M^{n-1}\times S^1_\pm,h')$ is the character of one dimensional representation $Z(M^{n-1},f)$ of group $\mathrm{Aut}(f)$ defined above, and applied orthogonality property of characters. Only characters of trivial representations survive after taking the sums over the elements of the group. Finally, since $Z$ is an invertible spin-TQFT, we have:
\begin{equation}
 Z(M^{n-1}\times S^1_\pm,\text{pr}^*f) = \dim Z^0(M^{n-1},f)\pm \dim Z^1(M^{n-1},f) 
\end{equation} 
where in the right hand side one of the graded dimensions is zero and the other is $1$. In particular, the total dimension of the Hilbert space of the gauged spin-TQFT on $M^{n-1}$ is given by
\begin{equation}
 \dim Z_\text{gauged}(M^{n-1}) = |C_{M^{n-1}}|.
\end{equation} 

The value of the functor $Z_\text{gauged}$ on a bordism $M^n$ between $M_1^{n-1}$ and $M_2^{n-1}$ is then a linear map
\begin{equation}
 Z_\text{gauged}(M^{n}):\;\;
 \bigoplus_{[f_1]\in C_{M^{n-1}_1}}Z(M^{n-1}_1,f_1)
 \longrightarrow
 \bigoplus_{[f_2]\in C_{M^{n-1}_2}}Z(M^{n-1}_2,f_2)
 \label{gauged-hom}
\end{equation} 
given by the following expression:
\begin{equation}
 Z_\text{gauged}(M^{n})=\bigoplus_{[f_{1,2}]\in C_{M_{1,2}^{n-1}}}\;\; 
 \sum_{\footnotesize\begin{array}{c}
        [g] \\
        g|_{M_{1,2}^{n-1}}=f_{1,2}
       \end{array}}
\;
 \frac{Z(M^{n},g)}{|\mathrm{Aut}(g)|\,|\mathrm{Aut}(f_1)|}
 \label{gauged-bordism-value}
\end{equation} 
where
\begin{equation}
 Z(M^{n},g):\;\; Z(M^{n-1}_1,f_1)
 \longrightarrow
 Z(M^{n-1}_2,f_2).
\end{equation} 
The factor $1/|\mathrm{Aut}(f_1)|$ is needed so that the functor satisfies composition property (cf. \cite{Freed:1991bn}). The formulas (\ref{gauged-hom}) and (\ref{gauged-hom}) then can be considered as the definition of the gauging map (\ref{gauging-map}).

Invertible TQFTs with symmetry $H_n$, satisfying certain additional properties: \textit{reflection-positivity} and being \textit{extended}, were classified in \cite{FH} (also in \cite{Kapustin:2014tfa,Kapustin:2014dxa,Yonekura:2018ufj} from more physical perspective). From physics point of view these additional requirements are quite natural and expected in theories describing realistic quantum systems. Namely, reflection-positivity is a Wick-rotated version of unitarity and being extended corresponds to locality. In the case of $H_n=Spin(n)\times G$ the result can be formulated as follows\footnote{Note that the free part of the classification, conjecturally given by
\begin{equation}
	\text{Hom}(\Omega_{n+1}^{Spin}(BG),\Z),
\end{equation}
 contains Chern-Simons-like theories, which are not strictly topological.}:
\begin{equation}
 \text{Tor}\,\left\{
   \begin{array}{c}
     \text{deformation classes of $G$-equivariant} \\
     \text{reflection-positive invertible spin-TQFTs}
    \end{array}
 \right\}
 \cong
 \text{Hom}(\text{Tor}\,\Omega_n^{Spin}(BG),U(1))
 \label{equiv-spin-inv-classification}
\end{equation}
More general formulation of the result will be mentioned and used in Section \ref{sec:spin-twisted-bordism}. 
As was pointed out above, invertible TQFTs form an abelian group (the additional conditions are respected by the product) and the isomorphism above should be understood as an isomorphism between abelian groups. The right hand side is the Pontryagin dual to the torsion subgroup of spin-bordism group of $BG$, which is defined as follows:
\begin{equation}
	\Omega_n^{Spin}(BG) := 
	\left\{
	\begin{array}{c}
	   \text{pairs }(M^{n},f:M^{n}\rightarrow BG), \\
		M^n\text{ is spin $n$-manifold}
	\end{array}
	\right\}/\sim\text{ bordisms}
	\label{spin-bordism-group-def}
\end{equation}
where bordisms are understood as morphisms in the category $\mathrm{Bord}_{n+1}^{Spin}(BG)$ defined above. The set of equivalence classes has a natural abelian group structure under disjoint union operation. This abelian group is always discrete and finitely generated, that is isomorphic to the finite product of finite cyclic groups and copies of $\Z$.

The correspondence between $G$-equivariant TQFTs and the elements of the abelian group in the right hand side of (\ref{equiv-spin-inv-classification}) is realized as follows. The embedding of the torsion subgroup $\Tor\Omega_n^{Spin}(BG)$ into the full spin bordism group induces a surjective map between their Pontryagin duals.
Moreover, the connected elements of the group $\text{Hom}(\Omega_n^{Spin}(BG),U(1))$ map to the same elements of the Pontryagin dual to the torsion subgroup. Therefore the right hand side of (\ref{equiv-spin-inv-classification}) can be understood as the group of connected components:
\begin{equation}
	\text{Hom}(\text{Tor}\,\Omega_n^{Spin}(BG),U(1)) = \pi_0\text{Hom}(\Omega_n^{Spin}(BG),U(1))
\end{equation}

One can consider a $G$-equivariant spin-TQFT $Z^\mu$ corresponding to an element
\begin{equation}
	\mu \in \text{Hom}(\Omega_n^{Spin}(BG),U(1))
\end{equation}
so that TQFTs corresponding to elements in the same connected component can be continuously deformed into each other\footnote{Meaning that there are continuous maps from a path connecting points in $\text{Hom}(\Omega_n^{Spin}(BG),U(1))$ to all values of the TQFT functor.}. Such TQFT can be characterized by its values on closed manifolds as follows:
\begin{equation}
	Z^\mu(M^n,g)=\mu\big([(M^n,g)]\big)\; \in U(1) \subset \C
\end{equation}
where $[\,\cdot\,]$ denotes a class in the bordism group (\ref{spin-bordism-group-def}). Applying the gauging map described above one can consider a (generically non-invertible) spin-TQFT $Z^\mu_\text{gauged}$ labeled by the elements of the same set. Note that non-torsion elements only appear in $\Omega_n^{Spin}(BG)$ in dimensions $n=0\mod 4$.

\subsection{Relation to Dijkgraaf-Witten gauge theories and bosonic TQFTs}
\label{sec:relation-to-DW}

The TQFTs labeled by the elements $\mu \in \text{Hom}(\Omega_n^{Spin}(BG),U(1))$ which are constructed above can be understood as generalizations of Dijkgraaf-Witten  topological gauge theories \cite{Dijkgraaf:1989pz}, for both ungauged (``classical'') and gauged (``quantum'') versions. The Dijkgraaf-Witten theories are labeled by elements of $H^n(BG,U(1))$. The explicit relation between two families of TQFTs is the following. The map
\begin{equation}
 \begin{array}{rcl}
    \Omega^{Spin}_n(BG) & \longrightarrow & H_n(BG,\Z)\\
    {[(M,f)]} & \longmapsto & f_*[M]
 \end{array}
 \label{spin-hom-map}
\end{equation}
induces
\begin{equation}
  H^n(BG,U(1))\cong \mathrm{Hom}(H_n(BG,\Z),U(1))
  \rightarrow
  \mathrm{Hom}(\Omega_n^{Spin}(BG),U(1)).
  \label{DW-spin-map}
\end{equation} 
This maps Dijkgraaf-Witten theories to the theories constructed above. Note that the map is in general neither surjective nor injective. Non-injectivity of the map (\ref{DW-spin-map}) (which is equivalent to non-surjectivity of (\ref{spin-hom-map})) means that SPTs
(the un-gauged Dijkgraaf-Witten theories) labeled by different elements of $H^n(BG,U(1))$ can become equivalent when considered on smooth spin-manifolds.\footnote{
Of course, it is known that Dijkgraaf-Witten theories corresponding to different elements of group cohomology can become equivalent after dynamical gauging (e.g.\ \cite{Wang2014oya1404.7854} and References therein).
However, here we  mean a more surprising statement: There are identical TQFTs even before gauging (i.e.\ SPTs). 
After gauging, there might be additional identifications corresponding to field redefinitions.
In dimension $n>6$ Dijkgraaf-Witten theories labeled by different elements of $H^n(BG,U(1))$ can become equivalent even as non-spin TQFTs, cf.\ \cite{Kapustin:2014tfa}.
This is because starting from dimension $n=7$ the map
$$H^n(BG,U(1)) \to \mathrm{Hom}(\mathrm{Tor}(\Omega^{SO}_n(BG)),U(1) )$$
is not injective in general. This is because there are examples in
degree 7 when a homology class of $BG$ cannot be represented by an image
of a smooth manifold continuously mapped to $BG$. Since SPTs/invertible
TQFTs are classified by r.h.s.\ of the above equation, the TQFTs labeled by the elements of l.h.s.\ that
map to the same element realize equivalent TQFTs (more concretely, the
actions constructed via two different classes of $H^n(BG,U(1))$ will
have the same value on any smooth $n$-manifold).} 
However this will not happen in any examples that we consider in this paper. That is, in all examples in this article, the map (\ref{spin-hom-map}) is surjective, and, therefore, (\ref{DW-spin-map}) is injective. In this case one can consider $H^n(BG,U(1))$ as a subgroup of $\Hom(\Omega_n^{Spin}(BG),U(1))$. The TQFTs $Z^\mu$ and $Z^\mu_\text{gauged}$ that correspond to elements $\mu\in \mathrm{Hom}(\Omega_n^{Spin}(BG),U(1))$ that are in the image of (\ref{DW-spin-map}) are not proper spin-TQFTs, meaning that the values of the functor $Z^\mu$ do not actually depend on spin structures. Following physics terminology we will call such TQFTs \textit{bosonic}, while others will be called \textit{fermionic}. In this work, we are interested in presenting explicit examples of the latter.

Note that invertible $n$-dimensional $G$-equivariant (non-spin) TQFTs are actually classified by $\Hom(\Tor\Omega_n^{SO}(BG),U(1))$, instead of $H^n(BG,U(1))$, where $\Omega_n^{SO}(BG)$ is the ordinary oriented bordism group of $BG$ generated by pairs $(M,f:M\rightarrow BG)$ where $M$ is an oriented manifold. The relation to Dijkgraaf-Witten theories is given by the direct analog of (\ref{spin-hom-map})-(\ref{DW-spin-map}). However, in dimension $n\leq 4$ there is no difference, that is $\Hom(\Tor\Omega_n^{SO}(BG),U(1))\cong H^n(BG,U(1))$. In general instead of 
(\ref{spin-hom-map}) one should consider the map
\begin{equation}
 \Omega^{Spin}_n(BG)\longrightarrow \Omega^{SO}_n(BG)
\end{equation} 
realized by forgetting spin-structure. The dual map
\begin{equation}
 \Hom(\Omega^{SO}_n(BG),U(1))\longrightarrow \Hom(\Omega^{Spin}_n(BG),U(1))
\end{equation} 
provides a relation between invertible $G$-equivariant spin- (fermionic) and non-spin (bosonic) TQFTs.

\subsection{Definitions of invertible spin-TQFTs v.s.\ short-range entangled fSPTs}

\label{sec:def-fSPT} 

However, there is a caveat.
Some theorists na\"ively regard
``$G$-equivariant reflection-positive invertible spin-TQFTs'' as a definition of SPTs protected by global symmetry $G$.
In contrast, other theorists define SPTs as short-range entangled (SRE) states whose existence must be protected by nontrivial global symmetry $G$.
The second definition, in some sense, is more physical and suitable for the lattice-regularized condensed matter setting. 
In quantum system, two distinct condensed matter phases cannot be deformed into each other via local unitary transformations. 
While all SPTs can be deformed into each other via local unitary transformations if all symmetry is broken, distinct 
$G$-SPTs cannot be deformed into each other when $G$ is preserved.
In the second definition, 
based on the local unitary transformation classification of phases,
the invertible spin-TQFTs protected by no symmetry (except the fermion parity $\Z_2^f$), classified
by ${\text{Hom}(\text{Tor}\,\Omega_n^{Spin}(pt),U(1))}$, are actually 
long-range entangled (LRE) invertible topologically ordered states (instead of short-range entangled SPTs).\footnote{
See a recent discussion in Ref.~\cite{2017RMPWen1610.03911} and Ref.~\cite{Prakash2018ugo1804.11236}'s Section 5.4 along this statement, and References therein.
For example, the 2d Arf invariant or equivalently  the 1+1D Kitaev fermionic chain \cite{2001KitaevWire}, obtained from the generator of ${\text{Hom}(\text{Tor}\,\Omega_2^{Spin}(pt),U(1))}\cong \Z_2$,
is actually \emph{not} a short-range entangled fSPTs, \emph{but} instead a  
long-range entangled invertible fermionic topological order in 1+1D (in 2d).
}
Formally, in the second definition of SPTs, we need to mod out those LRE invertible spin-TQFTs which are invertible topological order states; 
so 
we can propose a definition of fermionic $G$-SPTs and their classification, mathematically, by modifying the (\ref{equiv-spin-inv-classification}) to
\begin{equation}
 \left\{
    \begin{array}{c}
     \text{deformation classes of fermionic SPTs} \\
     \text{with an internal symmetry $G$ (i.e.\ $G$-fSPTs)}
    \end{array}
 \right\}
 \cong
 \frac{\text{Hom}(\text{Tor}\,\Omega_n^{Spin}(BG),U(1))}{\text{Hom}(\text{Tor}\,\Omega_n^{Spin}(pt),U(1))}
 \label{equiv-SPT-inv-classification}
\end{equation}
Later in Section \ref{sec:computation} and \ref{sec:spin-twisted-bordism}, when we show, in various Tables, the data of bordism groups and the classification of fSPTs,
we always use the second definition (which is, physically, the precise definition of short-range entangled SPTs), as proposed in (\ref{equiv-SPT-inv-classification}).
Since the entangled structure and the locality for local unitary transformation is more sharply defined in spacetime dimensions $n \geq 2$,
we will only classify fSPT for dimensions $n \geq 2$, focusing on $n=2,3,4$ (namely 1+1D, 2+1D, and 3+1D), shown in our Tables.

\section{Some useful invariants of spin-manifolds}
\label{sec:spin-invariants}

In this section we review basic invariants and structures that one can consider on closed spin and pin$^\pm$ manifolds in low dimensions (see e.g.\ \cite{kirby1990pin} for details) and fix their notations. This will be useful later in explicit construction of TQFTs.

\subsection{1-manifolds}
\label{sec:1man}

The only one-dimensional connected closed manifold is $S^1$. There are two choices of spin-structure that are usually referred to as odd and even. We will denote the corresponding spin circles as $S^1_-$ and $S^1_+$ respectively. The circle with even spin structure is a boundary of a disk with the unique spin structure, while the circle with odd spin structure is the generator of $\Omega_1^{Spin}(\text{pt})\cong \Z_2$. We will denote the corresponding bordism invariant as $\eta \in \Z_2$, its value is determined as follows:
\begin{equation}
 \eta(M^1) =\left\{
 \begin{array}{cl}
0, & M^1=S^1_+, \\
1, & M^{1}=S^1_-, \\
\sum_i \eta(N_i), & M^1=\sqcup_i N_i.
 \end{array}
 \right.
\end{equation}

\subsection{2-manifolds}
\label{sec:2man}

Consider first an oriented 2-manifold $\Sigma$. Spin structures then are in one-to-one correspondence with quadratic forms\footnote{Physically the value of $q(a)$ corresponds to periodicity condition on spinors along the Poincar\'e dual 1-cycle: $1$ for periodic and $0$ for anti-periodic} 
\begin{equation}
 \tilde{q}:H^1(\Sigma,\Z_2) \rightarrow \Z_2
\end{equation} 
such that
\begin{equation}
 \tilde{q}(a+b)-\tilde{q}(a)-\tilde{q}(b) = \int_\Sigma a\cup b\,.
\end{equation} 
The 2-dimensional spin bordism group is $\Omega_2^{Spin}(\text{pt})\cong \Z_2$ and the corresponding bordism invariant is Arf invariant:
\begin{equation}
 \begin{array}{rcl}
      \Omega_2^{Spin}(\text{pt})& \stackrel{~}\longrightarrow & \Z_2 \\
      {[\Sigma]} & \longmapsto & \text{Arf}(\Sigma) :=\sum_{i=1}^g \tilde{q}(a_i)\tilde{q}(b_i)\mod 2
 \end{array}
\end{equation} 
where $\{a_i,b_i\}_{i=1}^g$ is any symplectic basis in $H^1(\Sigma,\Z_2)$.

Consider now non-orientable a 2-manifold $\Sigma$. It always admits a pin$^-$ structure. Similarly to the spin case, pin$^-$ structures are in one-to-one correspondence with quadratic enhancement
\begin{equation}
 {q}:H^1(\Sigma,\Z_2) \rightarrow \Z_4
\end{equation} 
such that
\begin{equation}
{q}(a+b)-{q}(a)-{q}(b) = 2\int_\Sigma a\cup b\ \mod 4.
\end{equation} 
In particular:
\begin{equation}
{q}(a)=\int_\Sigma a\cup a \mod 2.
\end{equation} 
The pin$^-$-bordism group is $\Omega_2^{Pin^-}(\text{pt})\cong \Z_8$ and the isomorphism is explicitly given by Arf-Brown-Kervaire invariant:
\begin{equation}
 \begin{array}{rcl}
      \Omega_2^{Pin^-}(\text{pt})& \stackrel{~}\longrightarrow & \Z_8 \\
      {[\Sigma]} & \longmapsto & \text{ABK}(\Sigma) 
 \end{array}
\end{equation} 
One of the definitions of the Arf-Brown-Kervaire invariant valued mod 8 is given in terms of the following Gauss sum:
\begin{equation}
 \exp\{\pi i \text{ABK}(\Sigma)/4\} := \frac{1}{\sqrt{|H^1(\Sigma,\Z_2)|}}\sum_{a\in H^1(\Sigma,\Z_2)}
e^{\pi i q(a)/2}.
\label{ABK-definition}
\end{equation} 
When surface $\Sigma$ is orientable this reduces to the previous case with ${q}=2\tilde{q} \mod 4$, {$\text{ABK}=4\text{Arf}\mod 8$.

\subsection{3-manifolds}
\label{sec:3man}

Let $M^3$ be an oriented closed 3-manifold. Because of odd dimension there is no usual intersection pairing as in the two dimensional case above. However, one can instead define a linking pairing on the torsion part of the first homology\footnote{There is an analogous definition of linking pairing on $\mathrm{Tor}\,H_m(M^{2m+1},\Z)$ for any odd dimensional manifold.}:
\begin{equation}
\begin{array}{rccl}
 \ell k: &\mathrm{Tor}\,H_1(M^3,\Z) \otimes \mathrm{Tor}\,H_1(M^3,\Z)
 & \longrightarrow & \Q/\Z \\
 & {[\alpha]}\otimes [\beta]
 & \longmapsto & \frac{\#(\alpha \cap \tilde{\beta})}{n} \mod 1
\end{array}
\end{equation} 
where $\tilde{\beta}$ is a 2-chain such that $\partial \tilde{\beta}=n\cdot \beta$ for some $n\in \Z$ (such $n$ always exist because $[\beta]$ is torsion).

A spin-structure on $M^3$ again allows to define a quadratic refinement of the linking pairing:
\begin{equation}
\begin{array}{c}
 \gamma: \Tor H_1(M_3,\Z) \longrightarrow \Q/\Z \\
 \\
 \gamma(\alpha+\beta)-\gamma(\alpha)-\gamma(\beta)=\ell k(a,b) \mod 1
\end{array}
\end{equation} 
 The value of $\gamma$ can be geometrically defined as follows. Take a smooth embedding $\iota:S^1\rightarrow M^3$ representing a torsion element $\alpha\in \Tor H_1(M^3,\Z_2)$. Framings on  $K=\iota(S^1)$, i.e.\ trivializations of the normal bundle, are in one-to-one correspondence to rational numbers $q$ such that $q=\ell k(\alpha,\alpha) \mod 1$ (because framings corresponds to a choice of push-off $K$ into the boundary of the tabular neighborhood).  Given a spin structure on $M^3$ there is a subset of \textit{even} framings defined as follows. Framing on the normal bundle together with spin structure on $M^3$ fixes a spin structure on $K$. Then one can define a subset of \textit{even framings} such that $\eta(K)=0$, that is $K$ is a spin-boundary. Such framings have a fixed value of $q \mod 2$. Then one defines $\gamma(a)=q/2\mod 1$.

Spin structure on $M^3$ also allows one to define a symmetric function 
\begin{equation}
 \begin{array}{c}
  \delta:H^1(M_3,\Z_2)\times H^1(M_3,\Z_2) \longrightarrow \Z_4
 \end{array}
\end{equation} 
which is an enhancement of
\begin{equation}
 \begin{array}{rcl}
   H^1(M_3,\Z_2)\otimes H^1(M_3,\Z_2) \otimes H^1(M_3,\Z_2) & \longrightarrow & \Z_2 \\
   a \otimes b \otimes c & \longmapsto & \int_{M^3}a\cup b\cup c
 \end{array}
\end{equation} 
in the sense that
\begin{equation}
 \delta(a,b+c)-\delta(a,b)-\delta(a,c)=2
 \int_{M^3}a\cup b\cup c
 \label{delta-refinement}
\end{equation} 
The values of $\delta$ can be defined as follows. Let $\text{PD}(a)$ be a smooth, possibly non-orientable, surface which represents a class in $H_2(M^3,\Z_2)$ Poincar\'e dual to $a\in H^1(M^3,\Z_2)$ (it always exists). A spin structure on $M^3$ induces canonically a pin$^{-}$ structure on $\text{PD}(a)$. This is follows from the fact that $TM^3=T\text{PD}(a)\oplus \det T\text{PD}(a)$ due to orientability of $TM^3$, and from the fact that there is one-to-one correspondence between pin$^-$ structures on $V$ and spin-structures on $V\oplus \det V$. Then 
\begin{equation}
 \delta(a,b)={q}_{\text{PD}(a)}(b)\;\in \Z_4
\end{equation} 
where $q_{\text{PD}(a)}$ is a quadratic enhancement corresponding to the pin$^-$ structure on $\text{PD}(a)$ considered above, and $b$ is implicitly assumed to be restricted on $\text{PD}(a)$. 

Moreover, there is an enhancement of $\delta$, given by
\begin{equation} \label{eq:betaZ8}
	\begin{array}{rccl}
		\beta: & H^1(M^3,\Z_2) & \longrightarrow & \Z_8 \\
		& a & \longmapsto & \mathrm{ABK}(\Sigma_a)
	\end{array}
\end{equation}
so that
\begin{equation}
	\beta(a+b)=\beta(a)+\beta(b)+2\delta(a,b).
\end{equation}

\subsubsection{Abelian spin-Chern-Simons theory}

The quadratic enhancement $\gamma$ defined above can be used to explicitly write expression for the partition function of abelian spin-Chern-Simons theories, a known simple family of spin-TQFTs. The (``classical'') data needed to define a spin-Chern-Simons TQFT is a symmetric bilinear form 
\begin{equation}
 K:\Z^L\otimes \Z^L \;\longrightarrow \Z. 
\end{equation} 
or, equivalently, a lattice. On the physical level the partition function on a closed oriented spin 3-manifold $M^3$ is given by the path integral
\begin{equation} \label{eq:KmatCS}
 Z^K_\text{spin-CS}(M^3)=\int\limits_{U(1)\text{ connections}/\sim\text{ gauge}} DA\;\exp{(\frac{i}{8\pi}\int_{M^3}\sum_{i,j}K^{ij}A_idA_j)}.
\end{equation} 
The expression in the exponent is actually ill defined because connection 1-form $A_i$ is not globally defined for non-trivial bundles. To avoid this problem one has to take a spin 4-manifold $M^4$ such that $M^3=\partial M^4$ with induced spin-structure\footnote{We use the usual normalization of connection 1-form/curvature such that the first Chern class is $c_1=F/(2\pi)=dA/(2\pi)$.}:
\begin{equation}
  \exp{(\frac{i}{8\pi}\int_{M^3}\sum_{i,j}K^{ij}A_idA_j)}
:=
  \exp{(\frac{i}{8\pi}\int_{M^4}\sum_{i,j}K^{ij}F_i\,F_j)}
\end{equation} 
which is independent of the choice of spin-manifold $M^4$ because of integrality of $K$ and the fact that intersection form on spin-manifolds is even. If the spin-structure was not required, $K$ would have to be even (i.e.\ $K_{ii}=0\mod 2,\;\forall i$), which is the quantization condition on the level matrix for the ordinary, non-spin (``bosonic'') Chern-Simons theory.
 If the spin-structure is required, we have the spin (``fermionic'') Chern-Simons theory.
 
Different $K$ can actually give equivalent spin-TQFTs. The classification of non-equivalent abelian spin-Chern-Simons theories was done in \cite{Belov:2005ze}.

Even though in general path integral over the space of connections (modulo gauge transformations) is ill-defined, in this case the action is quadratic in connection 1-forms and the path integral can be defined and computed formally\footnote{Alternatively, one can also mathematically define spin-Chern-Simons TQFT via spin-generalization (see e.g.\ \cite{kirby19913,blanchet1992invariants,beliakova2014spin}) of Reshetikhin-Turaev \cite{reshetikhin1991invariants} construction where the input data is the spin modular tensor category of representations of lattice vertex operator algebra associated to $K$.}. The path integral reduces to the sum over critical points of the action, that is flat connections. To simplify the formulas let us assume that $M_3$ is a rational homology sphere, that is $H_1(M^3,\Z)=\Tor H_1(M^3,\Z)$. The moduli space of $U(1)^L$ flat connections is then a finite set
\begin{equation}
 \mathcal{M}_\text{flat}=\Hom(\pi_1(M^3),U(1)^L)/U(1)^L = \Hom(H_1(M^3,\Z),U(1)^L)
 \stackrel{\ell k}\cong H_1(M^3,\Z)^L
\end{equation} 
where the last isomorphism is explicitly given by the composition of linking pairing with the exponential map:
\begin{equation}
 \begin{array}{rcl}
   H_1(M^3,\Z) & \stackrel{\sim}\longrightarrow & \Hom(H_1(M^3,\Z),U(1)) \\
   \alpha & \longmapsto & \exp 2\pi i\, \ell k(\alpha,\cdot).  
 \end{array}
\end{equation} 
Assuming this correspondence between $U(1)$ flat connections and elements of $H_1(M_3,\Z)$, it is easy to see that the usual Chern-Simons invariant (valued mod 1) of $U(1)$ flat connection reads $\ell k(a,a)\mod 1$, while its spin-version is $\gamma(a) \mod 1$. The partition function of level $K$ spin-Chern-Simons theory with appropriate normalization\footnote{Such that $Z_\text{spin-CS}^K(S^2\times S^1)=1$.} is then given by
\begin{equation}
 Z_\text{spin-CS}^K(M^3)=
 \sum_{\alpha_i\in H_1(M_3,\Z)}
 \frac{\exp\left\{
 2\pi i \sum_{i=1}^L K_{ii}\gamma(\alpha_i)+
 2\pi i \sum_{i<j} K_{ij}\,\ell k(\alpha_i,\alpha_j)
 \right\}}{\sqrt{\det(iK)}\,|H_1(M_3,\Z)|^{L/2}}.
\end{equation}

As we will see later some of the TQFTs $Z^\mu_\text{gauged},\; \mu\in\Hom(\Omega_3(BG),U(1))$ are equivalent to abelian spin-Chern-Simons theories for certain choices of $K$. Consider in particular the case of 
\begin{equation}
	K=\left(
	\begin{array}{cc}
		0 & p \\
		p & k
	\end{array}
	\right),\;\;p,k>0.
\end{equation}
The sum above then can be partially performed explicitly:
\begin{equation}
	Z_\text{spin-CS}^{(0\,p;\,p\,k)}(M^3)=
	\sum_{\alpha_1,\alpha_2\,\in\, H_1(M^3,\Z)}
	\frac{e^{
	2\pi i \,k\,\gamma(\alpha_1)+2\pi i\,p\,\ell k (\alpha_1,\alpha_2)
	}}{p\,|H_1(M^3,\Z)|}=
	\frac{1}{p}\sum_{\footnotesize\begin{array}{c}
	 \alpha_1\in H_1(M^3,\Z), \\
	 p\,\ell k(\alpha_1,\alpha_2)=0\mod 1\,, \;  \forall \alpha_2
	\end{array}
	}e^{2\pi i k\,\gamma(\alpha_1)}.
\end{equation}
The set over which the sum is performed can be identified with the first cohomology with $\Z_p$ coefficients:
\begin{equation}
	H^1(M^3,\Z_p)\;\cong \; 
	\{ 
 \alpha\in H_1(M^3,\Z)\;|\;
  e^{2\pi i p\,\ell k(\alpha,\alpha')}=1, \;  \forall \alpha'
	\}.
\end{equation}
The isomorphism is given by the universal coefficient theorem ($H^1(M^3,\Z_p)\cong \Hom(H_1(M^3,\Z),\Z_p)$) and the canonical embedding $\Hom(H_1(M^3,\Z),\Z_p)\hookrightarrow\Hom(H_1(M^3,\Z),U(1))\stackrel{\ell k}{\cong}H_1(M^3,\Z)$. Let us denote by $\hat{a}$ an element of $H_1(M^3,\Z_p)$ corresponding $a\in H^1(M^3,\Z_p)$. Then we can define a function
\begin{equation}
	\begin{array}{cccc}
	  \hat\gamma: & H^1(M^3,\Z_p) & \longrightarrow & \Z_{2p} \\
	  & a & \longmapsto & 2p\gamma(\hat{a}).
	\end{array}
\end{equation}
so that
\begin{equation}
	Z_\text{spin-CS}^{(0\,p;\,p\,k)}(M^3)=
	\frac{1}{p}\sum_{a\in H^1(M^3,\Z_p)}e^{\frac{\pi i k}{p}\,\hat\gamma(a)}.
\end{equation}
The function $\hat{\gamma}$ can be understood as a quadratic refinement of 
the bilinear form
\begin{equation}
 \begin{array}{ccl}
 H^1(M^3,\Z_p) \otimes H^1(M^3,\Z_p)  & \longrightarrow & \Z_p \\
 a\otimes b  & \longmapsto & \int_{M^3} a\,\mathcal{B} b
 \end{array}
\end{equation} 
where $\mathcal{B}$ is the Bockstein homomorphism $H^1(M^3,\Z_p)\rightarrow H^2(M^3,\Z_p)$ corresponding to the following short exact sequence of coefficients:
\begin{equation}
 0\rightarrow \Z_p \stackrel{p\,\cdot}\longrightarrow \Z_{p^2} \stackrel{\mod p}\longrightarrow \Z_p \rightarrow 0.
\end{equation} So that
\begin{equation}
 \hat{\gamma}(a) = \int_{M^3} a\,\mathcal{B} a \mod p.
\end{equation} 

One can argue that $\hat\gamma$ is a bordism invariant, that is can be considered as homomorphism
\begin{equation}
	\hat{\gamma}:\Omega^{Spin}_3(B\Z_p)\;\longrightarrow \Z_{2p}
\end{equation}
where we identify an element of $H^1(M^3,\Z_p)$ with a homotopy class of a map $M^3\rightarrow B\Z_p$ in the usual way. We then arrive at the following relation between the spin-Chern-Simons TQFT and spin-TQFTs associated to elements of the spin-cobordism group of $\Z_{p}$:
\begin{equation}
	Z_\text{spin-CS}^{(0\,p;\,p\,k)} = 
	Z_\text{gauged}^{{\hat\gamma}^*(k)}
\end{equation}
where $k$ is considered an element of $\Hom(\Z_{2p},U(1))\cong \Z_{2p}$ and ${\hat{\gamma}}^*(k)\in \Hom (\Omega_3^{Spin}(B\Z_p),U(1))$ is its pullback with respect to the map $\hat\gamma$ above.

\section{Spin-bordism groups and computations}
\label{sec:computation}

By the Pontryagin-Thom construction, 
\begin{equation}
	\Omega^{Spin}_n(X)=\pi_n(MSpin\wedge X_+)
\end{equation}
where, as usual, $\pi_n$ denotes the $n$-th stable homotopy group of a spectrum, $MSpin$ is a Thom spectrum associated to stable spin-structure\footnote{Note that in the case of spin structure Thom spectrum is equivalent to Madsen-Tillmann spectrum: $MSpin \cong MTSpin$.}.  We use the standard notations in homotopy theory: $X_+$ is a pointed space with a disjoint basepoint added, $\wedge$ denotes the smash product. 
For example, $\Omega^{Spin}_d(pt)$ are computed by Anderson-Brown-Peterson (\cite{abp}). 

We are interested in the case $X=BG$, a classifying space of a finite abelian group. In this work, we compute some simple examples where $G$ is a finite abelian group. At odd torsion $MSpin\cong MSO$. Since we are interested in constructing spin-TQFTs which non-trivially depend on spin-structure (i.e.\ not ``bosonic''), we will only consider some simple examples when $G$ itself is 2-torsion. We can then use the Adams spectral sequence for computation:
\begin{equation}
	E_2^{s,t}=\Ext_{\A}^{s,t}(\H^*(MSpin\wedge X_+),\Z_2)\Rightarrow\pi_{t-s}(MSpin\wedge X_+)_2^{\wedge}=\Omega^{Spin}_{t-s}(X).
\end{equation}
where $\H^*(-)$ stands for mod 2 cohomology and $\A$ is for Steenrod algebra. The abelian groups $\Ext^{s,t}_\A(M,N)$ are understood as groups of extensions of length $s$ between modules over Steenrod algebra $N[t]$ and $M$, where $[t]$ denotes the shift of grading by $t$. Notations $A_2^\wedge$ stands for 2-completion of the abelian group $A$. We refer to \cite{beaudry2018guide} as a brief review of necessary definitions and techniques in stable homotopy theory that requires a minimal prior background.

At 2-torsion and in degree $n<8$ the Thom spectrum $MSpin$ is equivalent to $ko$, the connective version (stable homotopy groups in negative degree are zero) of the real K-theory $KO$. Therefore, if the abelian group $G$ has only 2-torsion (which means that $G$ is a product of finite abelian groups of the form $\Z_{2^m}$), one have
\begin{equation} \label{eq:bordism-ko}
 \Omega_n^{Spin}(BG)\cong ko_n(BG).
\end{equation} 
This relation can be used to immediately give answer for $\Omega_n^{Spin}(BG)$ for certain cases of $G$, where $ko_n(BG)$ were already calculated in the literature (\cite{yu1995connective,bruner2010connective}). We consider one of such cases in the next subsection.

We would like to remark that such statement about classification of \textit{interacting} fermionic SPTs with finite symmetry $G$ via the real connective K-theory ($ko$) should not be confused with a different statement about classification of certain \textit{free} fermionic SPTs with via real or complex periodic K-theory ($K$ or $KO$) given in \cite{Kitaev2009mg0901.2686}.

Below we present the calculations of bordism group $\Omega_n^{Spin}(BG)$ which classifies the invertible fermionic spin-TQFTs with the fermionic parity symmetry $\Z_2^f$
and internal symmetry $G$. 
The invertible fermionic spin-TQFTs include all the interacting fermionic SPTs (short-ranged entangled states) and some invertible fermionic topological orders (long-ranged entangled states).  To recall the definitions, see (\ref{equiv-spin-inv-classification}) and (\ref{equiv-SPT-inv-classification}).
Here the $n$ corresponds to the spacetime dimension $n$ in physical systems.

\subsection{$\Omega^{Spin}_{n}(B\Z_2^k)$}
\label{sec:Z2n}
\subsubsection{Computation}
When $G=\Z_2^k$ one can use the known results in the literature
\begin{theorem}\cite{yu1995connective} Let $(B\Z_2)^k$ denote the $k$-fold smash product of the classifying space of $\Z_2$. Then
\begin{equation*}
  ko_n((B\Z_2)^k)=  \begin{cases}
        \Z_2\oplus C_{k,n} & \text{for } n=8l+1, k<4l+2\\
                           & \text{for } n=8l+2, k<4l+3\\
        \Z_{2^{4l+4-k}}\oplus C_{k,n} , & \text{for }n=8l+3, k< 4l+4\\
        \Z_{2^{4l+5-k}}\oplus C_{k,n} , & \text{for }n=8l+7, k<4l+5\\
         C_{k,n} , &otherwise
\end{cases}
\end{equation*}
where $C_{k,n}$ is a $\F_2$-vector space, whose dimension is the coefficient of $t^n$ in the series 
\begin{equation}
 P_k(t)=\frac{t^k}{(1-t^4)(1+t^3)(1-t)^{k-1}}
\end{equation} 
with
\begin{equation*}
  Q_k(t)=  \begin{cases}
        t^{k-1} & \text{for } k\equiv 0,1 \text{ mod }4\\
      t^{k-2}(1+t-t^2+t^3-t^5) & \text{for } k\equiv 2 \text{ mod }4\\
         t^{k-3}(1+t^3-t^5) & \text{for } k\equiv 3 \text{ mod }4
         \end{cases}
\end{equation*}
\end{theorem} 
Since we have $B^{\times k}= (\vee_{{k \choose 1}} B) \bigvee (\vee_{{k \choose 2}}B^{\wedge 2})\bigvee\cdots\bigvee  (\vee_{{k \choose k}}B^{\wedge k})$, from above theorem, we have the following 
\begin{theorem}
\begin{equation}
 \Omega^{Spin}_3(B\Z_2^k)=\Z_8^k\oplus\Z_4^{\frac{k^2-k}{2}}\oplus\Z_2^{\frac{k^3-3k^2+2k}{6}},
 \label{spin-3-Z2k}
\end{equation} 
\begin{equation}
 \Omega^{Spin}_4(B\Z_2^k)=\Z\oplus\Z_2^{\frac{k^4+2k^3+11k^2-14k}{24}}.
 \label{spin-4-Z2k}
\end{equation} 
\end{theorem}
As described in section \ref{sec:relation-to-DW} there is a natural map from spin-bordism groups to integer homology groups (\ref{spin-hom-map}) which provides a relation with Dijkgraaf-Witten theories. Therefore it is instructive to compare (\ref{spin-3-Z2k})-(\ref{spin-4-Z2k}) with:
\begin{equation}
 H_3(B\Z_2^k,\Z)=\Z_2^{k+\frac{k^2-k}{2}+\frac{k^3-3k^2+2k}{6}},
\end{equation} 
\begin{equation}
 H_4(B\Z_2^k,\Z)=\Z\oplus\Z_2^{\frac{k^4+2k^3+11k^2-14k}{24}}.
\end{equation} 
In particular we see that $\Tor\Omega^{Spin}_4(B\Z_2^k)\cong H_4(B\Z_2^k,\Z)$. Therefore in this case we do not get any fermionic TQFTs (according to the terminology explained in section \ref{sec:relation-to-DW}).

\subsubsection{Bordism invariants}
\label{sec:Z2-inv}

Let us explicitly describe the corresponding bordism invariants. In dimension 3, the isomorphism above is explicitly given by:
\begin{equation}
\begin{array}{rcl}
 \Omega^{Spin}_3(B\Z_2^k)
 &\stackrel{\sim}{\longrightarrow}&
 \Z_8^k\oplus\Z_4^{\frac{k^2-k}{2}}\oplus\Z_2^{\frac{k^3-3k^2+2k}{6}},\\
 \\
 (M^3,g_1,\ldots g_k)& \longmapsto &
 \underline{\oplus_{i=1}^n\beta(g_i)}\bigoplus 
 \underline{\oplus_{i\neq j}\delta(g_i,g_j)} \bigoplus
 \oplus_{i\neq j\neq k,\; i\neq k} \int_{M^3}g_ig_jg_k
\end{array}
\end{equation} 
where $\beta$, $\delta$ are functions defined in section \ref{sec:3man}. In the right hand side $g_i:M^3\rightarrow BG$ are treated as the elements of $H^1(M^3,\Z_2)$ and the multiplication is performed with respect to the usual cup product. The non-underlined bordism invariants, considered as elements of $\Hom(\Omega_3^{Spin}(BG),U(1))$ (by embedding a cyclic group in $U(1)$), belong to the subgroup $H^3(BG,U(1))$ embedded via the map (\ref{DW-spin-map})\footnote{Note that via Yoneda lemma there is one-to-one correspondence between cohomology operations
\begin{equation}
 H^1(\,\cdot\,,G)\longrightarrow H^n(\,\cdot\,,U(1))
\end{equation} 
and elements of $H^n(BG,U(1))$.
}.

The underlined bordism invariants, considered as elements of $\Hom(\Omega_3^{Spin}(BG),U(1))$, are fermionic (in the terminology explained in in section \ref{sec:relation-to-DW}), but provide refinement of bosonic elements from $H^3(BG,U(1))$. That is, a certain power of them gives an element from subgroup $H^3(BG,U(1))$. For example, from (\ref{delta-refinement}) it follows that $\delta_{g_i,g_j}=\int_{M^3}g_i^2g_j \mod 2$. 

In dimension 4 we have:
\begin{equation}
\begin{array}{rcl}
 \Omega^{Spin}_4(B\Z_2^k)
 &\stackrel{\sim}{\longrightarrow}&
 \Z\oplus\Z_2^{\frac{k^4+2k^3+11k^2-14k}{24}},\\
 \\
 (M^4,g_1,\ldots g_k)& \longmapsto &
 \frac{p_1(TM^4)}{48} \bigoplus \oplus_{i,j,k,l}\int_{M^4}g_ig_jg_kg_l \;/\;\{g_i^4,\;g_i^2g_j^2,
 \;g_ig_jg_k^2 + g_jg_kg_i^2+g_kg_ig_j^2\}
\end{array}
\end{equation} 
where $p_1$ is the first Pontryagin class and the denominator of the quotient contains identically vanishing polynomials in elements of $H^1(M^4,\Z_2)$ of degree 4. Note that in dimension $4$ none of the invariants actually depends on the spin structure. Therefore their gauging does not give proper spin-TQFTs. Moreover, $\Hom(\Tor\Omega^{Spin}_4(B\Z_2^k),U(1))\cong H^4(B\Z_2^k,U(1))$ and the gauged TQFTs coincide with Dijkgraaf-Witten TQFTs \cite{Putrov:2016qdo}.

Hence we have the following theorem:
\begin{theorem}
\begin{equation}
\begin{tabular}{c c c c}
\hline
$n$ & $\Omega^{Spin}_n(B(\Z_2^k))$ & $H_n(B(\Z_2^k),\Z)$  & 
$\text{fermionic SPTs classes}$
\\
\hline
0& $\Z$ &  $\Z$   &  \\
1& $\Z_2^{k+1}$ & $\Z_2^k$  & \\
2& $\Z_2^{1+k+\binom{k}{2}}$ & $\Z_2^{\binom{k}{2}}$   & $\Z_2^{k+\binom{k}{2}}$\\
3 & $\Z_8^k\times\Z_4^{\binom{k}{2}}\times\Z_2^{\binom{k}{3}}$ & $\Z_2^{k+\binom{k}{2}+\binom{k}{3}}$ & $\Z_8^k\times\Z_4^{\binom{k}{2}}\times\Z_2^{\binom{k}{3}}$ \\
4 & $\Z\times\Z_2^{\frac{k^4+2k^3+11k^2-14k}{24}}$ & $\Z_2^{\frac{k^4+2k^3+11k^2-14k}{24}}$ & $\Z_2^{\frac{k^4+2k^3+11k^2-14k}{24}}$  \\ 
\hline
\end{tabular}
\end{equation}
where in the second column, for comparison, we list the known homology groups. Note that $\frac{k^4+2k^3+11k^2-14k}{24}=2 \binom{k}{2}+2\binom{k}{3}+\binom{k}{4}$.
\end{theorem}

\subsection{$\Omega^{Spin}_n(B(\Z_2\times\Z_4))$}
\subsubsection{Computation}
As was mentioned in the beginning of the section, the computation involves no odd torsion and we can use the Adams spectral sequence 
\begin{multline}
	E_2^{s,t}=\Ext_{\A}^{s,t}(\H^*(MSpin\wedge(B(\Z_2\times\Z_4))_+),\Z_2)\Rightarrow
	\\
	\pi_{t-s}(MSpin\wedge(B(\Z_2\times\Z_4))_+)_2^{\wedge}=\Omega^{Spin}_{t-s}(B(\Z_2\times\Z_4)).
\end{multline}
The mod 2 cohomology of Thom spectrum $MSpin$ is
\begin{equation}
\label{MSpin-coh}
	\H^*(MSpin)=\A\otimes_{\A(1)}\{\Z_2\oplus M\}
\end{equation}
where $M$ is a graded $\A(1)$-module with the degree $i$ homogeneous part $M_i=0$ for $i<8$. Here $\A$ stands for Steenrod algebra and $\A(1)$ stands for $\F_2$-algebra generated by $Sq^1$
and $Sq^2$.
Thus, for $t-s<8$, we can identify the $E_2$-page with 
\begin{equation}
	\Ext_{\A(1)}^{s,t}(\H^*(B(\Z_2\times\Z_4)),\Z_2).
\end{equation}
The mod 2 cohomology is the following: $\H^*(B\Z_2)=\Z_2[a]$ where $|a|=1$, $\H^*(B\Z_4)=\Z_2[y]\otimes\Lambda(x)$ where $|x|=1$, $|y|=2$, $Sq^1(y)=Sq^1(x)=0$. The differential on the second page are the following: $d_2(y)=xh_0^2$, $d_2(y\alpha)=xyh_0^3$, $d_2(ya^3+y^2a)=(xa^3+xya)h_0^2$. We use the standard notation in the stable homotopy theory where $h_0$ denotes an element of\footnote{There is natural action
\begin{equation}
 \Ext^{s,t}_{\A(1)}(\Z_2,\Z_2)\otimes \Ext^{s',t'}_{\A(1)}(M,\Z_2)
 \rightarrow \Ext^{s+s',t+t'}_{\A(1)}(M,\Z_2)
\end{equation} 
realized by
} $\Ext_{\A(1)}^{1,1}(\Z_2,\Z_2)$ corresponding to extension $\Z_2[1]\rightarrow \H^*(\mathbb{RP}^2)[-1] \rightarrow \Z_2$.  The $\A(1)$-module structure of $\H^*(B(\Z_2\times\Z_4))$ and the $E_2$ page are depicted in Figures \ref{fig:A1-Z2Z4} and \ref{fig:E2-Z2Z4}.
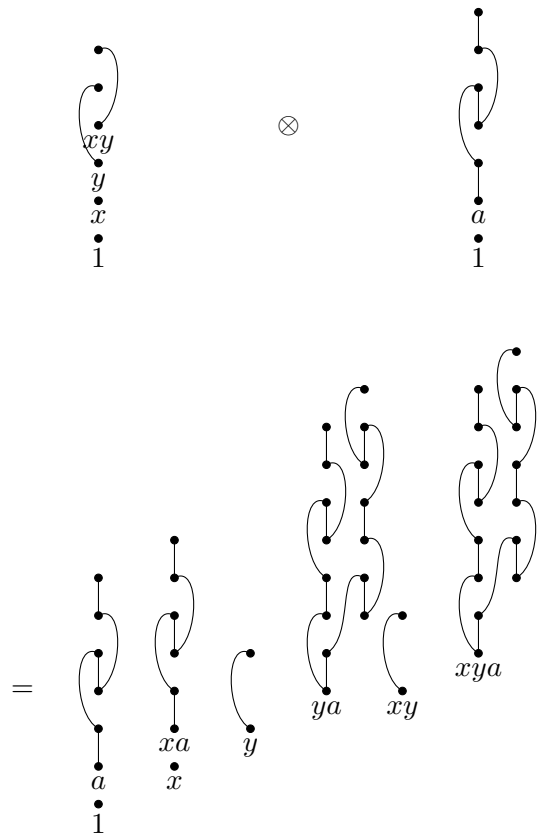
\begin{figure}[!h]
\begin{center}
\begin{tikzpicture}[scale=0.5]
\node [below] at (0,10) {1};
\draw[fill] (0,10) circle(.1);

\node [below] at (0,11) {$x$};
\draw[fill] (0,11) circle(.1);

\node [below] at (0,12) {$y$}; 
\draw[fill] (0,12) circle(.1);

\node [below] at (0,13) {$xy$};
\draw[fill] (0,13) circle(.1);
\draw[fill] (0,14) circle(.1);
\draw (0,12) to [out=150,in=150] (0,14);
\draw[fill] (0,15) circle(.1);
\draw (0,13) to [out=30,in=30] (0,15);

\node [below] at (10,10) {1};
\draw[fill] (10,10) circle(.1);

\node [below] at (10,11) {$a$};
\draw[fill] (10,11) circle(.1);
\draw[fill] (10,12) circle(.1);
\draw (10,11) -- (10,12);
\draw[fill] (10,13) circle(.1);
\draw[fill] (10,14) circle(.1);
\draw (10,13) -- (10,14);
\draw (10,12) to [out=150,in=150] (10,14);
\draw[fill] (10,15) circle(.1);
\draw (10,13) to [out=30,in=30] (10,15);
\draw[fill] (10,16) circle(.1);
\draw (10,15) -- (10,16);

\node at (5,13) {$\otimes$};

\node at (-2,-2) {$=$};

\node [below] at (0,-5) {1};
\draw[fill] (0,-5) circle(.1);

\node [below] at (0,-4) {$a$};
\draw[fill] (0,-4) circle(.1);
\draw[fill] (0,-3) circle(.1);
\draw (0,-4) -- (0,-3);
\draw[fill] (0,-2) circle(.1);
\draw[fill] (0,-1) circle(.1);
\draw (0,-2) -- (0,-1);
\draw (0,-3) to [out=150,in=150] (0,-1);
\draw[fill] (0,0) circle(.1);
\draw (0,-2) to [out=30,in=30] (0,0);
\draw[fill] (0,1) circle(.1);
\draw (0,0) -- (0,1);

\node [below] at (2,-4) {$x$};
\draw[fill] (2,-4) circle(.1);

\node [below] at (2,-3) {$xa$};
\draw[fill] (2,-3) circle(.1);
\draw[fill] (2,-2) circle(.1);
\draw (2,-3) -- (2,-2);
\draw[fill] (2,-1) circle(.1);
\draw[fill] (2,0) circle(.1);
\draw (2,-1) -- (2,0);
\draw (2,-2) to [out=150,in=150] (2,0);
\draw[fill] (2,1) circle(.1);
\draw (2,-1) to [out=30,in=30] (2,1);
\draw[fill] (2,2) circle(.1);
\draw (2,1) -- (2,2);

\node [below] at (4,-3) {$y$};
\draw[fill] (4,-3) circle(.1);
\draw[fill] (4,-1) circle(.1);
\draw (4,-3) to [out=150,in=150] (4,-1);

\node [below] at (6,-2) {$ya$};
\draw[fill] (6,-2) circle(.1);
\draw[fill] (6,-1) circle(.1);
\draw (6,-2) -- (6,-1);
\draw[fill] (6,0) circle(.1);
\draw (6,-2) to [out=150,in=150] (6,0);
\draw[fill] (6,1) circle(.1);
\draw (6,0) -- (6,1);
\draw[fill] (6,2) circle(.1);
\draw[fill] (6,3) circle(.1);
\draw (6,2) -- (6,3);
\draw (6,1) to [out=150,in=150] (6,3);
\draw[fill] (6,4) circle(.1);
\draw (6,2) to [out=30,in=30] (6,4);
\draw[fill] (6,5) circle(.1);
\draw (6,4) -- (6,5);
\draw[fill] (7,0) circle(.1);
\draw[fill] (7,1) circle(.1);
\draw (7,0) -- (7,1);
\draw (6,-1) to [out=30,in=150] (7,1);
\draw[fill] (7,2) circle(.1);
\draw (7,0) to [out=30,in=30] (7,2);
\draw[fill] (7,3) circle(.1);
\draw (7,2) -- (7,3);
\draw[fill] (7,4) circle(.1);
\draw[fill] (7,5) circle(.1);
\draw (7,4) -- (7,5);
\draw[fill] (7,6) circle(.1);
\draw (7,3) to [out=30,in=30] (7,5); 
\draw (7,4) to [out=150,in=150] (7,6);

\node [below] at (8,-2) {$xy$};
\draw[fill] (8,-2) circle(.1);
\draw[fill] (8,0) circle(.1);
\draw (8,-2) to [out=150,in=150] (8,0);

\node [below] at (10,-1) {$xya$};
\draw[fill] (10,-1) circle(.1);
\draw[fill] (10,0) circle(.1);
\draw (10,-1) -- (10,0);
\draw[fill] (10,1) circle(.1);
\draw (10,-1) to [out=150,in=150] (10,1);
\draw[fill] (10,2) circle(.1);
\draw (10,1) -- (10,2);
\draw[fill] (10,3) circle(.1);
\draw[fill] (10,4) circle(.1);
\draw (10,3) -- (10,4);
\draw (10,2) to [out=150,in=150] (10,4);
\draw[fill] (10,5) circle(.1);
\draw (10,3) to [out=30,in=30] (10,5);
\draw[fill] (10,6) circle(.1);
\draw (10,5) -- (10,6);
\draw[fill] (11,1) circle(.1);
\draw[fill] (11,2) circle(.1);
\draw (11,1) -- (11,2);
\draw (10,0) to [out=30,in=150] (11,2);
\draw[fill] (11,3) circle(.1);
\draw (11,1) to [out=30,in=30] (11,3);
\draw[fill] (11,4) circle(.1);
\draw (11,3) -- (11,4);
\draw[fill] (11,5) circle(.1);
\draw[fill] (11,6) circle(.1);
\draw (11,5) -- (11,6);
\draw[fill] (11,7) circle(.1);
\draw (11,4) to [out=30,in=30] (11,6); 
\draw (11,5) to [out=150,in=150] (11,7);

\end{tikzpicture}
\end{center}
\caption{The $\A(1)$-module structure of $\H^*(B(\Z_2\times\Z_4))$}
\label{fig:A1-Z2Z4}
\end{figure}

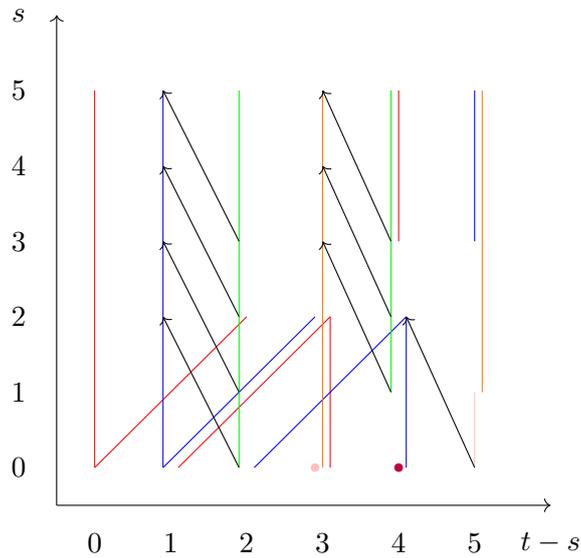
\begin{figure}[!h]
\begin{center}
\begin{tikzpicture}
\node at (0,-1) {0};
\node at (1,-1) {1};
\node at (2,-1) {2};
\node at (3,-1) {3};
\node at (4,-1) {4};
\node at (5,-1) {5};
\node at (6,-1) {$t-s$};
\node at (-1,0) {0};
\node at (-1,1) {1};
\node at (-1,2) {2};
\node at (-1,3) {3};
\node at (-1,4) {4};
\node at (-1,5) {5};
\node at (-1,6) {$s$};

\draw[->] (-0.5,-0.5) -- (-0.5,6);
\draw[->] (-0.5,-0.5) -- (6,-0.5);

\draw[color=red] (0,0) -- (0,5);
\draw[color=red] (0,0) -- (2,2);
\draw[color=red] (1.1,0) -- (3.1,2);
\draw[color=red] (3.1,2) -- (3.1,0);
\draw[color=red] (4,3) -- (4,5);
\draw[color=blue] (0.9,0) -- (0.9,5);
\draw[color=blue] (0.9,0) --(2.9,2);
\draw[color=blue] (2.1,0) -- (4.1,2);
\draw[color=blue] (4.1,2) -- (4.1,0);
\draw[color=blue] (5,3) -- (5,5);
\draw[color=green] (1.9,0) -- (1.9,5);
\draw[color=green] (3.9,1) -- (3.9,5);

\draw[color=orange] (3,0) -- (3,5);
\draw[color=orange] (5.1,1) -- (5.1,5);

\draw[color=pink,fill=pink] (2.9,0) circle(0.05);

\draw[color=pink] (5,0) -- (5,1);

\draw[color=purple,fill=purple] (4,0) circle(0.05);

\draw[->] (1.9,0) -- (0.9,2);
\draw[->] (1.9,1) -- (0.9,3);
\draw[->] (1.9,2) -- (0.9,4);
\draw[->] (1.9,3) -- (0.9,5);

\draw[->] (3.9,1) -- (3,3);
\draw[->] (3.9,2) -- (3,4);
\draw[->] (3.9,3) -- (3,5);

\draw[->] (5,0) -- (4.1,2);
\end{tikzpicture}
\end{center}
\caption{The $E_2$ page of the Adams spectral sequence for $\Z_2\times\Z_4$}
\label{fig:E2-Z2Z4}
\end{figure}

Hence we have the following theorem:
\begin{theorem}
\begin{equation}
\begin{tabular}{c c c c}
\hline
$n$ & $\Omega^{Spin}_n(B(\Z_2\times\Z_4))$ & $H_n(B(\Z_2\times\Z_4),\Z)$ & 
$\text{fermionic SPTs classes}$
\\
\hline
0& $\Z$ & $\Z$  &  \\
1& $\Z_2^2\times\Z_4$ & $\Z_2\times \Z_4$ & \\
2& $\Z_2^4$ & $\Z_2$ & $\Z_2^3$\\
3 & $\Z_2^3\times\Z_8^2$ & $\Z_2^2\times \Z_4$ & $\Z_2^3\times\Z_8^2$\\
4 & $\Z\times\Z_2\times \Z_4$ & $\Z_2^2$ & $\Z_2\times {\Z_4}$ \\ 
\hline
\end{tabular}
\end{equation}
where in the last column, for comparison, we list the known homology groups.
\end{theorem}

\subsubsection{Bordism invariants and manifold generators}
The bordism groups are explicitly realized as follows:
\begin{equation}
	\Omega_n^{Spin}(B(\Z_2\times\Z_4))=\{\text{spin $n$-manifolds }M^n\text{ with maps }f:M\to B\Z_4,g:M\to B\Z_2\}/\sim.
\end{equation}

Equivalently, $f\in H^1(M^n,\Z_4)$, $g\in H^1(M^n,\Z_2)$. The functions $f$ and $g$ can be used to  pull back the generators of the mod 2 cohomology of classifying spaces to $M^n$:
\begin{eqnarray*}
H^*(B\Z_4,\Z_2)&\stackrel{f^*}{\To}& H^*(M,\Z_2)\\
x&\mapsto& f^*(x)=f\mod2\\
y&\mapsto& f^*(y)
\end{eqnarray*}
\begin{eqnarray*}
H^*(B\Z_2,\Z_2)&\stackrel{g^*}{\To}& H^*(M,\Z_2)\\
a&\mapsto& g^*(a)=g
\end{eqnarray*}

Consider in particular the dimension $n=4$. Reading out from the $E_2$ page of Adams spectral sequence, the bordism group $\Z_4\oplus\Z_2$ is mapped to $\Z_2\oplus\Z_2$ by sending an element $(M^4, f\in H^1(M,\Z_4), g\in H^1(M^4,\Z_2))$ to the invariants $f^*(x) \cup g^*(a^3)$ and $f^*(xy) \cup g^*(a)$, we can see that
\begin{theorem}
 $(S^1\times\RP^3, f, g)$ generates $\Z_4$ and $(S^1\times L(4,1), h,\ell )$ generates $\Z_2$, where $L(4,1)\cong S^3/\Z_4$ is the Lens space, $f$ is the generator of $H^1(S^1,\Z_4)$, $g$ is the generator of $H^1(\RP^3,\Z_2)$,  $h$ is the generator of $H^1(L(4),\Z_4)$, and $\ell$ is the generator of $H^1(S^1,\Z_2)$.
\end{theorem}

It follows that complete bordism invariant in dimension $n<5$ read
\begin{equation}
\label{eq:topZ4Z2}
\begin{array}{rcl}
(M^1,f,g) & \stackrel{~}{\longmapsto} & \doubleunderline{\eta(M^1)} \oplus \int_{M^1} g \oplus \int_{M^1} f \\
\\
(M^2,f,g) & \stackrel{~}{\longmapsto} & \doubleunderline{\mathrm{Arf}(M^2)} \oplus \doubleunderline{\tilde{q}(f^*(x))} \oplus \doubleunderline{\tilde{q}(g)} \oplus \int_{M^2} g\,(f^*(x)) \\
\\
(M^3,f,g) &  \stackrel{~}{\longmapsto} & \int_{M^3}f^*(y)g \oplus \doubleunderline{\mathrm{Arf}(\mathrm{PD}(f^*(x)))} \oplus \doubleunderline{\tilde{q}_{\mathrm{PD}(f^*(x))}(g)} \oplus \underline{\beta(g)} \oplus \underline{\hat{\gamma}(f)} \\
\\
(M^4,f,g) & \stackrel{~}{\longmapsto} & \frac{p_1(TM^4)}{48} \oplus \int_{M^4} f^*(xy)\,g^*(a)
\oplus \underline{\delta_{\text{PD}(f)}(g,g)}
\end{array}
\end{equation}
where, as before, functions ${q}$, $\beta$, $\delta$ are the ones defined in section \ref{sec:spin-invariants}. Below we elaborate on the notations and the definition of the bordism invariants listed above. The expressions written in terms of $\text{PD}(f)$ or $\text{PD}(f^*(x))$ assume that the Poincar\'e duals of $f$ and $f^*(x)\equiv f\mod 2$ can be represented by an embedding\footnote{In fact, for the purposes of inducing spin structure from the ambient space, as described below, it is enough to have an immersion.} of a smooth oriented\footnote{Note that if $f^*(x)\equiv f\mod 2\neq 0$ and a smooth representative $N^{n-1}$ of the Poincar\'e dual of $f$ exists, it
will be necessarily orientable, because $[\text{PD}(f)]\in H_{n-1}(M^n,\Z_4)$ can be obtained by the pushforward of $[N^{n-1}]$ under the embedding. On the other hand, if $f\mod 2=0$ this implies that $f\in H^1(M^n,\Z_4)$ is the image of some $g'\in H^1(M^n,\Z_2)$ under the
  canonical map induced by the non-trivial homomorphism $\Z_2\to \Z_4$. Such homomorphism also induces a homomorphism $\Omega_n^{Spin}(B\Z_4\times B\Z_2)\rightarrow \Omega_n^{Spin}(B\Z_2\times B\Z_2)$ and the corresponding bordism invariant for $G=\Z_2\times \Z_4$ should reduce (i.e.\ pulled back) to the one of the invariants in Section \ref{sec:Z2-inv}. A naive argument shows that it should be $\int_{M_4}g'g^3$.} 
manifold $N^{n-1}$ of codimension one in $M^n$. In particular, this is automatically the case when $f=h \mod 4$ for some element $h\in H^1(M^n,\Z)$. In this case it is known that the Poincar\'e dual to $h$ can be always represented by a smooth oriented submanifold of codimension one inside $M^n$ and one can take $N^{n-1}$ to be this submanifold. An embedding of an oriented $N^{n-1}$, together with the orientation of the ambient space $M^n$ provides a trivialization of the normal bundle. The spin structure on $M^n$ then induces a spin structure on $N^{n-1}$. In the expressions above $\text{PD}(f)$ or $\text{PD}(f^*(x))$ are used to denote the spin manifold $N^{n-1}$.

 As in Section \ref{sec:Z2-inv} non-underlined elements (understood as elements of $\Hom(\Omega_n^{Spin}(BG),U(1))$) are bosonic (i.e.\ belong to $H^n(BG,U(1))$ subgroup), and elements underlined with a single line are fermionic that provide refinement of the bosonic elements from $H^n(BG,U(1))$. In particular, from the formulae of Section \ref{sec:spin-invariants} we have:
 \begin{equation}
 \beta(g) \mod 2 = \int_{M^3}g^3,
 \end{equation}
 \begin{equation}
	 \hat{\gamma}(f) \mod 4 =\int_{M^4}f\mathcal{B}f,
	\end{equation}
 \begin{equation}
	 \delta_{\text{PD}(f)}(g,g)\mod 2 = \int_{M^4}f^*(x)g^3.
	\end{equation}
  The elements in (\ref{eq:topZ4Z2}) underlined with a double line are fermionic and do not refine any elements of $H^n(BG,U(1))$.

\subsection{$\Omega^{Spin}_n(B\Z_4^2)$}
\label{sec:Z4Z4}
\subsubsection{Computation}
As in the previous case the computation involves no odd torsion and we can use the Adams spectral sequence:
\begin{equation}
	E_2^{s,t}=\Ext_{\A}^{s,t}(\H^*(MSpin\wedge(B\Z_4^2)_+),\Z_2)\Rightarrow\pi_{t-s}(MSpin\wedge(B\Z_4^2)_+)_2^{\wedge}=\Omega^{Spin}_{t-s}(B\Z_4^2).
\end{equation}
Using again the expression (\ref{MSpin-coh}) for mod 2 cohomology of $MSpin$, for $t-s<8$, we can identify the $E_2$-page with 
\begin{equation}
\Ext_{\A(1)}^{s,t}(\H^*(B\Z_4^2),\Z_2).
\end{equation}
The mod 2 cohomology and the differential are the following: $\H^*(B\Z_4)=\Z_2[y]\otimes\Lambda(x)$ where $|x|=1$, $|y|=2$, $Sq^1(y)=Sq^1(x)=0$. $d_2(y_i)=x_ih_0^2$, $d_2(y_i\alpha)=x_iy_ih_0^3$ for $i=1,2$, $d_2(y_1x_2)=d_2(x_1y_2)=x_1x_2h_0^2$, $d_2(y_1y_2)=x_1y_2h_0^2+y_1x_2h_0^2$, $d_2(x_1y_1y_2)=x_1y_1x_2h_0^2$, $d_2(y_1x_2y_2)=x_1x_2y_2h_0^2$ where $x_1,y_1$ and $x_2,y_2$ are generators of two copies of $\H^*(B\Z_4)$.

The $\A(1)$-module structure of $\H^*(B\Z_4^2)$ and the $E_2$ page are presented in Figures \ref{fig:A1-Z4Z4} and \ref{fig:E2-Z4Z4} respectively.

\begin{figure}[!h]
\begin{center}
\begin{tikzpicture}[scale=0.5]

\node [below] at (0,10) {1};
\draw[fill] (0,10) circle(.1);

\node [below] at (0,11) {$x_1$};
\draw[fill] (0,11) circle(.1);

\node [below] at (0,12) {$y_1$};
\draw[fill] (0,12) circle(.1);

\node [below] at (0,13) {$x_1y_1$};
\draw[fill] (0,13) circle(.1);
\draw[fill] (0,14) circle(.1);
\draw (0,12) to [out=150,in=150] (0,14);
\draw[fill] (0,15) circle(.1);
\draw (0,13) to [out=30,in=30] (0,15);

\node [below] at (10,10) {1};
\draw[fill] (10,10) circle(.1);

\node [below] at (10,11) {$x_2$};
\draw[fill] (10,11) circle(.1);

\node [below] at (10,12) {$y_2$};
\draw[fill] (10,12) circle(.1);

\node [below] at (10,13) {$x_2y_2$};
\draw[fill] (10,13) circle(.1);
\draw[fill] (10,14) circle(.1);
\draw (10,12) to [out=150,in=150] (10,14);
\draw[fill] (10,15) circle(.1);
\draw (10,13) to [out=30,in=30] (10,15);

\node at (5,13) {$\otimes$};

\node at (-2,-2) {$=$};

\node [below] at (0,-5) {1};
\draw[fill] (0,-5) circle(.1);

\node [below] at (-0.5,-4) {$x_1$};
\draw[fill] (-0.5,-4) circle(.1);

\node [below] at (0.5,-4) {$x_2$};
\draw[fill] (0.5,-4) circle(.1);

\node [left] at (0,-3) {$x_1x_2$};
\draw[fill] (0,-3) circle(.1);

\node [below] at (1,-3) {$y_1$};
\draw[fill] (1,-3) circle(.1);
\draw[fill] (1,-1) circle(.1);
\draw (1,-3) to [out=150,in=150] (1,-1);

\node [below] at (2,-3) {$y_2$};
\draw[fill] (2,-3) circle(.1);
\draw[fill] (2,-1) circle(.1);
\draw (2,-3) to [out=150,in=150] (2,-1);

\node [below] at (3,-2) {$x_1y_1$};
\draw[fill] (3,-2) circle(.1);
\draw[fill] (3,0) circle(.1);
\draw (3,-2) to [out=150,in=150] (3,0);

\node [below] at (4,-3) {$x_2y_2$};
\draw[->][color=red] (4,-3) -- (4,-2);
\draw[fill] (4,-2) circle(.1);
\draw[fill] (4,0) circle(.1);
\draw (4,-2) to [out=150,in=150] (4,0);

\node [below] at (5,-2) {$x_1y_2$};
\draw[fill] (5,-2) circle(.1);
\draw[fill] (5,0) circle(.1);
\draw (5,-2) to [out=150,in=150] (5,0);

\node [below] at (6,-3) {$x_2y_1$};
\draw[->][color=red] (6,-3) -- (6,-2);
\draw[fill] (6,-2) circle(.1);
\draw[fill] (6,0) circle(.1);
\draw (6,-2) to [out=150,in=150] (6,0);

\node [below] at (7,-1) {$y_1y_2$};
\draw[fill] (7,-1) circle(.1);
\draw[fill] (7,1) circle(.1);
\draw (7,-1) to [out=150,in=150] (7,1);

\node [below] at (8,-2) {$x_1y_1x_2$};
\draw[->][color=red] (8,-2) -- (8,-1);
\draw[fill] (8,-1) circle(.1);
\draw[fill] (8,1) circle(.1);
\draw (8,-1) to [out=150,in=150] (8,1);

\node [below] at (9,-3) {$x_1x_2y_2$};
\draw[->][color=red] (9,-3) -- (9,-1);
\draw[fill] (9,-1) circle(.1);
\draw[fill] (9,1) circle(.1);
\draw (9,-1) to [out=150,in=150] (9,1);

\node [below] at (10,0) {$x_1y_1y_2$};
\draw[fill] (10,0) circle(.1);
\draw[fill] (10,2) circle(.1);
\draw (10,0) to [out=150,in=150] (10,2);

\node [below] at (11,-1) {$x_2y_1y_2$};
\draw[->][color=red] (11,-1) -- (11,0);
\draw[fill] (11,0) circle(.1);
\draw[fill] (11,2) circle(.1);
\draw (11,0) to [out=150,in=150] (11,2);

\end{tikzpicture}
\end{center}
\caption{The $\A(1)$-module structure of $\H^*(B\Z_4^2)$}
\label{fig:A1-Z4Z4}
\end{figure}
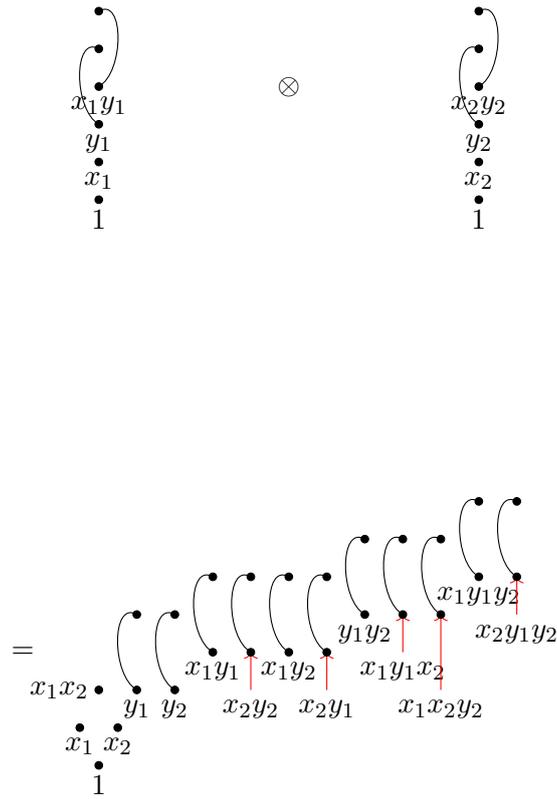

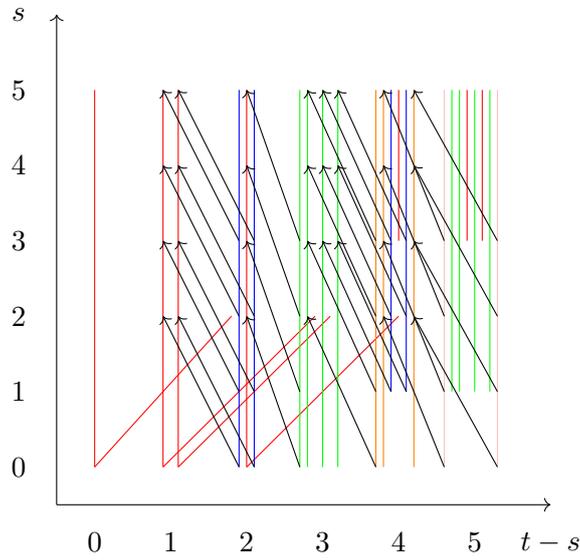
\begin{figure}[!h]
\begin{center}
\begin{tikzpicture}
\node at (0,-1) {0};
\node at (1,-1) {1};
\node at (2,-1) {2};
\node at (3,-1) {3};
\node at (4,-1) {4};
\node at (5,-1) {5};
\node at (6,-1) {$t-s$};
\node at (-1,0) {0};
\node at (-1,1) {1};
\node at (-1,2) {2};
\node at (-1,3) {3};
\node at (-1,4) {4};
\node at (-1,5) {5};
\node at (-1,6) {$s$};

\draw[->] (-0.5,-0.5) -- (-0.5,6);
\draw[->] (-0.5,-0.5) -- (6,-0.5);

\draw[color=red] (0,0) -- (0,5);
\draw[color=red] (0,0) -- (1.8,2);
\draw[color=red] (4,3) -- (4,5);
\draw[color=red] (0.9,0) -- (0.9,5);
\draw[color=red] (0.9,0) --(2.9,2);
\draw[color=red] (4.9,3) -- (4.9,5);
\draw[color=red] (1.1,0) -- (1.1,5);
\draw[color=red] (1.1,0) -- (3.1,2);
\draw[color=red] (5.1,3) -- (5.1,5);
\draw[color=red] (2,0) -- (2,5);
\draw[color=red] (2,0) -- (4,2);

\draw[color=blue] (1.9,0) -- (1.9,5);
\draw[color=blue] (2.1,0) -- (2.1,5);
\draw[color=blue] (3.9,1) -- (3.9,5);
\draw[color=blue] (4.1,1) -- (4.1,5);

\draw[color=green] (3,0) -- (3,5);
\draw[color=green] (2.8,0) -- (2.8,5);
\draw[color=green] (2.7,0) -- (2.7,5);
\draw[color=green] (3.2,0) -- (3.2,5);
\draw[color=green] (5,1) -- (5,5);
\draw[color=green] (4.8,1) -- (4.8,5);
\draw[color=green] (4.7,1) -- (4.7,5);
\draw[color=green] (5.2,1) -- (5.2,5);

\draw[color=orange] (3.8,0)--(3.8,5);
\draw[color=orange] (3.7,0)--(3.7,5);
\draw[color=orange] (4.2,0)--(4.2,5);

\draw[color=pink] (4.6,0)--(4.6,5);
\draw[color=pink] (5.3,0)--(5.3,5);

\draw[->] (1.9,0) -- (0.9,2);
\draw[->] (1.9,1) -- (0.9,3);
\draw[->] (1.9,2) -- (0.9,4);
\draw[->] (1.9,3) -- (0.9,5);

\draw[->] (2.1,0) -- (1.1,2);
\draw[->] (2.1,1) -- (1.1,3);
\draw[->] (2.1,2) -- (1.1,4);
\draw[->] (2.1,3) -- (1.1,5);

\draw[->] (2.7,0) -- (2,2);
\draw[->] (2.7,1) -- (2,3);
\draw[->] (2.7,2) -- (2,4);
\draw[->] (2.7,3) -- (2,5);

\draw[->] (3.7,0) -- (2.8,2);
\draw[->] (3.7,1) -- (2.8,3);
\draw[->] (3.7,2) -- (2.8,4);
\draw[->] (3.7,3) -- (2.8,5);

\draw[->] (3.9,1) -- (3,3);
\draw[->] (3.9,2) -- (3,4);
\draw[->] (3.9,3) -- (3,5);

\draw[->] (4.1,1) -- (3.2,3);
\draw[->] (4.1,2) -- (3.2,4);
\draw[->] (4.1,3) -- (3.2,5);

\draw[->] (4.6,0) -- (3.8,2);
\draw[->] (4.6,1) -- (3.8,3);
\draw[->] (4.6,2) -- (3.8,4);
\draw[->] (4.6,3) -- (3.8,5);

\draw[->] (5.3,0) -- (4.2,2);
\draw[->] (5.3,1) -- (4.2,3);
\draw[->] (5.3,2) -- (4.2,4);
\draw[->] (5.3,3) -- (4.2,5);

\end{tikzpicture}
\end{center}
\caption{The $E_2$ page of the Adams spectral sequence for $\Z_4^2$}
\label{fig:E2-Z4Z4}
\end{figure}

Hence we have the following theorem
\begin{theorem}
\begin{equation}
\begin{tabular}{c c c c}
\hline
$n$ & $\Omega^{Spin}_n(B\Z_4^2)$& $H_n(B\Z_4^2,\Z)$ & 
$\text{fermionic SPTs classes}$
\\
\hline
0& $\Z$ & $\Z$  &    \\
1& $\Z_2\times\Z_4^2$ & $\Z_4^2$ & \\
2& $\Z_2^3\times\Z_4$& $\Z_4$ & $\Z_2^2\times\Z_4$\\
3 & $\Z_2^3\times\Z_4\times\Z_8^2$ & $\Z_4^3$ & $\Z_2^3\times\Z_4\times\Z_8^2$\\
4 & $\Z\times\Z_4^2\times \Z_2$ & $\Z_4^2$ & $\Z_4^2\times {\Z_2}$\\ 
\hline
\end{tabular}
\end{equation}
\end{theorem}

\subsubsection{Bordism invariants and manifold generators for $G=\Z_4^2$}
The bordism groups are explicitly realized as follows:
\begin{equation}
\Omega_n^{Spin}(B(\Z_4\times\Z_4))=\{\text{spin $n$-manifolds }M^n\text{ with maps }f_1:M^n\to B\Z_4,\,f_2:M^n\to B\Z_4\}/\sim.
\end{equation}
Equivalently, $f_1\in H^1(M^n,\Z_4)$, $f_2\in H^1(M^n,\Z_4)$. As before, the functions $f_1$ and $f_2$ can be used to  pull back the generators of the mod 2 cohomology of the classifying spaces to $M^n$:
\begin{eqnarray*}
 H^*(B\Z_4,\Z_2)&\stackrel{f_1^*}{\To}& H^*(M,\Z_2)\\
x&\mapsto& f_1^*(x)=f_1\mod2\\
y&\mapsto& f_1^*(y)
\end{eqnarray*}
\begin{eqnarray*}
H^*(B\Z_4,\Z_2)&\stackrel{f_2^*}{\To}& H^*(M,\Z_2)\\
x&\mapsto& f_2^*(x)=f_2\mod2\\
y&\mapsto& f_2^*(y)
\end{eqnarray*}
Reading out from the $E_2$ page of Adams spectral sequence, the complete bordism invariants in dimension $n<5$ read as follows
\begin{equation}
\begin{array}{rcl}
(M^1,f_1,f_2) & \stackrel{~}{\longmapsto} & \doubleunderline{\eta(M^1)} \oplus \int_{M^1} f_1 \oplus \int_{M^1} f_2 \\
\\
(M^2,f_1,f_2) & \stackrel{~}{\longmapsto} & \doubleunderline{\mathrm{Arf}(M^2)} \oplus \doubleunderline{\tilde{q}(f_1^*(x))} \oplus \doubleunderline{\tilde{q}(f_2^*(x))} \oplus \int_{M^2} f_1\,f_2 
\\
\\
(M^3,f_1,f_2) &  \stackrel{~}{\longmapsto} & \doubleunderline{\mathrm{Arf}(\mathrm{PD}(f_1^*(x)))} \oplus \doubleunderline{\mathrm{Arf}(\mathrm{PD}(f_2^*(x)))} \oplus \doubleunderline{\tilde{q}_{\mathrm{PD}(f_1^*(x))}(f_2^*(x))} \oplus \int_{M^3}f_1\,\mathcal{B}f_2 \oplus \underline{\hat\gamma( f_1)} \oplus \underline{\hat{\gamma}(f_2)} \\
\\
(M^4,f_1,f_2) & \stackrel{~}{\longmapsto} & \frac{p_1(TM^4)}{48} \oplus \int_{M^4} f_1\,f_2\,\mathcal{B}f_2 \oplus \int_{M^4} f_2\,f_1\,\mathcal{B}f_1
\oplus \doubleunderline{\mathrm{Arf}(\mathrm{PD}(f_1^*(x))\cap \mathrm{PD}(f_2^*(x)))}
\end{array}
\end{equation}
where $\mathcal{B}:H^1(M^n,\Z_4)\rightarrow H^2(M^n,\Z_4)$ is the Bockstein homomorphism w.r.t.\ to the short exact sequence $\Z_4\rightarrow \Z_{16}\rightarrow \Z_4$. The codimension one submanifolds $\mathrm{PD}(f_1^*(x))$ and $\mathrm{PD}(f_2^*(x))$ are chosen such that they intersect transversally along a smooth 2-manifold. The spin structure on $\mathrm{PD}(f_1^*(x))\cap \mathrm{PD}(f_2^*(x))$ is induced as follows. The pair of normal vectors to $\mathrm{PD}(f_1^*(x))$ and $\mathrm{PD}(f_2^*(x))$ define a trivialization of the normal bundle to $\mathrm{PD}(f_1^*(x))\cap \mathrm{PD}(f_2^*(x))$. As described before, together with the spin structure on the ambient space $M^4$ this unambiguously defines a spin-structure on  $\mathrm{PD}(f_1^*(x))\cap \mathrm{PD}(f_2^*(x))$.

It follows that in dimension 4 the generators of the bordism group are given by the following
\begin{theorem}
$(S^1\times L(4,1),f,g )$ and $(S^1\times L(4,1),h,\ell)$ generates two $\Z_4$ individually and $(S^1\times S^1\times S^1\times S^1,k,j)$ generates $\Z_2$, where $L(4,1)=S^3/Z_4$ is the Lens space, $f$ is the generator of $H^1(S^1,\Z_4)$, $g$ is the generator of $H^1( L(4),\Z_4)$,  $h$ is the generator of $H^1(L(4),\Z_4)$, $\ell$ is the generator of $H^1(S^1,\Z_4)$, $k$ is the generator of the first $H^1(S^1,\Z_4)$, $j$ is the generator of the second $H^1(S^1,\Z_4)$ and there are odd spin structures on the last two $S^1$.
\end{theorem}

\subsection{$\Omega^{Spin}_n(B(\Z_2^2\times\Z_4))$}
\subsubsection{Computation}
We can  again use the Adams spectral sequence, since the computation involves no odd torsion:
\begin{multline}
 E_2^{s,t}=\Ext_{\A}^{s,t}(\H^*(MSpin\wedge(B(\Z_2^2\times\Z_4))_+),\Z_2)\Rightarrow\pi_{t-s}(MSpin\wedge(B(\Z_2^2\times\Z_4))_+)_2^{\wedge}=\Omega^{Spin}_{t-s}(B(\Z_2^2\times\Z_4)).
\end{multline} 
Using (\ref{MSpin-coh}), for $t-s<8$, we can identify the $E_2$-page with 
\begin{equation}
 \Ext_{\A(1)}^{s,t}(\H^*(B(\Z_2^2\times\Z_4)),\Z_2).
\end{equation} 
The cohomology rings of the classifying spaces are the following: $\H^*(B\Z_2)=\Z_2[a]$ where $|a|=1$, $\H^*(B\Z_4)=\Z_2[y]\otimes\Lambda(x)$ where $|x|=1$, $|y|=2$, $Sq^1(y)=Sq^1(x)=0$. The second page differential acts as follows: $d_2(y)=xh_0^2$, $d_2(y\alpha)=xyh_0^3$, $d_2(ya_i^3+y^2a_i)=(xa_i^3+xya_i)h_0^2$ for $i=1,2$, where $a_i$ are generators of the two copies of $\H^*(B\Z_2)$.

The $\A(1)$-module structure of $\H^*(B(\Z_2^2\times\Z_4))$ and the $E_2$ page are shown in Figures \ref{fig:A1-Z2Z2Z4} and \ref{fig:E2-Z2Z2Z4}.

\begin{figure}[!h]
\begin{center}
\begin{tikzpicture}[scale=0.5]
\node [below] at (0,0) {1};
\node [below] at (0,1) {$a_1$};
\draw[fill] (0,0) circle(.1);
\draw[fill] (0,1) circle(.1);
\draw[fill] (0,2) circle(.1);
\draw (0,1) -- (0,2);
\draw[fill] (0,3) circle(.1);
\draw[fill] (0,4) circle(.1);
\draw (0,2) to [out=150,in=150] (0,4);
\draw (0,3) -- (0,4);
\draw[fill] (0,5) circle(.1);
\draw (0,3) to [out=30,in=30] (0,5);
\draw[fill] (0,6) circle(.1);
\draw (0,5) -- (0,6);

\node [below] at (2,1) {$x$};
\node [below] at (2,2) {$xa_1$};
\draw[fill] (2,1) circle(.1);
\draw[fill] (2,2) circle(.1);
\draw[fill] (2,3) circle(.1);
\draw (2,2) -- (2,3);
\draw[fill] (2,4) circle(.1);
\draw[fill] (2,5) circle(.1);
\draw (2,3) to [out=150,in=150] (2,5);
\draw (2,4) -- (2,5);
\draw[fill] (2,6) circle(.1);
\draw (2,4) to [out=30,in=30] (2,6);
\draw[fill] (2,7) circle(.1);
\draw (2,6) -- (2,7);

\node [below] at (4,2) {$y$};
\draw[fill] (4,2) circle(.1);
\draw[fill] (4,4) circle(.1);
\draw (4,2) to [out=150,in=150] (4,4);

\node [below] at (5,3) {$ya_1$};
\draw[fill] (5,3) circle(.1);
\draw[fill] (5,4) circle(.1);
\draw (5,3) -- (5,4);
\draw[fill] (5,5) circle(.1);
\draw (5,3) to [out=150,in=150] (5,5);
\draw[fill] (5,6) circle (.1);
\draw (5,5) -- (5,6);
\draw[fill] (5,7) circle(.1);
\draw[fill] (5,8) circle(.1);
\draw (5,7) -- (5,8);
\draw (5,6) to [out=150,in=150] (5,8);
\draw[fill] (5,9) circle(.1);
\draw (5,7) to [out=30,in=30] (5,9);
\draw[fill] (5,10) circle(.1);
\draw (5,9) -- (5,10);
\draw[fill] (6,5) circle(.1);
\draw[fill] (6,6) circle(.1);
\draw (6,5) -- (6,6);
\draw (5,4) to [out=30,in=150] (6,6);
\draw[fill] (6,7) circle(.1);
\draw (6,5) to [out=30,in=30] (6,7);
\draw[fill] (6,8) circle(.1);
\draw (6,7) -- (6,8);
\draw[fill] (6,9) circle(.1);
\draw[fill] (6,10) circle(.1);
\draw (6,8) to [out=30,in=30] (6,10);
\draw (6,9) -- (6,10);
\draw[fill] (6,11) circle(.1);
\draw (6,9) to [out=150,in=150] (6,11);

\node [below] at (7,3) {$xy$};
\draw[fill] (7,3) circle(.1);
\draw[fill] (7,5) circle(.1);
\draw (7,3) to [out=150,in=150] (7,5);

\node [below] at (8,4) {$xya_1$};
\draw[fill] (8,4) circle(.1);
\draw[fill] (8,5) circle(.1);
\draw (8,4) -- (8,5);
\draw[fill] (8,6) circle(.1);
\draw (8,4) to [out=150,in=150] (8,6);
\draw[fill] (8,7) circle (.1);
\draw (8,6) -- (8,7);
\draw[fill] (8,8) circle(.1);
\draw[fill] (8,9) circle(.1);
\draw (8,8) -- (8,9);
\draw (8,7) to [out=150,in=150] (8,9);
\draw[fill] (8,10) circle(.1);
\draw (8,8) to [out=30,in=30] (8,10);
\draw[fill] (8,11) circle(.1);
\draw (8,10) -- (8,11);
\draw[fill] (9,6) circle(.1);
\draw[fill] (9,7) circle(.1);
\draw (9,6) -- (9,7);
\draw (8,5) to [out=30,in=150] (9,7);
\draw[fill] (9,8) circle(.1);
\draw (9,6) to [out=30,in=30] (9,8);
\draw[fill] (9,9) circle(.1);
\draw (9,8) -- (9,9);
\draw[fill] (9,10) circle(.1);
\draw[fill] (9,11) circle(.1);
\draw (9,9) to [out=30,in=30] (9,11);
\draw (9,10) -- (9,11);
\draw[fill] (9,12) circle(.1);
\draw (9,10) to [out=150,in=150] (9,12);

\node [below] at (10,1) {$a_2$};
\draw[fill] (10,1) circle(.1);
\draw[fill] (10,2) circle(.1);
\draw (10,1) -- (10,2);
\draw[fill] (10,3) circle(.1);
\draw[fill] (10,4) circle(.1);
\draw (10,2) to [out=150,in=150] (10,4);
\draw (10,3) -- (10,4);
\draw[fill] (10,5) circle(.1);
\draw (10,3) to [out=30,in=30] (10,5);
\draw[fill] (10,6) circle(.1);
\draw (10,5) -- (10,6);

\node [below] at (12,2) {$a_1a_2$};
\draw[fill] (12,2) circle(.1);
\draw[fill] (12,3) circle(.1);
\draw (12,2) -- (12,3);
\draw[fill] (12,5) circle(.1);
\draw (12,3) to [out=150,in=150] (12,5);
\draw[fill] (12,6) circle(.1);
\draw (12,5) -- (12,6);
\draw[fill] (13,3) circle(.1);
\draw[fill] (13,4) circle(.1);
\draw (13,3) -- (13,4);
\draw (12,2) to [out=30,in=150] (13,4);
\draw (13,4) to [out=150,in=30] (12,6);
\draw[fill] (13,5) circle(.1);
\draw (13,3) to [out=30,in=30] (13,5);
\draw[fill] (13,6) circle(.1);
\draw (13,5) -- (13,6);
\draw[fill] (13,7) circle(.1);
\draw[fill] (13,8) circle(.1);
\draw (13,7) -- (13,8);
\draw (13,6) to [out=150,in=150] (13,8);
\draw[fill] (13,9) circle(.1);
\draw (13,7) to [out=30,in=30] (13,9);

\node [below] at (15,4) {$a_1a_2^3$};
\draw[fill] (15,4) circle(.1);
\draw[fill] (15,5) circle(.1);
\draw (15,4) -- (15,5);
\draw[fill] (15,6) circle(.1);
\draw (15,4) to [out=150,in=150] (15,6);
\draw[fill] (15,7) circle(.1);
\draw (15,6) -- (15,7);
\draw[fill] (16,7) circle(.1);
\draw (15,5) to [out=30,in=150] (16,7);
\draw[fill] (16,8) circle(.1);
\draw (16,7) -- (16,8);
\draw (15,6) to [out=30,in=150] (16,8);
\draw[fill] (16,9) circle(.1);
\draw (15,7) to [out=30,in=150] (16,9);
\draw[fill] (16,10) circle(.1);
\draw (16,9) -- (16,10);
\draw (16,8) to [out=30,in=30] (16,10);

\node [below] at (17,4) {$a_1^3a_2$};
\draw[fill] (17,4) circle(.1);
\draw[fill] (17,5) circle(.1);
\draw (17,4) -- (17,5);
\draw[fill] (17,6) circle(.1);
\draw (17,4) to [out=150,in=150] (17,6);
\draw[fill] (17,7) circle(.1);
\draw (17,6) -- (17,7);
\draw[fill] (18,7) circle(.1);
\draw (17,5) to [out=30,in=150] (18,7);
\draw[fill] (18,8) circle(.1);
\draw (18,7) -- (18,8);
\draw (17,6) to [out=30,in=150] (18,8);
\draw[fill] (18,9) circle(.1);
\draw (17,7) to [out=30,in=150] (18,9);
\draw[fill] (18,10) circle(.1);
\draw (18,9) -- (18,10);
\draw (18,8) to [out=30,in=30] (18,10);

\node [below] at (19,2) {$xa_2$};
\draw[fill] (19,2) circle(.1);
\draw[fill] (19,3) circle(.1);
\draw (19,2) -- (19,3);
\draw[fill] (19,4) circle(.1);
\draw[fill] (19,5) circle(.1);
\draw (19,3) to [out=150,in=150] (19,5);
\draw (19,4) -- (19,5);
\draw[fill] (19,6) circle(.1);
\draw (19,4) to [out=30,in=30] (19,6);
\draw[fill] (19,7) circle(.1);
\draw (19,6) -- (19,7);

\node [below] at (21,3) {$xa_1a_2$};
\draw[fill] (21,3) circle(.1);
\draw[fill] (21,4) circle(.1);
\draw (21,3) -- (21,4);
\draw[fill] (21,6) circle(.1);
\draw (21,4) to [out=150,in=150] (21,6);
\draw[fill] (21,7) circle(.1);
\draw (21,6) -- (21,7);
\draw[fill] (22,4) circle(.1);
\draw[fill] (22,5) circle(.1);
\draw (22,4) -- (22,5);
\draw (21,3) to [out=30,in=150] (22,5);
\draw (22,5) to [out=150,in=30] (21,7);
\draw[fill] (22,6) circle(.1);
\draw (22,4) to [out=30,in=30] (22,6);
\draw[fill] (22,7) circle(.1);
\draw (22,6) -- (22,7);
\draw[fill] (22,8) circle(.1);
\draw[fill] (22,9) circle(.1);
\draw (22,8) -- (22,9);
\draw (22,7) to [out=150,in=150] (22,9);
\draw[fill] (22,10) circle(.1);
\draw (22,8) to [out=30,in=30] (22,10);

\node [below] at (24,3) {$ya_2$};
\draw[fill] (24,3) circle(.1);
\draw[fill] (24,4) circle(.1);
\draw (24,3) -- (24,4);
\draw[fill] (24,5) circle(.1);
\draw (24,3) to [out=150,in=150] (24,5);
\draw[fill] (24,6) circle (.1);
\draw (24,5) -- (24,6);
\draw[fill] (24,7) circle(.1);
\draw[fill] (24,8) circle(.1);
\draw (24,7) -- (24,8);
\draw (24,6) to [out=150,in=150] (24,8);
\draw[fill] (24,9) circle(.1);
\draw (24,7) to [out=30,in=30] (24,9);
\draw[fill] (24,10) circle(.1);
\draw (24,9) -- (24,10);
\draw[fill] (25,5) circle(.1);
\draw[fill] (25,6) circle(.1);
\draw (25,5) -- (25,6);
\draw (24,4) to [out=30,in=150] (25,6);
\draw[fill] (25,7) circle(.1);
\draw (25,5) to [out=30,in=30] (25,7);
\draw[fill] (25,8) circle(.1);
\draw (25,7) -- (25,8);
\draw[fill] (25,9) circle(.1);
\draw[fill] (25,10) circle(.1);
\draw (25,8) to [out=30,in=30] (25,10);
\draw (25,9) -- (25,10);
\draw[fill] (25,11) circle(.1);
\draw (25,9) to [out=150,in=150] (25,11);

\node [below] at (27,4) {$ya_1a_2$};
\draw[fill] (27,4) circle(.1);
\draw[fill] (27,5) circle(.1);
\draw (27,4) -- (27,5);
\draw[fill] (27,6) circle(.1);
\draw (27,4) to [out=150,in=150] (27,6);
\draw[fill] (27,7) circle(.1);
\draw (27,6) -- (27,7);
\draw[fill] (28,7) circle(.1);
\draw (27,5) to [out=30,in=150] (28,7);
\draw[fill] (28,8) circle(.1);
\draw (28,7) -- (28,8);
\draw (27,6) to [out=30,in=150] (28,8);
\draw[fill] (28,9) circle(.1);
\draw (27,7) to [out=30,in=150] (28,9);
\draw[fill] (28,10) circle(.1);
\draw (28,9) -- (28,10);
\draw (28,8) to [out=30,in=30] (28,10);

\node [below] at (29,4) {$xya_2$};
\draw[fill] (29,4) circle(.1);
\draw[fill] (29,5) circle(.1);
\draw (29,4) -- (29,5);
\draw[fill] (29,6) circle(.1);
\draw (29,4) to [out=150,in=150] (29,6);
\draw[fill] (29,7) circle (.1);
\draw (29,6) -- (29,7);
\draw[fill] (29,8) circle(.1);
\draw[fill] (29,9) circle(.1);
\draw (29,8) -- (29,9);
\draw (29,7) to [out=150,in=150] (29,9);
\draw[fill] (29,10) circle(.1);
\draw (29,8) to [out=30,in=30] (29,10);
\draw[fill] (29,11) circle(.1);
\draw (29,10) -- (29,11);
\draw[fill] (30,6) circle(.1);
\draw[fill] (30,7) circle(.1);
\draw (30,6) -- (30,7);
\draw (29,5) to [out=30,in=150] (30,7);
\draw[fill] (30,8) circle(.1);
\draw (30,6) to [out=30,in=30] (30,8);
\draw[fill] (30,9) circle(.1);
\draw (30,8) -- (30,9);
\draw[fill] (30,10) circle(.1);
\draw[fill] (30,11) circle(.1);
\draw (30,9) to [out=30,in=30] (30,11);
\draw (30,10) -- (30,11);
\draw[fill] (30,12) circle(.1);
\draw (30,10) to [out=150,in=150] (30,12);
 
\end{tikzpicture}
\end{center}
\caption{The $\A(1)$-module structure of $\H^*(B(\Z_2^2\times\Z_4))$}
\label{fig:A1-Z2Z2Z4}
\end{figure}
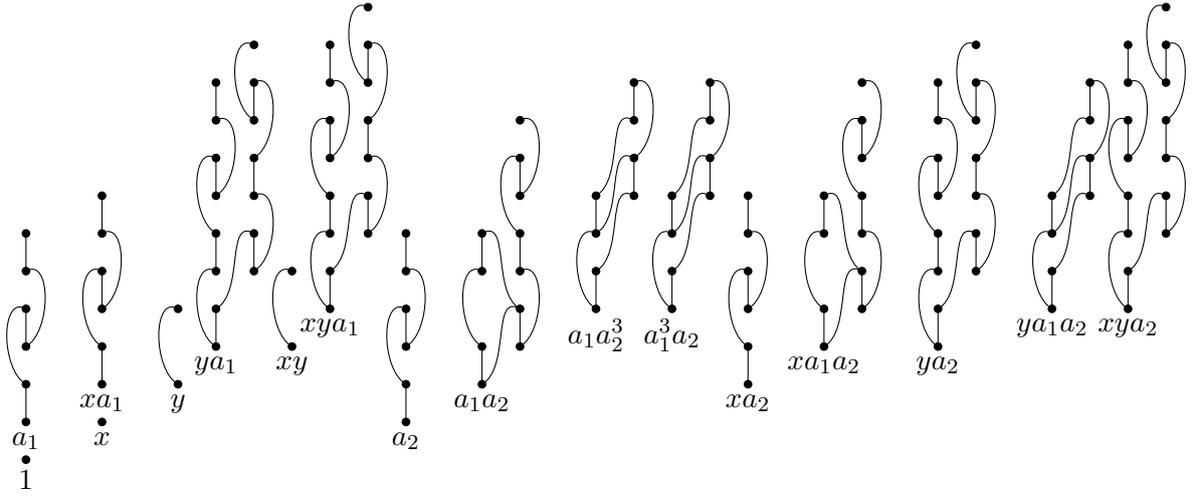

\begin{figure}[!h]
\begin{center}
\begin{tikzpicture}
\node at (0,-1) {0};
\node at (1,-1) {1};
\node at (2,-1) {2};
\node at (3,-1) {3};
\node at (4,-1) {4};
\node at (5,-1) {5};
\node at (6,-1) {$t-s$};
\node at (-1,0) {0};
\node at (-1,1) {1};
\node at (-1,2) {2};
\node at (-1,3) {3};
\node at (-1,4) {4};
\node at (-1,5) {5};
\node at (-1,6) {$s$};

\draw[->] (-0.5,-0.5) -- (-0.5,6);
\draw[->] (-0.5,-0.5) -- (6,-0.5);

\draw (0,0) -- (0,5);
\draw (0,0) -- (1.8,2);
\draw (0.9,0) -- (0.9,5);
\draw (0.9,0) -- (2.9,2);
\draw (1,0) -- (3,2);
\draw (3,2) -- (3,0);
\draw (1.1,0) -- (3.1,2);
\draw (3.1,2) -- (3.1,0);
\draw (1.9,0) -- (1.9,5);
\draw (2,0) -- (4,2);
\draw (4,2) -- (4,0);
\draw (2.1,0) -- (4.1,2);
\draw (4.1,2) -- (4.1,0);
\draw (2.2,0) -- (3.2,1);
\draw (3.2,1) -- (3.2,0);
\draw[fill] (2.8,0) circle(0.05);
\draw[fill] (2.9,0) circle(0.05);
\draw (3.3,0) -- (4.2,1);
\draw (4.2,1) -- (4.2,0);
\draw (3.4,0) -- (3.4,5);
\draw (4,3) -- (4,5);
\draw (4.3,1) -- (4.3,5);
\draw[fill] (3.7,0) circle(0.05);
\draw[fill] (3.8,0) circle(0.05);
\draw[fill] (3.9,0) circle(0.05);
\draw[fill] (4.3,0) circle(0.05);
\draw[fill] (4.4,0) circle(0.05);
\draw (5,0) -- (5,1);
\draw (5.1,0) -- (5.1,1);
\draw (5,3) -- (5,5);
\draw (5.2,1) -- (5.2,5);

\draw[->,color=red] (1.9,0) -- (0.9,2);
\draw[->,color=red] (1.9,1) -- (0.9,3);
\draw[->,color=red] (1.9,2) -- (0.9,4);
\draw[->,color=red] (1.9,3) -- (0.9,5);

\draw[->,color=red] (4.3,1) -- (3.4,3);
\draw[->,color=red] (4.3,2) -- (3.4,4);
\draw[->,color=red] (4.3,3) -- (3.4,5);

\draw[->,color=red] (5,0) -- (4,2);
\draw[->,color=red] (5.1,0) -- (4.1,2);
\end{tikzpicture}
\end{center}
\caption{The $E_2$ page of the Adams spectral sequence for $\Z_2^2\times\Z_4$}
\label{fig:E2-Z2Z2Z4}
\end{figure}
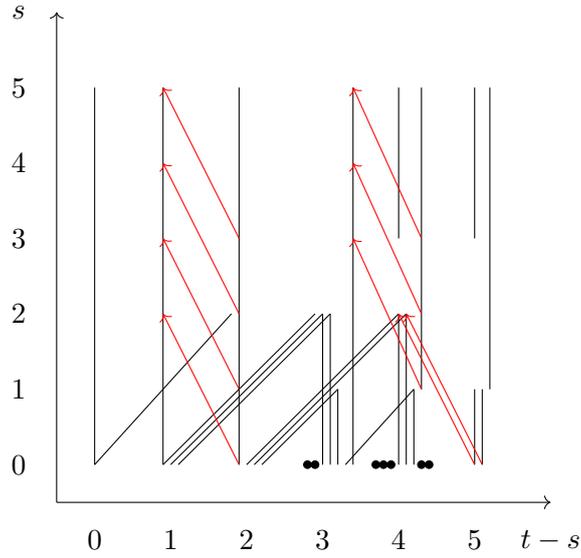

Hence we have the following
\begin{theorem} \label{thm:Z2Z2Z4}
\begin{equation}
\begin{tabular}{c c c c}
\hline
$n$ & $\Omega^{Spin}_n(B(\Z_2^2\times\Z_4))$& $H_n(B(\Z_2^2\times\Z_4),\Z_4)$ & 
$\text{fermionic SPTs classes}$
\\
\hline
0& $\Z$& $\Z$ &  \\
1& $\Z_2^3\times\Z_4$& $\Z_2^2\times \Z_4$ & \\
2& $\Z_2^7$& $\Z_2^3$ & $\Z_2^6$\\
3 & $\Z_2^6\times\Z_4\times\Z_8^3$& $\Z_2^6\times\Z_4$ & $\Z_2^6\times\Z_4\times\Z_8^3$ \\
4 & $\Z\times\Z_2^5\times {\Z_4^3}$ & $\Z_2^8$  & $\Z_2^5\times {\Z_4^3}$\\ 
\hline
\end{tabular}
\end{equation}
\end{theorem}

\subsubsection{Bordism invariants}
The bordism groups are explicitly realized as follows:
\begin{equation}
\Omega_n^{Spin}(B(\Z_2^2\times\Z_4))=\{\text{spin $n$-manifolds }M^n\text{ with maps }g_{1,2}:M\to B\Z_2,f:M\to B\Z_4\}/\sim.
\end{equation}
The complete bordism invariant in dimension $n<5$ read as follows
\begin{equation}
\begin{array}{rcl}
(M^1,f,g_1,g_2) & \stackrel{~}{\longmapsto} & \doubleunderline{\eta(M^1)} \oplus \int_{M^1} g_1 \oplus \int_{M^1} g_2 \oplus \int_{M^1}f \\
\\
(M^2,f,g_1,g_2) & \stackrel{~}{\longmapsto} & \doubleunderline{\mathrm{Arf}(M^2)} \oplus \doubleunderline{\tilde{q}(g_1)} \oplus \doubleunderline{\tilde{q}(g_2)} \oplus \doubleunderline{\tilde{q}(f^*(x))} \oplus \int_{M^2} g_1\,g_2\oplus \int_{M^2} g_1\,f^*(x)\oplus \int_{M^2} g_2\,f^*(x) \\
\\
(M^3,f,g_1,g_2) &  \stackrel{~}{\longmapsto} & 
\int_{M^3}g_1\,f^*(y) \oplus \int_{M^3}g_2\,f^*(y) \oplus \int_{M^3}g_1\,g_2\,f^*(x)\oplus 
\\
\\ & &
\; \doubleunderline{\mathrm{Arf}(\mathrm{PD}(f^*(x)))} \oplus \doubleunderline{\tilde{q}_{\mathrm{PD}(f^*(x))}(g_1)}\oplus \doubleunderline{\tilde{q}_{\mathrm{PD}(f^*(x))}(g_2)} \oplus
\\
\\ & &
 \; \underline{\delta(g_1,g_2)} \oplus \underline{\beta(g_1)} \oplus \underline{\beta(g_2)} \oplus \underline{\hat{\gamma}(f)}
\\
\\
(M^4,f,g_1,g_2) & \stackrel{~}{\longmapsto} & \frac{p_1(TM^4)}{48} \oplus \int_{M^4}g_1\,g_2^3 \oplus \int_{M^4}g_2\,g_1^3 
\oplus \int_{M^4}g_1\,f^*(xy)\oplus \int_{M^4}g_2\,f^*(xy)
\oplus \int_{M^4}g_1\,g_2\,f^*(y) \oplus 
\\
\\
& & \; \underline{\delta_{\mathrm{PD}(f)}(g_1,g_1)}\oplus 
\underline{\delta_{\mathrm{PD}(f)}(g_2,g_2)}\oplus 
\underline{\delta_{\mathrm{PD}(f)}(g_1,g_2)}
\end{array}
\end{equation}

\subsection{$\Omega^{Spin}_n(B(\Z_2\times\Z_4^2))$}
\subsubsection{Computation}
As in the previous sections we use the Adams spectral sequence
\begin{multline}
 E_2^{s,t}=\Ext_{\A}^{s,t}(\H^*(MSpin\wedge(B(\Z_2\times\Z_4^2))_+),\Z_2)\Rightarrow 
 \\
 \pi_{t-s}(MSpin\wedge(B(\Z_2\times\Z_4^2))_+)_2^{\wedge}=
 \Omega^{Spin}_{t-s}(B(\Z_2\times\Z_4^2)).
\end{multline} 
and the fact (\ref{MSpin-coh}). For $t-s<8$, we can then identify the $E_2$-page with 
\begin{equation}
 \Ext_{\A(1)}^{s,t}(\H^*(B(\Z_2\times\Z_4^2)),\Z_2).
\end{equation} 

The cohomology ring of the classifying space is given by:
$\H^*(B\Z_2)=\Z_2[a]$ where $|a|=1$, $\H^*(B\Z_4)=\Z_2[y]\otimes\Lambda(x)$ where $|x|=1$, $|y|=2$, $Sq^1(y)=Sq^1(x)=0$. The second page differential acts as follows:
$d_2(y_i)=x_ih_0^2$, $d_2(y_i\alpha)=x_iy_ih_0^3$, $d_2(y_ia^3+y_i^2a)=(x_ia^3+x_iy_ia)h_0^2$ for $i=1,2$. $d_2(y_1x_2)=d_2(x_1y_2)=x_1x_2h_0^2$, $d_2(y_1y_2)=x_1y_2h_0^2+y_1x_2h_0^2$, $d_2(x_1y_1y_2)=x_1y_1x_2h_0^2$, $d_2(y_1x_2y_2)=x_1x_2y_2h_0^2$, where $x_1,y_1$ and $x_2,y_2$ are generators of two copies of $\H^*(B\Z_4)$.

The $\A(1)$-module structure of $\H^*(B(\Z_2\times\Z_4^2))$ and the $E_2$ page are shown in Figures \ref{fig:A1-Z2Z4Z4} and \ref{fig:E2-Z2Z4Z4} where we use the known result for $\Z_4^2$ in Figure \ref{fig:E2-Z4Z4}.

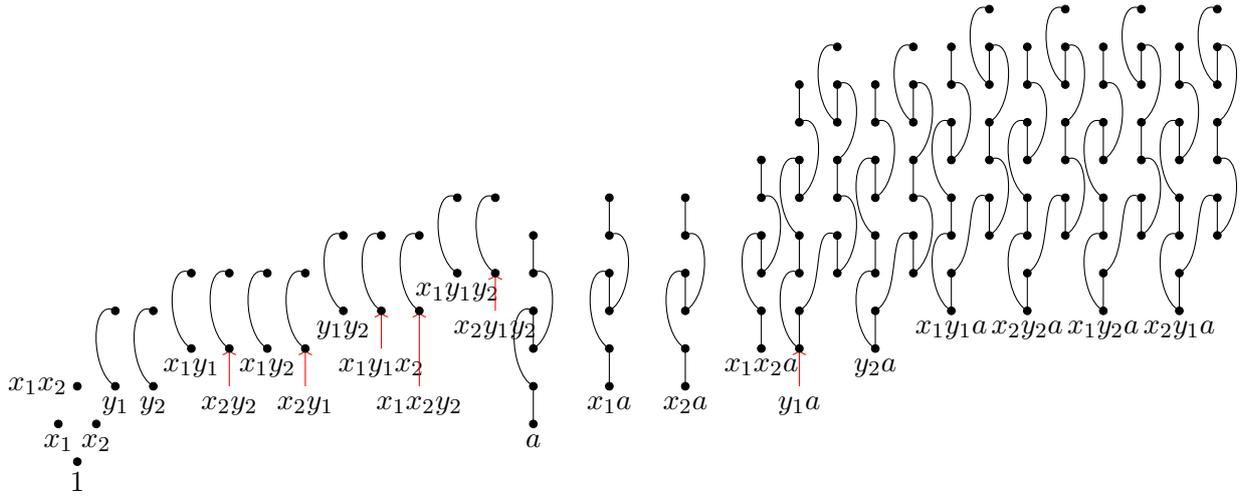
\begin{figure}[!h]
\begin{center}
\begin{tikzpicture}[scale=0.5]
\node [below] at (0,0) {1};
\draw[fill] (0,0) circle(.1);

\node [below] at (-0.5,1) {$x_1$};
\draw[fill] (-0.5,1) circle(.1);

\node [below] at (0.5,1) {$x_2$};
\draw[fill] (0.5,1) circle(.1);

\node [left] at (0,2) {$x_1x_2$};
\draw[fill] (0,2) circle(.1);

\node [below] at (1,2) {$y_1$};
\draw[fill] (1,2) circle(.1);
\draw[fill] (1,4) circle(.1);
\draw (1,2) to [out=150,in=150] (1,4);

\node [below] at (2,2) {$y_2$};
\draw[fill] (2,2) circle(.1);
\draw[fill] (2,4) circle(.1);
\draw (2,2) to [out=150,in=150] (2,4);

\node [below] at (3,3) {$x_1y_1$};
\draw[fill] (3,3) circle(.1);
\draw[fill] (3,5) circle(.1);
\draw (3,3) to [out=150,in=150] (3,5);

\node [below] at (4,2) {$x_2y_2$};
\draw[->][color=red] (4,2) -- (4,3);
\draw[fill] (4,3) circle(.1);
\draw[fill] (4,5) circle(.1);
\draw (4,3) to [out=150,in=150] (4,5);

\node [below] at (5,3) {$x_1y_2$};
\draw[fill] (5,3) circle(.1);
\draw[fill] (5,5) circle(.1);
\draw (5,3) to [out=150,in=150] (5,5);

\node [below] at (6,2) {$x_2y_1$};
\draw[->][color=red] (6,2) -- (6,3);
\draw[fill] (6,3) circle(.1);
\draw[fill] (6,5) circle(.1);
\draw (6,3) to [out=150,in=150] (6,5);

\node [below] at (7,4) {$y_1y_2$};
\draw[fill] (7,4) circle(.1);
\draw[fill] (7,6) circle(.1);
\draw (7,4) to [out=150,in=150] (7,6);

\node [below] at (8,3) {$x_1y_1x_2$};
\draw[->][color=red] (8,3) -- (8,4);
\draw[fill] (8,4) circle(.1);
\draw[fill] (8,6) circle(.1);
\draw (8,4) to [out=150,in=150] (8,6);

\node [below] at (9,2) {$x_1x_2y_2$};
\draw[->][color=red] (9,2) -- (9,4);
\draw[fill] (9,4) circle(.1);
\draw[fill] (9,6) circle(.1);
\draw (9,4) to [out=150,in=150] (9,6);

\node [below] at (10,5) {$x_1y_1y_2$};
\draw[fill] (10,5) circle(.1);
\draw[fill] (10,7) circle(.1);
\draw (10,5) to [out=150,in=150] (10,7);

\node [below] at (11,4) {$x_2y_1y_2$};
\draw[->][color=red] (11,4) -- (11,5);
\draw[fill] (11,5) circle(.1);
\draw[fill] (11,7) circle(.1);
\draw (11,5) to [out=150,in=150] (11,7);

\node [below] at (12,1) {$a$};
\draw[fill] (12,1) circle(.1);
\draw[fill] (12,2) circle(.1);
\draw (12,1) -- (12,2);
\draw[fill] (12,3) circle(.1);
\draw[fill] (12,4) circle(.1);
\draw (12,2) to [out=150,in=150] (12,4);
\draw (12,3) -- (12,4);
\draw[fill] (12,5) circle(.1);
\draw (12,3) to [out=30,in=30] (12,5);
\draw[fill] (12,6) circle(.1);
\draw (12,5) -- (12,6);

\node [below] at (14,2) {$x_1a$};
\draw[fill] (14,2) circle(.1);
\draw[fill] (14,3) circle(.1);
\draw (14,2) -- (14,3);
\draw[fill] (14,4) circle(.1);
\draw[fill] (14,5) circle(.1);
\draw (14,3) to [out=150,in=150] (14,5);
\draw (14,4) -- (14,5);
\draw[fill] (14,6) circle(.1);
\draw (14,4) to [out=30,in=30] (14,6);
\draw[fill] (14,7) circle(.1);
\draw (14,6) -- (14,7);

\node [below] at (16,2) {$x_2a$};
\draw[fill] (16,2) circle(.1);
\draw[fill] (16,3) circle(.1);
\draw (16,2) -- (16,3);
\draw[fill] (16,4) circle(.1);
\draw[fill] (16,5) circle(.1);
\draw (16,3) to [out=150,in=150] (16,5);
\draw (16,4) -- (16,5);
\draw[fill] (16,6) circle(.1);
\draw (16,4) to [out=30,in=30] (16,6);
\draw[fill] (16,7) circle(.1);
\draw (16,6) -- (16,7);

\node [below] at (18,3) {$x_1x_2a$};
\draw[fill] (18,3) circle(.1);
\draw[fill] (18,4) circle(.1);
\draw (18,3) -- (18,4);
\draw[fill] (18,5) circle(.1);
\draw[fill] (18,6) circle(.1);
\draw (18,4) to [out=150,in=150] (18,6);
\draw (18,5) -- (18,6);
\draw[fill] (18,7) circle(.1);
\draw (18,5) to [out=30,in=30] (18,7);
\draw[fill] (18,8) circle(.1);
\draw (18,7) -- (18,8);

\node [below] at (19,2) {$y_1a$};
\draw[->][color=red] (19,2) -- (19,3);
\draw[fill] (19,3) circle(.1);
\draw[fill] (19,4) circle(.1);
\draw (19,3) -- (19,4);
\draw[fill] (19,5) circle(.1);
\draw (19,3) to [out=150,in=150] (19,5);
\draw[fill] (19,6) circle (.1);
\draw (19,5) -- (19,6);
\draw[fill] (19,7) circle(.1);
\draw[fill] (19,8) circle(.1);
\draw (19,7) -- (19,8);
\draw (19,6) to [out=150,in=150] (19,8);
\draw[fill] (19,9) circle(.1);
\draw (19,7) to [out=30,in=30] (19,9);
\draw[fill] (19,10) circle(.1);
\draw (19,9) -- (19,10);
\draw[fill] (20,5) circle(.1);
\draw[fill] (20,6) circle(.1);
\draw (20,5) -- (20,6);
\draw (19,4) to [out=30,in=150] (20,6);
\draw[fill] (20,7) circle(.1);
\draw (20,5) to [out=30,in=30] (20,7);
\draw[fill] (20,8) circle(.1);
\draw (20,7) -- (20,8);
\draw[fill] (20,9) circle(.1);
\draw[fill] (20,10) circle(.1);
\draw (20,8) to [out=30,in=30] (20,10);
\draw (20,9) -- (20,10);
\draw[fill] (20,11) circle(.1);
\draw (20,9) to [out=150,in=150] (20,11);

\node [below] at (21,3) {$y_2a$};
\draw[fill] (21,3) circle(.1);
\draw[fill] (21,4) circle(.1);
\draw (21,3) -- (21,4);
\draw[fill] (21,5) circle(.1);
\draw (21,3) to [out=150,in=150] (21,5);
\draw[fill] (21,6) circle (.1);
\draw (21,5) -- (21,6);
\draw[fill] (21,7) circle(.1);
\draw[fill] (21,8) circle(.1);
\draw (21,7) -- (21,8);
\draw (21,6) to [out=150,in=150] (21,8);
\draw[fill] (21,9) circle(.1);
\draw (21,7) to [out=30,in=30] (21,9);
\draw[fill] (21,10) circle(.1);
\draw (21,9) -- (21,10);
\draw[fill] (22,5) circle(.1);
\draw[fill] (22,6) circle(.1);
\draw (22,5) -- (22,6);
\draw (21,4) to [out=30,in=150] (22,6);
\draw[fill] (22,7) circle(.1);
\draw (22,5) to [out=30,in=30] (22,7);
\draw[fill] (22,8) circle(.1);
\draw (22,7) -- (22,8);
\draw[fill] (22,9) circle(.1);
\draw[fill] (22,10) circle(.1);
\draw (22,8) to [out=30,in=30] (22,10);
\draw (22,9) -- (22,10);
\draw[fill] (22,11) circle(.1);
\draw (22,9) to [out=150,in=150] (22,11);

\node [below] at (23,4) {$x_1y_1a$};
\draw[fill] (23,4) circle(.1);
\draw[fill] (23,5) circle(.1);
\draw (23,4) -- (23,5);
\draw[fill] (23,6) circle(.1);
\draw (23,4) to [out=150,in=150] (23,6);
\draw[fill] (23,7) circle (.1);
\draw (23,6) -- (23,7);
\draw[fill] (23,8) circle(.1);
\draw[fill] (23,9) circle(.1);
\draw (23,8) -- (23,9);
\draw (23,7) to [out=150,in=150] (23,9);
\draw[fill] (23,10) circle(.1);
\draw (23,8) to [out=30,in=30] (23,10);
\draw[fill] (23,11) circle(.1);
\draw (23,10) -- (23,11);
\draw[fill] (24,6) circle(.1);
\draw[fill] (24,7) circle(.1);
\draw (24,6) -- (24,7);
\draw (23,5) to [out=30,in=150] (24,7);
\draw[fill] (24,8) circle(.1);
\draw (24,6) to [out=30,in=30] (24,8);
\draw[fill] (24,9) circle(.1);
\draw (24,8) -- (24,9);
\draw[fill] (24,10) circle(.1);
\draw[fill] (24,11) circle(.1);
\draw (24,9) to [out=30,in=30] (24,11);
\draw (24,10) -- (24,11);
\draw[fill] (24,12) circle(.1);
\draw (24,10) to [out=150,in=150] (24,12);

\node [below] at (25,4) {$x_2y_2a$};
\draw[fill] (25,4) circle(.1);
\draw[fill] (25,5) circle(.1);
\draw (25,4) -- (25,5);
\draw[fill] (25,6) circle(.1);
\draw (25,4) to [out=150,in=150] (25,6);
\draw[fill] (25,7) circle (.1);
\draw (25,6) -- (25,7);
\draw[fill] (25,8) circle(.1);
\draw[fill] (25,9) circle(.1);
\draw (25,8) -- (25,9);
\draw (25,7) to [out=150,in=150] (25,9);
\draw[fill] (25,10) circle(.1);
\draw (25,8) to [out=30,in=30] (25,10);
\draw[fill] (25,11) circle(.1);
\draw (25,10) -- (25,11);
\draw[fill] (26,6) circle(.1);
\draw[fill] (26,7) circle(.1);
\draw (26,6) -- (26,7);
\draw (25,5) to [out=30,in=150] (26,7);
\draw[fill] (26,8) circle(.1);
\draw (26,6) to [out=30,in=30] (26,8);
\draw[fill] (26,9) circle(.1);
\draw (26,8) -- (26,9);
\draw[fill] (26,10) circle(.1);
\draw[fill] (26,11) circle(.1);
\draw (26,9) to [out=30,in=30] (26,11);
\draw (26,10) -- (26,11);
\draw[fill] (26,12) circle(.1);
\draw (26,10) to [out=150,in=150] (26,12);

\node [below] at (27,4) {$x_1y_2a$};
\draw[fill] (27,4) circle(.1);
\draw[fill] (27,5) circle(.1);
\draw (27,4) -- (27,5);
\draw[fill] (27,6) circle(.1);
\draw (27,4) to [out=150,in=150] (27,6);
\draw[fill] (27,7) circle (.1);
\draw (27,6) -- (27,7);
\draw[fill] (27,8) circle(.1);
\draw[fill] (27,9) circle(.1);
\draw (27,8) -- (27,9);
\draw (27,7) to [out=150,in=150] (27,9);
\draw[fill] (27,10) circle(.1);
\draw (27,8) to [out=30,in=30] (27,10);
\draw[fill] (27,11) circle(.1);
\draw (27,10) -- (27,11);
\draw[fill] (28,6) circle(.1);
\draw[fill] (28,7) circle(.1);
\draw (28,6) -- (28,7);
\draw (27,5) to [out=30,in=150] (28,7);
\draw[fill] (28,8) circle(.1);
\draw (28,6) to [out=30,in=30] (28,8);
\draw[fill] (28,9) circle(.1);
\draw (28,8) -- (28,9);
\draw[fill] (28,10) circle(.1);
\draw[fill] (28,11) circle(.1);
\draw (28,9) to [out=30,in=30] (28,11);
\draw (28,10) -- (28,11);
\draw[fill] (28,12) circle(.1);
\draw (28,10) to [out=150,in=150] (28,12);

\node [below] at (29,4) {$x_2y_1a$};
\draw[fill] (29,4) circle(.1);
\draw[fill] (29,5) circle(.1);
\draw (29,4) -- (29,5);
\draw[fill] (29,6) circle(.1);
\draw (29,4) to [out=150,in=150] (29,6);
\draw[fill] (29,7) circle (.1);
\draw (29,6) -- (29,7);
\draw[fill] (29,8) circle(.1);
\draw[fill] (29,9) circle(.1);
\draw (29,8) -- (29,9);
\draw (29,7) to [out=150,in=150] (29,9);
\draw[fill] (29,10) circle(.1);
\draw (29,8) to [out=30,in=30] (29,10);
\draw[fill] (29,11) circle(.1);
\draw (29,10) -- (29,11);
\draw[fill] (30,6) circle(.1);
\draw[fill] (30,7) circle(.1);
\draw (30,6) -- (30,7);
\draw (29,5) to [out=30,in=150] (30,7);
\draw[fill] (30,8) circle(.1);
\draw (30,6) to [out=30,in=30] (30,8);
\draw[fill] (30,9) circle(.1);
\draw (30,8) -- (30,9);
\draw[fill] (30,10) circle(.1);
\draw[fill] (30,11) circle(.1);
\draw (30,9) to [out=30,in=30] (30,11);
\draw (30,10) -- (30,11);
\draw[fill] (30,12) circle(.1);
\draw (30,10) to [out=150,in=150] (30,12);

\end{tikzpicture}
\end{center}
\caption{The $\A(1)$-module structure of $\H^*(B(\Z_2\times\Z_4^2))$}
\label{fig:A1-Z2Z4Z4}
\end{figure}

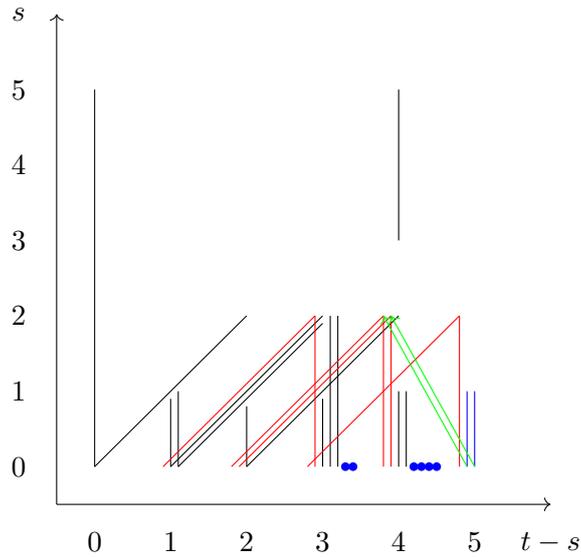
\begin{figure}[!h]
\begin{center}
\begin{tikzpicture}
\node at (0,-1) {0};
\node at (1,-1) {1};
\node at (2,-1) {2};
\node at (3,-1) {3};
\node at (4,-1) {4};
\node at (5,-1) {5};
\node at (6,-1) {$t-s$};
\node at (-1,0) {0};
\node at (-1,1) {1};
\node at (-1,2) {2};
\node at (-1,3) {3};
\node at (-1,4) {4};
\node at (-1,5) {5};
\node at (-1,6) {$s$};

\draw[->] (-0.5,-0.5) -- (-0.5,6);
\draw[->] (-0.5,-0.5) -- (6,-0.5);

\draw (0,0) -- (0,5);
\draw (0,0) -- (2,2);
\draw (1,0) --(1,0.9);
\draw (1,0) -- (3,2);
\draw (1.1,0) -- (1.1,1);
\draw (1.1,0) -- (3,1.9);
\draw (2,0) -- (2,0.8);
\draw (2,0) -- (4,2);
\draw (3,0) -- (3,0.9);
\draw (3.1,0) -- (3.1,2);
\draw (3.2,0) -- (3.2,2);
\draw (4,0) -- (4,1);
\draw (4.1,0) -- (4.1,1);
\draw (4,3) -- (4,5);
\draw[color=red] (0.9,0) -- (2.9,2);
\draw[color=red] (2.9,2) -- (2.9,0);
\draw[color=red] (1.8,0) -- (3.8,2);
\draw[color=red] (3.8,2) -- (3.8,0);
\draw[color=red] (1.9,0) -- (3.9,2);
\draw[color=red] (3.9,2) -- (3.9,0);
\draw[color=red] (2.8,0) -- (4.8,2);
\draw[color=red] (4.8,2) -- (4.8,0);
\draw[fill=blue,color=blue] (3.3,0) circle(0.05);
\draw[fill=blue,color=blue] (3.4,0) circle(0.05);

\draw[fill=blue,color=blue] (4.2,0) circle(0.05);
\draw[fill=blue,color=blue] (4.3,0) circle(0.05);
\draw[fill=blue,color=blue] (4.4,0) circle(0.05);
\draw[fill=blue,color=blue] (4.5,0) circle(0.05);
\draw[color=blue] (4.9,0) -- (4.9,1);
\draw[color=blue] (5,0) -- (5,1);

\draw[->,color=green] (4.9,0) -- (3.8,2);
\draw[->,color=green] (5,0) -- (3.9,2);
\end{tikzpicture}
\end{center}
\caption{The $E_2$ page of the Adams spectral sequence for $\Z_2\times\Z_4^2$}
\label{fig:E2-Z2Z4Z4}
\end{figure}

Hence we have the following 
\begin{theorem}
\begin{equation}
\begin{tabular}{c c c c}
\hline
$n$ & $\Omega^{Spin}_n(B(\Z_2\times\Z_4^2))$ & $H_n(B(\Z_2\times\Z_4^2),\Z)$ & 
$\text{fermionic SPTs classes}$
\\
\hline
0& $\Z$ & $\Z$ &  \\
1& $\Z_2^2\times\Z_4^2$& $\Z_2\times\Z_4^2$ & \\
2& $\Z_2^6\times\Z_4$& $\Z_2^2\times\Z_4$ & $\Z_2^5\times\Z_4$\\
3 & $\Z_2^{8}\times\Z_4\times\Z_8^3$& $\Z_2^4\times\Z_4^3$ & $\Z_2^{8}\times\Z_4\times\Z_8^3$\\
4 & $\Z\times\Z_2^6\times {\Z_4^4}$& $\Z_2^6\times\Z_4^2$ & $\Z_2^6\times {\Z_4^4}$\\ 
\hline
\end{tabular}
\end{equation}
\end{theorem}

\subsubsection{Bordism invariants}
The bordism groups are explicitly realized as follows:
\begin{equation}
\Omega_n^{Spin}(B(\Z_2\times\Z_4^2))=\{\text{spin $n$-manifolds }M^n\text{ with maps }g:M\to B\Z_2,f_{1,2}:M\to B\Z_4\}/\sim.
\end{equation}
The complete bordism invariant in dimension $n<5$ read as follows
\begin{equation}
\begin{array}{rcl}
(M^1,f_1,f_2,g) & \stackrel{~}{\longmapsto} & \doubleunderline{\eta(M^1)} \oplus \int_{M^1} g \oplus \int_{M^1} f_1\oplus \int_{M^1} f_2 \\
\\
(M^2,f_1,f_2,g) & \stackrel{~}{\longmapsto} & \doubleunderline{\mathrm{Arf}(M^2)} \oplus \doubleunderline{\tilde{q}(f^*_1(x))} \oplus \doubleunderline{\tilde{q}(f^*_2(x))} \oplus \doubleunderline{\tilde{q}(g)} \oplus \int_{M^2} g\,f_1^*(x)\oplus \int_{M^2} g\,f_2^*(x) \oplus \int_{M^2} f_1\,f_2 \\
\\
(M^3,f_1,f_2,g) &  \stackrel{~}{\longmapsto} & 
\int_{M^3}g\,f_1^*(x)\,f_2^*(x) \oplus \int_{M^3}g\,f_1^*(y)\oplus \int_{M^3}g\,f_2^*(y) \oplus 
\\ \\ & & \;
\doubleunderline{\mathrm{Arf}(\mathrm{PD}(f_1^*(x)))} \oplus \doubleunderline{\mathrm{Arf}(\mathrm{PD}(f^*_2(x)))} \oplus \doubleunderline{\tilde{q}_{\mathrm{PD}(f^*_1(x))}(g)}
\oplus 
\doubleunderline{\tilde{q}_{\mathrm{PD}(f^*_2(x))}(g)} \oplus 
\\ \\ & & \;
\doubleunderline{\tilde q_{\mathrm{PD}(f^*_1(x))}(f_2^*(x))}\oplus\int_{M^3}f_1\,\mathcal{B}f_2 \oplus \underline{\hat\gamma( f_1)} \oplus \underline{\hat{\gamma}(f_2)}\oplus \underline{\beta(g)} \\
\\
(M^4,f_1,f_2,g) & \stackrel{~}{\longmapsto} & \frac{p_1(TM^4)}{48}
\oplus \int_{M^4} g\,f_1^*(xy) \oplus \int_{M^4} g\,f_2^*(xy)
\oplus \int_{M^4} g\,f_1^*(x)f_2^*(y) \oplus \int_{M^4} g\,f_2^*(x)f_1^*(y)
\oplus
\\ \\ & & \;
\doubleunderline{\mathrm{Arf}(\mathrm{PD}(f^*_1(x))\cap \mathrm{PD}(f^*_2(x)))}
\oplus \int_{M^4} f_1f_2\mathcal{B}f_2 \oplus \int_{M^4} f_2f_1\mathcal{B}f_1 \oplus
\\ \\ & & \;
\doubleunderline{\tilde q_{\mathrm{PD}(f^*_1(x))\cap \mathrm{PD}(f^*_2(x))}(g)}
\oplus 
\underline{\delta_{\mathrm{PD}(f_1)}(g,g)}
\oplus \underline{\delta_{\mathrm{PD}(f_2)}(g,g)}
\end{array}
\end{equation}

\subsection{$\Omega^{Spin}_n(B\Z_4^3)$}
\label{sec:Z4Z4Z4}
\subsubsection{Computation}
As before, we can use the Adams spectral sequence 
\begin{multline}
  E_2^{s,t}=\Ext_{\A}^{s,t}(\H^*(MSpin\wedge(B\Z_4^3)_+),\Z_2)\Rightarrow\pi_{t-s}(MSpin\wedge(B\Z_4^3)_+)_2^{\wedge}=\Omega^{Spin}_{t-s}(B\Z_4^3).
\end{multline} 
and, for $t-s<8$, we can identify the $E_2$-page with 
\begin{equation}
 \Ext_{\A(1)}^{s,t}(\H^*(B\Z_4^3),\Z_2).
\end{equation} 
The mod 2 cohomology ring and the second page differential are determined by the following formulas. $\H^*(B\Z_4)=\Z_2[y]\otimes\Lambda(x)$ where $|x|=1$, $|y|=2$, $Sq^1(y)=Sq^1(x)=0$. $d_2(y_i)=x_ih_0^2$, $d_2(y_i\alpha)=x_iy_ih_0^3$ for $i=1,2,3$, $d_2(y_1x_2)=d_2(x_1y_2)=x_1x_2h_0^2$, $d_2(y_1y_2)=x_1y_2h_0^2+y_1x_2h_0^2$, $d_2(x_1y_1y_2)=x_1y_1x_2h_0^2$, $d_2(y_1x_2y_2)=x_1x_2y_2h_0^2$.
$d_2(x_1y_2x_3)=d_2(x_2y_1x_3)=d_2(x_1x_2y_3)=x_1x_2x_3h_0^2$, $d_2(y_1y_2x_3)=x_1y_2x_3h_0^2+x_2y_1x_3h_0^2$, $d_2(y_1y_3x_2)=x_1x_2y_3h_0^2+x_3y_1x_2h_0^2$, $d_2(y_2y_3x_1)=x_3y_2x_1h_0^2+x_2y_3x_1h_0^2$.

The $\A(1)$-module structure of $\H^*(B\Z_4^3)$ is depicted in Figure 
\ref{fig:A1-Z4Z4Z4}.

\begin{figure}[!h]
\begin{center}
\begin{tikzpicture}[scale=0.4]
\node [below] at (0,0) {1};
\draw[fill] (0,0) circle(.1);

\node [below] at (0,1) {$x_2$};
\draw[fill] (0,1) circle(.1);

\node [below] at (-1,1) {$x_1$};
\draw[fill] (-1,1) circle(.1);

\node [below] at (1,1) {$x_3$};
\draw[fill] (1,1) circle(.1);

\node [below] at (0,2) {$x_1x_3$};
\draw[fill] (0,2) circle(.1);

\node [left] at (-1,2) {$x_1x_2$};
\draw[fill] (-1,2) circle(.1);

\node [above] at (1,2) {$x_2x_3$};
\draw[fill] (1,2) circle(.1);

\node [above] at (0,3) {$x_1x_2x_3$};
\draw[fill] (0,3) circle(.1);

\node [below] at (2,2) {$y_1$};
\draw[fill] (2,2) circle(.1);
\draw[fill] (2,4) circle(.1);
\draw (2,2) to [out=150,in=150] (2,4);

\node [below] at (3,2) {$y_2$};
\draw[fill] (3,2) circle(.1);
\draw[fill] (3,4) circle(.1);
\draw (3,2) to [out=150,in=150] (3,4);

\node [below] at (4,2) {$y_3$};
\draw[fill] (4,2) circle(.1);
\draw[fill] (4,4) circle(.1);
\draw (4,2) to [out=150,in=150] (4,4);

\node [below] at (5,3) {$x_1y_1$};
\draw[fill] (5,3) circle(.1);
\draw[fill] (5,5) circle(.1);
\draw (5,3) to [out=150,in=150] (5,5);

\node [below] at (6,2) {$x_2y_2$};
\draw[->][color=red] (6,2) -- (6,3);
\draw[fill] (6,3) circle(.1);
\draw[fill] (6,5) circle(.1);
\draw (6,3) to [out=150,in=150] (6,5);

\node [below] at (7,3) {$x_3y_3$};
\draw[fill] (7,3) circle(.1);
\draw[fill] (7,5) circle(.1);
\draw (7,3) to [out=150,in=150] (7,5);

\node [below] at (8,2) {$x_1y_2$};
\draw[->][color=red] (8,2) -- (8,3);
\draw[fill] (8,3) circle(.1);
\draw[fill] (8,5) circle(.1);
\draw (8,3) to [out=150,in=150] (8,5);

\node [below] at (9,3) {$x_1y_3$};
\draw[fill] (9,3) circle(.1);
\draw[fill] (9,5) circle(.1);
\draw (9,3) to [out=150,in=150] (9,5);

\node [below] at (10,2) {$x_2y_1$};
\draw[->][color=red] (10,2) -- (10,3);
\draw[fill] (10,3) circle(.1);
\draw[fill] (10,5) circle(.1);
\draw (10,3) to [out=150,in=150] (10,5);

\node [below] at (11,3) {$x_2y_3$};
\draw[fill] (11,3) circle(.1);
\draw[fill] (11,5) circle(.1);
\draw (11,3) to [out=150,in=150] (11,5);

\node [below] at (12,2) {$x_3y_1$};
\draw[->][color=red] (12,2) -- (12,3);
\draw[fill] (12,3) circle(.1);
\draw[fill] (12,5) circle(.1);
\draw (12,3) to [out=150,in=150] (12,5);

\node [below] at (13,3) {$x_3y_2$};
\draw[fill] (13,3) circle(.1);
\draw[fill] (13,5) circle(.1);
\draw (13,3) to [out=150,in=150] (13,5);

\node [below] at (14,4) {$y_1y_2$};
\draw[fill] (14,4) circle(.1);
\draw[fill] (14,6) circle(.1);
\draw (14,4) to [out=150,in=150] (14,6);

\node [below] at (15,3) {$y_1y_3$};
\draw[->][color=red] (15,3) -- (15,4);
\draw[fill] (15,4) circle(.1);
\draw[fill] (15,6) circle(.1);
\draw (15,4) to [out=150,in=150] (15,6);

\node [below] at (16,4) {$y_2y_3$};
\draw[fill] (16,4) circle(.1);
\draw[fill] (16,6) circle(.1);
\draw (16,4) to [out=150,in=150] (16,6);

\node [below] at (17,2) {$x_1x_2y_1$};
\draw[->][color=red] (17,2) -- (17,4);
\draw[fill] (17,4) circle(.1);
\draw[fill] (17,6) circle(.1);
\draw (17,4) to [out=150,in=150] (17,6);

\node [below] at (18,3) {$x_1x_2y_2$};
\draw[->][color=red] (18,3) -- (18,4);
\draw[fill] (18,4) circle(.1);
\draw[fill] (18,6) circle(.1);
\draw (18,4) to [out=150,in=150] (18,6);

\node [below] at (19,4) {$x_1y_1x_3$};
\draw[fill] (19,4) circle(.1);
\draw[fill] (19,6) circle(.1);
\draw (19,4) to [out=150,in=150] (19,6);

\node [below] at (20,2) {$x_2y_2x_3$};
\draw[->][color=red] (20,2) -- (20,4);
\draw[fill] (20,4) circle(.1);
\draw[fill] (20,6) circle(.1);
\draw (20,4) to [out=150,in=150] (20,6);

\node [below] at (21,3) {$x_1y_2x_3$};
\draw[->][color=red] (21,3) -- (21,4);
\draw[fill] (21,4) circle(.1);
\draw[fill] (21,6) circle(.1);
\draw (21,4) to [out=150,in=150] (21,6);

\node [below] at (22,4) {$x_2y_1x_3$};
\draw[fill] (22,4) circle(.1);
\draw[fill] (22,6) circle(.1);
\draw (22,4) to [out=150,in=150] (22,6);

\node [below] at (23,2) {$x_1x_2y_3$};
\draw[->][color=red] (23,2) -- (23,4);
\draw[fill] (23,4) circle(.1);
\draw[fill] (23,6) circle(.1);
\draw (23,4) to [out=150,in=150] (23,6);

\node [below] at (24,3) {$x_1x_3y_3$};
\draw[->][color=red] (24,3) -- (24,4);
\draw[fill] (24,4) circle(.1);
\draw[fill] (24,6) circle(.1);
\draw (24,4) to [out=150,in=150] (24,6);

\node [below] at (25,4) {$x_2x_3y_3$};
\draw[fill] (25,4) circle(.1);
\draw[fill] (25,6) circle(.1);
\draw (25,4) to [out=150,in=150] (25,6);

\node [below] at (26,5) {$x_1y_1y_2$};
\draw[fill] (26,5) circle(.1);
\draw[fill] (26,7) circle(.1);
\draw (26,5) to [out=150,in=150] (26,7);

\node [below] at (27,2) {$x_2y_1y_2$};
\draw[->][color=red] (27,2) -- (27,5);
\draw[fill] (27,5) circle(.1);
\draw[fill] (27,7) circle(.1);
\draw (27,5) to [out=150,in=150] (27,7);

\node [below] at (28,3) {$y_1y_2x_3$};
\draw[->][color=red] (28,3) -- (28,5);
\draw[fill] (28,5) circle(.1);
\draw[fill] (28,7) circle(.1);
\draw (28,5) to [out=150,in=150] (28,7);

\node [below] at (29,4) {$x_1x_2x_3y_1$};
\draw[->][color=red] (29,4) -- (29,5);
\draw[fill] (29,5) circle(.1);
\draw[fill] (29,7) circle(.1);
\draw (29,5) to [out=150,in=150] (29,7);

\node [below] at (30,5) {$x_1x_2x_3y_2$};
\draw[fill] (30,5) circle(.1);
\draw[fill] (30,7) circle(.1);
\draw (30,5) to [out=150,in=150] (30,7);

\node [below] at (31,2) {$x_1y_1y_3$};
\draw[->][color=red] (31,2) -- (31,5);
\draw[fill] (31,5) circle(.1);
\draw[fill] (31,7) circle(.1);
\draw (31,5) to [out=150,in=150] (31,7);

\node [below] at (32,3) {$x_2y_2y_3$};
\draw[->][color=red] (32,3) -- (32,5);
\draw[fill] (32,5) circle(.1);
\draw[fill] (32,7) circle(.1);
\draw (32,5) to [out=150,in=150] (32,7);

\node [below] at (33,4) {$x_1y_2y_3$};
\draw[->][color=red] (33,4) -- (33,5);
\draw[fill] (33,5) circle(.1);
\draw[fill] (33,7) circle(.1);
\draw (33,5) to [out=150,in=150] (33,7);

\node [below] at (34,5) {$x_2y_1y_3$};
\draw[fill] (34,5) circle(.1);
\draw[fill] (34,7) circle(.1);
\draw (34,5) to [out=150,in=150] (34,7);

\node [below] at (35,2) {$x_3y_1y_3$};
\draw[->][color=red] (35,2) -- (35,5);
\draw[fill] (35,5) circle(.1);
\draw[fill] (35,7) circle(.1);
\draw (35,5) to [out=150,in=150] (35,7);

\node [below] at (36,3) {$x_3y_2y_3$};
\draw[->][color=red] (36,3) -- (36,5);
\draw[fill] (36,5) circle(.1);
\draw[fill] (36,7) circle(.1);
\draw (36,5) to [out=150,in=150] (36,7);

\node [below] at (37,4) {$x_1x_2x_3y_3$};
\draw[->][color=red] (37,4) -- (37,5);
\draw[fill] (37,5) circle(.1);
\draw[fill] (37,7) circle(.1);
\draw (37,5) to [out=150,in=150] (37,7);

\end{tikzpicture}
\end{center}
\caption{The $\A(1)$-module structure of $\H^*(B\Z_4^3)$}
\label{fig:A1-Z4Z4Z4}
\end{figure}
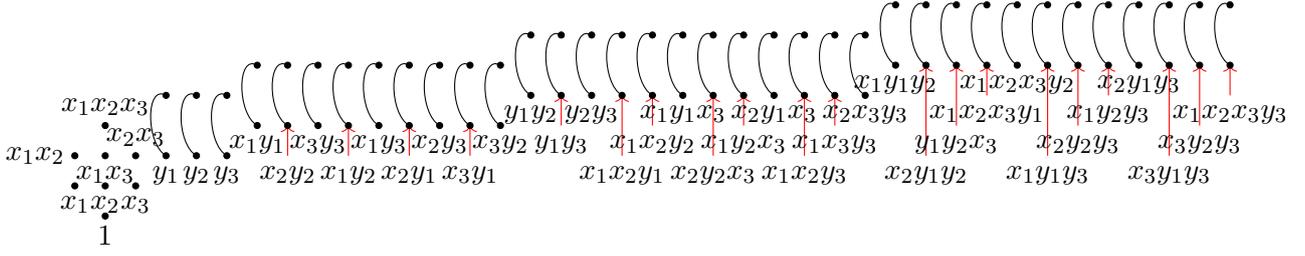

The calculation and the result are similar to the previous ones. Here we only present the result in dimension 4:
\begin{theorem}
\begin{equation}
 \Omega^{Spin}_4(B\Z_4^3)=\Z\times\Z_2^4\times\Z_4^8.
\end{equation} 
\end{theorem}
which can be compared with
\begin{equation}
 H_4(B\Z_4^3,\Z)=\Z_4^8.
\end{equation} 

\subsubsection{Bordism invariants}
\begin{equation}
\Omega_n^{Spin}(B(\Z_4^3))=\{\text{spin $n$-manifolds }M^n\text{ with maps }f_{1,2,3}:M\to B\Z_4\}/\sim.
\end{equation}
\begin{equation}
\begin{array}{rcl}
(M^4,f_1,f_2,f_3) & \stackrel{~}{\longmapsto} & \frac{p_1(TM^4)}{48} \oplus 
\int_{M^4} f_1f_2\mathcal{B}f_2 \oplus 
\int_{M^4} f_1f_3\mathcal{B}f_3 \oplus
\int_{M^4} f_2f_1\mathcal{B}f_1 \oplus 
\int_{M^4} f_2f_3\mathcal{B}f_3 \oplus
\\ \\ & & \;
\int_{M^4} f_3f_1\mathcal{B}f_1 \oplus 
\int_{M^4} f_3f_2\mathcal{B}f_2 \oplus
\int_{M^4} f_1f_2\mathcal{B}f_3 \oplus 
\int_{M^4} f_1f_3\mathcal{B}f_2 \oplus
\\ \\ & & \;
\doubleunderline{\mathrm{Arf}(\mathrm{PD}(f^*_1(x))\cap \mathrm{PD}(f^*_2(x)))}
\oplus 
\doubleunderline{\mathrm{Arf}(\mathrm{PD}(f^*_2(x))\cap \mathrm{PD}(f^*_3(x)))}
\oplus
\\ \\ & & \;
\doubleunderline{\mathrm{Arf}(\mathrm{PD}(f^*_1(x))\cap \mathrm{PD}(f^*_3(x)))}
\oplus 
\doubleunderline{\eta(\mathrm{PD}(f^*_1(x))\cap \mathrm{PD}(f^*_2(x))\cap \mathrm{PD}(f^*_3(x)))}
\end{array}
\end{equation}
Note that the last invariant can be equivalently written in a following not explicitly symmetric way: $\tilde{q}_{\mathrm{PD}(f^*_1(x))\cap \mathrm{PD}(f^*_2(x))}(f^*_3(x))$.

\subsection{$\Omega_n^{Spin}(BG)$ for general finite abelian $G$.}
We do not present a calculation for general finite abelian $G$, however the result is expected to be qualitatively similar. In particular, the odd torsion part will coincide with the odd torsion part the oriented bordism group $\Omega_n^{SO}(BG)$ and will not provide any fermionic spin-TQFTs (that is, TQFTs depending non-trivially on spin-structure). Moreover, for $n\leq 4$, $\Tor\Omega_n^{SO}(BG)=H_n(BG,\Z)$. Therefore one can restrict to the case of $G=\prod_i\Z_{2^{m_i}}$. Therefore, the result is expected to be of the similar form as in the examples considered above. 

\section{Fermionic topological invariants of links and surface links}
\label{sec:links}
An $n$-dimensional TQFT, when evaluated on a closed $n$-manifold gives a numerical invariant of this manifold, valued in\footnote{As $\Hom(\C,\C)\cong \C$}  $\C$. The TQFT structure also provides invariants of embedded submanifolds with respect to ambient isotopy. Namely, consider a possibly disjoint closed oriented $k$-manifold $Y^k$ embedded via a map $\iota$ into a $n$-manifold $M^n$.  Denote the image $\iota(Y^k)=L$ and its tabular neighborhood as $\mathcal{N}(L)$. The complement $M^n\setminus \mathcal{N}(L)$ then can be considered as a bordism from the empty manifold $\emptyset$ to $Y^k\times S^{n-k-1}$. The value of a TQFT $Z$ on it then can be considered as an element in the complex vector space\footnote{As $\Hom(\C,Z(Y^k\times S^{n-k-1}))\cong Z(Y^k\times S^{n-k-1})$.} $Z(Y^k\times S^{n-k-1})$. If one fixes a basis in this vector space, for each basis element there is a numeric invariant of $\iota:Y^k\hookrightarrow M^n$ under ambient isotopy.

A very well known example of this construction is Witten-Reshetikhin-Turaev 3d TQFTs which provide ``quantum'' invariants of both closed 3-manifolds and links. Similarly, the 3d and 4d spin-TQFTs $Z^{\mu}_\text{gauged}$ considered in this article give rise to invariants of links in 3-manifolds and surface links in 4-manifolds respectively. 

We will restrict our attention on the codimension 2 oriented submanifolds in $S^n$ where $n=3$ or $n=4$. In this case the normal bundle $NL$ to $L=\iota(Y^{n-2})$ is always trivial. This follows, for example from the triviality of the Euler class (see e.g. Corollary 11.4 in \cite{milnor2016characteristic}). Let us fix the framing of $L=\iota(Y^{n-2})$, that is a trivialization of the normal bundle. Since the normal bundle is two-dimensional, the choice of the framing is equivalent to a choice of non-vanishing section of the normal bundle (up to homotopy), that is a choice of normal vector at each point of $L$. Without loss of generality the end of the vector can lie on the boundary of the tabular neighborhood $\mathcal{N}$. The framing thus fixes a homeomorphism $Y^{n-2}\times S^1\cong \partial(S^n\setminus \mathcal{N}(L))$ by mapping a fixed point on $S^1$ to the end of the framing vector. The framing also provides a spin-structure on $Y^{n-2}\times S^1$ induced from the unique spin-structure on $S^n$. Namely, the trivialization of of normal bundle $NL$ together with spin structure on $S^n$, via the decomposition $NL\oplus TL = TS^n$, fixes a spin structure on $TL\stackrel{d\iota}\cong TY^{n-2}$ \cite{kirby1990pin}. Spin structure induced from $S^n$ on the $S^1$ factor in $S^1\times Y^{n-1}$ is even (i.e.\ bounding), since it is identified with a circle bounding a small disk surrounding $Y^{n-2}$. Together they give a product spin structure on $Y^{n-2}\times S^1$. Thus a framed oriented codimension 2 submanifold (possible disconnected) $L$ embedded in $S^n$ gives a well defined element of $[S^n\setminus \mathcal{N}(L)]\in \Hom(\emptyset,Y^{n-1}\times S^1)$ in the spin-bordism category $\mathrm{Bord}_{n}^{Spin}$ and one can consider the evaluation of the spin-TQFT functor on it:
\begin{equation}
Z^\mu_\text{gauged}(S^n\setminus \mathcal{N}(L)) \in Z^\mu_\text{gauged}(Y^{n-2}\times S^1)
\label{link-bordism}
\end{equation} 
for different values of $\mu\in \Hom(\Omega^{Spin}_n(BG),U(1))$.

There is always distinguished choice of the ``zero'' framing on each connected component $L_i$ of $L=\sqcup_i L_i$ in the following sense. The space of homotopy classes of framings on the normal bundle to $L_i$ is non-canonically isomorphic to $H^1(L_i,\Z)$. However there is a distinguished isomorphism for which the element $a\in H^1(L_i,\Z)$ corresponding to the given framing is determined by $a([\ell])=\ell k(\ell',L_i)$, where $\ell$ is a smooth curve on $L_i$ representing a class $[\ell]\in H_1(L_i,\Z)$ and $\ell'$ is its push-off in $S^n$ towards the framing vector. For framing corresponding to the zero element in $H^1(L_i,\Z)$ there is always exist oriented Seifert hyper-surface $V^{n-1}_i$ smoothly embedded in $S^n$ such that $\partial V^{n-2}_i=L_i$ and the framing vector field on $\Sigma$ is the normal vector to $L_i$ inside $V_i$ (see e.g. \cite{sato1984cobordisms}). For such choice of framing $L_i$, with the spin-structure induced from $S^n$, represents a trivial element in $\Omega_{n-2}^{Spin}(\text{pt})$.

Note that knowledge of invariants of (surface) links (\ref{link-bordism}) in $S^n$ allows easy calculation of the invariants $Z^\mu_\text{gauged}(M^n)$ on closed manifolds $M^n$ constructed via surgery, by using the functoriality of $Z^\mu_\text{gauged}$.

In what follows we consider a few examples of particular choices $\mu\in \Hom(\Omega^{Spin}_n(BG),U(1))$ in more detail, extending and clarifying some of the statements appeared in \cite{Putrov:2016qdo}. All other choices are analogous.  Note that the cases when $\mu$ is ``bosonic'' (i.e.\ belongs to the $H^n(BG,U(1))$ subgroup) were already considered in great detail in \cite{Putrov:2016qdo}. We provide a list of examples the invariant of (surface) links produced by 3- and 4-dimensional fermionic finite group gauge theories in Table \ref{table:link-sum}.}
\begin{table}[h!]
\footnotesize
    \hspace{-1.8em}\begin{tabular}{| c | c|c | c | c|c|c|}
    \hline
   Dim & $\begin{array}{c} \text{symmetry group} \\ G\times \Z_2^f\end{array}$ & $\begin{array}{c} \text{spin TQFTs}\\ 
   \text{from gauging fSPTs}:\\ \text{Action } \end{array}$  & Values &   $\begin{array}{c} \text{Link}\\ \text{invariants} \end{array}$  \\ \hline\hline
3d (2+1D) & $\Z_2 \times \Z_2^f $ & 
$\frac{\pi}{4}\beta(g)$ & $\Z_8$ &  $\begin{array}{c}
\text{Arf/ABK} \\ \text{ABK}(V_1) \end{array}$
\\ \hline   
3d (2+1D) & $ (\Z_2)^2 \times \Z_2^f$ &
$\frac{\pi}{2} \delta(g_1,g_2)$
 & $\Z_4$ & $\begin{array}{c} \text{unoriented} \\ \text{Sato-Levine}  \\ {q_{V_1}(V_1\cap V_2)}\end{array}$
 \\ \hline  \hline  
4d (3+1D) & $\Z_4\times \Z_2 \times \Z_2^f $ & 
$\frac{\pi}{2}\delta_{\text{PD}(f)}(g,g)$ 
& $\Z_4$ & $\text{ABK}(V_1\cap V_2) \mod 4$
\\ \hline   
4d (3+1D) & $(\Z_4)^2 \times \Z_2^f $  & 
$\pi \text{Arf} (\text{PD}(f_1\mod 2)\cap \text{PD}(f_2\mod 2))$ 
 & $\Z_2$ & $\begin{array}{c}
\text{Sato-Levine} \\ \text{Arf}(V_1\cap V_2) \end{array}$
\\ \hline
4d (3+1D) & $(\Z_4)^3 \times \Z_2^f $  & 
$\pi \eta(\text{PD}(f_1\mod 2)\cap \text{PD}(f_2\mod 2)\cap \text{PD}(f_3\mod 2))$ 
 & $\Z_2$ & $\eta(V_1\cap V_2 \cap V_3)$
\\
 \hline   
    \end{tabular}
\caption{
A list of some of the (surface) link invariants produced by fermionic finite group gauge theories, which is meant to be an update and improvement of the previous study in
Ref.~\cite{Putrov:2016qdo}'s Section 8 and its Table 3. Here, as in the main text, $V_i$ denote Seifert surfaces (volumes) of the (surface) link components. In the next subsections we elaborate on how the spin-structure is induced on their intersections.}
\label{table:link-sum}
\end{table}

\subsection{$G=\Z_2$, in $3$ dimensions}
\label{sec:5-1}

Consider a framed oriented knot\footnote{The case of the multi-component link is analogous.} $L$  in $M^3=S^3$, that is an embedding $\iota:S^1\hookrightarrow S^3$. As described above the unique spin-structure on $S^3$ together with framing defines a spin structure on  $T^2\cong \partial (S^3\setminus\mathcal{N}(L))$. The framings of $L\hookrightarrow S^3$ are in one-to-one correspondence with integers numbers. The correspondence is given by the self-linking number of $L$:
\begin{equation}
\ell k(L,L):=\ell k (L,L') \;\;\in \Z
\end{equation} 
where $L'$ is a push-off of $L$ towards the framing vector. The framing with $\ell k(L,L)=0$ is often called Seifert framing. In this case the framing vector can be realized as a normal vector to $L$ pointing inwards an \textit{oriented} Seifert surface (which always exist) $\Sigma$ smoothly embedded in  $S^3$ such that $\partial \Sigma = L$. The spin structure induced on $L\cong S^1$ is then even, since $L$ is the spin-boundary of oriented $\Sigma$ which has a spin-structure canonically induced from $S^3$. Any non-zero even framing (i.e.\ $\ell k(L,L)\in 2\Z$) can be realized by taking the normal vector pointing inwards an \textit{unorientable} surface $\Sigma$, such that $\partial\Sigma = L$. Such surfaces can be realized by taking band-connected sums of an orientable Seifert surfaces with a M\"obius band. Each such sum changes framing by $2$. Any even framing than induces an even spin structure on $L\cong S^1$ and, as a spin-manifold $\partial (S^3\setminus\mathcal{N}(L))=T^2_{++}\cong S^1_+\times S^1_+$ where two circles are identified with the meridian and longitude cycles\footnote{Meridian cycle is a small cycle surrounding the knot and longitude is the cycle given by the push-off $L'$ towards the framing vector.}. For odd framings Seifert surface, orientable or not, that is compatible with the framing does not exist. The spin structure on the knot complement is then $\partial (S^3\setminus\mathcal{N}(L))\cong S^1_+\times S^1_-$. This follows, for example, from the realization of spin-structures on $T^2$ as quadratic forms on $H_1(T^2,\Z_2)$ and the fact that changing framing by one corresponds to the action
\begin{equation}
\begin{array}{rcl}
H_1(T^2,\Z_2) & \longrightarrow & H_1(T^2,\Z_2), \\
{[m]} & \longmapsto & {[m]}, \\
{[\ell]} & \longmapsto & {[\ell]+[m]}. \\
\end{array}
\end{equation} 
where $[m]$ and $[\ell]$ are representatives of the median and longitude of $L$ respectively.

Let us take $\mu$ to be the generator of $\Hom (\Omega_3^{Spin}(B\Z_2),U(1))\cong \Z_8$:
\begin{equation}
\begin{array}{crcl}
\mu: & \Omega_3^{Spin}(BG) & \longrightarrow & U(1), \\
& [M^3,f] & \longmapsto & \exp{\frac{\pi i}{4} \beta(f)} \equiv \exp{\frac{\pi i}{4}\mathrm{ABK}[\text{PD}(f)]}.
\end{array}
\label{ABK-TQFT-closed}
\end{equation}

Let us choose an even framing. The value of the spin-TQFT functor on $T^2_{++}$ is the following complex 3-dimensional space (see general formula (\ref{gauged-hilbert-space}) and the calculation in Section 5.2 of \cite{Wang:2018edf})
\begin{equation}
Z^\mu_{\text{gauged}}(T^2_{++}) = Z^\mu(T^2_{++},0) \oplus Z^\mu(T^2_{++},[\ell]) \oplus Z^\mu(T^2_{++},[m])
\label{Z2-2-torus-hilbert}
\end{equation} 
where we specify the elements $f\in H^1(T^2_{++},\Z_2)$ (i.e.\ the homotopy class of the map $f:T^2_{++} \rightarrow B\Z_2$) by their Poincar\'e duals in $H_1(T^2_{++},\Z_2)$. Since $Z^\mu$ is an invertible TQFT, each component in the sum is a one-dimensional vector space. However, in order to obtain numerical invariants of knots one still has to fix a basis, that is provide unit maps
\begin{equation}
\C \longrightarrow  Z^\mu(T^2_{++},f)
\end{equation} 
for each component in (\ref{Z2-2-torus-hilbert}). This can be realized by using the value of the TQFT on the complement of an unknot $L=U$ with trivial framing (that is a solid torus $D^2\times S^1$):
\begin{equation}
Z_\text{gauged}^\mu(S^3\setminus \mathcal{N}(U))\;=\; 1\oplus 1 \oplus 0
\end{equation} 
Note that the value in $Z^\mu(T^2_{++},[m])$ vanishes because the  map $f=[m]:T^2\rightarrow B\Z_2$ cannot be extended to interior of the solid torus and the corresponding term in (\ref{gauged-bordism-value}) is absent. The choice of the basis in $Z^\mu(T^2_{++},[m])$ can be done by swapping meridian and the longitude. 

On closed 3-manifolds the value of the invertible $\Z_2$-equivariant TQFT $Z^\mu$ is given by the right hand side of (\ref{ABK-TQFT-closed}). It can be extended to a general element of the bordism $(M^3,g)\in \Hom((M_1^{2},f_1),(M_2^{2},f_2))$ in $\mathrm{Bord}^{Spin}_3(B\Z_2)$ as follows\footnote{The identification of the map between one-dimensional complex spaces with $\C$ itself requires a choice of basis for each $Z^\mu(M^2,f)$, that is a linear map $\C\rightarrow Z^\mu(M^2,f)$. Due to the monoidal property $Z^\mu(M_1^2\sqcup M^2_2,f_1\sqcup f_2)=\Z^\mu(M_1^2,f_1)\otimes Z^\mu(M_2^2,f_2)$ and existence of a canonical bordism between disjoint union and connected sum, it is sufficient to consider only for genus one Riemann surfaces. For the case when $M^2=S^1_+\times S^1_+$ and $f$ is such that $(M^2,f)$ represents a zero class in $\Omega_2^{Spin}(B\Z_2)\cong \Z_2^2$ the basis has been fixed above. All other null-bordant pairs $(M^2,f)$ can be related by the mapping class group action on $T^2$. For pairs $(M^2,f)$ which represent non-trivial elements in $\Omega_2^{Spin}(B\Z_2)$ one can first obtain a map $\C\rightarrow Z^\mu(M^2,f)\otimes Z^\mu(M^2,f)$ by cutting a 3-torus $T^3$ in half. This fixes a basis in $Z^\mu(M^2,f)$ up to a sign. Since such $(M,f)$ only appear in pairs in the boundary of the bordism $(M^3,g)$, different choices do not affect the choice of the isomorphism $Z^\mu(M^3,g)\cong \C$.}:
\begin{equation}
Z^\mu(M^3,g)=\exp{\frac{\pi i}{4}\mathrm{ABK}[\text{PD}(g)]}:
\qquad Z^\mu(M_1^2,f_1)\rightarrow Z^\mu(M_2^2,f_2)
\label{Inv-3d-Z2-functor}
\end{equation}
where now $\text{PD}(g)$ is a smooth surface (in general non-orientable and with boundary) inside $M^3$ representing an element in the relative homology $H_2(M^3,M_1^2 \sqcup M_2^2;\Z_2)$ Poincar\'e dual to $g\in H^1(M^3,\Z_2)$. The condition $g|_{M_i^2}=f_i \in H^1(M^2_i,\Z_2)$ is equivalent to the requirement that the boundary of $\text{PD}(g)$ is a collection of smooth curves on $M_1^2$ and $M_2^2$ representing Poincar\'e duals to $f_i$. The spin structure on $M^3$ induces a spin structure on $\text{PD}(g)$ analogously to the closed case discussed in Section \ref{sec:3man}. In the formula above $ABK[\Sigma]$ denotes the extension of the $\Z_8$ valued Arf-Brown-Kervaire invariant to pin$^-$ surface $\Sigma$ with boundary (see e.g. \cite{kirby2004local}). As in the case of closed $\Sigma$ discussed in Section \ref{sec:2man},  pin$^-$ structures are in one-to-one correspondence with quadratic enhancements $q:H_1(\Sigma,\Z_2)\cong H^1(\Sigma,\partial \Sigma; \Z_2)\rightarrow \Z_4$ of the intersection pairing. The ABK invariant is then defined by a formula similar to (\ref{ABK-definition}).  The functoriality property of (\ref{Inv-3d-Z2-functor}) then follows from the additivity of ABK invariant under gluing surfaces with pin$^-$ structure.

The value of the spin-TQFT on the complement of a general oriented knot $L$ with even framing inside $S^3$ is given by
\begin{equation}
Z_\text{gauged}^\mu(S^3\setminus \mathcal{N}(L))\;=\;
1\oplus e^{\frac{\pi i}{4}\text{ABK}(\Sigma)} \oplus 0
\;\;\in Z^\mu_{\text{gauged}}(T^2_{++})
\label{ABK-knot-complement}
\end{equation}
where $\Sigma$ is a possibly unorientable surface embedded in $S^3$ with $\partial \Sigma=L$ and inducing the given framing on $L$. We used the fact that, as in the case of unknot, there is a unique $g\in H^1(S^3\setminus \mathcal{N}(L),\Z_2)$ with a fixed $f=g|_{\partial (S^3\setminus \mathcal{N}(L))}\in H^1(T^2,\Z_2)$. The pin$^-$ structure on $\Sigma$ is induced from the unique spin-structure on $S^3$. The values of the corresponding quadratic enhancement $q_{\Sigma}:H_1(\Sigma,\Z_2)\rightarrow \Z_4$  geometrically can be realized as follows \cite{kirby2004local}. Take a smooth curve in $\Sigma$ that represents an element $\alpha \in H^1(\Sigma,\Z_2)$. Then $q(\alpha)$ is the number of half-twists mod 4 of the thin band embedded in $\Sigma$, that is a tabular neighborhood of the curve.

The mod 8 valued invariant of a knot $L$ in (\ref{ABK-knot-complement}) appeared in \cite{kirby2004local} and is simply related to more usual Arf invariant of $L$ (which is defined using an oriented Seifert surface $\Sigma$):
\begin{equation}
\mathrm{ABK}(L)=4\,\mathrm{Arf}(L_0)+\frac{\ell k(L,L)}{2}\mod 8
\end{equation} 
where $L_0$ is the same knot but with zero framing. The relation can be shown by changing $\Sigma$ to an orientable surface by taking a band-connected sum with appropriate number of M\"obius bands. A simple example of a knot $L_0$ with $\mathrm{Arf}(L_0)=1 \mod 2$ is given by the trefoil ($3_1$ in the classification table).

\subsection{$G=\Z_2\times \Z_2$, in $3$ dimensions 
}
\label{sec:5-2}

\begin{figure}[h!]
\centering
\includegraphics{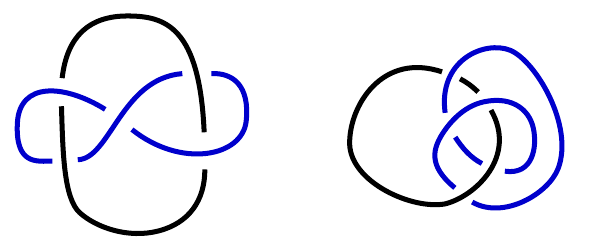}
\caption{Left: Whitehead link, a link with the vanishing linking number. The value of unoriented Sato-Levine invariant is $2\mod 4$. Note that the value of the ordinary Sato-Levine invariant is $1\in \pi_3(S^2)\cong \Z$. Right: A link with even intersection number with value $1\mod 4$ of unoriented Sato-Levine invariant.}
\label{fig:SL-3d-generators}
\end{figure}

Let us take $\mu$ to be a generator of a $\Z_4$ subgroup in $\Hom(\Omega_3^{Spin}(BG),U(1))$ (see Section \ref{sec:Z2n}):
\begin{equation}
\begin{array}{crcl}
\mu: & \Omega_3^{Spin}(BG) & \longrightarrow & U(1), \\
& [M^3,f_1,f_2] & \longmapsto & \exp{\frac{\pi i}{2} \delta(f_1,f_2)}.
\end{array}
\label{delta-TQFT-closed}
\end{equation}
In this case the interesting situation is when $L=L_1\sqcup L_2$ is a two component framed oriented link in $S^3$ such that each component has even framing. The case of larger number of components is similar. Denote by $\Sigma_{1,2}$ possibly unorientable surfaces such that $\partial \Sigma_i = L_i$ and they induce framings on $L_{1,2}$ respectively. We also assume that $\Sigma_1\cap L_2=\emptyset$, $\Sigma_2\cap L_1=\emptyset$ and $\Sigma_1$ intersect with $\Sigma_2$ transversely. Such surfaces exist if and only if the link is \textit{proper}, that  is $\ell k(L_1,L_2)=0\mod 2$ \cite{saito1993unoriented}. The unique spin structure on $S^3$ induces a pin$^-$ structure on each $\Sigma_i$ which is represented by a quadratic function $q_{\Sigma_i}:H_1(\Sigma_i,\Z_2)\rightarrow \Z_4$. The same reasoning as in the previous section can be used to show that:
\begin{equation}
Z^\mu_\text{gauged}(S^3\setminus \mathcal{N}(L))|_{Z^\mu(T^2_{++},[\ell]\oplus 0)\otimes Z^\mu(T^2_{++},0\oplus [\ell])}=
\left\{
\begin{array}{ll}
\exp{\frac{\pi i }{2}q_{\Sigma_1}([\Sigma_1\cap \Sigma_2])}, & L\text{ is proper}, \\
0,& \text{ otherwise}.
\end{array}
\right.
\end{equation}
where we only listed the projection on the functor on the one-dimensional subspace of $Z^\mu_\text{gauged}(T^2_{++}\sqcup T^2_{++})$ in the decomposition (\ref{gauged-hilbert-space}) that gives a non-trivial invariant of the link $L$. Geometrically $q_{\Sigma_1}([\Sigma_1\cap \Sigma_2])$ counts the number of half-twists of the thin band in $\Sigma_1$ containing the curve $\Sigma_1\cup \Sigma_2$. Since $\Sigma_1$ and $\Sigma_2$ intersect transversely, $q_{\Sigma_1}([\Sigma_1\cap \Sigma_2])=q_{\Sigma_2}([\Sigma_1\cup \Sigma_2])$. Such mod 4 valued invariant of proper links is known as unoriented Sato-Levine invariant \cite{saito1993unoriented}. The examples of 2-component proper links with non-trivial values of unoriented Sato-Levine invariant are shown in Figure \ref{fig:SL-3d-generators}.

\subsection{$G=\Z_4\times \Z_4$, in $4$ dimensions 
}

\label{sec:5-3}

\begin{figure}[h!]
\centering
\includegraphics{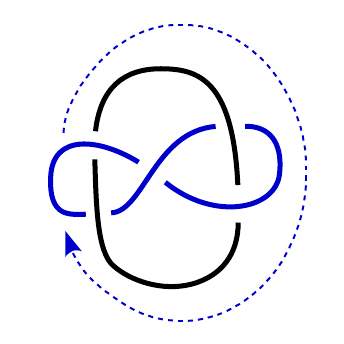}
\caption{A twisted spun of Whitehead link. This 2-component surface link $L\subset S^4$ can be realized inside (topologically trivially) embedded $D^3\times S^1\subset S^4$ where $D^3$ is a solid ball. For each $t\in [0,2\pi)$ parametrizing $S^1$, $L\cap (D^3\times \{t\})$ is a copy of a Whitehead link where one component is a fixed circle and the other component is rotated by angle $t$ around the axis orthogonal to that circle and passing through its center, as shown in the picture. This construction can be understood as 4-dimensional generalization of a braid representation of a link.}
\label{fig:whitehead-twisted-spun}
\end{figure}

Take $\mu$ to be a generator of a $\Z_2$ subgroup in $\Hom(\Omega_4^{Spin}(BG),U(1))$ (see Section \ref{sec:Z2n}):
\begin{equation}
\begin{array}{crcl}
\mu: & \Omega_3^{Spin}(BG) & \longrightarrow & U(1), \\
& [M^3,f_1,f_2] & \longmapsto & (-1)^{\mathrm{Arf}(\mathrm{PD}(f_1^*(x))\cap \mathrm{PD}(f_2^*(x)))}.
\end{array}
\end{equation}
Let $L=L_1\sqcup L_2$ be a two-component oriented surface-link in $S^4$, so that $L_i$ is the image of a closed oriented Riemann surface $\Sigma_i$ under the embedding map $\iota:\,Y^2\equiv \Sigma_1\sqcup \Sigma_2\hookrightarrow S^4$. Choose zero framing on both components $L_i$ as described in the beginning of the section. Then there exist three-dimensional Seifert volumes $V_i$ such that $\partial V_i =\Sigma_i$ and each $\Sigma_i$ with the induced spin structure is a spin-boundary. Moreover, let us assume that $L$ is \textit{semiboundary}, that is, by definition, there exist Seifert volumes such that $V_1\cap L_2=\emptyset$ and $V_2\cap L_1=\emptyset$ \cite{sato1984cobordisms}. One can choose $V_1,\,V_2$ to intersect transversally so that $V_1\cap V_2$ is an oriented Riemann surface. The normal bundle to $V_1\cap V_2$ in $S^4$ has a natural framing given by the two normal vectors pointing inward $V_i$. Given this framing, the spin structure on $S^4$ induces a spin structure on $V_1\cap V_2$.  Then, by an argument similar to the one in the three-dimensional case,
\begin{equation}
Z^\mu_\text{gauged}(S^4\setminus \mathcal{N}(L))|_{Z^\mu(\Sigma_1\times S^1_+,[\Sigma_1]\oplus 0)\otimes Z^\mu(\Sigma_2\times S^1_+,0\oplus [\Sigma_2])}=
(-1)^{\mathrm{Arf}(V_1\cap V_2)}, \qquad L\text{ is semi-boundary}.
\end{equation}
This $\Z_2$-valued invariant of a semi-boundary link is equal to the Sato-Levine invariant of a two-component semi-boundary surface link in $S^4$ \cite{sato1984cobordisms} (see also \cite{cochran1984invariant} for an alternative realization of the same invariant, very close to the one described here) which follows from the canonical isomorphism between $\Omega_2^{Spin}(\text{pt})\cong \Z_2$ and two-dimensional framed bordism group $\Omega_2^\text{fr}(\text{pt})\cong \pi_2^S$, which is in turn, by Pontryagin-Thom, isomorphic to the second stable homotopy group of spheres. An example of a surface link with non-trivial value of the invariant is shown in Figure \ref{fig:whitehead-twisted-spun}.

\subsection{$G=\Z_4\times \Z_4 \times \Z_4$, in $4$ dimensions 
}

\label{sec:5-4}

\begin{figure}[h!]
\centering
\includegraphics{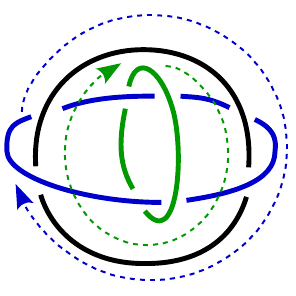}
\caption{A twisted spun of Borromean rings. This 3-component surface link $L\subset S^4$ can be realized inside (topologically trivially) embedded $D^3\times S^1\subset S^4$ where $D^3$ is a solid ball. For each $t\in [0,2\pi)$ parametrizing $S^1$, $L\cap (D^3\times \{t\})$ is a copy of a Borromean rings link where one component is a fixed circle and the other two components are rotated by angle $t$ around the axis orthogonal to that circle and passing through its center, as shown in the picture. This construction can be understood as 4-dimensional generalization of a braid representation of a link.}
\label{fig:borromean-rings-twisted-spun}
\end{figure}

Take $\mu$ to be a generator of a $\Z_2$ subgroup in $\Hom(\Omega_4^{Spin}(BG),U(1))$ (see Section \ref{sec:Z4Z4Z4}):
\begin{equation}
\begin{array}{crcl}
\mu: & \Omega_3^{Spin}(BG) & \longrightarrow & U(1), \\
& [M^3,f_1,f_2,f_3] & \longmapsto & (-1)^{\tilde{q}_{\mathrm{PD}(f^*_1(x_1))\cap \mathrm{PD}(f^*_2(x_2))}(f^*_3(x_3))}.
\end{array}
\end{equation}
Let $L=L_1\sqcup L_2\sqcup L_3$ be now a three-component oriented surface-link in $S^4$, where each $L_i$ is the image of a closed oriented Riemann surface $\Sigma_i$ under embedding map $\iota:\,Y^2\equiv \Sigma_1\sqcup \Sigma_2\sqcup \Sigma_3\hookrightarrow S^4$. Choose again zero framing on each $L_i$ and assume  that $L$ is \textit{semi-boundary}, that is, for each component $L_i$ there exist a Seifert volume $V_i$ such that it induces the framing and $V_i\cap(L\setminus L_i) =\emptyset$.  One can choose $V_1,\,V_2,\, V_3$ all intersect transversally so that $V_1\cap V_2$ is an oriented Riemann surface and $V_1\cap V_2\cap V_3$ is a smooth curve in it. As in the previous subsection, the spin-structure on $S^4$ induces a spin-structure on $V_1\cap V_2$ described by a certain quadratic form $q_{V_1\cap V_2}: H_1(V_1\cap V_2)\rightarrow \Z_2$. Moreover, there is a natural framing on the 1-manifold $V_1\cap V_2 \cap V_3$ embedded in $S^4$ given by the two three normal vectors pointing inward $V_i$. Given this framing, the spin structure on $S^4$ induces a spin structure on $V_1\cap V_2\cap V_3$.  Then
\begin{multline}
Z^\mu_\text{gauged}(S^4\setminus \mathcal{N}(L))|_{Z^\mu(\Sigma_1\times S^1_+,[\Sigma_1]\oplus 0 \oplus 0)\otimes Z^\mu(\Sigma_2\times S^1_+,0\oplus [\Sigma_2]\oplus 0)\otimes Z^\mu(\Sigma_3\times S^1_+,0\oplus 0 \oplus [\Sigma_3])}=\\
(-1)^{\tilde{q}_{V_1\cap V_2}(V_1\cap V_2\cap V_3)}\equiv (-1)^{\eta(V_1\cap V_2\cap V_3)}, \qquad L\text{ is semiboundary}.
\end{multline}
where $\eta$ is the invariant of spin 1-manifolds described in Section \ref{sec:1man}. Note that using the isomorphism $\Omega_1^{Spin}(\text{pt})\cong \Omega_1^\text{fr}(\text{pt})\cong \pi_1^S \Z_2$, one can define the value as the class of $V_1\cap V_1\cap V_3$ in $\Omega_1^\text{fr}(\text{pt})$, similarly to the original Sato-Levine invariant.  An example of a surface link with non-trivial value of the invariant is shown in Figure \ref{fig:borromean-rings-twisted-spun}.

\subsection{Categorification of invariants of links and 3-manifolds}
Before we proceed, let us briefly review the general definition of Sato-Levine invariants \cite{sato1984cobordisms} of higher dimensional links. Let $L^{n-2}=L_1^{n-2}\sqcup L_2^{n-2}$ be a pair of disjoint  oriented dimension $(n-2)$ manifolds embedded in $\mathbb{R}^n$ (one can replace it with $S^n$ by adding a point at infinity). Each $L_i^{n-2},\,i=1,2$ is not necessarily connected. There always exist Seifert volumes $V^{n-1}_i$ such that $\partial V^{n-1}_i=L_i$. By definition the link $L$ is \textit{semiboundary} if one can find $V_i$ such that $V_i\cap L_j=\emptyset,i\neq j$. Then the $n$-dimensional Sato-Levine invariant is defined as follows\footnote{The more usual notation is $\beta$, however we are already using this symbol for a different invariant.}:
\begin{equation}
	\begin{array}{rrcc}
	\text{SL}_n: & \frac{\left\{\begin{array}{c}
	\text{semiboundary}\\
	\text{links in $\mathbb{R}^n$}
	\end{array}
	\right\}}{\sim\text{ ambient isotopy}} & \longrightarrow & \pi_n(S^{2}), \\
	\\
	&
	L^{n-2}=L_1^{n-2} \sqcup L_2^{n-2} & \longmapsto &
	\begin{array}{c}
	[V_1^{n-2}\cap V_2^{n-2}] \\
	\text{ via Pontryagin-Thom}
	\end{array}
	\end{array}
\end{equation}
where, as in  Section \ref{sec:5-3}, we use vectors tangential to $V_i$ to define a framing on the normal bundle to $V_1^{n-1}\cap V_2^{n-1}$. Given this dimension $(n-2)$ submanifold in $\mathbb{R}^n$ with a framing of the normal bundle, Pontryagin-Thom construction\footnote{For completeness, let us remind it. Let $N^{n-m}$ be a codimension $n$ submanifold in $\mathbb{R}^n$ with a framing on the normal bundle. The corresponding continuous map $S^n\to S^{m}$ then can be explicitly constructed as follows. First let us identify the source $S^n$ with the ambient $\mathbb{R}^n$ with added point at infinity, and the target $S^n$ as $D^n/\partial D^n$ where $D^n$ is a unit ball. Pick a tabular neighborhood $\mathcal{T}(N^{n-m})\subset \mathbb{R}^n$ of $N^{n-m}$. The framing on the normal bundle then provides an explicit isomorphism $\mathcal{T}(N^{n-m})\cong N^{n-m}\times D^{m}$ where $D^{m}$ is the $m$-dimensional unit ball. The map $S^n\to S^m$ is then given by taking all the points outside of the tabular neighborhood to $\partial D^m$ (collapsed to a single point in $S^m$) and the points inside to be projected on $D^m$.} then provides an element in the $n$-homotopy group of $S^2$.

Similarly, one can define \textit{stable} Sato-Levine invariant:
\begin{equation}
	\widetilde{\text{SL}}_n: \;\frac{\left\{\begin{array}{c}
	\text{semiboundary}\\
	\text{links in $\mathbb{R}^n$}
	\end{array}
	\right\}}{\sim\text{ ambient isotopy}} \longrightarrow  \pi_{n-2}^s=\lim_{k\to\infty}\pi_{n+k}(S^{2+k}) 
\end{equation}
where we replace the ambient $\mathbb{R}^n$ with $\mathbb{R}^{n+k}$, and $L$ with $L\times \text{pt}$ and then take $k\to\infty$.

As was explained Section \ref{sec:5-3}, the 4d spin-TQFT obtained by gauging $G=\Z_4\times \Z_4$-equivariant invertible spin-TQFT with action $\text{Arf}(\text{PD}(f_1\mod 2)\cap \text{PD}(f_2\mod 2))$ provides a realization of $\widetilde{\text{SL}}_4$ invariant of surface links. By a similar argument one can show that 3d spin-TQFT obtained by gauging invertible $G$-equivariant spin-TQFT with the same $G$ and action $\eta(\text{PD}(f_1\mod 2)\cap \text{PD}(f_2\mod 2))$ 
(where $\eta$ is the invariant of $\Omega_1^{Spin}(\text{pt})\cong \Z_2$) provides a realization of $\widetilde{\text{SL}}_3$ invariant of surface links. In both cases one can see this from the fact that $\Omega^{Spin}_{n}(\text{pt})\cong \pi^s_{n}$ for $n\leq 2$. Namely, if the invariants $\widetilde{\text{SL}}_3$ are understood to be valued in $\{\pm 1\}$ (i.e.\ multiplicative realization of $\Z_2$):

\begin{equation}
	\widetilde{\text{SL}}_3(L_1^1\sqcup L_2^1) =(-1)^{\eta(V_1^2\cap V_2^2)},
\end{equation}
\begin{equation}
	\widetilde{\text{SL}}_4(L_1^1\sqcup L_2^1) =(-1)^{\text{Arf}(V_1^3\cap V_2^3)},
\end{equation}
where we used the notations for Seifert surfaces/volumes as before.

\begin{figure}[h!]
\centering
\includegraphics[scale=1.4]{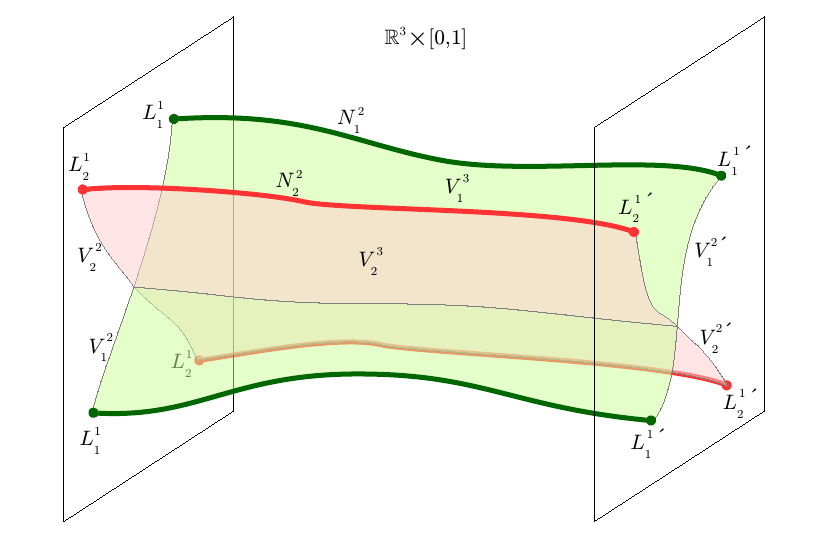}
\caption{A schematic picture of a morphism in the category $\text{LinkBord}_4^\text{sb}$ where, for visualization purpose, all dimensions are reduced by one.}
\label{fig:link-bord}
\end{figure}

Now we would like to claim that, in a certain sense, $\widetilde{\text{SL}}_4$ categorifies $\widetilde{\text{SL}}_3$ . In order to make this statement meaningful one has to extend $\widetilde{\text{SL}}_4$ from invariant of (semiboundary) surface links in $\mathbb{R}^4$ to the functor from the category of (semiboundary) link-bordisms to the category of complex $\Z_2$-vector spaces $\text{Vect}^{\Z_2}_\mathbb{C}$:
\begin{equation}
	\widetilde{\mathrm{SL}}_4:\;\text{LinkBord}_4^\text{sb}\rightarrow \text{Vect}^{\Z_2}_\mathbb{C}.
	\label{SL4-functor}
\end{equation}
The objects in the category $\text{LinkBord}_4^\text{sb}$ are semiboundary links in $\mathbb{R}^3$:
\begin{equation}
	\text{Ob}(\text{LinkBord}_4^\text{sb})=\left\{\begin{array}{c}
	\text{semiboundary}\\
	\text{links $L_1^{1}\sqcup L_2^{1}$ in $\mathbb{R}^3$}
	\end{array}
	\right\}
\end{equation}
The morphisms are pairs of 2-manifolds embedded in $\mathbb{R}^3\times [0,1]$ with boundaries coinciding with the links sitting in $\mathbb{R}^3\times \{0,1\}$ and satisfying semiboundary property (see Fig.~\ref{fig:link-bord}):
\begin{equation}
	\text{Hom}(L_1^{1}\sqcup L_2^{1},{L_1^{1}}'\sqcup {L_2^{1}}')=
	\left\{\begin{array}{c}
		\text{$N_1^{2}\sqcup N_2^{2}\subset \mathbb{R}^3\times [0,1]$, s.t.\ $\partial N_i^2\subset \mathbb{R}^3\times \{0,1\}$,}\\
		\;\partial N_i^2\cap \mathbb{R}^3 \times \{0\} =L_i^1,\;\;\partial N_i^2\cap \mathbb{R}^3 \times \{1\} ={L_i^1}'
		\\
		\exists V^3_i,\text{ s.t.\ }\partial V^3_i=N_i^2\cup V_i^2\cup {V_i^2}',\; V^3_i\cap N_j^2=\emptyset,\,i\neq j
		\end{array}
		\right\}/\sim\text{diffeo}.
		\label{link-bord-mor}
\end{equation}
The value of the functor (\ref{SL4-functor}) on objects is given by
\begin{equation}
 \widetilde{\text{SL}}_4(L_1^1\sqcup L_2^1)=
 \left\{
 \begin{array}{cl}
 \C[0],& \widetilde{\text{SL}}_3(L_1^1\sqcup L_2^1)=1,\\
 \C[1],& \widetilde{\text{SL}}_3(L_1^1\sqcup L_2^1)=-1,
 \end{array}
 \right.
 = Z^\text{Arf}(V^2_1\cap V^2_1)
 \label{SL4-Arf-obj}
\end{equation}
where $\C[n]$ denotes a one-dimensional complex vector space with $\Z_2$ grading $n$, and $Z^\text{Arf}$ is the non-trivial (fully extendable) invertible 2d spin-TQFT such that its value on a closed spin-surface $\Sigma$ is $Z^\text{Arf}(\Sigma)=(-1)^{\text{Arf}(\Sigma)}$. Such TQFT was considered in detail in \cite{Debray:2018wfz} (see also \cite{gunningham2016spin}). The value of $\widetilde{\text{SL}}_4$ on  morphisms (using the conventions in (\ref{link-bord-mor})) is
\begin{equation}
	\widetilde{\text{SL}}_4(N_1^2\sqcup N_2^2)=
	Z^{\text{Arf}}(V_1^3\cap V_2^3).
	\label{SL4-Arf-mor}
\end{equation}
where $V_1^3\cap V_2^3$ is a surface in $\mathbb{R}^3\times [0,1]$ with induced spin-structure and boundary components lying in $\mathbb{R}^3\times \{0\}$ and $\mathbb{R}^3\times \{1\}$. From the definition of semi-boundary link bordisms (\ref{link-bord-mor}) the r.h.s.\ of (\ref{SL4-Arf-mor}) indeed provides a $\Z_2$-linear map between the $\Z_2$-vector spaces associated to the objects via  (\ref{SL4-Arf-obj}) which is functorial. 

Then the functor $\widetilde{\text{SL}}_4$ categorifies link invariant $\widetilde{\text{SL}}_3$ in the sense that
\begin{equation}
	\widetilde{\text{SL}}_3(L_1^1\sqcup L_2^1)=
	\dim \widetilde{\text{SL}}_4^{0}(L_1^1\sqcup L_2^1)-\dim \widetilde{\text{SL}}_4^{1}(L_1^1\sqcup L_2^1)
\end{equation}
where, as before, the upper indices $0,1$ denote the components of the vector space with particular grading. This categorification of link invariants can be understood as a toy version of Khovanov homology \cite{khovanov1999categorification}, which categorifies Jones polynomial invariant of links (valued in $\Z[q,q^{-1}]$), and is also functorial with respect to link bordisms. Moreover, $\widetilde{\text{SL}}_3$ invariant has a close relation to the Jones polynomial invariant evaluated at $q=i$ (see e.g. \cite{kirby2004local})).

As was described above, $\widetilde{\text{SL}}_{3,4}$ can  be realized as $Z^{\mu_{3,4}}_\text{gauged}$ with $\mu_{3,4}$ being particular elements of $\text{Hom}(\text{Tor}\,\Omega^{Spin}_{3,4}(B\Z_4^2),U(1))$. Namely
\begin{equation}
	\begin{array}{c}
		\mu_4:=\text{Arf}(\text{PD}(f_1 \mod 2) \cap \text{PD}(f_2 \mod 2)),\\
		\mu_3:=\eta(\text{PD}(f_1 \mod 2) \cap \text{PD}(f_2 \mod 2)).
	\end{array}
\end{equation}
 It is natural to ask if $Z^{\mu_{4}}_\text{gauged}$ also categorifies $Z^{\mu_{3}}_\text{gauged}$. However, it is easy to see that categorification in the most naive sense, is not possible, because generically the value of $Z^{\mu_{3}}_\text{gauged}$ on a closed spin 3-manifold is rational (e.g. $Z^{\mu_3}_\text{gauged}(S^3)=1/16$), which cannot be interpreted as a (signed) sum of dimensions. That is, it is not possible to have a simple relation of the form $Z^{\mu_3}_\text{gauged}(\;\cdot\;)=Z^{4d}(\;\cdot \times S^1_\pm)$ for any 4d spin-TQFT $Z^{4d}$. But, as was shown in \cite{Wang:2018edf}, the 3d and 4d TQFTs $Z^{\mu_3}_\text{gauged}$ and $Z^{\mu_4}_\text{gauged}$ are still closely related. Namely:
\begin{equation}
	\begin{array}{rcl}
	Z^{\mu_4}_\text{gauged}(\;\cdot \times S^1_-) & = &
	(Z^{\mu_3}_\text{gauged})^{\oplus\,4}\oplus (Z^{\mu_3'}_\text{gauged})^{\oplus\,12}\\
	Z^{\mu_4}_\text{gauged}(\;\cdot \times S^1_+) & = &
	(Z^{0}_\text{gauged})^{\oplus\,4}\oplus (Z^{\mu_3+\mu_3'}_\text{gauged})^{\oplus\,12}
	\end{array}.
\end{equation}
where $\oplus$ is the direct sum operation\footnote{The crucial property of the direct sum operation on n-dimensional TQFTs is 
\begin{equation}
	(Z_1\oplus Z_2)(M^{n-1})=Z_1(M^{n-1})\oplus Z_2(M^{n-1}), \text{ if }\pi_0(M^{n-1})=1.
\end{equation}
The direct operation is then naturally extended to the TQFT values on disjoint $(n-1)$-dimensional manifolds and bordisms between them so that functoriality and symmetric monoidal property hold.
 } on TQFT functors and the summands are $\Z_4^2$ fermionic topological gauge theories corresponding to different elements of $\text{Hom}(\text{Tor}\,\Omega_3^{Spin}(B\Z_4^2),U(1))$. In particular, 
\begin{equation}
		\mu_3':=\text{Arf}(\text{PD}(f_1 \mod 2)).
\end{equation}

\section{Other bordism groups and computations: 
$\Omega_n^{\frac{Spin\times\Z_{2m}}{{\Z_2}}}$.}

\label{sec:spin-twisted-bordism}

Let $Spin(n)\times_{\Z_2}\Z_{2m}\equiv (Spin(n)\times\Z_{2m})/\Z_2$ where the quotient is with respect to the diagonal center $\Z_2$ subgroup. Similarly to spin manifolds, one can consider manifolds with $Spin \times_{\Z_2}\Z_{2m}$ tangential structure. That is, manifolds equipped with $H_n=Spin(n)\times_{\Z_2}\Z_{2m}$ principle bundle which is a lift of the $SO(n)$ orthonormal tangent frame bundle with respect to the extension $\Z_{2m}\rightarrow Spin(n)\times_{\Z_2}\Z_{2m}\rightarrow SO(n)$. 

The Pontryagin-Thom isomorphism provides a relation between the bordism groups of manifolds with (stable\footnote{In the cases when $H=Spin$ and $H=Spin\times_{\Z_2}\Z_{2m}$ the stable and unstable structures are equivalent, see \cite{kirby1990pin}.}) tangential structure $H$ and homotopy groups of the Madsen-Tillmann spectrum associated to tangential structure $H$:
\begin{equation}\label{MTH}
\Omega_n^H\equiv \Omega_n^H(\text{pt}) =\pi_nMTH\equiv\text{colim}_{k\to\infty}\pi_{n+k}MTH_k.
\end{equation}
On the other hand, in the work of Freed-Hopkins \cite{FH}, there is a 1:1 correspondence \footnote{Note that the free part of the classification contains Chern-Simons-like theories, which are not strictly topological.}
\begin{equation}
	\text{Tor}\,\left\{\begin{array}{ccc}\text{deformation classes of reflection positive}\\\text{invertible }n\text{-dimensional extended topological}\\\text{field theories with symmetry group }H_n\end{array}\right\}\cong[MTH,\Sigma^{n+1}I\Z]_{\text{tors}}.
\end{equation}
The abelian group $[MTH,\Sigma^{n+1}I\Z]$ is denoted by $TP_n(H)$ in \cite{FH}. In particular, $[MTH,\Sigma^{n+1}I\Z]_{\text{tors}}$ stands for the torsion part of homotopy classes of maps from spectrum $MTH$ to the $(n+1)$-th suspension of spectrum $I\Z$. The Anderson dual $I\Z$ is a spectrum that is the fiber of $I\C\to I\C^{\times}$
where $I\C(I\C^{\times})$ is the Brown-Comenetz dual spectrum defined by 
\begin{equation}
[X,I\C]=\Hom(\pi_0X,\C),
\end{equation}
\begin{equation}
[X,I\C^{\times}]=\Hom(\pi_0X,\C^{\times}).
\end{equation}
There is an exact sequence
\begin{equation}
0\to\Ext^1(\pi_nMTH,\Z)\to[MTH,\Sigma^{n+1}I\Z]\to\Hom(\pi_{n+1}MTH,\Z)\to0.
\end{equation}
The torsion part $[MTH,\Sigma^{n+1}I\Z]_{\text{tors}}$ is $\Ext^1((\pi_nMTH)_{\text{tors}},\Z)=\Hom((\pi_nMTH)_{\text{tors}},U(1))$.
This provides relation to the bordism groups in \eqref{MTH}.

Now let $H=Spin\times_{\Z_2}\Z_{2m}$.
Write $m=2^l\cdot n$ where $l\ge0$ and $k$ is odd, let $H'=Spin\times_{\Z_2}\Z_{2^{l+1}}$.
Then $H=H'\times\Z_k$ and $MTH=MTH'\wedge(B\Z_k)_+$, $\Omega_k^H=\Omega_k^{H'}(B\Z_k)$. The 2-torsion part of $\Omega_d^H$ is $\Omega_d^{H'}$ which is computed in \cite{campbell} for $l=1$, $d\le5$ and $l\ge2$, $d\le4$. The group $\Omega_5^H$ has been computed in \cite{hsieh}, but here we provide an alternative computation.

For $l=0$, $H'=Spin$, 
\begin{equation}
\Omega_d^{Spin}
=\left\{\begin{array}{llllll}\Z&d=0\\\Z_2&d=1\\\Z_2&d=2\\0&d=3\\\Z&d=4\\0&d=5\end{array}\right.,
\quad
\text{invertible fermionic TQFT classes}
=\left\{\begin{array}{llllll} \Z_2&d=1\\\Z_2&d=2\\ \Z&d=3\\ 0&d=4\\0&d=5\end{array}\right..
\end{equation}

For $l=1$, $H'=Spin\times_{\Z_2}\Z_4$,
\begin{equation}
\Omega_d^{Spin\times_{\Z_2}\Z_4}=\left\{\begin{array}{llllll}\Z&d=0\\\Z_4&d=1\\0&d=2\\0&d=3\\\Z&d=4\\\Z_{16}&d=5\end{array}\right.,
\quad
\text{fSPTs classes}=\left\{\begin{array}{llllll}0&d=2\\0&d=3\\ 0&d=4\\\Z_{16}&d=5\end{array}\right..
\end{equation}

For $l\ge2$, 
\begin{equation}\Omega_d^{Spin\times_{\Z_2}\Z_{2^{l+1}}}=\left\{\begin{array}{llllll}\Z&d=0\\\Z_{2^{l+1}}&d=1\\0&d=2\\\Z_{2^{l-1}}&d=3\\\Z&d=4\\\Z_{2^{l+3}}\times\Z_{2^{l-1}}&d=5\end{array}\right.,
\quad
\text{fSPTs classes}=\left\{\begin{array}{llllll}0&d=2\\\Z_{2^{l-1}}&d=3\\0 &d=4\\\Z_{2^{l+3}}\times\Z_{2^{l-1}}&d=5\end{array}\right..
\end{equation}
where the $d=5$ case is new.

Write $n=p_1^{k_1}\cdots p_r^{k_r}$ where $p_i$ are odd primes and $k_i\ge1$.
The $p_i$-torsion part of $\Omega_d^H$ is $\Omega_d^{SO}(B\Z_{p_i^{k_i}})_{p_i}^{\wedge}$.
\begin{equation}
\Ext_{\A_{p_i}}^{s,t}( H^*(MSO,\Z_{p_i})\otimes H^*(B\Z_{p_i^{k_i}},\Z_{p_i}),\Z_{p_i})\Rightarrow\Omega_{t-s}^{SO}(B\Z_{p_i^{k_i}})_{p_i}^{\wedge}
\end{equation}
where $\A_{p_i}$ is the mod $p_i$ Steenrod algebra.

\begin{equation}
 H^*(B\Z_{p_i^{k_i}},\Z_{p_i})=\Lambda_{\Z_{p_i}}(a)\otimes\Z_{p_i}[b]
\end{equation}
where $b=\beta a$, $\beta$ is the Bockstein homomorphism associated to $0\to\Z_{p_i}\to\Z_{p_i^2}\to\Z_{p_i}\to0$, $a\in H^1(B\Z_{p_i^{k_i}},\Z_{p_i})$, $b\in H^2(B\Z_{p_i^{k_i}},\Z_{p_i})$. The dual of $\A_{p_i}=H^*(H\Z_{p_i},\Z_{p_i})$ is 
\begin{equation}
\A_{p_i*}=H_*(H\Z_{p_i},\Z_{p_i})=\Lambda_{\Z_{p_i}}(\tau_0,\tau_1,\dots)\otimes\Z_{p_i}[\xi_1,\xi_2,\dots]
\end{equation}
where $\tau_j=(P^{p_i^{j-1}}\cdots P^{p_i}P^1\beta)^*$ and $\xi_j=(P^{p_i^{j-1}}\cdots P^{p_i}P^1)^*$, $P^i$ being standard generators of the mod $p_i>2$ Steenrod algebra. Let $C=\Z_{p_i}[\xi_1,\xi_2,\dots]\subseteq\A_{p_i*}$, then 
\begin{equation}
H_*(MSO,\Z_{p_i})=C\otimes\Z_{p_i}[z_1,z_2,\dots]
\end{equation}
where $|z_k|=4k$ for $k\ne\frac{p_i^t-1}{2}$. The cohomology is then
\begin{equation}
H^*(MSO,\Z_{p_i})=(\Z_{p_i}[z_1,z_2,\dots])^*\otimes C^*
\end{equation}
where $C^*=\A_{p_i}/(\beta)$ and $(\beta)$ is the two-sided ideal of $\A_{p_i}$ generated by $\beta$.

If $p_i=3$, 
\begin{equation}
H^*(MSO,\Z_3)=C^*\oplus\Sigma^8C^*\oplus\cdots
\end{equation}
\begin{equation}
\cdots\To\Sigma^2\A_3\oplus\Sigma^6\A_3\oplus\cdots\To\Sigma\A_3\oplus\Sigma^5\A_3\oplus\cdots\To\A_3\To\A_3/(\beta)
\end{equation}
is an $\A_3$-resolution of $\A_3/(\beta)$.
\begin{figure}[!h]
\begin{center}
\begin{tikzpicture}
\node at (0,-1) {0};
\node at (1,-1) {1};
\node at (2,-1) {2};
\node at (3,-1) {3};
\node at (4,-1) {4};
\node at (5,-1) {5};
\node at (6,-1) {$t-s$};
\node at (-1,0) {0};
\node at (-1,1) {1};
\node at (-1,2) {2};
\node at (-1,3) {3};
\node at (-1,4) {4};
\node at (-1,5) {5};
\node at (-1,6) {$s$};

\draw[->] (-0.5,-0.5) -- (-0.5,6);
\draw[->] (-0.5,-0.5) -- (6,-0.5);

\draw (0,0) -- (0,5);
\draw (4,1) -- (4,5);

\draw[fill] (1,0) circle(0.05);
\draw[fill] (3,0) circle(0.05);
\draw (5,0) -- (5,1);

\end{tikzpicture}
\end{center}
\caption{Adams chart for $\Omega_*^{SO}(B\Z_{3^{k_i}})_3^{\wedge}$}
\end{figure}
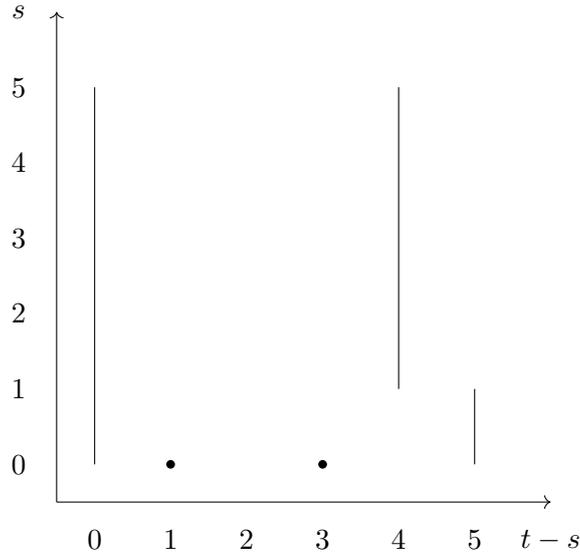
Therefore we arrive at the following
\begin{theorem}
\begin{equation}
\begin{tabular}{c c c}
\hline
$d$ & $\Omega_d^{SO}(B\Z_{3^{k_i}})_3^{\wedge}$ \\
\hline
0& $\Z$ \\
1& $\Z_3$ \\
2& $0$ \\
3 & $\Z_3$\\
4 & $\Z$ \\
5 & $\Z_9$\\ 
\hline
\end{tabular}
\end{equation}
\end{theorem}

The topological term of $\Omega_5^{SO}(B\Z_{3^{k_i}})_3^{\wedge}$ is $\mathfrak{P}(b)$ where $\mathfrak{P}$ is the Postnikov square operation $H^2(-,\Z_3)\rightarrow H^5(-,\Z_9)$.

If $p_i\ge5$, 
$$\H^*(MSO,\Z_{p_i})=C^*\oplus\Sigma^4C^*\oplus\cdots$$
$$\cdots\To\Sigma^2\A_{p_i}\oplus\Sigma^{2p_i}\A_{p_i}\oplus\cdots\To\Sigma\A_{p_i}\oplus\Sigma^{2p_i-1}\A_{p_i}\oplus\cdots\To\A_{p_i}\To\A_{p_i}/(\beta)$$
is an $\A_{p_i}$-resolution of $\A_{p_i}/(\beta)$.

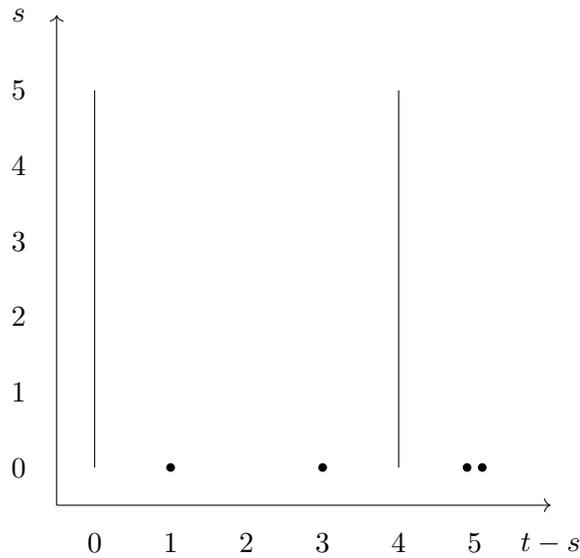
\begin{figure}[!h]
\begin{center}
\begin{tikzpicture}
\node at (0,-1) {0};
\node at (1,-1) {1};
\node at (2,-1) {2};
\node at (3,-1) {3};
\node at (4,-1) {4};
\node at (5,-1) {5};
\node at (6,-1) {$t-s$};
\node at (-1,0) {0};
\node at (-1,1) {1};
\node at (-1,2) {2};
\node at (-1,3) {3};
\node at (-1,4) {4};
\node at (-1,5) {5};
\node at (-1,6) {$s$};

\draw[->] (-0.5,-0.5) -- (-0.5,6);
\draw[->] (-0.5,-0.5) -- (6,-0.5);

\draw (0,0) -- (0,5);
\draw (4,0) -- (4,5);

\draw[fill] (1,0) circle(0.05);
\draw[fill] (3,0) circle(0.05);

\draw[fill] (4.9,0) circle(0.05);
\draw[fill] (5.1,0) circle(0.05);

\end{tikzpicture}
\end{center}
\caption{Adams chart for $\Omega_*^{SO}(B\Z_{p_i^{k_i}})_{p_i}^{\wedge}$}
\end{figure}

Therefore we have
\begin{theorem}
\begin{equation}
\centering
\begin{tabular}{c c}
\hline
$d$ & $\Omega_d^{SO}(B\Z_{p_i^{k_i}})_{p_i}^{\wedge}$\\
\hline
0& $\Z$\\
1& $\Z_{p_i}$\\
2& $0$\\
3 & $\Z_{p_i}$\\
4 & $\Z$\\
5 & $\Z_{p_i}^2$\\ 
\hline
\end{tabular}
\end{equation}
for $p_i\geq 5$.
\end{theorem}

The topological terms of $\Omega_5^{SO}(B\Z_{p_i^{k_i}})_{p_i}^{\wedge}$ are $ab^2$ and $a(\sigma\mod p_i)$ where $\sigma$ is the signature.

\section{Time-reversal, pin$^{\pm}$-TQFTs and non-orientable manifolds}
\label{sec:TR-Pin-TQFT}
As was already mentioned in Section \ref{sec:spin-TQFT}, the notion of spin-TQFT can be generalized to non-orientable manifolds. The analogue of spin structure is pin$^\pm$ structure, that is the lift of $O(n)$ principle bundle of orthonormal tangent frames to $Pin^\pm(n)$ principle bundle with respect to the extensions (\ref{pin-groups}). Physically the $O(1)$ subgroup of $O(n)$ plays role of the time reversal symmetry and often denoted as $O(1)=\Z_2^T$. It is extended by fermionic parity $\Z_2^f$ to the subgroups $Pin^+(1)=\Z_2^f\times \Z_2^T$ and $Pin^-(1)=\Z_4^{Tf}$ of $Pin^+(n)$ and $Pin^-(n)$ respectively, so that the diagram (\ref{pin-groups}) commutes. A pin$^\pm$ manifold is a manifold with a chosen pin$^\pm$ structure.

 The definition of a pin$^\pm$-TQFT is the same as the one for a spin-TQFT but with the spin bordism category replaced by the pin$^\pm$ bordism category, that is the category with objects being closed pin$^\pm$ manifolds and morphisms being bordisms between them equipped with pin$^\pm$ structure that is reduced to the pin$^\pm$ structure of manifolds at the boundary. The corresponding bordism group is also defined similarly to (\ref{spin-bordism-group-def}):

\begin{equation}
 	\Omega_n^{Pin^\pm}(BG) := 
 	\left\{
 	\begin{array}{c}
 	   \text{pairs }(M^{n},f:M^{n}\rightarrow BG), \\
 		M^n\text{ is pin$^\pm$ $n$-manifold}
 	\end{array}
 	\right\}/\sim\text{ bordisms}
 	\label{pin-bordism-group-def}
\end{equation}

The non-orientable version of (\ref{equiv-spin-inv-classification}) then reads
\begin{equation}
 \left\{
    \begin{array}{c}
     \text{deformation classes of $G$-equivariant} \\
     \text{reflection-positive invertible pin$^\pm$-TQFTs}
    \end{array}
 \right\}
 \cong
  \text{Hom}(\Omega_n^{Pin^\pm}(BG),U(1))\,.
 \label{equiv-pin-inv-classification}
\end{equation}

Compared to (\ref{equiv-spin-inv-classification}) picking the torsion subgroup is not required because the bordism groups are all torsion. Physically, such invertible TQFTs provide description of fSPT with time reversal (TR) symmetry (TR-fSPT). Note that, after Wick rotation from Euclidean to Lorentzian signature, $Pin^+$ and $Pin^-$ correspond respectively to $T^2=(-1)^F$ and $T^2=1$ relations between the generators of time-reversal symmetry and fermionic parity.

The gauging operation 
\begin{equation}
\begin{array}{ccc}
\left\{\begin{array}{c}
          \text{invertible $G$-equivariant}\\
          \text{pin$^\pm$-TQFTs}
        \end{array}
\right\} 
&
\longrightarrow
&
\left\{ \text{pin$^\pm$-TQFTs}
\right\} 
\\
\\
Z & \longmapsto & Z_\text{gauged}
\end{array}
\label{pin-gauging-map}
\end{equation} 
is realized completely analogous the gauging in spin case, described in detail in Section (\ref{sec:gauging}).

 Following the general Pontryagin-Thom isomorphism (cf. (\ref{MTH})), the pin$^{\pm}$ bordism groups of a topological space $X$ can be related to stable homotopy groups:
\begin{equation}
	\Omega_*^{Pin^\pm}(X)= \pi_*(MTPin^\pm\wedge X_{+})
\end{equation}
where $MTH$ denotes the Madsen-Tillmann spectrum corresponding to (stable) tangential $H$-structure\footnote{Unlike in the $H=Spin$ case it is not the same as Thom spectrum $MH$. However, $MTPin^\pm\cong MPin^\mp$.}, and, as before, $X_+$ is $X$ with a disjoint marked point added.  	
The reduced $Pin^{+/-}$ bordism groups of $X$ are given by
\begin{equation}
	\widetilde{\Omega}_*^{Pin^\pm}(X)= \pi_*(MTPin^\pm\wedge X).
\end{equation}
In particular, we have 		
\begin{equation}
\Omega_*^{Pin^\pm}(X)= \Omega_*^{Pin^\pm} \oplus \widetilde{\Omega}_*^{Pin^\pm}(X).	
\end{equation}
In what follows we provide the results of calculations of pin$^\pm$ bordism groups of $BG$, for a few simple finite abelian groups $G$.  

Since pin$^\pm$ bordism group are 2-torsion, we only need to consider the Adams spectral sequence at prime $2$: 
\begin{equation}
	E_2^{s,t}=\Ext_{\A}^{s,t}(\H^*(MPin^\pm\wedge X),\Z_2)\Rightarrow\pi_{s-t}(MTPin^\pm\wedge X)=\widetilde{\Omega}_{s-t}^{Pin^\pm}(X).
	\label{pin-adams}
\end{equation}
In particular, if $r=2^km$ with $m$ odd, 
\begin{equation}
\Omega_*^{Pin^+}(B\Z_r)=\Omega_*^{\Pin^+}(B\Z_{2^k})	
\end{equation}

\subsection{Pin$^{+}$ bordism groups}
For calculation we use the fact that \cite{FH}
\begin{equation}
	\Sigma^{-1}MTPin^+ \cong MSpin\wedge MTO_1
\end{equation}
Together with (\ref{MSpin-coh}) it implies that the second page of the Adams spectral sequence (\ref{pin-adams}) for $s-t<8$ can be identified with
\begin{equation}
	\Ext_{\A(1)}^{s,t}(\H^{*-1}(MTO_1 \wedge X),\Z_2)
\end{equation}

\subsubsection{$\Omega_*^{Pin^+}(B\Z_{2})$}

$$\Omega_*^{Pin^{+}}(B\Z_{2})= \Omega_*^{Pin^{+}} \oplus \widetilde{\Omega}_*^{Pin^{+}}(B\Z_{2})$$

\begin{figure}[!h]
	\begin{center}
		\begin{tikzpicture}[scale=0.5]
		\node [below] at (0,0) {$u$};
		\draw[fill] (0,0) circle(.1);
		\draw[fill] (0,1) circle(.1) ;
		\node [right] at (0,1) {$w_1$};
		\draw (0,0) -- (0,1);
		\draw[fill] (0,2) circle(.1);
		\draw (0,0) to [out=150,in=150] (0,2);
		\draw[fill] (0,3) circle(.1);
		\draw[fill] (0,4) circle(.1);
		\draw[fill] (0,5) circle(.1);
		\draw (0,2) -- (0,3);
		\draw (0,3) to [out=150,in=150] (0,5);
		\draw (0,4) -- (0,5);

		\node [below] at (3,1) {$a$};
		\draw[fill] (3,1) circle(.1);
		\draw[fill] (3,2) circle(.1);
		\draw (3,1) -- (3,2);
		\draw[fill] (3,3) circle(.1);
		\draw[fill] (3,4) circle(.1);
		\draw (3,3) -- (3,4);
		\draw (3,2) to [out=150,in=150] (3,4);
		\draw[fill] (3,5) circle(.1);
		\draw (3,3) to [out=30,in=30] (3,5);
		\draw[fill] (3,6) circle(.1);
		\draw (3,5) -- (3,6);

		\node at (1.5,3) {$\otimes$};
		
		\node at (5,3) {$=$};
		
		\node [below] at (6,1) {$ua$};
		\draw[fill] (6,1) circle(.1);
		\draw[fill] (6,2) circle(.1);
		\draw (6,1) -- (6,2);
		\draw[fill] (6,3) circle(.1);
		\draw (6,1) to [out=150,in=150] (6,3);
		\draw[fill] (6,4) circle(.1);
		\draw (6,3) -- (6,4);
		\draw[fill] (7,4) circle(.1);
		\draw (7,4) -- (7,5);
		\draw (6,2) to [out=30,in=140] (7,4);
		\draw[fill] (7,5) circle(.1);
		\draw (6,3) to [out=30,in=140] (7,5);
		\draw[fill] (7,6) circle(.1);
		\draw (6,4) to [out=30,in=140] (7,6);
		\draw[fill] (7,7) circle(.1);
		\draw (7,6) -- (7,7);
		\draw (7,5) to [out=30,in=30] (7,7);

		\node [below] at (10,2) {$w_1a$};
		\draw[fill] (10,2) circle(.1);
		\draw (10,2) -- (10,3);
		\draw[fill] (10,3) circle(.1);
		\draw[fill] (10,4) circle(.1);
		\draw[fill] (10,5) circle(.1);
		\draw (10,3) to [out=150,in=150] (10,5);
		\draw (10,4) -- (10,5);
		\draw[fill] (10,7) circle(.1);
		\draw[fill] (10,6) circle(.1);
		\draw (10,4) to [out=30,in=30] (10,6);
		\draw (10,6) -- (10,7);
		\draw[fill] (10,8) circle(.1);
		\draw[fill] (10,9) circle(.1);
		\draw (10,7) to [out=150,in=150] (10,9);
		\draw (10,8) -- (10,9);
		\draw (10,8) to [out=30,in=30] (10,10);

		\node [below] at (12,3) {$ua^3$};
		\draw[fill] (12,3) circle(.1);
		\draw[fill] (12,4) circle(.1);
		\draw (12,3) -- (12,4);
		\draw[fill] (12,5) circle(.1);
		\draw (12,3) to [out=150,in=150] (12,5);
		\draw[fill] (12,6) circle(.1);
		\draw (12,5) -- (12,6);
		\draw[fill] (13,6) circle(.1);
		\draw (13,6) -- (13,7);
		\draw (12,4) to [out=30,in=140] (13,6);
		\draw[fill] (13,7) circle(.1);
		\draw (12,5) to [out=30,in=140] (13,7);
		\draw[fill] (13,8) circle(.1);
		\draw (12,6) to [out=30,in=140] (13,8);
		\draw[fill] (13,9) circle(.1);
		\draw (13,8) -- (13,9);
		\draw (13,7) to [out=30,in=30] (13,9);

		\node [below] at (14,5) {$ua^5$};
		\draw[fill] (14,5) circle(.1);
		\draw[fill] (14,6) circle(.1);
		\draw (14,5) -- (14,6);
		\draw[fill] (14,7) circle(.1);
		\draw (14,5) to [out=150,in=150] (14,7);
		\draw[fill] (14,8) circle(.1);
		\draw (14,7) -- (14,8);
		\draw[fill] (15,8) circle(.1);
		\draw (15,8) -- (15,9);
		\draw (14,6) to [out=30,in=140] (15,8);
		\draw[fill] (15,9) circle(.1);
		\draw (14,7) to [out=30,in=140] (15,9);
		\draw[fill] (15,10) circle(.1);
		\draw (14,8) to [out=30,in=140] (15,10);
		\draw[fill] (15,11) circle(.1);
		\draw (15,10) -- (15,11);
		\draw (15,9) to [out=30,in=30] (15,11);		
		\end{tikzpicture}
	\end{center}
	\caption{The $\A(1)$-module structure of $\H^*(MTO_1)\otimes\H^*(B\Z_2)$.}
	\label{fig:Pin+Z2}
\end{figure}
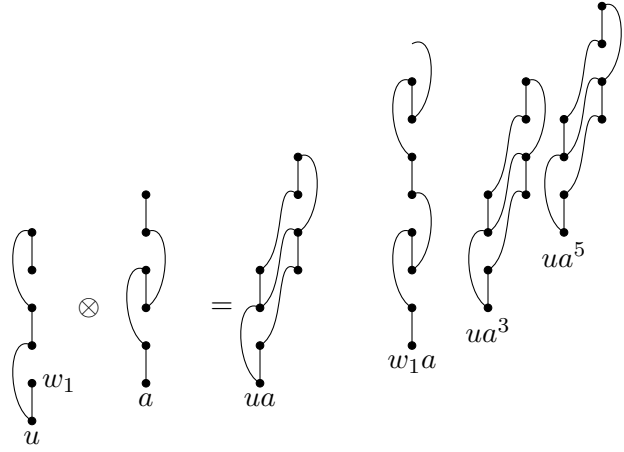

\begin{figure}[!h]
	\begin{center}
		\begin{tikzpicture}
		\node at (0,-1) {0};
		\node at (1,-1) {1};
		\node at (2,-1) {2};
		\node at (3,-1) {3};
		\node at (4,-1) {4};
		\node at (5,-1) {5};
		\node at (6,-1) {$t-s$};
		\node at (-1,0) {0};
		\node at (-1,1) {1};
		\node at (-1,2) {2};
		\node at (-1,3) {3};
		\node at (-1,4) {$s$};
		
		\draw[->] (-0.5,-0.5) -- (-0.5,4);
		\draw[->] (-0.5,-0.5) -- (6,-0.5);

		\draw[color=purple,fill=purple] (1,0) circle(0.05);
		\draw[color=purple,fill=purple] (3,0) circle(0.05);
		\draw[color=purple,fill=purple] (5,0) circle(0.05);
		\draw[color=purple,fill=purple] (5.1,0) circle(0.05);

		\draw[color=red] (2,0) -- (4,2);
		\draw[color=red] (4,2) -- (4,0);

		\end{tikzpicture}
	\end{center}
	\caption{The $E_2$ page of the Adams spectral sequence for $\widetilde{\Omega}_*^{Pin^{+}}(B\Z_{2})$}
	\label{fig:E2-Pin+Z2}
\end{figure}
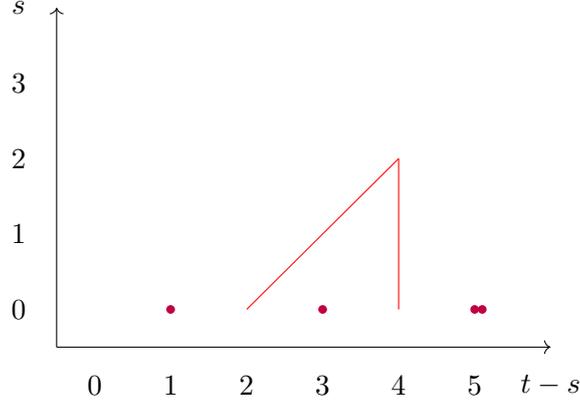

The $\A(1)$-module structure of $\H^*(MTO_1\wedge B\Z_2)$ is shown in Figure \ref{fig:Pin+Z2}.
The $E_2$-page of Adams spectral sequence is  shown	 in Figure \ref{fig:E2-Pin+Z2}.
There are no further differentials due to degree reasons.	Hence we have the following theorem:
\begin{theorem}
	\begin{equation}
	\begin{tabular}{c c }
	\hline
	$i$ & $\Omega^{\Pin^+}_i(B\Z_2)$ 
	\\
	\hline
	0& $\Z_2$ \\
	1& $\Z_2$ \\
	2& $\Z_2^2$ \\
	3 & $\Z_2^3$ \\
	4 & $\Z_{16}\times\Z_8$  \\ 
	\hline
	\end{tabular}
	\end{equation}
\end{theorem}

The corresponding bordism invariants of $(M^n,g:M^n\rightarrow B\Z_2)$ in dimensions $n=3$ and $n=4$: 
\begin{equation}
\begin{array}{rcl}
(M^3,g) &  \stackrel{~}{\longmapsto} & \int_{M^3}g^3 \oplus 
\text{Arf}(\text{PD}(w_1))\oplus 
\tilde{q}_{\text{PD}(w_1)}(g)
\\
\\
(M^4,g) & \stackrel{~}{\longmapsto} & \eta_{Pin^+}(M_4) \oplus \beta_{\text{PD}(w_1)}(g)
\end{array}
\end{equation}
where we used the fact that Poincar\'e dual to $w_1(TM^n)$ can be represented by an orientable codimension 1 submanifold with spin structure induced from the pin$^-$ structure on $M^n$ \cite{kirby1990pin}. The invariant $\eta_{Pin^+}(M_4)$ is $\Z_{16}$ valued eta-invariant of Dirac operator.

\begin{figure}[!h]
	\begin{center}
		\begin{tikzpicture}[scale=0.5]
		\node [below] at (0,0) {$u$};
		\node [right] at (0,1) {$w_1$};
		\node [right] at (0,2) {$w_1^2$};
		\draw[fill] (0,0) circle(.1);
		\draw[fill] (0,1) circle(.1) ;
		\node [right] at (0,1) {$w_1$};
		\draw (0,0) -- (0,1);
		\node [right] at (0,2) {$w_1^2$};
		\draw[fill] (0,2) circle(.1);
		\draw (0,0) to [out=150,in=150] (0,2);
		\draw[fill] (0,3) circle(.1);
		\draw[fill] (0,4) circle(.1);
		\draw[fill] (0,5) circle(.1);
		\draw (0,2) -- (0,3);
		\draw (0,3) to [out=150,in=150] (0,5);
		\draw (0,4) -- (0,5);
		\draw (0,4) to [out=30,in=30] (0,6);
		
		\node at (1.5,3) {$\otimes$};
		
		\node [below] at (3,1) {$x$};
		\draw[fill] (3,1) circle(.1);
		\draw[fill] (3,2) circle(.1);
		\node [right] at (3,2) {$y$};
		\draw[fill] (3,3) circle(.1);
		\node [right] at (3,3) {$xy$};
		\draw[fill] (3,4) circle(.1);
		\draw (3,2) to [out=150,in=150] (3,4);
		\draw[fill] (3,5) circle(.1);
		\draw (3,3) to [out=30,in=30] (3,5);

		\node at (5,3) {$=$};

		\node [below] at (6,1) {$ux$};
		
		\draw[fill] (6,1) circle(.1);
		\draw[fill] (6,2) circle(.1);
		\draw[fill] (6,3) circle(.1);
		\draw[fill] (6,4) circle(.1);
		\draw[fill] (6,5) circle(.1);
		\draw[fill] (6,6) circle(.1);
		\draw (6,1) -- (6,2);
		\draw (6,3) -- (6,4);
		\draw (6,5) -- (6,6);
		\draw (6,1) to [out=150,in=150] (6,3);
		\draw (6,4) to [out=150,in=150] (6,6);
		\draw (6,5) to [out=30,in=30] (6,7);

		\node [below] at (8,2) {$uy$};
		\draw[fill] (8,2) circle(.1);
		\draw[fill] (8,3) circle(.1);
		\draw[fill] (8,4) circle(.1);
		\draw[fill] (8,5) circle(.1);
		\draw[fill] (8,6) circle(.1);
		\draw[fill] (8,7) circle(.1);
		\draw[fill] (9,4) circle(.1);
		\draw[fill] (9,5) circle(.1);
		\draw[fill] (9,6) circle(.1);
		\draw[fill] (9,7) circle(.1);
		\draw[fill] (9,8) circle(.1);
		\draw[fill] (9,9) circle(.1);
		\draw (8,2) -- (8,3);
		\draw (8,4) -- (8,5);
		\draw (8,6) -- (8,7);
		\draw (9,4) -- (9,5);
		\draw (9,6) -- (9,7);
		\draw (9,8) -- (9,9);
		\draw (8,2) to [out=150,in=150] (8,4);
		\draw (8,5) to [out=150,in=150] (8,7);
		\draw (8,6) to [out=30,in=30] (8,8);
		\draw (8,3) to [out=30,in=140] (9,5);
		\draw (9,4) to [out=30,in=30] (9,6);
		\draw (9,7) to [out=30,in=30] (9,9);
		\draw (9,8) to [out=150,in=150] (9,10);

		\node [below] at (11,3) {$uxy$};
		\draw[fill] (11,3) circle(.1);
		\draw[fill] (11,4) circle(.1);
		\draw[fill] (11,5) circle(.1);
		\draw[fill] (11,6) circle(.1);
		\draw[fill] (11,7) circle(.1);
		\draw[fill] (11,8) circle(.1);
		\draw[fill] (12,5) circle(.1);
		\draw[fill] (12,6) circle(.1);
		\draw[fill] (12,7) circle(.1);
		\draw[fill] (12,8) circle(.1);
		\draw[fill] (12,9) circle(.1);
		\draw[fill] (12,10) circle(.1);
		\draw (11,3) -- (11,4);
		\draw (11,5) -- (11,6);
		\draw (11,7) -- (11,8);
		\draw (12,5) -- (12,6);
		\draw (12,7) -- (12,8);
		\draw (12,9) -- (12,10);
		\draw (11,3) to [out=150,in=150] (11,5);
		\draw (11,6) to [out=150,in=150] (11,8);
		\draw (11,7) to [out=30,in=30] (11,9);
		\draw (11,4) to [out=30,in=140] (12,6);
		\draw (12,5) to [out=30,in=30] (12,7);
		\draw (12,8) to [out=30,in=30] (12,10);
		\draw (12,9) to [out=150,in=150] (12,11); 
		\end{tikzpicture}
	\end{center}
	\caption{The $\A(1)$-module structure of $\H^*(MTO_1)\otimes\H^*(B\Z_4)$.}
	\label{fig:Pin+Z4}
\end{figure}
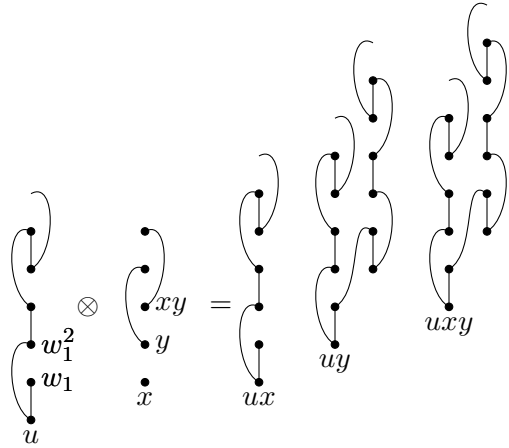

\begin{figure}[!h]
	\begin{center}
		\begin{tikzpicture}
		\node at (0,-1) {0};
		\node at (1,-1) {1};
		\node at (2,-1) {2};
		\node at (3,-1) {3};
		\node at (4,-1) {4};
		\node at (5,-1) {5};
		\node at (6,-1) {$t-s$};
		\node at (-1,0) {0};
		\node at (-1,1) {1};
		\node at (-1,2) {2};
		\node at (-1,3) {3};
		\node at (-1,4) {$s$};
		
		\draw[->] (-0.5,-0.5) -- (-0.5,4);
		\draw[->] (-0.5,-0.5) -- (6,-0.5);

		\draw[color=black,fill=black] (1,0) circle(0.05);
		\draw[color=black] (3,1) -- (5,3);
		\draw[color=black] (5,3) -- (5,0);
		
		\draw[color=green,fill=green] (2,0) circle(0.05);
		\draw[color=green,fill=green] (4,1) -- (4,0);
		
		\draw[color=blue,fill=blue] (3,0) circle(0.05);
		\draw[color=blue,fill=blue] (5.1,1) -- (5.1,0);
		
		\end{tikzpicture}
	\end{center}
	\caption{The $E_2$ page of the Adams spectral sequence for $\widetilde{\Omega}_*^{\Pin^{+}}(B\Z_{4})$}
	\label{fig:E2-Pin+Z4}
\end{figure}
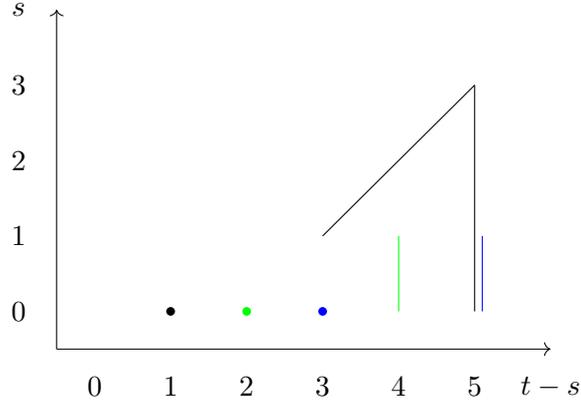

\subsubsection{$\Omega_*^{Pin^+}(B\Z_{4})$}

\begin{equation}
	\Omega_*^{Pin^{+}}(B\Z_{4})= \Omega_*^{Pin^{+}} \oplus \widetilde{\Omega}_*^{Pin^{+}}(B\Z_{4}).
\end{equation}
The $\A(1)$-module structure of $\H^*(MTO_1\wedge B\Z_4)$ is shown in Figure \ref{fig:Pin+Z4}.
The $E_2$-page of Adams spectral sequence is  shown in Figure \ref{fig:E2-Pin+Z4} .

Hence we have the following theorem:
\begin{theorem}
	\begin{equation}
	\begin{tabular}{c c c}
	\hline
	$i$ & $\Omega^{\Pin^+}_i(B\Z_4)$ 
	\\
	\hline
	0& $\Z_2$  \\
	1& $\Z_2$ \\
	2& $\Z_2^2$ \\
	3 & $\Z_2^3$ \\
	4 & $\Z_{16}\times\Z_8$ \\ 
	\hline
	\end{tabular}
	\end{equation}
\end{theorem}

The corresponding bordism invariants of $(M^n,g:M^n\rightarrow B\Z_4)$ in dimensions $n=3$ and $n=4$: 
\begin{equation}
\begin{array}{rcl}
(M^3,f) &  \stackrel{~}{\longmapsto} & \int_{M^3}f^*(xy) \oplus 
\text{Arf}(\text{PD}(w_1))\oplus 
\tilde{q}_{\text{PD}(w_1)}(f^*(x))
\\
\\
(M^4,f) & \stackrel{~}{\longmapsto} & \eta_{Pin^+}(M_4) \oplus \hat{\gamma}_{\text{PD}(w_1)}(f)
\end{array}
\end{equation}

\subsection{Pin$^{-}$ bordism groups}
For calculation we use the fact that \cite{FH}
\begin{equation}
	\Sigma^{+1}MTPin^+ \cong MSpin\wedge MO_1
\end{equation}
Together with (\ref{MSpin-coh}) it implies that the second page of the Adams spectral sequence (\ref{pin-adams}) for $s-t<8$ can be identified with
\begin{equation}
	\Ext_{\A(1)}^{s,t}(\H^{*+1}(MO_1 \wedge X),\Z_2)
\end{equation}

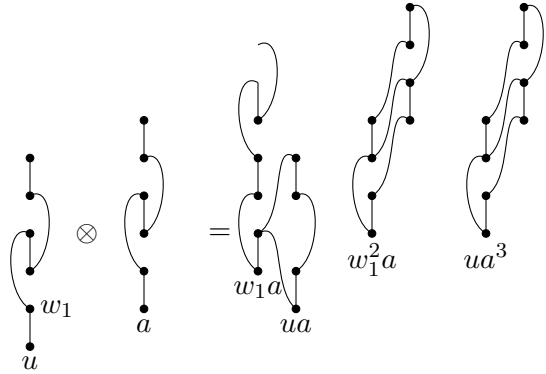
\begin{figure}[!h]
	\begin{center}
		\begin{tikzpicture}[scale=0.5]
		\node [below] at (0,0) {$u$};
		\node [right] at (0,1) {$w_1$};
		\draw[fill] (0,0) circle(.1);
		\draw[fill] (0,1) circle(.1) ;
		\draw[fill] (0,2) circle(.1);
		\draw[fill] (0,3) circle(.1);
		\draw[fill] (0,4) circle(.1);
		\draw[fill] (0,5) circle(.1);
		\draw (0,0) -- (0,1);
		\draw (0,2) -- (0,3);
		\draw (0,4) -- (0,5);
		\draw (0,1) to [out=150,in=150] (0,3);
		\draw (0,2) to [out=30,in=30] (0,4);

		\node [below] at (3,1) {$a$};
		\draw[fill] (3,1) circle(.1);
		\draw[fill] (3,2) circle(.1);
		\draw (3,1) -- (3,2);
		\draw[fill] (3,3) circle(.1);
		\draw[fill] (3,4) circle(.1);
		\draw (3,3) -- (3,4);
		\draw (3,2) to [out=150,in=150] (3,4);
		\draw[fill] (3,5) circle(.1);
		\draw (3,3) to [out=30,in=30] (3,5);
		\draw[fill] (3,6) circle(.1);
		\draw (3,5) -- (3,6);

		\node at (1.5,3) {$\otimes$};
		
		\node at (5,3) {$=$};
		
		\node [below] at (7,1) {$ua$};
		\node [below] at (6,2) {$w_1a$};
		\draw[fill] (7,1) circle(.1);
		\draw[fill] (7,2) circle(.1);
		\draw[fill] (7,4) circle(.1);
		\draw[fill] (7,5) circle(.1);
		\draw[fill] (6,2) circle(.1);
		\draw[fill] (6,3) circle(.1);
		\draw[fill] (6,4) circle(.1);
		\draw[fill] (6,5) circle(.1);
		\draw[fill] (6,6) circle(.1);
		
		\draw (6,2) -- (6,3);
		\draw (6,4) -- (6,5);
		\draw (6,6) -- (6,7);
		\draw (7,1) -- (7,2);
		\draw (7,4) -- (7,5);
		
		\draw (6,2) to [out=150,in=150] (6,4);
		\draw (6,5) to [out=150,in=150] (6,7);
		\draw (6,3) to [out=30,in=140] (7,1);
		\draw (6,3) to [out=30,in=140] (7,5);
		\draw (6,6) to [out=30,in=30] (6,8);
		\draw (7,2) to [out=30,in=30] (7,4);

		\node [below] at (9,3) {$w_1^2a$};
		\draw[fill] (9,3) circle(.1);
		\draw[fill] (9,4) circle(.1);
		\draw (9,3) -- (9,4);
		\draw[fill] (9,5) circle(.1);
		\draw (9,3) to [out=150,in=150] (9,5);
		\draw[fill] (9,6) circle(.1);
		\draw (9,5) -- (9,6);
		\draw[fill] (10,6) circle(.1);
		\draw (10,6) -- (10,7);
		\draw (9,4) to [out=30,in=140] (10,6);
		\draw[fill] (10,7) circle(.1);
		\draw (9,5) to [out=30,in=140] (10,7);
		\draw[fill] (10,8) circle(.1);
		\draw (9,6) to [out=30,in=140] (10,8);
		\draw[fill] (10,9) circle(.1);
		\draw (10,8) -- (10,9);
		\draw (10,7) to [out=30,in=30] (10,9);

		\node [below] at (12,3) {$ua^3$};
		\draw[fill] (12,3) circle(.1);
		\draw[fill] (12,4) circle(.1);
		\draw[fill] (12,5) circle(.1);
		\draw[fill] (12,6) circle(.1);
		\draw[fill] (13,6) circle(.1);
		\draw[fill] (13,7) circle(.1);
		\draw[fill] (13,8) circle(.1);
		\draw[fill] (13,9) circle(.1);
		\draw (12,3) -- (12,4);
		\draw (12,3) to [out=150,in=150] (12,5);
		\draw (12,5) -- (12,6);
		\draw (13,6) -- (13,7);
		\draw (12,4) to [out=30,in=140] (13,6);
		\draw (12,5) to [out=30,in=140] (13,7);
		\draw (12,6) to [out=30,in=140] (13,8);
		\draw (13,8) -- (13,9);
		\draw (13,7) to [out=30,in=30] (13,9);	
		\end{tikzpicture}
	\end{center}
	\caption{The $\A(1)$-module structure of $\H^*(MO_1)\otimes\H^*(B\Z_2)$}
	\label{fig:Pin-Z2}
\end{figure}

\begin{figure}[!h]
	\begin{center}
		\begin{tikzpicture}
		\node at (0,-1) {0};
		\node at (1,-1) {1};
		\node at (2,-1) {2};
		\node at (3,-1) {3};
		\node at (4,-1) {4};
		\node at (5,-1) {5};
		\node at (6,-1) {$t-s$};
		\node at (-1,0) {0};
		\node at (-1,1) {1};
		\node at (-1,2) {2};
		\node at (-1,3) {3};
		\node at (-1,4) {$s$};
		
		\draw[->] (-0.5,-0.5) -- (-0.5,4);
		\draw[->] (-0.5,-0.5) -- (6,-0.5);

		\draw[color=purple,fill=purple] (3,0) circle(0.05);
		\draw[color=purple,fill=purple] (3.1,0) circle(0.05);
		\draw[color=purple,fill=purple] (5,0) circle(0.05);

		\draw[color=black] (1,0) -- (2,1);
		\draw[color=black] (2,1) -- (2,0);

		\end{tikzpicture}
	\end{center}
	\caption{The $E_2$ page of the Adams spectral sequence for $\widetilde{\Omega}_*^{Pin^{-}}(B\Z_{2})$}
	\label{fig:E2-Pin-Z2}
\end{figure}
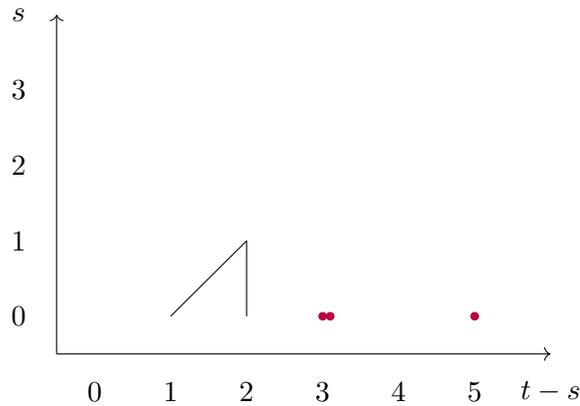

\subsubsection{$\Omega_*^{Pin^-}(B\Z_{2})$} 

\begin{equation}
	\Omega_*^{Pin^{-}}(B\Z_{2})= \Omega_*^{Pin^{-}} \oplus \widetilde{\Omega}_*^{Pin^{-}}(B\Z_{2})
\end{equation}
The $\A_1$-module structure of $\H^*(MO_1\wedge B\Z_2)$ is shown in Figure \ref{fig:Pin-Z2}.
The $E_2$-page of Adams spectral sequence is  shown	 in Figure \ref{fig:E2-Pin-Z2}.

There is no further differential due to degree reasons.	Hence we have the following theorem:
\begin{theorem}
	\begin{equation}
	\begin{tabular}{c c c}
	\hline
	$i$ & $\Omega^{Pin^-}_i(B\Z_2)$ 
	\\
	\hline
	0& $\Z_2$  \\
	1& $\Z_2^2$ \\
	2& $\Z_4\times\Z_8$ \\
	3 & $\Z_2^2$\\
	4 & $0$\\ 
	\hline
	\end{tabular}
	\end{equation}
\end{theorem}

The corresponding bordism invariants of $(M^n,g:M^n\rightarrow B\Z_2)$ in dimensions $n=2$ and $n=3$: 
\begin{equation}
\begin{array}{rcl}
(M^2,g) &  \stackrel{~}{\longmapsto} & \text{ABK}(M^2) \oplus 
{q}_{M^2}(g)
\\
\\
(M^3,g) & \stackrel{~}{\longmapsto} & \int_{M^3}w_1^2g \;\oplus\; \int_{M^3}g^3
\end{array}
\end{equation}

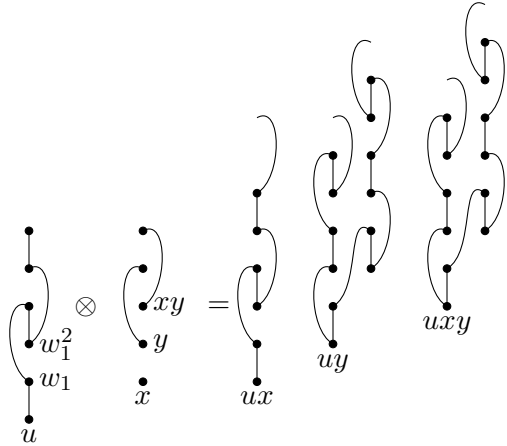
\begin{figure}[!h]
	\begin{center}
		\begin{tikzpicture}[scale=0.5]
		\node [below] at (0,0) {$u$};
		\node [right] at (0,1) {$w_1$};
		\node [right] at (0,2) {$w_1^2$};
		
		\draw[fill] (0,0) circle(.1);
		\draw[fill] (0,1) circle(.1) ;
		\draw[fill] (0,2) circle(.1);
		\draw[fill] (0,3) circle(.1);
		\draw[fill] (0,4) circle(.1);
		\draw[fill] (0,5) circle(.1);
		\draw (0,0) -- (0,1);
		\draw (0,2) -- (0,3);
		\draw (0,4) -- (0,5);
		\draw (0,1) to [out=150,in=150] (0,3);
		\draw (0,2) to [out=30,in=30] (0,4);
		
		\node [below] at (3,1) {$x$};
		\draw[fill] (3,1) circle(.1);
		\draw[fill] (3,2) circle(.1);
		\node [right] at (3,2) {$y$};
		\draw[fill] (3,3) circle(.1);
		\node [right] at (3,3) {$xy$};
		\draw[fill] (3,4) circle(.1);
		\draw (3,2) to [out=150,in=150] (3,4);
		\draw[fill] (3,5) circle(.1);
		\draw (3,3) to [out=30,in=30] (3,5);
		
		\node at (1.5,3) {$\otimes$};

		\node at (5,3) {$=$};

		\node [below] at (6,1) {$ux$};
		
		\draw[fill] (6,1) circle(.1);
		\draw[fill] (6,2) circle(.1);
		\draw[fill] (6,3) circle(.1);
		\draw[fill] (6,4) circle(.1);
		\draw[fill] (6,5) circle(.1);
		\draw[fill] (6,6) circle(.1);
		\draw (6,1) -- (6,2);
		\draw (6,3) -- (6,4);
		\draw (6,5) -- (6,6);
		\draw (6,2) to [out=150,in=150] (6,4);
		\draw (6,3) to [out=30,in=30] (6,5);
		\draw (6,6) to [out=30,in=30] (6,8);

		\node [below] at (8,2) {$uy$};
		\draw[fill] (8,2) circle(.1);
		\draw[fill] (8,3) circle(.1);
		\draw[fill] (8,4) circle(.1);
		\draw[fill] (8,5) circle(.1);
		\draw[fill] (8,6) circle(.1);
		\draw[fill] (8,7) circle(.1);
		\draw[fill] (9,4) circle(.1);
		\draw[fill] (9,5) circle(.1);
		\draw[fill] (9,6) circle(.1);
		\draw[fill] (9,7) circle(.1);
		\draw[fill] (9,8) circle(.1);
		\draw[fill] (9,9) circle(.1);
		\draw (8,2) -- (8,3);
		\draw (8,4) -- (8,5);
		\draw (8,6) -- (8,7);
		\draw (9,4) -- (9,5);
		\draw (9,6) -- (9,7);
		\draw (9,8) -- (9,9);
		\draw (8,2) to [out=150,in=150] (8,4);
		\draw (8,5) to [out=150,in=150] (8,7);
		\draw (8,6) to [out=30,in=30] (8,8);
		\draw (8,3) to [out=30,in=140] (9,5);
		\draw (9,4) to [out=30,in=30] (9,6);
		\draw (9,7) to [out=30,in=30] (9,9);
		\draw (9,8) to [out=150,in=150] (9,10);

		\node [below] at (11,3) {$uxy$};
		\draw[fill] (11,3) circle(.1);
		\draw[fill] (11,4) circle(.1);
		\draw[fill] (11,5) circle(.1);
		\draw[fill] (11,6) circle(.1);
		\draw[fill] (11,7) circle(.1);
		\draw[fill] (11,8) circle(.1);
		\draw[fill] (12,5) circle(.1);
		\draw[fill] (12,6) circle(.1);
		\draw[fill] (12,7) circle(.1);
		\draw[fill] (12,8) circle(.1);
		\draw[fill] (12,9) circle(.1);
		\draw[fill] (12,10) circle(.1);
		\draw (11,3) -- (11,4);
		\draw (11,5) -- (11,6);
		\draw (11,7) -- (11,8);
		\draw (12,5) -- (12,6);
		\draw (12,7) -- (12,8);
		\draw (12,9) -- (12,10);
		\draw (11,3) to [out=150,in=150] (11,5);
		\draw (11,6) to [out=150,in=150] (11,8);
		\draw (11,7) to [out=30,in=30] (11,9);
		\draw (11,4) to [out=30,in=140] (12,6);
		\draw (12,5) to [out=30,in=30] (12,7);
		\draw (12,8) to [out=30,in=30] (12,10);
		\draw (12,9) to [out=150,in=150] (12,11); 
		\end{tikzpicture}
	\end{center}
	\caption{The $\A(1)$-module structure of $\H^*(MO_1)\otimes\H^*(B\Z_4)$}
	\label{fig:Pin-Z4}
\end{figure}

\subsubsection{$\Omega_*^{Pin^-}(B\Z_{4})$}

\begin{equation}
	\Omega_*^{Pin^{-}}(B\Z_{4})= \Omega_*^{Pin^{-}} \oplus \widetilde{\Omega}_*^{Pin^{-}}(B\Z_{4})
\end{equation}
The $\A_1$-module structure of $\H^*(MO_1\wedge B\Z_4)$ is shown in Figure \ref{fig:Pin-Z4}.

\begin{figure}[!h]
	\begin{center}
		\begin{tikzpicture}
		\node at (0,-1) {0};
		\node at (1,-1) {1};
		\node at (2,-1) {2};
		\node at (3,-1) {3};
		\node at (4,-1) {4};
		\node at (5,-1) {5};
		\node at (6,-1) {$t-s$};
		\node at (-1,0) {0};
		\node at (-1,1) {1};
		\node at (-1,2) {2};
		\node at (-1,3) {3};
		\node at (-1,4) {$s$};
		
		\draw[->] (-0.5,-0.5) -- (-0.5,4);
		\draw[->] (-0.5,-0.5) -- (6,-0.5);

		\draw[color=black] (1,0) -- (3,2);
		\draw[color=black] (3,2) -- (3,0);
		
		\draw[color=green,fill=green] (2,0) circle(0.05);
		\draw[color=green,fill=green] (4,1) -- (4,0);
		
		\draw[color=blue,fill=blue] (3.1,0) circle(0.05);
		\draw[color=blue,fill=blue] (5,1) -- (5,0);
		
		\draw[->,color=red,fill=red] (4,0) -- (3,2);
		
		\end{tikzpicture}
	\end{center}
	\caption{The $E_2$ page of the Adams spectral sequence for $\widetilde{\Omega}_*^{\Pin^{-}}(B\Z_{4})$}
	\label{fig:E2-Pin-Z4}
\end{figure}
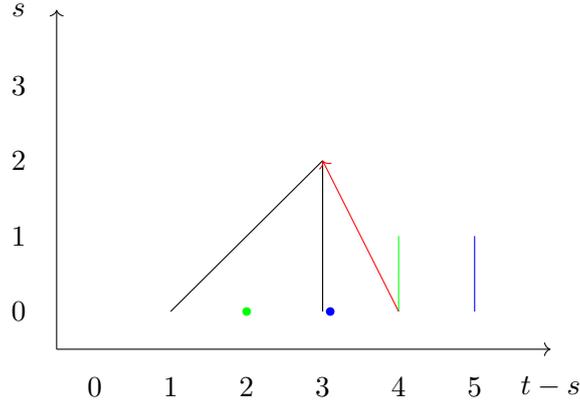

The $E_2$-page of Adams spectral sequence is shown in Figure \ref{fig:E2-Pin-Z4}.

Hence we have the following theorem:
\begin{theorem}
	\begin{equation}
	\begin{tabular}{c c }
	\hline
	$i$ & $\Omega^{Pin^-}_i(B\Z_4)$ 
	\\
	\hline
	0& $\Z_2$  \\
	1& $\Z_2^2$ \\
	2& $\Z_2^2\times\Z_8$  \\
	3 & $\Z_2\times\Z_4$ \\
	4 & $\Z_{2}$  \\ 
	\hline
	\end{tabular}
	\end{equation}
\end{theorem}

The corresponding bordism invariants of $(M^n,f:M^n\rightarrow B\Z_4)$ in dimension $n=3$: 
\begin{equation}
\begin{array}{rcl}
(M^3,f) & \stackrel{~}{\longmapsto} & \int_{M^3}f^*(xy) \;\oplus\; (\text{ABK}(\text{PD}(f^*(x)))\mod 4)

\end{array}
\end{equation}
Notice that $\tilde{\Omega}^{Spin}_4(B\Z_2\times B\Z_4) \to\Omega^{Pin^-}_3(B\Z_4)$ is an isomorphism. The map is a spin 4-manifold $M^4$ with $g: M\to B_2$ and $f:M\to B\Z_4$ is sent to an submanifold $N^3$ dual to $g$ with $N^3\to M^4\xrightarrow{f} B\Z_4$.


\section{Symmetric anomalous Spin-TQFTs as the boundary state of fermionic SPTs}
\label{sec:SbTQFT}

In this section we construct gapped boundary theories coupled with some of the fSPTs protected by finite group symmetry we have constructed.
The construction works for the two cases. The one is where the symmetry group $G$, or its subgroup large enough to trivialize the bordism invariant defining the fSPT, is spontaneously broken at the boundary. 
Another, more nontrivial, case is when, in 4-dimensions, the symmetry group $G$ has the form of  $\mathbb{Z}_4^2\times H$ and the bordism invariant can be formally written as ``$f_1^*(x)\cup f_2^*(x) \cup X$", where $f_{1,2}^*(x)$ is the modulo 2 reduction of the $\mathbb{Z}_4$ backgrounds $f_{1,2}$, and $X$ is a bordism invariant in $\mathrm{Hom}(\Omega^{Pin^-}_2(BH),U(1))$.
In Section~\ref{sec:spin-TQFT}, we have found three of such invariants. In the latter case no subgroup of $G$ is broken on the boundary.
The summary of this section can be found in Subsection~\ref{subsec:boundarysummary}.

What we would like to do is in precise the following.
A $\dd$ dimensional fSPT with symmetry $G$ is an invertible $G$-equivariant TQFT, which is in particular a functor $\Bordspin{\dd}(BG) \to \VectZtwo$, as stated in section \ref{sec:spin-TQFT}.
We want to find a topological boundary condition for some of fSPTs we have found.
That is, we want to find an enhancement of a given TQFT functor $\Bordspin{\dd}(BG) \to \VectZtwo$ to a functor $Z:\Bordspinbdry{\dd}(BG) \to \VectZtwo$, where $\Bordspinbdry{\dd}$ is the bordism category whose objects can have boundaries and morphisms can have corners.\footnote{Here we are not trying to define a full-fledged extended TQFT, which also encodes the set of possible boundary conditions, but we only describe a single boundary condition for a given fSPT.}
Further, since the fSPT is trivial when its background is ignored (i.e.\ maps to $BG$ are trivial),  physically we expect that we can obtain the boundary TQFT $Z^\partial:\Bordspin{\dd-1}\to\VectZtwo$ from the bulk-boundary TQFT $Z$. The non-equivariant version $Z_0:\Bordspinbdry{\dd}\to\VectZtwo$ of $Z$ goes through the category of spin null-bordant manifolds,\footnote{That is, $Z_0(M^\dd_1)$ and $Z_0(M^\dd_2)$ are the same if $\partial M^\dd_1$ is spin-diffeomorphic to $\partial M^\dd_2$. This is because we can drill out $M^\dd\setminus N_{\partial}$ where $N_{\partial}$ is the tubular neighborhood of the boundary, and then fill there with the $\dd$-dimensional ball without changing the value of $Z_0$, since the bulk theory is trivial.}
and we demand that this functor enhances to the boundary TQFT $Z^\partial:\Bordspin{\dd-1}\to\VectZtwo$.
The boundary TQFT cannot be trivial as long as the bulk fSPT is non-trivial.

Physically, it is expected that for any $\fSPT$ for a finite group there exists a boundary TQFT on which the $G$ symmetry is spontaneously broken. In such cases the vector space $Z^\partial(S^{\dd-2})$ is $|G|$ dimensional. While for $1+1$d boundary there is no other option, for higher dimensional boundary we would like to find a more non-trivial boundary TQFT where $Z^\partial(S^{\dd-2})$ is one dimensional, which will turns out to be possibly in many cases.

\subsection{General strategy}\label{subsec:boundary-gen}
In \cite{Kapustin:2014dxa} and more generally in \cite{Gaiotto:2017zba},
it is stated that a $G$-protected fSPT can be described by attaching intrinsic (i.e.\ non-equivariant) invertible TQFTs on the $G$-symmetry defects.
A good example of the statement appeared in \cite{Kapustin:2014dxa} is the 3d $\Z_2$ fSPT defined by $\beta(g)$ where $g$ is the $\Z_2$ background, since
\begin{equation}
	\beta(g) = \ABK(\PD(g)).
	\label{eq:betaABK}
\end{equation}
Physically, $\PD(g)$ is regarded as the subspace that the $\Z_2$ symmetry defect occupies, and therefore we regard \eqref{eq:betaABK} as putting the invertible TQFT defined by the $\ABK$ invariant on the $\Z_2$ symmetry defect.\footnote{This might seem contradicting since, while the $\ABK$ invariant is $\Z_8$ valued, the $\Z_2$ defect should vanish when two of them are stacked. This is actually not the case because when $\PD(g)$ is oriented $\ABK$ has order 2, and when $\PD(g)$ is unorientable it has non-vanishing self-intersection and therefore two of symmetry defect occupying the same unorientable homology class cannot be stacked in a parallel way.}
The pin$^-$ structure on $\PD(g)$ is induced by the spin structure of the total space. See \cite{kirby1990pin}. (It is also briefly explained in \cite{Wang:2018edf}).
Another, more intricate, example is the 4d $\Z_4\times \Z_4$ fSPT defined by the invariant
\begin{equation}
\Arf(f_1^*(x),f_2^*(x)),
\label{eq:ArfZ4Z4}
\end{equation}
where $f_1$ and $f_2$ are the $\Z_4$ backgrounds.
We regard this invariant as decorating the intersection of $\Z_4$ defects with the $\Arf$ invertible TQFT.

Now we want to put a fSPT on a manifold with boundary. We want to preserve the $G$ symmetry on the boundary (though it is spontaneously broken on boundary for a 3d fSPT case). In other words, we take the symmetry background $g$ in the cohomology group $H^1(M)$ (and not in the relative cohomology cohomology $H^1(M,\partial M)$), so that its dual $\PD(g)$ is in the relative homology group $H^1(M,\partial M)$. A representative of $\PD(g)$ can have its boundary in $\partial M$, which is $\dd-2$ dimensional.
If a symmetry defect $\PD(g)$ (or intersection of them) supports an intrinsic invertible TQFT, the value of the invertible TQFT on the defect (e.g.\ $\ABK(\PD(g))$) is not well-defined when the defect have a boundary.
A naive way to fix this problem is to extend the symmetry defect along $\partial{M}$ to close the boundary of the defect. This operation would define an element of $H_{\dd-1}(M)$ out of $\PD(g)\in H_{\dd-1}(M;\partial M)$. However, the way to close the boundary of the defect is not unique and the ambiguity is captured by $H_{\dd-1}(\partial M)$, because of the exact sequence
\begin{equation}
	H_{\dd-1}(\partial M) \to H_{\dd-1}(M) \stackrel{p}{\to} H_{\dd-1}(M, \partial M).
\end{equation}
Therefore, a way to define the bulk-boundary TQFT is to take the sum over $H_{\dd-1}(\partial M)$ (If $\PD(g)$ is not in the image of $p$, simply we set the partition function to be zero).
In this way, we arrive at the partition function, for instance, for the $\beta(g)$ fSPT on a spin manifold $M^3$ with boundaries:
\begin{equation}
	Z(M^3,g) = \sum_{c \in \partial^{-1}([\partial \PD(g)])} e^{\frac{2\pi\mathrm{i}}{8}\ABK(c\cup\PD(g))},	
	\label{eq:partfbeta2}
\end{equation}
where $\partial:H_*(\partial M^3,\partial \PD(g);\Z_2)\to H_{*-1}(\partial \PD(g);\Z_2)$ is the boundary map, and $[\partial \PD(g)]$ is the class in $H_1(\partial \PD(g);\Z_2)$ that is the image of the fundamental class of $\PD(g)$ under the boundary map $H_2(\PD(g),\partial \PD(g);\Z_2)\to H_1(\partial\PD(g);\Z_2)$.
The pin$^-$ structure on $\PD(c)\cap \PD(g)$ is induced by the spin structure of $M^3$.

For the invariant \eqref{eq:ArfZ4Z4}, we would like to propose the partition function on a spin manifold $M^4$ with boundary in a similar fashion to \eqref{eq:partfbeta2}. The construction is, however, more involved and thus we postpone the discussion till Subsection \ref{subsec:ArfZ4Z4}.

In the rest of the section, we construct the (1-)functor for the bulk-boundary TQFT for the 3d fSPT $\beta(g)$, and then also discuss bulk-boundary systems for some of 4d bulk fSPTs.

\subsection{Bulk-boundary TQFT for 2+1d $\Z_2$ fSPT $\beta(g)$} 
Let us construct the functor $Z:\Bordspinbdry{3}(B\Z_2)\to \VectZtwo$ for the bulk $\Z_2$ fSPT $\beta$.
We can promote the partition function \eqref{eq:partfbeta2} into the functor  $Z$ in the following way.
The value of $Z$ on an object $(M^2,g)\in \mathrm{Ob}(\Bordspinbdry{3})$, with a (homotopy class of) map $g:M^2 \to B\Z_2$, is
\begin{equation}
	Z(M^2,g) =
		\bigoplus_{c \in \partial^{-1}([\partial \PD(g)])} \left(\C v_{c} \otimes Z^{\ABK}(c\cup \PD(g))\right) 
\end{equation}
where $v_{c}$ is an indeterminate vector with $\Z_2$ degree zero corresponding to an element $c$ and $Z^{\ABK}:\text{Bord}_2^{Pin^-}\to\VectZtwo$ is the functor representing the pin$^-$ fSPT defined by the $\ABK$ invariant. Such (fully extended) 2d TQFT was considered in detail in \cite{Debray:2018wfz}.
The functor $Z$ evaluated on a morphism $(M^3,g)\in \mathrm{Hom}_{\Bordspinbdry{3}}((M^2_1,g),(M^2_2,g))$ is similarly constructed as:
\begin{equation}
	Z(M^3,g) =
	\bigoplus_{c \in \partial^{-1}([\partial \PD(g)])} \left(F_{c|_{M_1^2},c|_{M_2^2}} \otimes Z^{\ABK}(c\cup \PD(g))\right)
\end{equation}
where $F_{\phi,\psi}$ is the linear map sending $v_{\phi} \mapsto v_{\psi}$.

By setting $g=0$, we obtain the boundary TQFT $Z^\partial:\Bordspin{2}\to\VectZtwo$.
For example, the functor evaluated on the object $S^1$ gives
\begin{equation}
	\begin{split}
		Z^\partial(S^1) &=\bigoplus_{c\in H_1(S^1;\Z_2)}\C v_{c}\otimes Z^{\ABK}(\PD(\phi))\\
				&=\C v_{0}\oplus (\C v_{[S^1]}\otimes Z^{\ABK}(S^1)).
	\end{split}
\end{equation}
In particular, $Z^\partial(S^1_+) = \C^2$ and $Z^\partial(S^1_-) = \C^{1|1}$.\footnote{The one-dimensional $\Z_2$-graded vector space $Z^{\ABK}(S^{1}_-)$ has the odd $\Z_2$ degree, which can be understood from the partition function of $Z^{\ABK}$ on a torus.}
The ground state degeneracy of $Z^\partial(S^1_{+})$ is interpreted as the spontaneous symmetry breaking of the $\Z_2$ symmetry on the boundary.

The construction here can be easily generalize to arbitrary $G$-fSPT with boundary condition with spontaneously broken $G$ symmetry (or its subgroup large enough to trivialize the fSPT anomaly). Next, we would like to construct a more nontrivial boundary condition for a fSPT, that is the boundary condition with which $G$ symmetry is not spontaneously broken on boundary.

\subsection{Bulk-boundary TQFT for 3+1d $\Z_4 \times \Z_2$ fSPT $\Arf(\PD(f_1^*(x))\cap \PD(f_2^*(x)))$} 
\label{subsec:ArfZ4Z4}
\paragraph{Partition function}
We would like to generalize the above construction to the $\Z_4\times \Z_4$ fSPT defined by the invariant $\Arf(\PD(f_1')\cap \PD(f_2'))$ where $f_1,f_2$ are the two $\Z_4$ backgrounds and $f_i'= f_i^*(x)$ (the modulo 2 reductions of them). 
On boundary, we do not want neither $\Z_4$ symmetry to be broken. In particular, the partition function of the boundary theory $Z^\partial$ on $S^1\times S^2$ should be 1.

As said in Subsection \ref{subsec:boundary-gen}, the invariant $\Arf(\PD(f_1')\cap \PD(f_2'))$ is interpreted as decorating the intersection of the symmetry defects $\PD(f_1')$ and $\PD(f_2')$. The intersection of the defects $\PD(f_1')\cap \PD(f_2')$ intersects with the boundary $\partial M^4$ of the space time $M^4$ at a link $L = \PD(f_1')\cap \PD(f_2') \cap \partial M^4$. The orientation of $M^4$ induces the fundamental class $[L]\in H_1(L;\Z_2)$.

Imitating the previous section, naively one might think that we can define the boundary theory by summing over the preimage $\partial^{-1}([L])$ of the boundary map $\partial: H_2(\partial M,L;\Z_2) \to H_1(L;\Z_2)$. However, for a general element $c\in \partial^{-1}([L])$, the pin$^-$ structures induced on $c$ and $\PD(f_1')\cap\PD(f_2')$ are not compatible along $L$. Hence we cannot define the invariant $\ABK(c\cup_L (\PD(f_1')\cap\PD(f_2')))$ with a general $c\in\partial^{-1}([L])$. To avoid this problem, we propose to sum over the surfaces bounded by $L$ with a pin$^-$ compatible with $\PD(f_1')\cap \PD(f_2')$. This can be done as follows.

The spin structure on $\PD(f_1')\cap \PD(f_2')$ is induced form that of $M^4$ using the framing of the normal bundle with framing vectors tangential to $\PD(f_1')$ and $\PD(f_2')$. This also induces the framing of $L$ in $\partial M^4$ in the same way. Let $T(L)$ be the tubular neighborhood of $L$ in $\partial M^4$.
By pushing each component of $L$ along one of the framing vectors, we can define a map $H_1(L;\Z_2)\to H_1(\partial T(L);\Z_2)$. We denote the image of $[L]$ under this map by $\alpha \in H_1(\partial T(L);\Z_2)$. The homology class $\alpha$ does not depends on the choice of the framing vector at each component. 
We have the long exact sequence
\begin{equation}
	H_2(\partial T(L))\stackrel{a}{\to}H_2(\partial M^4 - T(L)) \stackrel{}{\to} H_2(\partial M^4 - T(L),\partial T(L)) \stackrel{\partial'}{\to}H_1(\partial T(L)) \stackrel{i}{\to} H_1(\partial M^4 - T(L)),
	\label{eq:exacthom}
\end{equation}
where all the coefficients of the cohomologies are $\Z_2$.
For each element $c \in \partial'^{-1}(\alpha)$, we define the element in $\hat{c}\in H_2(\partial M^4,L;\Z_2)$ by extending $c$ to $L$ along the framing vector\footnote{Note that because $\partial'(c)=\alpha \in H^1(T(L),\Z_2)$, in principle a smooth representative of $c$ ends on a cycle in $T(L)$ which represents an element in \textit{integral} homology $H_1(T(L),\Z)$ that can be different from the one given by the map $H_1(L,\Z)\rightarrow H_1(T(L),\Z)$ via pushoff towards a framing vector. However, by gluing the appropriate number of M\"obius strips to the boundary components of $c$ one can always fix this mismatch. This can be seen from the fact that framings at each component of $L$ are (non-canonically) in one-to-one correspondence with integers and gluing a single M\"obius strip changes the integer by 2 and that the values of framings mod 2 is fixed by $\alpha \in H^1(T(L),\Z_2)$. See \cite{kirby2004local}.} we used to define $\alpha$. By construction, the framing, and hence the spin structure, induced on $L$ from $c$ and from $\PD(f_1')\cap\PD(f_2')$ are the same.
Therefore, we can uniquely define the pin$^-$ structure on $\hat{c}\cup_L \PD(f_1')\cap\PD(f_2')$.

Now, we propose that the partition function on a 4-manifold $M^4$ with boundary $\partial M^4$ is
\begin{equation}
	Z(M^4,f_1,f_2) = \frac1{2^{|\pi_0(\partial M^4)|}}\sum_{c \in \partial'^{-1}(\alpha)} e^{\frac{2\pi\mathrm{i}}{8}\ABK(\hat{c}\cup_L (\PD(f_1)\cap\PD(f_2)))}.
	\label{eq:ArfZ4Z4part}
\end{equation}
The overall normalization factor $2^{|\pi_0(\partial M^4)|}$ is coming from the gauge redundancy $H^0(\partial M^4;\Z_2)$ on the boundary theory.
When $\partial M^4=\varnothing$, the partition function is just $(-1)^{\Arf(\PD(f_1)\cap\PD(f_2))}$ as desired. 
If $i(\alpha)=0$ (where $i$ is defined in \eqref{eq:exacthom}), the partition function is zero. Otherwise, the preimage $\partial'^{-1}(\alpha)$ is a $H_2(\partial M^4-T(L))/a(H_2(\partial T(L)))$-torsor, and thus we can non-canonically map the sum over $\partial'^{-1}(\alpha)$ to the sum over $H_2(\partial M^4-T(L))/a(H_2(\partial T(L)))$.
In particular, when the backgrounds are off, i.e.\ $f_1=f_2=0$, we have $\partial'^{-1}(\alpha) = H_2(\partial M)$, and the partition function is 
\begin{equation}
	Z^\partial(\partial M^4) = Z(M^4,f_1=0,f_2=0) = \frac1{2^{|\pi_0(\partial M^4)|}}\sum_{c \in H_2(\partial M^4;\Z_2)} e^{\frac{2\pi\mathrm{i}}{8}\ABK(c)},
\end{equation}
which is the partition function of the $\Z_2$ gauge theory with action $\frac{2\pi\mathrm{i}}{8}\beta(g)$ with $g=\PD(c)$ on $\partial M^4$.
The boundary theory can be replaced by $\Z_2$ gauge theory with action $\frac{2\pi\mathrm{i}}{8}k\ABK$ with $k=3,5,7$.

The interpretation of the system is that the lines $L$ behave the 't Hooft lines in the boundary theory, and $\alpha$ is the framing of the operator.
Although there is no naive 1-form symmetry coupled to the operator since the operator requires a framing, the intersection of two 0-from symmetry defects naturally caries a framing as explained, and therefore $f_1'\cup f_2'$ can couple with the 't Hooft line of the boundary $\Z_2$ gauge theory, with anomaly $\Arf(\PD(f_1')\cap\PD(f_2'))$.

\paragraph{Functor}
Next, we describe the functor $Z:\Bordspinbdry{4}(B\Z_4\times B\Z_4)\to \VectZtwo$ for the bulk $\Z_4\times \Z_4$ fSPT $\Arf(\PD(f_1')\cap\PD(f_2'))$ that is compatible with the partition function \eqref{eq:ArfZ4Z4part}.
The value of $Z$ on an object $(M^3,f_1,f_2)\in \mathrm{Ob}(\Bordspinbdry{4})$, with maps $f_1:M^3\to B\Z_4$ and $f:M^3 \to B\Z_4$, is
\begin{equation}
	Z(M^3,f_1,f_2) =
	\bigoplus_{c \in \partial''^{-1}(\beta)} \left(\C v_{c} \otimes Z^{\ABK}(\hat{c}\cup_{P} (\PD(f_1')\cap\PD(f_2'))\right) ,
	\label{eq:ArfZ4Z4vec}
\end{equation}
where $P = \partial M^3 \cap \PD(f_1') \cap \PD(f_2')$ are points with tubular neighborhood $T(P)$, $\beta$ is the element in $H_0(\partial T(P))$ induced from $[P]$, $\partial'': H_1(\partial M^3-T(P),\partial T(P);\Z_2)\to H_0(\partial T(P);\Z_2)$ is the boundary map in the exact sequence analogous to \eqref{eq:exacthom}.
$\hat{c} \in H_1(\partial M^3,P)$ is the element obtained by extending $c \in \partial''^{-1}(\beta)$ to $P$.

$Z$ evaluated on a morphism $(M^4,f_1,f_2)\in \mathrm{Hom}_{\Bordspinbdry{4}}((M^3_1,g_1),(M^3_2,g_2))$ is:
\begin{equation}
	Z(M^4,f_1,f_2) =
	\frac1{N_{M^4}}\bigoplus_{c \in \partial'^{-1}(\alpha)} \left(F_{c|_{M^3_1},c|_{M^3_2}} \otimes Z^{\ABK}(\hat{c}\cup_L (\PD(f_1')\cap\PD(f_2') ))\right) ,
\end{equation}
Here $\frac1{N_{M^4}}$ is the normalization factor coming from the gauge redundancy on boundary, which is essentially the same as the factors appeared in \eqref{gauged-bordism-value}. Since the restriction of $\alpha$ on $M_{1,2}^3$ coincides with $\beta$ appeared in \eqref{eq:ArfZ4Z4vec}, we have consistent restriction map from $\partial '^{-1}(\alpha)$ to $\partial''^{-1}(\beta)$ for each $M_1^3$ and $M_2^3$.

One can directly observe that, as a functor, the boundary TQFT $Z^\partial:\Bordspin{3}\to\VectZtwo$ is the $\Z_2$ gauge theory with the action $\frac{2\pi\mathrm{i}}{8} \beta(g)$, namely:
\begin{equation}
	Z^\partial = Z^{\beta}_\text{gauged},
\end{equation}
as expected above.

\subsection{Bulk-boundary TQFT for 3+1d $\Z_4^2 \times \Z_2$ fSPT $\tilde{q}_{\PD(f_1^*(x)\cup f_2^*(x))}(g)$} 
\label{subsec:Z42Z2}

In the previous two examples, the boundary TQFT $Z^{\partial}$ depends on the spin-structure on its argument, when all the backgrounds are set to be trivial. This is not always the case even if the fSPT in the bulk depends on the spin-structure. However, in such a case symmetry action on the boundary  TQFT depends on the spin-structure on the boundary.

Consider the 3+1d $\Z_4^2 \times \Z_2$ fSPT $\tilde{q}_{\PD(f_1^*(x)\cup f_2^*(x))}(g)$, where $f_{1,2}$ are the $\Z_4^2$ backgrounds and $g$ is the $\Z_2$ background.
As before, we can construct a bulk-boundary TQFT as
\begin{equation}
	Z(M^3,f_1,f_2,g) =
	\bigoplus_{c \in \partial''^{-1}(\beta)} \left(\C v_{c} \otimes Z^{q(g)}(\hat{c}\cup (\PD(f_1')\cap\PD(f_2')))\right) 
\end{equation}
and $Z$ evaluated on a morphism $(M^4,f,g)\in \mathrm{Hom}_{\Bordspinbdry{4}}((M^3_1,g_1),(M^3_2,g_2))$ is:
\begin{equation}
	Z(M^4,f_1,f_2) =
	\frac1{N_{M^4}}\bigoplus_{c \in \partial'^{-1}(\alpha)} \left(F_{c|_{M^3_1},c|_{M^3_2}} \otimes Z^{q(g)}(\hat{c}\cup (\PD(f_1')\cap\PD(f_2')))\right) .
\end{equation}
where $Z^{q(g)}:\text{Bord}_2^{Pin^-}(B\Z_2)\to \VectZtwo$ is the $\Z_2$ fSPT for the invariant $q(g)$, and $f_i' = f_i^*(x)$. Here $g$ is understood to be restricted on the argument of $Z^q(g)$.
Other notations are the same as those in the previous subsection.
The normalization factor $N_{M^4}$ is also the same as that in the previous example.

When all the backgrounds $f_1,f_2$ and $g$ are turned off, the boundary TQFT $Z^\partial$ is identified with the $\Z_2$ gauge theory $Z^{0}_\text{gauged}$ with the trivial action. Further, the boundary theory $Z^\partial$ can be promoted into a $\Z_2$ equivariant TQFT $Z^\partial:\Bordspin{3}(B\Z_2) \to \VectZtwo$ with $\Z_2$ background $g$.
The value on an object $(M^2,g)\in \mathrm{Ob}(\Bordspin{3}(B\Z_2))$ is
\begin{equation}
	Z^\partial(M^2,g) = \bigoplus_{c\in H_1(M^2)} \C v_{c} \otimes Z^{q(g)}(c).
\end{equation}
This $\Z_2$-graded vector space homogeneously have even degree (i.e.\ there is no fermionic states), but is non-trivially acted on by the global $\Z_2$ symmetry, due to the second factor.
For example, on the torus with the even spin structure $T^2_+$ (meaning that a fermion is anti-periodic along at least one of the directions of $T^2$), the vector space $Z^\partial(T^2_+,0)$  has 3 states neutral under the global $\Z_2$ symmetry (not to be confused with the $\Z_2$ of the $\Z_2$ grading, which is related to the fermion parity symmetry) and 1 charged state, and on the torus $T^2_- = T^2_{PP}$ with the odd spin-structure (meaning that a fermion is periodic along both directions of $T^2$), the vector space $Z^\partial(T^2_-,0)$ has 1 neutral state and 3 charged states.

\subsection{General structure and comments}
\label{subsec:boundarysummary}
Here we summarize the results we have found on 3+1d fSPTs so far in this section, and add a several discussions.
For a $G$-protected spin-SPT listed in Table \ref{tab:BBTQFTs}, we can construct bulk-boundary TQFT $Z:\Bordspinbdry{4} \to \VectZtwo$ as follows. Here, the first column was discussed in Subsection~\ref{subsec:ArfZ4Z4}, the second was in Subsection~\ref{subsec:Z42Z2}, and the third is a generalization of them.
In all the cases, the symmetry group $G$ has the form of $\mathbf{Z}_4^2\times H$, and the bordism invariant defining the fSPTs formally has the form of "$f_1^*(x)\cup f_2^*(x)\cup X$" with an invariant $X\in \Hom(\Omega_2^{\Pin^-},U(1))$.
For an object $(M^4,b)\in \mathrm{Ob}(\Bordspinbdry{4})$ with background $b:M^4\to BG$, the value of $Z$ is
\begin{equation}
	Z(M^3,b) =
	\bigoplus_{c \in \partial''^{-1}(\beta)} \left(\C v_{c} \otimes Z^{X}(\hat{c}\cup (\PD(f_1^*(x)\cap \PD(f_2^*(x))))\right) 
	\label{eq:genZ1}
\end{equation}
The notations here was introduced in Subsection \ref{subsec:ArfZ4Z4}.
Similarly, the value of $Z$ on a morphism $(M^4,b)\in \Hom((M^3_1,b_1),(M^3,b_2))$ is
\begin{equation}
	Z(M^4,b) =
	\frac1{N_{M^4}}\bigoplus_{c \in \partial'^{-1}(\alpha)} \left(F_{c|_{M^3_1},c|_{M^3_2}} \otimes Z^{X}(\hat{c}\cup (\PD(f_1^*(x)\cap \PD(f_2^*(x))))\right) 
	\label{eq:genZ2}
\end{equation}
As before, the value of the normalization constant $N_{M^4}$ is $2^{|\pi_0(\partial M^3_1)|+|\pi_0'(\partial M^4)|}$, where $\pi_0'(\partial M^4)$ is the set of the connected components of $\partial M^4$ that does not intersect with neither $M^3_1$ or $M^3_2$.

In all the cases listed in Table \ref{tab:BBTQFTs}, the boundary TQFT $Z^\partial$ is a $\Z_2$ gauge theory. Its action is $\nu \beta$, where $\nu$ is specified in Table \ref{tab:BBTQFTs}.
When $\nu$ is odd the ground state degeneracy on a torus $T^2$ is 3 (all bosonic for even spin-structure and all fermionic for odd spin-structure), and otherwise it is $4$ (all bosonic).

\begin{table}[t]
	\centering
	$
	\begin{tabu}{|c|c|c|c|c|c|c|}
	\hline
		G = \Z_4^2\times H&\text{fSPTs} &X  &\nu & \mathrm{D}(T^2) &\mathrm{D}^H(T_\text{PP}^2) &\mathrm{D}^H(T^2_{\text{other}})  \\
		\hline
		\Z_4^2\times \{1\}& k\Arf(\PD(f'_1)\cap \PD(f'_2))  & k'\ABK  & k' & 3 & \text{--} & \text{--} \\
		\Z_4^2\times \Z_2 & k\tilde{q}_{\PD(f'_1)\cap\PD(f'_2)}(g) & k'' q_{\PD(\phi_1)}(g)  & 0 & 4 & 3 & 1\\
		\Z_4^2\times \Z_4 &\eta(\PD(f'_1)\cap\PD(f'_2)\cap\PD(f'_3))&  k'' q_{\PD(\phi_1)}(f'_3) &  0 & 4 & 3 & 1 \\
		\hline
	\end{tabu}
	$
	\caption{Data defining bulk-boundary TQFT for $G$-protected spin-SPTs. $k$ is a coefficient in the action of the SPT, which is in $Z_2$, $X$ is to be substituted in \eqref{eq:genZ1} and \eqref{eq:genZ2}. 	$f_{1,2,3}$ denotes $\Z_4$ backgrounds, $g$ denotes a $\Z_2$ background. The modulo 2 reduction of $f$ or $f_i$ is denoted by $f'$ or $f_i'$.  $k',k''$ are arbitrary elements of $\Z_8$ and $\Z_4$, respectively, whose modulo 2 reduction are $k$. In all the cases the boundary TQFT is a $\Z_2$ gauge theory $Z^{\nu \beta}_\text{gauged}$, and $\nu$ is the coefficient in the action. The fifth column $\mathrm{D}(T^2)$ is the ground state degeneracy (GSD, abbreviated as $\mathrm{D}$) of the boundary TQFT. For more information about the $\Z_2$ gauge theory with odd $\nu$, see Sec.~8.1 of \cite{Putrov:2016qdo}. In the last two rows the boundary TQFT is the $\Z_2$ gauge theory with the trivial action tensored with the trivial $\{1,f\}$ fermion line, and the action of $H$ symmetry on the states depends on the spin structure on the space. The last two columns are the numbers of states acted nontrivially by $H$ on $T^2$ with double periodic (odd) and the other (even) spin structure, respectively. In the third row with $H=\Z_4$, the states are acted only by the odd generator of $\Z_4$. The $\Z_2$ subgroup of $H=\Z_4$ case acts trivially both on bulk and boundary, and the action of the quotient $\Z_2=\Z_4/\Z_2$ on the system is the same as the $H= \Z_2$ of the second column. The action of $\Z_4^2 = G/H$ part of the symmetry on the boundary TQFT is more subtle, see the main text. }
	\label{tab:BBTQFTs}
\end{table}

Let us make a comment about a physics interpretation of the construction.
As stated in Subsection~\ref{subsec:ArfZ4Z4}, the intersection of the symmetry defects acts like a $\Z_2$ one-form symmetry defect on the boundary $\Z_2$ gauge theory. In fact, the $\Z_2$ gauge theory with the trivial action $S=0\times \beta$ have electric and magnetic $\Z_2$ one-form symmetries, with a mixed anomaly $\int a_1\cup a_2$ between them, where $a_1,a_2$ are the backgrounds for the one-form symmetries.\footnote{This means that the $Z_\text{gauged}^{0\beta}$ can be a boundary theory of the invertible TQFT defined on orientable manifolds with structure maps $(a_1,a_2):M^4 \to K(\Z_2,2)\times K(\Z_2,2)$ that have the partition function $(-1)^{\int_{M^4}a_1\cup a_2}$.}
Roughly speaking, the equations \eqref{eq:genZ1} and \eqref{eq:genZ2} can be regarded as the theory obtained by ``substituting $X$" into the magnetic one-form symmetry background $a_2$ and substituting $f_1'\cup f_2'$ into the electric one-form symmetry background realizes the anomaly ``$\int f_1'\cup f_2' \cup X$", which is the rough structure of the precise invariants in Table~\ref{tab:BBTQFTs}.
In this sense, the construction of the boundary TQFT for the fSPT we presented is an analog of what is done in \cite{Kapustin:2014tfa} for bosonic SPTs.

However, it is not very precise to say we substituted $f_1'\cup f_2' $ into the magnetic one-form symmetry background of $Z_\text{gauged}^{\nu \ABK}$, since the 't Hooft loop of the theory requires a framing, which cannot be specified just by a 1-cycle or 2-cocycle, which is supposed to be the background field for a one-form symmetry.
Therefore, in some see, the theory have ''framed one-form symmetry", meaning that the background is a 1-cycle equipped with framing, and the construction \eqref{eq:genZ1} and \eqref{eq:genZ2} amounts to substituting $f_1'\cup f_2'$ into the background of ''framed one-form symmetry". It would be interesting to find a precise physics meaning of such a concept.

In this paper, the authors could not find a construction of a bulk-boundary TQFT for the bulk 4d $\Z_4\times \Z_2$ fSPT $\delta_{\PD(f)}(g,g)$. This invariant involves, unlike other invariants we have discussed in this section, the $\Z_4$ background $f$ itself, which is not modulo 2 reduction. This indicates we need a $\Z_4$ gauge theory on boundary. Indeed, in \cite{2018arXiv180408628F}, a boundary $\Z_4$ gauge theory is proposed. It would be interesting to try to reconstruct the system in the language we have used in this section. In addition, the boundary TQFTs for pin$^+$ bordism invariants $\beta_{\PD(w_1)}(g)$ and $\hat{\gamma}_{\PD(w_1)}(f)$ for the global symmetries $\Z_2$ and $\Z_4$ are also remained to be constructed. Although the former only involves the $\Z_2$ symmetry defects, the method developed in this section is not applicable since there is no known way to give a pin$^-$ structure inside a codimension-1 submanifold in the 3-dimensional pin$^+$ boundary manifold.

In this paper we have only considered fSPTs with finite group $G$. On the other hand, in \cite{Garcia-Etxebarria:2017crf} it is proposed that a particular 9d $\mathrm{Sp}(N)$ fSPT should have boundary TQFT. It would be intriguing if the method in this section can be generalized into the continuous group case.

\section{Crystalline fSPTs (fermionic Symmetry Protected Topological states) and  bordism groups $\Omega _n^{Spin}(B(\mathbb {Z}\times G_o))$:
Dimensional extension}

\label{sec:crystalline}

\begin{figure}[h!]
\centering
\begin{tikzpicture}
\begin{scope}[shift={(-2,0,0)}]
\draw[->] (0,0,0)--(1,0,0) node[right]{$y$};
\draw[->] (0,0,0)--(0,1,0) node[above]{$z$};
\draw[->] (0,0,0)--(0,0,1) node[below left]{$x$};
\node[] at (0,2,0) {$\uparrow$};
\node[] at (0,3,0) {$\uparrow$};
\node[] at (0,4,0) {$\uparrow$};
\node[] at (0,5,0) {$h$};
\end{scope}
\foreach \i in {1,...,5}{
    \draw[fill=white!80!black] (0,\i,0) -- (4,\i,0)--(4,\i,2)--(0,\i,2)--cycle;
    \node[right] at (4,\i,0) {$z_\i$};
    }
\end{tikzpicture}
\label{fig:layer-stacking}
\caption{From fSPTs to crystalline fSPTs via \emph{layer-stacking}. More precisely, the figure shows
a construction of 4d (3+1D) crystalline fSPTs from 3d (2+1D) fSPTs.
Several 3d fSPTs layers are placed along the $x$-$y$ plane, while the layers are stacked along the $z$-direction. 
Physical interpretations and mathematical derivations are given in Section \ref{sec:crystalline} and \ref{sec:cond-mat}.
The $h$ is an example of the $z$-directional gauge field to probe the background symmetry associated to a translational $\Z$-symmetry. It can be interpreted either as an element of $H^1(M^n,\Z)$, or, as a map $h:M\rightarrow B\Z = S^1$ considered up to homotopy.}
\end{figure}
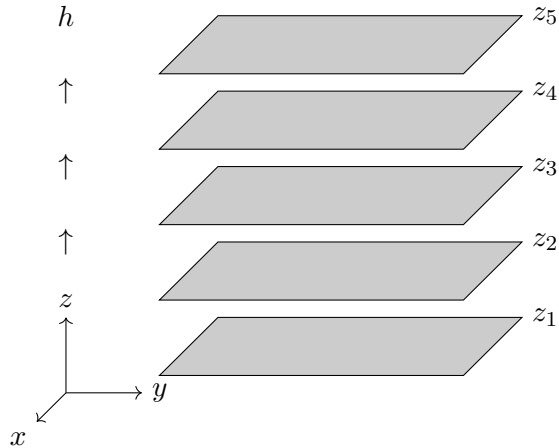

In the rest of the paper, for the sake of brevity, we denote the group that classifies fSPTs with symmetry $H$ as
\begin{equation}
	\Omega^{n}_{H,\text{Tor}}(BG) :=\text{Hom}(\text{Tor}\,\Omega^{H}_{n}(BG),U(1)),
\end{equation}
where $H$ is a tangential structure (e.g. $H=Spin,Pin^\pm,Spin\times_{\Z_2}\Z_{2m}$), and refer to $\Omega^{n}_{H,\text{Tor}}(BG)$ as the \textit{co}bordism group. Now we would like to re-organize spin-TQFT data obtained previously in Section \ref{sec:computation} 
and explore their relation to a different group, $\Omega^n_{Spin,\text{Tor}}(B(\mathbb {Z}\times G_o))$.
As explained in Figure \ref{fig:layer-stacking}, there is a physical application of a part of $\Omega^n_{Spin,\text{Tor}}(B(\mathbb {Z}\times G_o))$,  namely
the \emph{layer-stacking} of lower ($n-1$)d $G_o$-fSPTs to ($n$)d crystalline $\Z \times G_o$-SPTs along an extra dimension, 
where $G_o$ is interpreted as an internal onsite symmetry of ($n-1$)d fSPTs
(associated to $\Omega^{n-1}_{Spin,\text{Tor}}(BG_o)$, where the on-site internal symmetry $G_o$ represents the symmetry group $G$ considered previously in Section \ref{sec:computation}). 
Mathematically, 
we use the fact that a $n$-th bordism group associated to $n$d invertible spin-TQFTs with $\Z \times G_o$-symmetry
\bea  \label{eq:Kunneth1}
\Omega^n_{Spin,\text{Tor}}(B(\Z \times G_o))\cong\Omega^n_{Spin,\text{Tor}}(BG_o) \times \Omega^{n-1}_{Spin,\text{Tor}}(BG_o),
\eea 
contains explicitly the subgroup $\Omega_{n-1}^{Spin}(BG_o)$  associated to $(n-1)$d invertible spin-TQFTs with $G_o$-symmetry. The (canonical) isomorphism (\ref{eq:Kunneth1}) immediately follows from \footnote{Note that the classifying space $B\Z=S^1$. 
The formula (\ref{eq:Kunneth1-bord}) can be understood as a generalization of the particular case of the K\"unneth formula for integral homology: $H_n(B(\Z \times G_o))=H_n(BG_o) \times H_{n-1}(BG_o)$ to the spin-bordism generalized homology. The isomorphism (\ref{eq:Kunneth1-bord}) can be derived, for example, from the Adams spectral sequence for the bordism groups considered earlier in the paper. It follows from the fact that 
\begin{equation}
	H^*(S^1\times BG_o,\Z_2)\cong H^*(BG_o,\Z_2)\times H^{*-1}(BG_o,\Z_2)
\end{equation}
and that the are no non-trivial differentials in the Adams spectral sequence coming from $H^*(BG_o,\Z_2)$ and $H^{*-1}(BG_o,\Z_2)$ terms.
} 
\begin{equation}
	\label{eq:Kunneth1-bord}
	\Omega_n^{Spin}(B(\Z \times G_o))\cong\Omega_n^{Spin}(BG_o) \times \Omega_{n-1}^{Spin}(BG_o),
\end{equation}
that can be geometrically realized as follows:
\begin{equation}
	(M^n,h:M^n\rightarrow BZ,f:M^n\rightarrow BG_o) \longmapsto
	(M^n,h) \times (\text{PD}(h),f|_{\text{PD}(h)})
\end{equation}
where $\text{PD}(h)$ is a smooth oriented codimension 1 submanifold (which always exist) of $M^n$ representing a Poincar\'e dual to $h$. The spin structure on $\text{PD}(h)$ is induced from the spin structure on $M^n$ as previously described. 

Thus, when we interpret $\Z$ as a translation symmetry along an extra dimension, all the spin-TQFTs associated to $\Omega_{n-1}^{Spin}(BG_o)$
survive and can be stacked into crystalline fSPTs in one higher dimension.
If the crystalline symmetry is more general as $G_c$ instead of simply as $\Z$, and if $G_c$ commutes with the internal symmetry $G_o$,
{then we can evaluate instead
$
\Omega_n^{Spin}(B(G_c \times G_o))
$.
{More generally, 
one can consider $G_c$ and $G_o$ to be non-commutative, e.g.
$\Omega_n^{Spin}(B(G_c \ltimes G_o))$, but we will not need this 
in this article, since $\Omega_n^{Spin}(B(\Z \times G_o))$ already provides a non-trivial relations between fSPTs in $n$d and $(n-1)$d, see Section \ref{sec:relation-fSPTs}
for the details.}
In the next section, we will focus on $\Omega_n^{Spin}(B(\Z \times G_o))$ and relate topological terms 
between $n$d and ($n-1$)d associated to
$\Omega_{n-1}^{Spin}(BG_o)$.
}

\subsection{Computations of $\Omega _n^{Spin}(B(\mathbb {Z}\times G_o))$
for a finite abelian $G_o$}
\label{subsec:crystalline-cobordism}
Using the analogue of K\"unneth formula (\ref{eq:Kunneth1}), we can obtain the cobordism groups $\Omega^n_{Spin,\text{Tor}}(B(\mathbb {Z}\times G_o))$,
and associate this data to crystalline fermionic SPTs classes.
In the data below, the crystalline fermionic SPTs classes in $n$d are directly given by $\Omega^{n-1}_{Spin,\text{Tor}}(BG_o)$ again thanks to (\ref{eq:Kunneth1}).

We remark that the 2d Arf invariant spin-TQFT (1+1D Kitaev chain) is not an SRE fSPTs but is an LRE invertible fermionic topological order (see Section \ref{sec:def-fSPT},  eqs.(\ref{equiv-spin-inv-classification}) and (\ref{equiv-SPT-inv-classification})).
However, stacking LRE 1+1D Kitaev chains into a 2+1D system protected by $\Z$-translational symmetry, it becomes an SRE 2+1D crystalline fSPTs.\footnote{Thanks to the local unitary transformation, this 2+1D crystalline fSPTs  can be deformed to a trivial tensor product state once we break the $\Z$-translational symmetry.}
Therefore, all 2d invertible spin-TQFTs with $G_o$ symmetry can contribute to 3d crystalline $(\mathbb {Z}\times G_o)$-fSPTs. 

We only list crystalline fSPTs classes for dimensions $n=3$ and $4$, since we have more clear definitions of fSPTs with internal onsite symmetry $G_o$ in 1+1D (2d) or above,
which then can be stacked along one extra dimension to crystalline fSPTs in 2+1D (3d) and 3+1D (4d). See below.

{
\begin{theorem}
\label{thm:Z2Z}
\begin{equation}
\begin{tabular}{c c c c}
\hline
$n$ & $\Omega^{Spin}_n(B((\Z_2)\times\Z))$ & $H_n(B((\Z_2)\times\Z))$   & 
$\begin{array}{c}\text{crystalline fermionic SPTs classes}\\ \text{via }{\Omega^{n-1}_{Spin,\mathrm{Tor}}(B(\Z_2))}\end{array}$
\\
\hline
0& $\Z$ & $\Z$  \\
1& $\Z\times\Z_2^2$ & $\Z\times\Z_2$ \\
2&  $\Z_2^4$& $\Z_2$&   \\
3 & $\Z_2^2\times\Z_8$ &$\Z_2$ &  $\Z_2^2$\\
4 & $\Z\times {\Z_8}$&  $\Z_2$ &  $\Z_8$\\ 
\hline
\end{tabular}
\end{equation}
\end{theorem}
}

{
\begin{theorem}
\label{thm:Z4Z}
\begin{equation}
\begin{tabular}{c c c c}
\hline
$n$ & $\Omega^{Spin}_n(B((\Z_4)\times\Z))$ & $H_n(B((\Z_4)\times\Z))$   & 
$\begin{array}{c}\text{crystalline fermionic SPTs classes}\\ \text{via }{\Omega^{n-1}_{Spin,\mathrm{Tor}}(B(\Z_4))}\end{array}$
\\
\hline
0& $\Z$ & $\Z$ \\
1& $\Z \times \Z_2\times\Z_4$ & $\Z\times\Z_4$\\
2& $\Z_2^3 \times\Z_4$ & $\Z_4$  & \\
3 & $\Z_2^3\times\Z_8$ & $\Z_4$ & $\Z_2$\\
4 & $\Z\times {\Z_2}\times\Z_8$ &$\Z_4$  & ${\Z_2}\times\Z_8$\\ 
\hline
\end{tabular}
\end{equation}
\end{theorem}
}

{
\begin{theorem}
\label{thm:Z2Z2Z}
\begin{equation}
\begin{tabular}{c c c c}
\hline
$n$ & $\Omega^{Spin}_n(B((\Z_2^2)\times\Z))$ & $H_n(B((\Z_2^2)\times\Z))$  & 
$\begin{array}{c}\text{crystalline fermionic SPTs classes}\\ 
\text{via }{\Omega^{n-1}_{Spin,\mathrm{Tor}}(B(\Z_2^2))}\end{array}$
\\
\hline
0& $\Z$ &  $\Z$ \\
1& $\Z\times\Z_2^3$ & $\Z\times\Z_2^2$ \\
2& $\Z_2^7$ & $\Z_2^3$ &  \\
3 & $\Z_2^4\times\Z_4\times\Z_8^2$ & $\Z_2^4$ & $\Z_2^4$\\
4 & $\Z\times\Z_2^2\times\Z_4\times\Z_8^2$ & $\Z_2^5$ & $\Z_4\times\Z_8^2$ \\ 
\hline
\end{tabular}
\end{equation}
\end{theorem}
}

{
\begin{theorem}
\label{thm:ZZ2Z4}
\begin{equation}
\begin{tabular}{c c c c}
\hline
$n$ & $\Omega^{Spin}_n(B((\Z_2 \times \Z_4)\times\Z))$ & $H_n(B((\Z_2 \times \Z_4)\times\Z))$  & 
$\begin{array}{c}\text{crystalline fermionic SPTs classes}\\ 
\text{via }{\Omega^{n-1}_{Spin,\mathrm{Tor}}(B(\Z_2 \times \Z_4))}\end{array}$
\\
\hline
0& $\Z$ &  $\Z$ \\
1& $\Z\times\Z_2^2\times\Z_4$ & $\Z\times\Z_2\times\Z_4$ \\
2& $\Z_2^6\times\Z_4$ & $\Z_2^2\times\Z_4$ & \\
3 & $\Z_2^7\times\Z_8^2$ & $\Z_2^3\times\Z_4$ & $\Z_2^4$ \\
4 & $\Z\times\Z_2^4\times\Z_4\times\Z_8^2$ & $\Z_2^4\times\Z_4$ & $\Z_2^3\times\Z_8^2$ \\ 
\hline
\end{tabular}
\end{equation}
\end{theorem}
}

{
\begin{theorem}
\label{thm:Z4Z4Z}
\begin{equation}
\begin{tabular}{c c c c}
\hline
$n$ & $\Omega^{Spin}_n(B((\Z_4^2)\times\Z))$ & $H_n(B((\Z_4^2)\times\Z))$   & 
$\begin{array}{c}\text{crystalline fermionic SPTs classes}\\ 
\text{via }{\Omega^{n-1}_{Spin,\mathrm{Tor}}(B( (\Z_4)^2))}\end{array}$
\\
\hline
0& $\Z$ & $\Z$  \\
1& $\Z\times\Z_2\times\Z_4^2$ & $\Z\times\Z_4^2$\\
2& $\Z_2^4\times\Z_4^3$ & $\Z_4^3$  &  \\
3 & $\Z_2^6\times\Z_4^2\times\Z_8^2$ & $\Z_4^4$ & $\Z_2^3\times\Z_4$ \\
4 & $\Z\times\Z_2^4\times\Z_4^3\times\Z_8^2$ & $\Z_4^5$ & $\Z_2^3\times\Z_4\times\Z_8^2$\\ 
\hline
\end{tabular}
\end{equation}
\end{theorem}
}

\subsection{Relations between fSPTs and crystalline-fSPTs:\\
$\Omega^n_{Spin,\mathrm{Tor}}(B(\Z_N \times G_o))$ and $\Omega^{n-1}_{Spin, \mathrm{Tor}}(B(G_o))$ $\subset$ $\Omega^n_{Spin, \mathrm{Tor}}(B(\Z \times G_o))$}

\label{sec:relation-fSPTs}

We now relate the crystalline-$(\Z \times G_o)$-fSPTs in $n$d associated to $\Omega^{n-1}_{Spin,\mathrm{Tor}}(BG_o)$
(within the bordism group $\Omega^n_{Spin, \mathrm{Tor}}(B(\Z \times G_o))$ thanks to (\ref{eq:Kunneth1}))
to the fSPTs protected by an internal onsite symmetry $(\Z_N \times G_o)$, obtained from $\Omega^n_{Spin, \mathrm{Tor}}(B(\Z_N \times G_o))$.
Namely, we study the map  
\bea \label{eq:map-crystal-SPT}
\Omega^n_{Spin, \mathrm{Tor}}(B(\Z_N \times G_o)) \to \Omega^n_{Spin,\mathrm{Tor}}(B(\Z \times G_o)),
\eea
dual to the map between the corresponding bordism groups:
\begin{equation}
	\begin{array}{rcl}
	\Omega_n^{Spin}(B(\Z \times G_o)) &\longrightarrow & \Omega_n^{Spin}(B(\Z_N \times G_o)), \\
	(M^n,h:M^n\to BZ,f:M^n\to BG_o) &\longmapsto & (M^n,h\mod N,f)
	\end{array}
\end{equation}
We also use the map 
\bea \label{eq:map-crystal-SPT-2}
\Omega_{n-1}^{Spin,\mathrm{Tor}}(B(G_o)) \to \Omega_n^{Spin,\mathrm{Tor}}(B(\Z_N \times G_o)).
\eea
obtained by composing the embedding via (\ref{eq:Kunneth1}) with (\ref{eq:Kunneth1-bord}).

In Table \ref{table:ZxGo}, we consider the map (\ref{eq:map-crystal-SPT}) in the case of spacetime dimension $n=4$ and 
for $G_o=\Z_2$ or $\Z_4$, while $\Z_N=\Z_2$ or $\Z_4$.
In Table \ref{table:ZxGo2}, we consider the map of (\ref{eq:map-crystal-SPT})
for $G_o=(\Z_2)^2, \Z_2 \times \Z_4$ or $(\Z_4)^2$, while $\Z_N=\Z_2$ or $\Z_4$.
We denote the topological terms/bordism invariants in terms of the notations introduced in Section \ref{sec:computation},
and also the more informal notations used in our previous work \cite{Wang:2018edf}.
Furthermore, we denote $f:M\to B\Z_4$, $g:M\to B\Z_2$, and $h:M\to B\Z$ are the
maps from the manifold $M$ to the classifying space defining background gauge fields for the corresponding groups.

\begin{table}[!h]
\centering
\noindent
\xymatrix{
 & 
{\begin{array}{@{}l@{}}
\underline{(\Z_2^2)\text{-fSPTs}}: 
 \Omega^4_{Spin,\mathrm{Tor}}(B\Z_2^2)=\Z\times\Z_2^8
 \end{array}} 
 \ar[ld]
&\\
{\begin{array}{@{}l@{}}
\Omega^4_{Spin,\mathrm{Tor}}(B(\Z\times\Z_2))
\\
\underbrace{= \boxed{\Z_8}}_{\Omega^3_{Spin,\mathrm{Tor}}(B(\Z_2))= \boxed{\Z_8}}
\end{array}} \;{:} 
& 
{\left\{\begin{array}{lll}  \Z_8:& {\beta_{\text{PD}(h)}(g)=h\cup (g\cup\text{ABK})}  \end{array}\right.}
\\
&
{\begin{array}{@{}l@{}@{}}
\underline{(\Z_2\times\Z_4)\text{-fSPTs}}{:}\\[2mm]
\Omega^4_{Spin,\mathrm{Tor}}(B(\Z_2\times\Z_4))
\\
=\Z_2\times\boxed{\Z_4}
\end{array}}  \;{:} 
{\left\{\begin{array}{lll} \Z_4:& {\delta_{\text{PD}(f)}(g,g)}
\\ 
&
={f\cup(g\cup \text{ABK})}   \quad \text{[Ab]}\end{array}\right.}
\ar[lu]_{\text{Yes}}\ar[ld]^{\text{No}}
\\
{\begin{array}{@{}l@{}}
\Omega^4_{Spin,\mathrm{Tor}}(B(\Z \times \Z_4))
\\
\underbrace{=\boxed{\Z_2
\times \Z_8}}_{\Omega^3_{Spin,\mathrm{Tor}}(B(\Z_4))= \boxed{\Z_2
\times \Z_8}}
\end{array}}: 
&
{\left\{\begin{array}{lll} 
\Z_2: 
&  
{\text{Arf}(\text{PD}(f\mod 2)\cap\text{PD}(h\mod 2))} 
\\
& 
=(h\mod 2)\cup (f\mod 2) \cup \text{Arf} \;
\\ 
\Z_8:& {\hat{\gamma}_{\text{PD}(h)}(f)}
\\
&
= { \text{invert. $U(1)_1$ spin-CS with U(1) broken to $\Z_4$} \mid_{\text{PD}(h)} } 
\end{array}\right.}
\\
&
\ar[lu]_{\text{Yes}}
\hspace{-3em}
{\begin{array}{@{}l@{}}
\underline{(\Z_4^2)\text{-fSPTs}}{:}
\\
\Omega^4_{Spin,\mathrm{Tor}}(B(\Z_4^2))
\\
=\Z_4^2 \times\boxed{\Z_2}
\\
\end{array}}:
{\left\{\begin{array}{lll} 
\Z_2:
& 
{\text{Arf}(\text{PD}(f_1 \mod 2)\cap\text{PD}(f_2\mod 2))} 
\\
&
= (f_1\mod 2)\cup (f_2\mod 2)\cup \text{Arf} \;\quad\quad \text{[NAb]}
\end{array}\right.}
}
\hspace*{25mm}
\caption{
\fontsize{10}{11}\selectfont
We consider the map (\ref{eq:map-crystal-SPT})  from fSPTs with symmetries $G_o\times \Z_N$ ($G_o=\Z_2$ or $\Z_4$, $N=2$ or $4$) to
$\Omega^4_{Spin,\mathrm{Tor}}(B(\Z \times G_o))$,
which contains the crystalline fSPTs from $\Omega^{3}_{Spin,\mathrm{Tor}}(BG_o)$ subgroup.
The left hand side of the arrows shows the classes associated to fSPTs with $\Z$-symmetry,
the right hand side of arrows shows the classes associated to fSPTs with finite abelian group symmetries.
The group-valued classification whose generators associated to ``intrinsically fermionic'' (spin-TQFTs) are ``boxed.''
The arrows dressed with labels ``Yes'' mean that all fermionic states survive under the map; thus the maps labeled ``Yes'' are injective. 
The arrow dressed with a label ``No'' means that fermionic states do not survive under the map from the left to the right classes; thus the maps labeled ``No'' are \emph{not} injective. We also list topological invariants using a more informal notation used in \cite{Wang:2018edf}. For example, $h\cup (g\cup\text{ABK})$ actually means $\text{ABK}(\text{PD}(h)\cap \text{PD}(g))$ where we first induce spin structure on $\text{PD}(g)$ and then pin$^-$ structure on $\text{PD}(h)\cap \text{PD}(g)$ as described previously in the main text.
}
\label{table:ZxGo}
\end{table}

\restoregeometry

\newgeometry{left=1.3cm, right=1.3cm, top=1.5cm, bottom=1.5cm}
%
\begin{table}[!h]
\centering
\noindent
\fontsize{9}{8}\selectfont
{
\xymatrix{
 & \underline{(\Z_2^3)\text{-fSPTs}}: \Omega^4_{Spin,\mathrm{Tor}}(B\Z_2^3)=\Z\times\Z_2^8 
 \ar[ld]
 &\\
 \vspace{-2ex}
{\begin{array}{@{}l@{}}
\Omega^4_{Spin,\mathrm{Tor}}(B(\Z \times\Z_2^2))\\
\underbrace{
=\Z_2^2\times\boxed{\Z_4\times\Z_8^2}}_{
\Omega^3_{Spin,\mathrm{Tor}}(B(\Z_2^2))=\boxed{\Z_4\times\Z_8^2}}
\end{array}} \;{:} 
& 
{\left\{\begin{array}{lll} \Z_4:& 
{\delta_{\text{PD}(h)}(g_1,g_2)} 
=h \cup (g_1\cup g_2\cup\tilde{\eta})   \\  
\Z_8: &  {\beta_{\text{PD}(h)}(g_1)} 
=h \cup (g_1\cup\text{ABK}) \\ 
\Z_8:& {\beta_{\text{PD}(h)}(g_2)} 
=h\cup (g_2\cup\text{ABK}) \end{array}\right.}
&
\\
&
\ar[lu]_-{\text{Yes}}
\ar[ld]^-{\text{Partially}}
{\begin{array}{@{}l@{}}
\underline{(\Z_2^2\times\Z_4)\text{-fSPTs}}:\\[2mm]
\Omega^4_{Spin,\mathrm{Tor}}(B(\Z_2^2\times\Z_4)) \\
=\Z_2^5\times\boxed{\Z_4^3}
\end{array}}  \;{:}
{\left\{\begin{array}{lll} 
\Z_4:& {\delta_{\text{PD}(f)}(g_1,g_1)}  \\&
=f\cup(g_1\cup\text{ABK})  \quad \;[\text{Ab}]\\  
\Z_4: &  {\delta_{\text{PD}(f)}(g_2,g_2)} \\&
=f\cup(g_2\cup\text{ABK}) \quad \;[\text{Ab}]\\ 
\Z_4:& {\delta_{\text{PD}(f)}(g_1,g_2)} \\&
{=f\cup (g_1\cup g_2\cup\tilde{\eta})} \quad [{\text{NAb}}] \end{array}\right.}
&
\\
{\begin{array}{@{}l@{}}
\Omega^4_{Spin,\mathrm{Tor}}(B(\Z \times\Z_2\times\Z_4))\\
\underbrace{=\Z_2^2
\times\boxed{\Z_2^2\times\Z_4\times\Z_8^2}}_{
\Omega^3_{Spin,\mathrm{Tor}}(B(\Z_2\times\Z_4))= \Z_2\times\boxed{\Z_2^2\times\Z_8^2}}
\end{array}}: 
&
{\left\{\begin{array}{lll} \Z_2:& {\text{Arf}(\text{PD}(f\mod 2)\cap\text{PD}(h\mod 2))} \\ &
=(h \mod 2)\cup (f\mod 2)\cup\text{Arf}  \\  
\Z_2: &  {\tilde{q}_{\text{PD}(f\mod 2)\cap\text{PD}(h\mod 2)}(g)} \\&
{=(h\mod 2)\cup (f\mod 2)\cup g\cup\eta}\\ 
\Z_8:&{\beta_{\text{PD}(h)}(g)} 
=h\cup (g\cup\text{ABK}) \\
\Z_8:& {\hat{\gamma}_{\text{PD}(h)}(f)}
= { \text{$\Z_4\subset U(1)_1$-spin-CS} \mid_{\text{PD}(h)} }\\
\Z_4:&{\delta_{\text{PD}(f)}(g,g)} 
=\underbrace{f\cup (g\cup \text{ABK})}_{\text{[Non-stacking]}}\\  
\end{array}\right.}
&\\
&
\ar[lu]_-{\text{Yes}}
\ar[ld]_-{\text{Partially}}
\hspace{-3em}
{\begin{array}{@{}l@{}}
\underline{(\Z_2\times\Z_4^2)\text{-fSPTs}}:\\[2mm]
\Omega^4_{Spin,\mathrm{Tor}}(B(\Z_2\times\Z_4^2))\\
=\Z_2^4\times\Z_4^2\\
\times\boxed{\Z_2^2\times\Z_4^2}
\end{array}}:
{\left\{\begin{array}{lll} \Z_2:& {\text{Arf}(\text{PD}(f_1\mod 2)\cap\text{PD}(f_2\mod 2))} \\&=(f_1\mod 2)\cup (f_2\mod 2)\cup\text{Arf} \quad\quad [\text{NAb}] \\  
\Z_2: &  {\tilde{q}_{\text{PD}(f_1\mod 2)\cap\text{PD}(f_2\mod 2)}(g)} \\&=(f_1\mod 2)\cup (f_2\mod 2)\cup g\cup\eta \quad \; [{\text{NAb}}]\\ 
\Z_4:&{\delta_{\text{PD}(f_1)}(g,g)} =f_1\cup (g\cup \text{ABK}) \quad\quad \quad [\text{Ab}]\\
\Z_4:&{\delta_{\text{PD}(f_2)}(g,g)} =f_2\cup (g\cup \text{ABK}) \quad\quad\quad [\text{Ab}]\\
\end{array}\right.}
&
\\
{\begin{array}{@{}l@{}}
\Omega^4_{Spin,\mathrm{Tor}}(B(\Z\times \Z_4^2))\\
\underbrace{=\Z_4^3
\times\boxed{\Z_2^4\times\Z_8^2}}_{
\Omega^3_{Spin,\mathrm{Tor}}(B(\Z_4^2))
= \Z_4 
\times\boxed{\Z_2^3\times\Z_8^2}}
\end{array}} \;:
&
\hspace{-4em}
{\left\{\begin{array}{lll} 
\Z_2: &  \overbrace{\text{Arf}(\text{PD}(f_1\mod 2)\cap\text{PD}(f_2\mod 2))}^{\text{[Non-stacking]}} \\&
=(f_1\mod 2)\cup (f_2\mod 2)\cup\text{Arf}\\
\Z_2:&{\text{Arf}(\text{PD}(f_1\mod 2)\cap\text{PD}(h\mod 2))} \\&
=(h\mod 2)\cup (f_1\mod 2)\cup\text{Arf}\\
\Z_2:&{\text{Arf}(\text{PD}(f_2\mod 2)\cap\text{PD}(h\mod 2))} \\&
=(f_2\mod 2)\cup (h\mod 2)\cup\text{Arf}\\
\Z_2:& {\tilde{q}_{\text{PD}(f_1\mod 2)\cap\text{PD}(h\mod 2)}(f_2\mod 2)} \\&
=(f_1\mod 2)\cup (f_2\mod 2)\cup (h\mod 2)\cup\eta\\  
\Z_8:&{\hat{\gamma}_{\text{PD}(h)}(f_1)} = {\text{$\Z_4\subset U(1)_1$-spin-CS} \mid_{\text{PD}(h)} }.\\
\Z_8:&{\hat{\gamma}_{\text{PD}(h)}(f_2)} = { \text{$\Z_4\subset U(1)_1$-spin-CS} \mid_{\text{PD}(h)} }. 
\end{array}\right.}
&\\
&
\ar[lu]^-{\text{Yes}}  
\hspace{-5em}
{\begin{array}{@{}l@{}}
\underline{(\Z_4^3)\text{-fSPTs}}:\\[2mm]
\Omega^4_{Spin,\mathrm{Tor}}(B\Z_4^3)\\
=\Z_4^8 \times\boxed{\Z_2^4}
\end{array}}:
{\left\{\begin{array}{lll} \Z_2:& {\text{Arf}(\text{PD}(f_1\mod 2)\cap\text{PD}(f_2\mod 2))} \\&=(f_1\mod 2)\cup (f_2\mod 2)\cup\text{Arf}  \quad\quad\quad\quad [{\text{NAb}}] \\  
\Z_2: &  {\text{Arf}(\text{PD}(f_2\mod 2)\cap\text{PD}(f_3\mod 2))} \\&=(f_2\mod 2)\cup (f_3\mod 2)\cup\text{Arf}  \quad\quad\quad\quad [{\text{NAb}}]\\ 
\Z_2:&{\text{Arf}(\text{PD}(f_1\mod 2)\cap\text{PD}(f_3\mod 2))} \\&=(f_1\mod 2)\cup (f_3\mod 2)\cup\text{Arf} \quad\quad \quad\quad [{\text{NAb}}]\\
\Z_2:&{\tilde{q}_{\text{PD}(f_1\mod 2)\cap\text{PD}(f_2\mod 2)}(f_3\mod 2)} \\&=(f_1\mod 2)\cup (f_2\mod 2)\cup (f_3\mod 2)\cup\eta \quad\;\, [{\text{NAb}}]\\
\end{array}\right.}
&
}
\caption{
\fontsize{10}{11}\selectfont
We consider the map (\ref{eq:map-crystal-SPT})  from fSPTs with symmetries $G_o\times \Z_N$ ($G_o=(\Z_2)^2, \Z_2 \times \Z_4$ or $(\Z_4)^2$, $N=2$ or $4$) to
$\Omega^4_{Spin,\mathrm{Tor}}(B(\Z \times G_o))$,
which contains the crystalline fSPTs from $\Omega^{3}_{Spin,\mathrm{Tor}}(BG_o)$ subgroup.
The left hand side of arrows shows the classes associated to fSPTs with $\Z$-symmetry,
the right hand side of arrows shows the classes associated to fSPTs with finite abelian group symmetry.
The group-valued classifications whose generators associated to intrinsically fermionic  (spin-TQFTs) are boxed.
The arrows dressed with labels ``Yes'' mean that all fermionic states can completely map from the left to the right classes; thus the maps labeled ``Yes'' are injective. 
The arrows dressed with labels ``Partially'' mean that some fermionic states do not survive under the map; thus the maps labeled ``Partially'' are \emph{not} injective.
We \emph{warn} the readers, however, the maps from 4d fSPTs $\Omega^4_{Spin,\mathrm{Tor}}(B(\Z_N \times G_o))$ to crystalline 4d fSPTs $\Omega^{3}_{Spin,\mathrm{Tor}}(B(G_o))$ (a subgroup of $\Omega^4_{Spin,\mathrm{Tor}}(B(\Z \times G_o))$) are different, see Sec.~\ref{sec:cond-mat} for discussions.
 In particular, those topological invariants brace-labeled with \emph{non-stacking} can\emph{not} be obtained from a layer-stacking construction as a crystalline SPTs.
}
\label{table:ZxGo2}
}
\end{table}

\restoregeometry

\noindent
\underline{\emph{Fermionic topological terms. Stacking vs. Non-stacking}:}  In Table \ref{table:ZxGo} and \ref{table:ZxGo2}, we list down the classifications associated to $\Omega^n_{Spin, \mathrm{Tor}}(B(\Z \times G_o))$ or $\Omega^{n-1}_{Spin,\mathrm{Tor}}(B(G_o))$. The boxed classification groups are such that their generators are \emph{intrinsically fermionic} (i.e.\ they are \emph{not} 
generated by \emph{bosonic} Dijkgraaf-Witten topological term). There we only list down their intrinsically fSPT invariants.
Since $\Omega^n_{Spin, \mathrm{Tor}}(B(\Z \times G_o)) \supset \Omega^{n-1}_{Spin,\mathrm{Tor}}(B(G_o))$,
those fermionic topological terms that occur in the former (the right hand side of (\ref{eq:map-crystal-SPT})) but not in the latter  
(the left hand side of  (\ref{eq:map-crystal-SPT-2})'s) can\emph{not} be obtained from a 3d-layer-stacking construction as a crystalline 4d-SPT ---
We brace-labeled those topological terms with ``non-stacking'' label in Table \ref{table:ZxGo2}. The \emph{non-stacking} terms lack
the dependence on the $h:M^n\to B\Z$ (equivalently $h\in H^1(M^n,\Z)$) gauge field.

In Tables \ref{table:ZxGo} and \ref{table:ZxGo2},  ``$\Z_4\subset U(1)_1$-spin-CS'' 
means an  invertible (that is, depending on background gauge fields) $U(1)$ spin-CS theory with minimal non-zero level (i.e.\ level 1, in a proper normalization) with $U(1)$ broken to $\Z_4$ subgroup, see Sec.~\ref{sec:Z2Z4-234} for detailed discussions.

Interestingly, as shown in Table \ref{table:ZxGo} and \ref{table:ZxGo2}, 
it is \emph{not} possible to always map injectively from 
$ \Omega_4^{Spin}(B(\Z_2 \times G_o))\to \Omega_4^{Spin}(B(\Z \times G_o))$, see the arrows 
labeled with ``No'' or ``Partially.''\footnote{
For example, it is \emph{not} possible to obtain the $\Z_8$ classes generated by ${\hat{\gamma}_{\text{PD}(h)}(f)}$
in $\Omega^4_{Spin,\mathrm{Tor}}(B(\mathbb Z \times \mathbb Z_4))$ from the $\Z_4$ classes generated by  
${\delta_{\text{PD}(f)}(g,g)} =f\cup(g\cup \text{ABK})$
in $\Omega_4^{Spin}(B(\mathbb Z_2\times\mathbb Z_4))$, via obtaining a new the $\Z_2$ gauge field $g$ 
from the  $\Z$ gauge field $h$ by mod 2 reduction.
We can prove it is impossible by contradiction. 
 Suppose it is possible, then
\begin{equation}
\delta_{\text{PD}(f)}(h \mod 2, h \mod 2) \overset{?}{=} \hat{\gamma}_{\text{PD}(h)}(f)
 \mod 4
  \label{non-stacking-ex0}
\end{equation}
for any $f \in H^1(M^4,\Z_4)$, $h \in H^1(M^4,\Z)$ and any spin $M^4$. But then
one can take $M^4=S^1 \times M^3$, $h$ to be the generator of $H^1(S^1,\Z)$, and $f$ to
be in $H^1(M^3,\Z_4)$. Then the right hand side of the above formula (\ref{non-stacking-ex0}) becomes $\int_{M^3} f \mathcal{B} f$  
(where $B: H^1(M^4,\Z_4) \to H^2(M^4,\Z_4)$ is Bockstein morphism) and in general
is not zero, while the left hand side is identically zero in this setup (because
self-intersection of PD($h\mod 2$) inside PD($f$) is trivial), so we
have a contradiction. The
formula (\ref{non-stacking-ex0}) is false.}
However, it is \emph{always} allowed to map injectively from the $\Omega^4_{Spin,\mathrm{Tor}}(B(\Z_4 \times G_o)) \to \Omega^4_{Spin,\mathrm{Tor}}(B(\Z \times G_o))$.
One formal way to understand this fact is because Poincar\'e duals
(which physically correspond to domain walls) for elements of
$H^1(M,\Z_2)$ are in general non-orientable manifolds, while for both
$H^1(M,\Z_4)$ and $H^1(M,\Z)$ are oriented ($\Z_4$-orientation is equivalent to $\Z$-orientation). This is why one can reproduce
fSPTs with $\Z_4$, but not $\Z_2$, symmetry from the fSPTs with $\Z$ symmetry.

In Table \ref{table:ZxGo} and \ref{table:ZxGo2}, we put labels [Ab] (abelian) or [NAb] (non-abelian)
on the right-hand-side of some invertible spin TQFTs with $G$-symmetry (or $G$-fSPTs). 
What we mean there  is that the corresponding spin-TQFTs with
 dynamically gauged $G$, are  [Ab] (abelian) or [NAb] (non-abelian).
The criteria for determining, whether spin TQFTs with  dynamically gauged $G$ symmetry in 3d, 4d or above is [Ab] or [NAb],
have been given in Ref.~\cite{Wang:2018edf}'s Sec.~1.2.
 We briefly remind the readers our definition in Section \ref{sec:nAbTQFT}.

\subsection{Abelian vs. Non-Abelian gauged spin-TQFTs: Criteria} \label{sec:nAbTQFT}

In this work, the criteria we define for determining if the topological gauge theories, of finite gauge group $G$, 
are non-abelian [NAb] instead of abelian [Ab] for spacetime dimensions $n \geq 3$, are the following:

\noindent
\underline{\emph{Criteria}:} If (and only if, in our work), the partition function $ Z_\text{gauged}(M^n)$ of the gauged spin-TQFT defined in eqns. (\ref{gauged-TQFT-closed})
computed on a $n$-torus ($M=M^{n-1}\times S^1_+=T^{n-1}\times S^1_+=T^n$) is reduced to a smaller value from a particular power of the order finite gauge group, namely $|G|^{n-1}$,
then the  gauged spin-TQFT is non-abelian [NAb].

Physically, the partition function $ Z_\text{gauged}(T^n)$ represents the dimensions of Hilbert space, or equivalently the ground state degeneracy (GSD) on a $(n-1)$-torus 
(see various examples computed in \cite{Wang:2018edf}). If $ Z_\text{gauged}(T^n) <|G|^{n-1}$ is reduced,
this implies that the certain extended surface operator (physically, the open ends of a surface have anyonic extended/string excitations attached)
 have the \emph{quantum dimension}  $d_\alpha$ (associated to the local or non-local Hilbert space of such anyonic extended/string excitations $\alpha$) 
larger than 1, namely $d_\alpha > 1$. This non-abelian property can be seen, for example, 
from the quantum representation of the mapping class group MCG($T^{n-1}$)=SL($n,\mathbb{Z}$), in which case the modular $\mathcal{S}$-matrix (a generator of SL($n,\mathbb{Z}$))
will have at least one entry that satisfies $(\mathcal{S}_{0 \alpha}/\mathcal{S}_{0 0}) >1$.\footnote{See for example, Ref.~\cite{Wang2014oya1404.7854} on the computation of 
 modular $\mathcal{S}^{xyz}$-matrix (a generator of modular SL($3,\mathbb{Z}$) data)
 that shows this non-abelian property with $(\mathcal{S}_{0 \alpha}/\mathcal{S}_{0 0}) >1$ for certain non-abelian TQFTs and a certain anyonic string excitation from a surface operator labeled by $\alpha$.
 More examples of non-abelian TQFTs are given in Ref.~\cite{Wang:2018edf} and in Table \ref{table:ZxGo} and \ref{table:ZxGo2}.}
The  non-abelian property also
implies that a certain 3-loop braiding process will have non-abelian unitary matrix acting on the eigenstate-vector of the Hilbert space after the completion of adiabatic evolution of braiding process. 

From Tables \ref{table:ZxGo} and \ref{table:ZxGo2}, we learn, example by example,  that the fact that the dynamically gauged spin-TQFTs defined by $ Z_\text{gauged}$  
become non-abelian spin-TQFTs, when the original un-gauged invertible $G$-spin-TQFT theories (with abelian $G$) involve either of the following: 
\begin{enumerate}
\item an odd multiple of Arf invariant (generating $\Z_2$), 
\item an odd multiple of ABK (generating $\Z_8$),
\item an odd multiple of $\eta$ (generating $\Z_2$ class) 
\item an odd multiple of $\tilde \eta$ (generating $\Z_4$ class, see Ref.~\cite{Wang:2018edf}'s Sec.~5.)
\end{enumerate}
In physics, this means that the above non-abelian TQFTs must  
 induce lower dimensional 2d spin-TQFT (such as based on dimensional reduction \cite{Wang:2018edf} or compactification)
that involve the Kitaev's Majorana fermionic chain \cite{2001KitaevWire}, which corresponds to  2d Arf or ABK invariants.
We give more accounts on this phenomena in the next Section \ref{sec:cond-mat}.

To summarize Section \ref{sec:crystalline}, 
the physical
meaning of $\Z$ global symmetry is a discrete translation symmetry in one extra spatial dimension. 
Applications to realistic systems of $\Z \times G_0$-crystalline fSPTs make sense when we 
consider the spin bordism group of $G_0$ up to dimension $d-1$.
In Section \ref{sec:cond-mat}, we would like to explain all of the above in a more down to earth way 
and in a setting suitable for condensed matter community.

\section{Interpretation of the results 
in 
quantum matter
and more
}
\label{sec:cond-mat}

Now we reorganize our results into an alternative understanding: in terms of the setting and the language of current developments of condensed matter and topological 
quantum matter, and also in comparison to the tools developed by physics setting (versus the mathematical settings).

We will relate various fSPTs protected by internal onsite symmetry, to crystalline-fSPTs protected by the translational symmetry of infinite $\Z$ integer symmetry (i.e.\ space group symmetry).
We had presented the summary of new cobordism calculations involving the $B\Z=S^1$
and the classifications of  crystalline-fSPTs,
altogether in Sec.~\ref{subsec:crystalline-cobordism}.
Note that we explain the first example in more details in
Sec.~\ref{sec:commentZ2Z4}, while we go through later examples rather quickly based on the similar strategy. 

Then we will discuss more general 
crystalline-fSPTs (c-fSPT). We first clarify the differences and the meanings of the crystalline symmetry and the internal symmetry in Sec.~\ref{subsec:c-sym-int-sym}.
Then we include various examples of crystalline symmetries and their complete classifications via cobordism approach: 
\begin{itemize}
\item
Reflection/Mirror c-fSPT in Sec.~\ref{subsec:Reflection/Mirror}.
\item Inversion c-fSPT in Sec.~\ref{subsec:Inversion}.
\item Rotation c-fSPT in Sec.~\ref{subsec:Rotation}.
\end{itemize}
Then we discuss the partially gauging of fSPTs to obtain 
fermionic SETs (Symmetry Enriched Topologically ordered states) in Sec.~\ref{subsec:fSET}. 
These fermionic SETs are fermionic \emph{gauge} theories but enriched by the additional \emph{global} symmetry protection.

\subsection{4d $\Z_4 \times \Z_2$ and crystalline-$\Z \times \Z_2$-fSPTs }

\label{sec:commentZ2Z4}

Here we particularly aim to understand the 4d cobordism/SPT invariant ${\delta_{\mathrm{PD}(f)}(g,g)}$
given in \eqn{eq:topZ4Z2} and in Table \ref{table:ZxGo}, where
we have the spin $n$-manifolds $M^d$ with maps $f:M\to B\Z_4$, $g:M\to B\Z_2$ (equivalently, $f \in H^1(M, \Z_4)$ and $g \in H^1(M, \Z_2)$). 
Following the informal physical notation in \cite{Wang:2018edf},
this corresponds to the partition function\footnote{As discussed in Section \ref{sec:computation}, in such expression we assume that there exists a representation of Poincar\'e dual of $f$ by an immersed manifold $\text{PD}(f)$. }
\begin{equation}  \label{eq:a1a2ABK}
\exp(\frac{i \pi k}{2}  \int_{M^4}(f \cup (g\cup \text{ABK})) \equiv  
\exp(\frac{i \pi k}{2} \int_{\text{PD}(f)} g|_{\text{PD}(f)} \cup \text{ABK}),
\end{equation} 
where $k \in \Z_4$ and we focus on the minimal generator with $k=1$ below.

\subsubsection{Construction from 2d to 3d to 4d}

\label{sec:Z2Z4-234}

How do we construct this 4d $\Z_4 \times \Z_2$-fSPTs starting from lower dimensions, say from 2d and 3d?

\underline{\emph{Decorated domain wall construction}}:
We can start from the 2d $\Z_2^T$-fSPTs (with $\Z_2^T \times \Z_2^f$ symmetry) classified by $\Omega_2^{Pin^-}(pt)=\Z_8$, whose generator is the ABK invariant (see a recent exploration \cite{Debray2018wfz1803.11183} and reference therein) ---
Physically this is the so-called 2d Kitaev's fermionic chain \cite{2001KitaevWire, 0904.2197FK} with each isolated 0+1D edge on an open chain has a Majorana zero mode.
Now we can place the ABK invariants (i.e.~Kitaev chains) on all the 2d $\Z_2$-symmetry-breaking domain walls of 
3$d$ 
$\Z_2$-fSPTs (with $\Z_2 \times \Z_2^f$ symmetry)
classified by Pontryagin dual group to $\Omega_3^{Spin}(B \Z_2)=\Omega_2^{Pin^-}(pt)=\Z_8$. 
Then we can restore the full $\Z_2 \times \Z_2^f$ symmetry by, in condensed matter language, proliferating or condensing the 2d domain wall. This is known as the \emph{decorated domain walls proliferation construction} of SPTs, proposed by \cite{ChenLu1303.4301}.\footnote{See a field theory derivation of \emph{decorated domain walls proliferation construction} of SPTs and topological terms related to bosonic TQFTs and Dijkgraaf-Witten gauge theory 
in Ref.~\cite{GuWW2015lfa1503.01768}. Similar construction for fermionic SPTs is discussed in Ref.~\cite{Gaiotto:2017zba}}
 Although the 3$d$ $\Z_2$-fSPTs is placed on an oriented spin 3-manifold, the 2d domain wall can have an induced pin$^-$ structure thanks to Smith isomorphism \cite{kirby1990pin,Kapustin:2014dxa}:
 $\Omega_3^{Spin}(B \Z_2) \to  \Omega_2^{Pin^-}(pt)$. This construction matches 
the 3d partition function $\exp(\frac{i \pi k'}{4}  \int_{M^3} (a\cup \text{ABK}))$ with $k'\in \Z_8$ \cite{Putrov:2016qdo}.

 Continuing from 3d $\Z_2$-fSPTs, we can further construct both
 the 4d crystalline-$\Z \times \Z_2$-fSPTs and 4d $\Z_4 \times \Z_2$-fSPTs by stacking 3d systems layer by layer.
 In that case, we need to have $k'=2$ as 2-copies of ABK or Kitaev chain proliferated-domain walls in order to construct 
$k'=2 \in \Z_8$ of 3d $\Z_2$-fSPTs. 
Physically, a single $k'=2$ 3d $\Z_2$-fSPTs has two copies of $p_x+i p_y$ and $p_x-i p_y$ superconductors,\footnote{
Hereby the chiral and anti-chiral-$p$-wave superconductors, or the
$p_x\pm ip_y$ superconductors, we mean the Cooper pairing of two fermions are in the $p_x\pm ip_y$-orbital of $p$-orbital pairing states 
($p$ in terms of the angular momentum $\ell=1$ in the spherical harmonics, or the $p$ in the $s,p,d,f$, etc. of atomic orbitals). 
The pairing function results in the superconductor order parameter $\Delta(\mathbf{k})\propto k_x\pm ik_y$, where $k_x,k_y$ are spatial momentum in the $x,y$-directions.
}
 so called the chiral and anti-chiral-$p$-wave superconductors. The 2d boundary of each chiral-$p$-wave TSC (living on a 3-manifold) 
hosts a conformal field theory (CFT) described by a single free left moving the real Majorana-Weyl fermion with chiral central charge $c_-=1/2$.
So for a single $k'=2$ 3d $\Z_2$-fSPTs boundary, there are left-chiral central charge $c_L=2 \times 1/2= 1$ and 
right-chiral central charge $c_R=2 \times 1/2= 1$, combining into a 1+1D Dirac spinor fermion (with a total $c_-= c_L - c_R=0$). 
The corresponding mathematical 3d bulk spin-TQFTs (as dynamically gauged fSPTs) and 2d boundary theories are given in Section 8 and Table 2 of \cite{Putrov:2016qdo}.
Although a total central charge of 2d edge states $c_-= c_L - c_R=0$ is non-chiral, but the global $\Z_2$-symmetry assignment is actually 
\emph{chiral, anomalous} and \emph{non-onsite} on the 2d edge modes.

\underline{\emph{Layer-stacking}}:
Now we can, along the spatial $z$-direction, stack integral layers of $k'=2$ of 3d $\Z_2$-fSPTs placed at the spatial $x$-$y$-direction, following the idea of \cite{2018arXiv180408628F}.
In this way, we can construct a 4d crystalline-$\Z \times \Z_2$-fSPTs placed in the spatial $x$-$y$-$z$ space, 
protected by the internal $\Z_2$-symmetry and the spatial lattice translational symmetry $\Z$ along the $z$-direction.
How do we classify the interacting 4d crystalline-$\Z \times \Z_2$-fSPTs for this particular construction?
Mathematically this question can be answered by computing $\Omega^{Spin}_n(B((\Z_2)\times\Z))$, which we show in Theorem \ref{thm:Z2Z}.
Physically this question can be answered by asking how many (say, a number of $\tilde k$) layers of this
$k'=2$ of 3d $\Z_2$-fSPTs do we need in order to fully gap out the boundary gapless modes without breaking any global symmetries?
Namely,  how many layers we need to add with non-perturbative interactions among 
these $\tilde k$ copies of $k'=2$ of 2d boundary CFTs, in order to fully generate the energy gap for a symmetric topological gapped boundary?
Our answer is $\tilde k=4$.
Because the $\tilde k \cdot k' = 4 \cdot 2 =8$, 
since 8 layers of 3d $\Z_2$-fSPTs indeed can be fully gapped, thanks to their classification $\Omega_3^{Spin}(B \Z_2)=\Omega_2^{Pin^-}(pt)=\Z_8$ 
(physically, derived from gapping the boundary Majorana modes).
By adding the interactions among the neighbor 4 layers ($\tilde k=4$) along $z$-direction, we actually break the $\Z$-translation down to $4\Z$-translation symmetry,
which again is still a lattice translation symmetry (redefined the $4\Z$ as a new integer $\Z$ by rescaling the translational lattice constant by 4 times).
So physically we predict that there must be at least $\Z_{\tilde k}= \Z_4$-class for 
4d crystalline-$\Z \times \Z_2$-fSPTs thus also for $\Omega^{Spin}_4(B((\Z_2)\times\Z))$.
Indeed the answer agrees between our math result (Theorem \ref{thm:Z2Z}'s 4d $\Z_8$ classification\footnote{Which is 
generated by  $\int_{M^4} (g \cup \text{ABK}) \cup  (h \mod 8)$ with the $\Z_2$ gauge field $g$ and the $\Z$ gauge field of $h$ along the $z$-direction.} contains the $\Z_{\tilde k}= \Z_4$ normal subgroup) 
and physics arguments.

Moreover, we can view this 4d crystalline-$\Z \times \Z_2$-fSPTs for the $\nu =1 \in \Z_{\tilde k}= \Z_4$-class
as another 4d fSPTs protected only by internal symmetries, without the need of translational 
symmetry.\footnote{This means we can also re-interpret the role of crystalline-$\Z$ lattice translation as a new internal symmetry instead.} 
Which additional internal symmetry is required? The answer is constrained by
what the internal symmetry can still be preserved when we add non-perturbative interactions among
 the $z$-directional neighbors of 4d crystalline-$\Z \times \Z_2$-fSPTs (4 layers of  $k'=2$ of 3d $\Z_2$-fSPTs).
 For 8 Majorana zero modes in 1d, or 8 left-moving + 8 right-moving (8L+8R) chiral Majorana-Weyl fermions in 2d, 
 Ref.~\cite{0904.2197FK} shows that, under quartic-interaction fermion gapping terms, there is an internal symmetry of $SO(7) \subset SO(8)$ rotating between Majorana modes 
  that can still be preserved (also there is the obvious fermion parity $\Z_2^f$-symmetry preserved).
 Apparently, the $SO(7)$ rotates between 8L+8R Majorana-Weyl fermions in 2d, 
 while we wish to keep the subgroup of $SO(7)$ acting \emph{only} among each 2L+2R Majorana-Weyl fermions,
as a proper internal symmetry. Thus, we can take the intersection between the 2-flavor symmetry of left and right $SO(2)_L \times SO(2)_R  \subset SO(8)$ and the  $SO(7)$.
The finite group $\Z_4 \times \Z_2$-internal symmetry of 4d $\Z_4 \times \Z_2$-fSPTs is within the overlapped subgroup.
Therefore,  we derive physically the relation of constructions between 4d crystalline-$\Z \times \Z_2$-fSPTs and 4d $\Z_4 \times \Z_2$-fSPTs.
Formally, 
Table \ref{table:ZxGo}'s map labeled ``Yes'' 
for $\Omega^4_{Spin,\mathrm{Tor}}(B(\Z_4 \times \Z_2)) \to \Omega^4_{Spin,\mathrm{Tor}}(B(\Z \times \Z_2))$ is given by
\bea \label{eq:map-8-to-4}
\exp( \frac{2 \pi i}{8}  k' {\beta_{\text{PD}(h)}(g)} ) =\exp( \frac{2 \pi i}{4}  \tilde k \,{\delta_{\text{PD}(f)}(g,g)} ), 
\eea
with $f=h\mod 4$ and $k' \in \Z_8$, $\tilde{k} \in \Z_4$. The correspondence is that $\tilde{k} =0,1,2,3 \mod 4$ is mapped to $k'=0,2,4,6 \mod 8$. 
Namely,  the map is
\bea
\Z_4 \ni \tilde{k}\mapsto k' =2\tilde{k} \in \Z_8.
\eea
For this reason, we see that  $\tilde{k} \in \Z_4$ from 2d-to-3d-to-4d construction starts from an even number of Kitaev chains. The corresponding gauged 4d spin-TQFT will be abelian [Ab].

Although 4d crystalline-$\Z \times \Z_4$-fSPTs contains a $\Z_8$ classification subgroup (Theorem \ref{thm:Z4Z}), it is distinct from the $\Z_8$ subgroup of crystalline-$\Z \times \Z_2$-fSPTs. 
The reason is the following. The former $\Z_8$-classes are associated to the 3d $\Z_4$-fSPTs, which can be described by ungauged (i.e.\ depending on a background gauge field) $U(1)$-spin-CS theory given in (\ref{eq:KmatCS}) with a single $U(1)$ broken down to $\Z_4$ subgroup and $K=(p)$, $p\in \Z_8$.
The corresponding $\Z_8$-classes of dynamically gauged 3d $\Z_4$-fSPTs become $\Z_8$-classes of gauged spin-CS theory
with $U(1)^2$ gauge group and $K$-matrix $(\begin{smallmatrix} 0 & 4 \\ 4 & p\end{smallmatrix})$ with $p \in \Z_8$, potentially with an additional fully gapped fermionic sector (a fermionic trivial tensor product state).
This spin-CS theory is therefore distinct from the ungauged and gauged theory of (\ref{eq:a1a2ABK}).
Thus, it turns out that we \emph{cannot}
 relate the constructions between 4d crystalline-$\Z \times \Z_4$-fSPTs and 4d $\Z_4 \times \Z_2$-fSPTs.
This explains the fact that in Table \ref{table:ZxGo}'s the map is labeled ``No'' 
for $\Omega^4_{Spin,\mathrm{Tor}}(B(\Z_2 \times \Z_4)) \to \Omega^4_{Spin,\mathrm{Tor}}(B(\Z \times \Z_4))$.

\subsubsection{Symmetric anomalous gapped boundary TQFT}

We would like to comment the symmetric gapped boundary topological orders (TQFT) of SPTs (see a systematic study on this subject in  \cite{Wang2017loc1705.06728}).
Recently, Ref.~\cite{2018arXiv180408628F} derives a 3d anomalous $\Z_4$-gauge theory preserving the full global symmetry and living on the boundary of 4d $\Z_4 \times \Z_2$-fSPTs.
Namely Ref.~\cite{2018arXiv180408628F}'s 3d anomalous $\Z_4$-gauge theory can capture all the 't Hooft anomalies of the boundary of 4d $\Z_4 \times \Z_2$-fSPTs.

\underline{\emph{Symmetry Extension and Dimensional Reconstruction from 2d to 3d to 4d}}:
We first apply some simple arguments to construct a potential symmetry-preserving gapped boundary state.
Our strategy is to firstly follow the so-called symmetry extension construction developed in Ref.~\cite{Wang2017loc1705.06728}.
It is known that for \emph{any} bosonic types of 't Hooft anomalies of ordinary 0-form finite group global symmetries,
there \emph{always exists} a symmetry-extension constructed gapped boundary when spacetime dimensions $\geq 2$, proven in \cite{Wang2017loc1705.06728}, 
also later confirmed in a more mathematical setting in  \cite{2017arXiv1712.09542Yuji}.
 However, we are facing now the fermionic types of 't Hooft anomalies.
It is not guaranteed that symmetry extension \cite{Wang2017loc1705.06728} necessarily apply to all these cases.
Following Sec.~\ref{sec:Z2Z4-234}, we can reduce the problem from 4d to 2d,
by asking how to obtain symmetry extended gapped boundary for 2 layers of Kitaev chains ($k' = 2$ class in  $\Omega^2_{Pin^-,\text{Tor}}(\text{pt})=\Z_8$).
The attempt was made recently in Ref.~\cite{Prakash2018ugo1804.11236} (Section 4.3),
which finds that the finite symmetry group of a finite extension, say a group $N$ (usually a normal subgroup), 
must \emph{not commute} with the fermion parity symmetry $\Z_2^f$.
Thus $N$ does \emph{not commute} with the $n$-dimensional spacetime symmetry $Spin(n)$ group  \cite{Prakash2018ugo1804.11236}!
Similar outcome happens to 2 layers of 3d $\Z_2$-fSPTs ($k' = 2$ class in $\Omega^3_{Spin,\mathrm{Tor}}(B \Z_2)=\Z_8$),
  where the finite symmetry group $N$ of a finite extension must not commute with the fermion parity symmetry $\Z_2^f$.
Going from 2d to 3d back to 4d, for 4d $\Z_4 \times \Z_2$-fSPTs, if symmetry extension gapped boundary construction applies, 
its finite extension $N$ still cannot commute with the fermion parity symmetry $\Z_2^f$.
This indicates that in putative $N$ gauge theory (after gauging the extension $N$), the anyon (at the ends of open line operators) can be \emph{permuted} under the
$\Z_4 \times \Z_2$-symmetry transformation. This 2d-to-3d-to-4d argument leads to agreement with the anyon-permuting anomalous global symmetry, on the 3d boundary of 4d SPTs, emphasized in 
\cite{2018arXiv180408628F}!

\subsection{4d $\Z_4 \times \Z_4$ and crystalline-$\Z \times \Z_4$-fSPTs} 

\underline{\emph{Construction from 2d to 3d to 4d}}:
To obtain 4d crystalline-$\Z \times \Z_4$-fSPTs in Table \ref{table:ZxGo}, we can stack the 3d $\Z_4$-fSPTs which have $\Omega^3_{Spin,\text{Tor}}(B (\Z_4))=\Z_8 \times \Z_2$-classes.
The $\Z_8$-sub-classes are can be realized as the level 1 $U(1)$-spin-CS theory given in (\ref{eq:KmatCS}) with
a symmetric bilinear form $K=(1)$ and $U(1)$ broken to $\Z_4$ subgroup
(See the construction in the last paragraph of Sec.~\ref{sec:Z2Z4-234}.)
The generator of the $\Z_2$ subgroup can be obtained from the decoration and proliferation of 2d Kitaev chain into the $\Z_4$-symmetry breaking domain wall of 3d $\Z_4$-fSPTs.
Thus, in the above construction, we can gain all 4d crystalline-$\Z \times \Z_4$-fSPTs from lower dimensions via the \emph{stacking} and \emph{proliferation} of condensed matter methods.
 
In fact, the $\Z_2$-classes in $\Omega^2_{Spin,\text{Tor}}(\text{pt})$ are directly mapped to the $\Z_2$-classes in $\Omega^3_{Spin,\text{Tor}}(B \Z_4)$,
and then to the $\Z_2$-classes in $\Omega_{Spin,\text{Tor}}^4(B((\Z_4)\times\Z))$, which are also the image of the $\Z_2$-classes in $\Omega_{Spin,\text{Tor}}^4(B((\Z_4)^2))$ w.r.t.\ the map (\ref{eq:map-crystal-SPT}).
All these $\Z_2$ sub-classifications are related to the fact that two layers of 1+1D Kitaev chains (i.e.\ 2 Arf) becomes trivial.
Physically, it means that the two nearest-neighbored layers of Kitaev chains (in 2d, 3d or 4d) can be fully gapped by adding interactions but without breaking any symmetry.

\subsection{4d $\Z_2^2 \times \Z_4$, crystalline-$\Z \times (\Z_2)^2$ and $\Z \times \Z_2 \times \Z_4$-fSPTs}

\label{sec:commentZ22Z4}

\underline{\emph{Construction from 2d to 3d to 4d for $\Z \times (\Z_2)^2$-fSPTs}}:
We first construct the intrinsically fermionic  4d $\Z \times (\Z_2)^2$-fSPTs, which have $\Z_8^2 \times \Z_4$ classes, shown in Table \ref{table:ZxGo2}.
Both $\Z_8$ subgroups obviously can be constructed from the $\Z_8$-classes of Kitaev chain (ABK invariants)
 from 2d to 3d to 4d via  \emph{stacking} and \emph{proliferation}  as shown in Sec.~\ref{sec:Z2Z4-234}, which we will not repeat again.
{
The generator of $\Z_4$ classes has an interesting 4d topological term ${\delta_{\text{PD}(h)}(g_1,g_2)} =h\cup (g_1\cup g_2\cup\tilde{\eta})$,
which can be constructed by stacking (see $h$) along the $z$-direction a $\Z_4$ generator 
$g_1\cup g_2\cup\tilde{\eta}$ (from $\Z_4 \in \Omega^{Spin}_3(B((\Z_2)^2))$). 
Each 3d slice of the gauged $g_1\cup g_2\cup\tilde{\eta}$ is a 3d non-abelian spin-TQFT, due to the ground state degeneracy (GSD) reduction on the 3-torus. 
(See Sec.~\ref{sec:nAbTQFT} and Ref.\cite{Wang:2018edf}'s Sec.~5's calculation on this GSD.)
Thus, the gauged version of this minimal generator of $\Z_4$ class of 4d fSPT is also a non-abelian [NAb] spin-TQFT.
Therefore we construct above all fermionic states of $\Z_8^2 \times \Z_4$ classes of 4d $\Z \times (\Z_2)^2$-fSPTs from the 2d-to-3d-to-4d procedure.
}
 
Next, given 4d $\Z_2^2 \times \Z_4$-fSPTs, we construct 4d  $\Z \times (\Z_2)^2$-fSPTs.
Namely, we would like to know the fermionic states mapped in Table \ref{table:ZxGo}
from the ${\Z_4^3}$-classes of $\Omega^4_{Spin,\text{Tor}}(B(\Z_2^2\times\Z_4))$ to the
${\Z_4\times\Z_8^2}$-classes of
$\Omega^4_{Spin,\text{Tor}}(B(\Z \times\Z_2^2))$.
The classes in the  $\Z_4\to\Z_4$ subgroups are obviously and directly related to the 3d  theory described earlier associated to $g_1\cup g_2\cup\tilde{\eta}$.
The remaining classes are related by to copies of the map $\Z_4\to\Z_8$, the same as (\ref{eq:map-8-to-4}), where
$g$ in (\ref{eq:map-8-to-4}) can be replaced to either $g_1$ or $g_2$ to get two different $\Z_8$ subgroups.
Then for $k' \in \Z_8$ and $\tilde{k} \in \Z_4$ as in (\ref{eq:map-8-to-4}), the correspondence again is $k =0,1,2,3 \mod 4$ mapped to $\tilde k=0,2,4,6 \mod 8$. 
The $(\Z_4)^2$-classes of spin-TQFTs obtained from dynamical gauged fSPTs, are again abelian [Ab].

\underline{\emph{Construction from 2d to 3d to 4d for $\Z \times \Z_2 \times \Z_4$-fSPTs}}:

We first construct the intrinsically fermionic 4d $\Z \times  \Z_2 \times \Z_4$-fSPTs that are crystalline fSPTs, which has $\Z_8^2 \times \Z_2^2$ classes 
(from $\Omega^3_{Spin,\text{Tor}}(B(\Z_2\times\Z_4))=\Z_8^2 \times \Z_2^2$, box labeled), shown in Table \ref{table:ZxGo2}.
\begin{itemize}

\item One $\Z_8$ classification subgroup (involving ABK invariant) obviously can be constructed from the $\Z_8$-class generated by Kitaev chain from 2d to 3d to 4d via  \emph{stacking} and \emph{proliferation}.
The generator ${\beta_{\text{PD}(h)}(g)} =h\cup (g\cup\text{ABK})$ is explicitly obtained from 
stacking the generator of $\Z_8$-classes of 3d $\Z_2$-fSPTs (given by a topological term $g\cup\text{ABK}$) along the
$z$ direction.

\item The generator of another $\Z_8$ classification subgroup (involving a spin-CS) can be constructed from the $\Z_8$ class of a $K=(1)$
ungauged $U(1)$-spin-CS theory, given in (\ref{eq:KmatCS}), 
  with $U(1)$ broken to $\Z_4$
 as shown in Sec.~\ref{sec:Z2Z4-234}, which we will not repeat again.
 
 \item The non-trivial $\Z_2$-class 
${\text{Arf}(\text{PD}(f\mod 2)\cap\text{PD}(h\mod 2))} 
=(h\mod 2)\cup(f\mod 2))\cup\text{Arf}$
can be constructed via the 2d-3d-4d procedure: from the 2d Kitaev chain via 
 Arf invariant generating $\Omega^2_{Spin,\text{Tor}}(pt)=\Z_2$, we can map it to the generator of $\Omega^3_{Spin,\text{Tor}}(B \Z_4)=\Z_2$, then to
the generator of $\Z_2$ subclass in $\Omega^4_{Spin,\text{Tor}}(B (\Z \times \Z_4))$. This fSPTs is already discussed in Sec.~\ref{sec:Z2Z4-234}.
 
 \item Another non-trivial $\Z_2$ class  ${\tilde{q}_{\text{PD}(f\mod 2)\cap\text{PD}(h\mod 2)}(g)}$ ${=(h\mod 2)\cup (f\mod 2)\cup g\cup\eta}$ involves
all $\Z$-, $\Z_4$- and $\Z_2$-gauge fields. Consider the 1d intersection of the $\Z$- $\Z_4$- and $\Z_2$-symmetry-breaking domain walls with trapped  $\eta$ term 
(i.e.\ fermionic mode trapped at the 1d (=0+1D) intersection of domain wall decorated by the induced spin structure \cite{GuWW2015lfa1503.01768}) and 
then proliferate it to restore the full symmetry. This gives rise to the desired 4d fSPTs.
 \end{itemize}

 Next we would like to relate the above $\Z_8^2 \times \Z_2^2$ classes of intrinsically fermionic 4d $\Z \times  \Z_2 \times \Z_4$-crystalline 
 fSPTs (l.h.s.) to the ${\Z_4^3}$ classes of 4d $\Z_2^2 \times \Z_4$-fSPTs (r.h.s., within $\Omega^4_{Spin,\text{Tor}}(B(\Z_2^2\times\Z_4))$).
Depend on the meaning of $\Z_2$ gauge field $g$ (treated as $g_1$ or $g_2$), we can use the same map  
as (\ref{eq:map-8-to-4}),
taking $\tilde{k} \in \Z_4$ to $k'=2\tilde{k} \in \Z_8$. 
Moreover, we can map between one of the $\Z_4$ of r.h.s. and the $\Z_2$ of r.h.s. is given by:
\bea
\exp( \frac{2 \pi i }{2} p \,\tilde{q}_{\text{PD}(f\mod 2) \cap \text{PD}(h\mod 2)} (g)) = \exp( \frac{2 \pi i }{4} p' \,  \delta_{\text{PD}(f\mod 2)} (g,g') ), 
\eea
with $g'=h\mod 2$ and $p \in \Z_2$, $p' \in \Z_4$. We find that $p = p' \mod 2$, so the map $\Z_4\to \Z_2$ is given by mod 2 reduction. Alternatively, we can say that the topological term $\delta_{\text{PD}(f\mod 2)} (g,g')$ maps to $2\tilde{q}_{\text{PD}(\mod 2) \cap \text{PD}(h\mod 2 )}(g)$.
In summary, we have the r.h.s. fSPTs map to the l.h.s. fSPTs:
 \bea
 \Z_4 \times \Z_4 \times \Z_4 \ni (k'_1, k'_2, p')  \mapsto (0,\; p,\; k_1,\; 0)= (0 ,\; p' \text{ mod } 2, \; 2 k'_1,\; 0)  \in \Z_2 \times \Z_2 \times \Z_8 \times  \Z_8  . 
\eea
This explains the map (labeled ``Partially'') from 4d $\Z_2^2 \times \Z_4$- 
fSPTs to the 4d $\Z\times \Z_2 \times \Z_4$-crystalline-fSPTs.

\subsection{4d $\Z_2 \times \Z_4^2$, crystalline-$\Z \times \Z_2 \times \Z_4$ and $\Z \times (\Z_4)^2$-fSPTs}

\label{sec:commentZ2Z42}

We had constructed the intrinsically fermionic 4d $\Z \times  \Z_2 \times \Z_4$-fSPTs that are crystalline fSPTs with $\Z_8^2 \times \Z_2^2$ classes in
Sec.~\ref{sec:commentZ22Z4} from the 2d-to-3d-to-4d stacking procedure. By writing down the topological terms,
in Table \ref{table:ZxGo2}, we can see explicitly the injective map (labeled ``Yes'') between  
 the 4d $\Z_2 \times \Z_4^2$-fSPTs and 4d $\Z \times  \Z_2 \times \Z_4$-crystalline fSPTs via
 \begin{multline}
\Z_2 \times \Z_2 \times \Z_4 \times \Z_4 \ni ( p_1, p_2, k'_1, k'_2) 
 \mapsto ( p_1, p_2, k_1, k_2)= (p_1, p_2, 2k'_1, 2k'_2)  \in \Z_2 \times \Z_2 \times \Z_8 \times  \Z_8. \; 
\end{multline} 
 
 \underline{\emph{Construction from 2d to 3d to 4d for $\Z \times  (\Z_4)^2$-fSPTs}}:
We now construct the intrinsically fermionic 4d $\Z \times (\Z_4)^2$-fSPTs, which form $\Z_2^3 \times \Z_8^2$ classification, shown in Table \ref{table:ZxGo2}.
The $\Z_2^3$-classes of 4d fSPTs involve either the 2d Arf invariant or the 1d $\eta$ invariant. All these can be obtained from the decoration and proliferation of the domain-wall trapping 1+1D Kitaev chain or the 0+1D fermionic mode (with an induced spin structure along an $S^1$). We shall not repeat these constructions.

 Next we construct the map between the above $\Z_2^3 \times \Z_8^2$ classes of intrinsically fermionic 4d $\Z \times  (\Z_4)^2$-crystalline 
 fSPTs (l.h.s.) and the ${\Z_2^2 \times (\Z_4)^2}$ classes of 4d $\Z_2 \times \Z_4^2$-fSPTs (r.h.s., within $\Omega^4_{Spin,\text{Tor}}(B(\Z_2 \times \Z_4^2))$).
In Table \ref{table:ZxGo2}, we can see explicitly the map (labeled ``Partially'') between  
 the 4d $\Z_2 \times \Z_4^2$-fSPTs and 4d $\Z \times    (\Z_4)^2$-crystalline fSPTs via
\bea
 \Z_2 \times \Z_2 \times \Z_4 \times \Z_4 \ni ( p_1, p_2, k'_1, k'_2)
 \mapsto ( p_1, 0, p_2, 0, 0)  \in \Z_2 \times \Z_2 \times \Z_2 \times \Z_8 \times  \Z_8. 
\eea

\subsection{4d $\Z_4^3$ and crystalline-$\Z \times (\Z_4)^2$-fSPTs}

\label{sec:trans-cfSPT-5}

We had constructed the intrinsically fermionic 4d $\Z \times (\Z_4)^2$-fSPTs that are crystalline fSPTs with $\Z_2^3 \times \Z_8^2$ classes in
Sec.~\ref{sec:commentZ2Z42}.
By writing down the topological terms,
in Table \ref{table:ZxGo2}, we see explicitly the surjective map (labeled ``Yes'') between  
 the 4d $(\Z_4)^3$-fSPTs and 4d $\Z \times  (\Z_4)^2$-crystalline fSPTs via
 \begin{multline}
 ( p, p_1, p_2,  p_{123} )\in  \Z_2 \times \Z_2 \times \Z_2 \times \Z_2 
 \to ( p, p_1, p_2,  p_{123},0,0 )  \in \Z_2 \times \Z_2 \times \Z_2 \times \Z_2 \times  \Z_8 \times  \Z_8. \; 
\end{multline} 
Strictly speaking the first $p \in \Z_2$, the coefficient multiplying 
${\text{Arf}(\text{PD}(f_1\mod 2)\cap\text{PD}(f_2\mod 2))}={(f_1\mod 2)\cup (f_2\mod 2)\cup\text{Arf}}$,
is from a \emph{non-stacking} 4d SPTs, since it does not involve the $h$-gauge field.
This generator, however, maps to the generator of the first and the same $\Z_2$ group in the 4d $(\Z_4)^3$-fSPTs.
Note that here all dynamically gauged theories become non-abelian [NAb].

\subsection{Crystalline Symmetry $G_c$ v.s. Internal Symmetry $G_o$}

\label{subsec:c-sym-int-sym}

In previous sections, we mainly consider our spin TQFTs as either the fermionic SPTs (fSPT) protected by onsite internal global symmetry $G=G_o$,
or considering their fermionic finite group $G$ gauged theory. In this section \ref{subsec:c-sym-int-sym} and later on, we will map our results and classifications obtained from cobordism theory
to other new types of fSPTs protected by crystalline global symmetry $G=G_c$. These crystalline fSPTs (c-fSPT) not only can include
the translational symmetry of the space group (discussed in Sec.~\ref{sec:crystalline} and Sec.~\ref{sec:commentZ2Z4} to ~\ref{sec:trans-cfSPT-5}), but also other crystalline symmetries.
Crystalline symmetries, of great interest and applicable to condensed matter system and topological phases  \cite{2011PhRvLFu}, include:
\begin{itemize}
\item
Space group, in crystallography, which means the symmetry of the underlying lattice and crystal. In 3 dimensional spatial lattice, 
there are 230 possible space groups (e.g. \cite{Thorngren2018wwt1612.00846, 2017Po1703.00911} and references therein). 
Each element contains a collection of symmetry operations, like translation, glide, skew axis, etc.
\item Point group (e.g. \cite{PRX1604.08151, 2017PhRvB1705.09243} and references therein on their SPTs) 
which corresponds to the symmetries or isometries that keep at least one point fixed on the underlying lattice and crystal.
Point group is the quotient group of {the space group} by the {translational symmetry group}.
Here we will focus on the reflection/mirror c-fSPT in Sec.~\ref{subsec:Reflection/Mirror},
inversion c-fSPT in Sec.~\ref{subsec:Inversion},
and the rotation c-fSPT in Sec.~\ref{subsec:Rotation}.
\end{itemize}

Part of our work is inspired by Ref.~\cite{Thorngren2018wwt1612.00846}'s 
Crystalline Equivalence Principle: 
``Euclidean crystalline-SPTs with symmetry group $G$ are in 1-to-1 correspondence with SPTs protected by the internal symmetry $G$, 
where if $G$ has an orientation-reversing, it is mapped to an anti-unitary symmetry in the internal symmetry.''
We would like to apply their principles to include reflection/mirror, inversion and rotation SPTs, and
 compute explicit classifications from cobordism theory approach --- the
 cobordism theory will also suggest the appropriate relations between the topological terms of c-fSPTs and internal symmetric fSPTs.
 Moreover, we can compare with another independent approach on lattice models from the recent Ref.~\cite{Cheng2018aaz1810.12308} on interacting rotational fSPTs,
 which we find in agreement. Additional comments on other recent works can be found Sec.~\ref{sec:recent}.

\subsection{Reflection/Mirror SPTs}

\label{subsec:Reflection/Mirror}

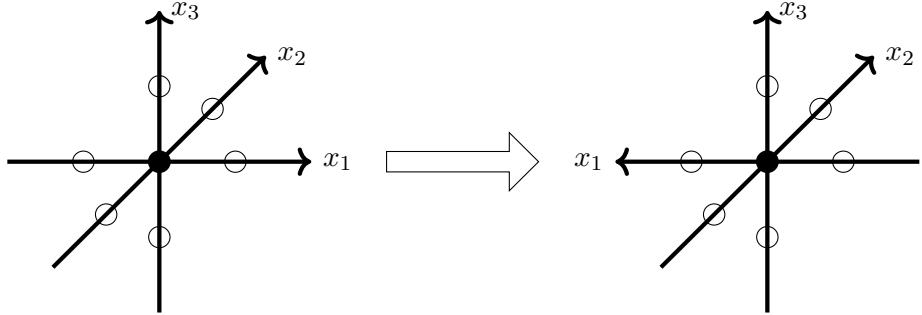
\begin{figure}[!h]
\centering
\begin{tikzpicture}
\draw[->, ultra thick] (-2.,0) -- (2.,0) node[right]{$x_1$};
\draw[->, ultra thick] (0,-2.) -- (0,2.) node[right]{$x_3$};  
\draw[->, ultra thick] (-1.4,-1.4) -- (1.4,1.4) node[right]{$x_2$}; 
 \node[circle, fill=black, inner sep=0pt, minimum size=3.mm] at (0,0){}; 
    \node[circle, minimum width=8pt, draw, inner sep=0pt] at (1,0){}; 
      \node[circle, minimum width=8pt, draw, inner sep=0pt] at (-1,0){}; 
         \node[circle, minimum width=8pt, draw, inner sep=0pt] at (.7,.7){}; 
            \node[circle, minimum width=8pt, draw, inner sep=0pt] at (-.7,-.7){}; 
               \node[circle, minimum width=8pt, draw, inner sep=0pt] at (0,1){}; 
      \node[circle, minimum width=8pt, draw, inner sep=0pt] at (0,-1){}; 
         \node[draw, single arrow,
              minimum height=20mm, minimum width=5mm,
              double arrow head extend=2mm,
              anchor=east, rotate=0] at (5,0) {};
\draw[->, ultra thick] (2.+8,0) -- (-2.+8,0)  node[left]{$x_1$};   
\draw[->, ultra thick] (0+8,-2.) -- (0+8,2.) node[right]{$x_3$}; 
\draw[->, ultra thick] (-1.4+8,-1.4) -- (1.4+8,1.4) node[right]{$x_2$}; 
 \node[circle, fill=black, inner sep=0pt, minimum size=3.mm] at (8,0){}; 
     \node[circle, minimum width=8pt, draw, inner sep=0pt] at (1+8,0){}; 
    \node[circle, minimum width=8pt, draw, inner sep=0pt] at (-1+8,0){}; 
         \node[circle, minimum width=8pt, draw, inner sep=0pt] at (.7+8,.7){}; 
            \node[circle, minimum width=8pt, draw, inner sep=0pt] at (-.7+8,-.7){}; 
               \node[circle, minimum width=8pt, draw, inner sep=0pt] at (0+8,1){}; 
      \node[circle, minimum width=8pt, draw, inner sep=0pt] at (0+8,-1){}; 
 \end{tikzpicture}
  \caption{Reflection or mirror symmetry. 
 The $x_j$, $j=1,2,3$ (more generally $j=1,2,...,d-1$) illustrate the local spatial coordinates with respect to an origin (the filled circle \CIRCLE) 
   which can be understood as a lattice site.
   The unfilled circles $\bigcirc$ mean some quantum degrees of freedom (e.g. complex fermion [a pair of real Majorana fermion modes], 
   iso-spin or boson) associated to a local site.
}
    \label{fig:Reflection}
\end{figure}

Reflection symmetry transformation means one of spatial coordinates (say $x_1$) is sent to its negative value (Fig.~\ref{fig:Reflection}), say with respect to the origin:
\bea
x_1 \to -x_1,
\eea
while others on the mirror hyperplane (say $x_j$, $j=2,...,d-1$) remain the same.

\begin{itemize}
\item
The local \emph{internal} symmetry is applied to all local sites, each within a local region around the unfilled circles $\bigcirc$.
\item
The  \emph{spatial} crystalline symmetry (like space group or point group) is applied to as the symmetry transformations between local sites, implemented \emph{between} the unfilled circles $\bigcirc$.
\end{itemize}

\begin{table}[h!]
\centering
\begin{tabular}{c   | l l  |  l  l }
\hline
dim  $\backslash$ $G_c \times G_o$  & $r^2=+1$ &  c-fSPTs classes & $r^2=(-1)^F$ &  c-fSPTs classes  \\
\hline
1+1D (2d) & $\Z_4^{Tf}$ & $\Omega_{Pin^+,\text{Tor}}^{2}(pt)=\Z_2.$ & $\Z_2^T\times\Z_2^f$ & $\Omega_{Pin^-,\text{Tor}}^{2}(pt)=\Z_8.$   \\
\hline
2+1D (3d) & $\Z_4^{Tf}$  & $\Omega_{Pin^+,\text{Tor}}^{2}(pt)=\Z_2.$ & $\Z_2^T\times\Z_2^f$ &  $\Omega_{Pin^-,\text{Tor}}^{3}(pt)=0.$    \\
\hline
3+1D (4d) & $\Z_4^{Tf}$  & $\Omega_{Pin^+,\text{Tor}}^{4}(pt)=\Z_{16}.$ & $\Z_2^T\times\Z_2^f$  &  $\Omega_{Pin^-,\text{Tor}}^{4}(pt)=0.$ \\ 
\hline
$(d-1)+1$D ($d$d) & $\Z_4^{Tf}$  & $\Omega_{Pin^+,\text{Tor}}^{d}(pt)$ & $\Z_2^T\times\Z_2^f$  &  $\Omega_{Pin^-,\text{Tor}}^{d}(pt)$ \\ 
$d  \geq 2$& $\Z_4^{Tf}  \times G_o$  & $\Omega_{Pin^+,\text{Tor}}^{d}(BG_o)$& $\Z_2^T\times\Z_2^f  \times G_o$ & $\Omega_{Pin^-,\text{Tor}}^{d}(BG_o)$   \\
\hline
\end{tabular}
\caption{The classification of reflection/mirror c-fSPTs for dimensions 1+1D, 2+1D, 3+1D and others.
The $r$ is the generator of reflection/mirror symmetry.
Below the column of crystalline symmetry ``$r^2=+1$'' or ``$r^2=(-1)^F$'', 
we list down the corresponding internal symmetry (in Minkowski signature, where TR operator is anti-unitary and $T^2=(\pm1)^F$ corresponds to $Pin^\mp$ \cite{Kapustin:2014dxa}) by a map from the crystalline symmetry (which map is explained in the main text).
Below the column of ``c-fSPTs classes'', we list down the corresponding the bordism groups and the group classification of c-fSPTs.
In the last row we list down the $G_c \times G_o$-fSPTs where $G_c$ is the reflection/mirror symmetry and $G_o$ is an additional internal symmetry.
The corresponding bordism groups can be  obtained from our approach in Sec.~\ref{sec:TR-Pin-TQFT}.}
\label{table:crystalline-internal-1}
\end{table}

In Table \ref{table:crystalline-internal-1}, we list down the complete classification of reflection/mirror c-fSPTs.
The $r$ is the general of reflection/mirror symmetry (up to the fermion parity, it is a $\Z_2$ group).
For fermionic system, there is a choice of $r^2=+1$  or $r^2=(-1)^F$. The latter means that $r^2=(-1)^F$ is locked with the fermion parity symmetry. 
The $(-1)^F$ is +1 or -1 for an even or odd number of fermions. The $-1$ is obviously related to the fermion statistics, rotating a fermion by $2 \pi$ gives rise to a $-1$ sign.

Since the reflection ($x_1 \to -x_1$) and the time-reversal symmetry ($t \to -t$) is the same for the reflection symmetry of Euclidean field theory,
we can map the ``$r^2=+1$'' or ``$r^2=(-1)^F$'' to ${Pin^+}$ or ${Pin^-}$ Euclidean field theory. This leads to the result summarized in Table \ref{table:crystalline-internal-1}.


\subsection{Inversion SPTs}

\label{subsec:Inversion}

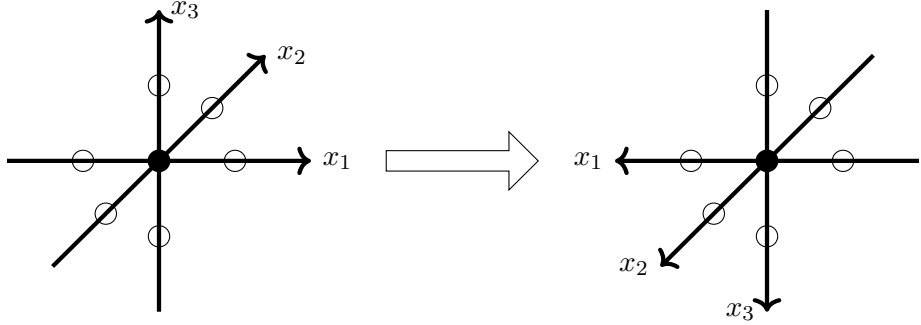
\begin{figure}[!h]
\centering
\begin{tikzpicture}
\draw[->, ultra thick] (-2.,0) -- (2.,0) node[right]{$x_1$};
\draw[->, ultra thick] (0,-2.) -- (0,2.) node[right]{$x_3$};
\draw[->, ultra thick] (-1.4,-1.4) -- (1.4,1.4) node[right]{$x_2$};
 \node[circle, fill=black, inner sep=0pt, minimum size=3.mm] at (0,0){}; 
   \node[circle, minimum width=8pt, draw, inner sep=0pt] at (1,0){}; 
      \node[circle, minimum width=8pt, draw, inner sep=0pt] at (-1,0){}; 
         \node[circle, minimum width=8pt, draw, inner sep=0pt] at (.7,.7){}; 
            \node[circle, minimum width=8pt, draw, inner sep=0pt] at (-.7,-.7){}; 
               \node[circle, minimum width=8pt, draw, inner sep=0pt] at (0,1){}; 
      \node[circle, minimum width=8pt, draw, inner sep=0pt] at (0,-1){}; 
         \node[draw, single arrow,
              minimum height=20mm, minimum width=5mm,
              double arrow head extend=2mm,
              anchor=east, rotate=0] at (5,0) {};
\draw[->, ultra thick] (2.+8,0) -- (-2.+8,0)  node[left]{$x_1$};
\draw[->, ultra thick]  (1.4+8,1.4) -- (-1.4+8,-1.4)  node[left]{$x_2$}; 
\draw[->, ultra thick] (0+8,2.) -- (0+8,-2.) node[left]{$x_3$}; 
 \node[circle, fill=black, inner sep=0pt, minimum size=3.mm] at (8,0){}; 
    \node[circle, minimum width=8pt, draw, inner sep=0pt] at (1+8,0){}; 
    \node[circle, minimum width=8pt, draw, inner sep=0pt] at (-1+8,0){}; 
         \node[circle, minimum width=8pt, draw, inner sep=0pt] at (.7+8,.7){}; 
            \node[circle, minimum width=8pt, draw, inner sep=0pt] at (-.7+8,-.7){}; 
               \node[circle, minimum width=8pt, draw, inner sep=0pt] at (0+8,1){}; 
      \node[circle, minimum width=8pt, draw, inner sep=0pt] at (0+8,-1){}; 
 \end{tikzpicture}
   \caption{Inversion.    See the caption to Fig.~\ref{fig:Reflection} for conventions.}
    \label{fig:Inversion}
\end{figure}

Inversion symmetry transformation means all spatial coordinates ($x_j$, $j=1,2,...,d-1$) are inverted (Fig.~\ref{fig:Inversion}), say respect to the origin (the black dot at $x_j=0$):
\bea
x_j \to -x_j,
\eea

\begin{table}[h!]
\centering
\begin{tabular}{c   | l l  |  l  l}
\hline
dim  $\backslash$ $G_c \times G_o$& $I^2=+1$  & c-fSPTs classes & $I^2=(-1)^F$  & c-fSPTs classes\\
\hline
1+1D (2d) & $\Z_4^{Tf}$ &  $\Omega_{Pin^+,\text{Tor}}^{2}(pt)=\Z_2.$  & $\Z_2^T\times\Z_2^f$ & $\Omega_{Pin^-,\text{Tor}}^{2}(pt)=\Z_8.$ \\
\hline
2+1D (3d) &  $\Z_4^f$  &  $\Omega_{(Spin \times \Z_4)/\Z_2,\text{Tor}}^{3}(pt)=0.$ & $\Z_2\times\Z_2^f$ &  $\Omega_{Spin,\text{Tor}}^{3}(B(\Z_2  )) =\Z_8.$  \\
\hline
3+1D (4d) & $\Z_4^{Tf}$  & $\Omega_{Pin^+,\text{Tor}}^{4}(pt)=\Z_{16}.$ & $\Z_2^T\times\Z_2^f$ & $\Omega_{Pin^-,\text{Tor}}^{4}(pt)=0.$  \\
\hline
$(d-1)+1$D ($d$d) &  $\Z_4^f$  &  $\Omega_{(Spin \times \Z_4)/\Z_2,\text{Tor}}^{d}(pt)$ & $\Z_2\times\Z_2^f$ &  $\Omega_{Spin,\text{Tor}}^{d}(B(\Z_2  ))$  \\
 $d \geq 2, d \in \text{odd}$ & $\Z_4^f \times G_0$ &    $\Omega_{(Spin \times \Z_4)/\Z_2,\text{Tor}}^{d}(BG_o)$& $\Z_2\times\Z_2^f\times G_0$ & $\Omega_{Spin,\text{Tor}}^{d}(B(\Z_2 \times G_o))$ \\
 \hline
$(d-1)+1$D ($d$d) & $\Z_4^{Tf}$  & $\Omega_{Pin^+,\text{Tor}}^{d}(pt)$ & $\Z_2^T\times\Z_2^f$  &  $\Omega_{Pin^-,\text{Tor}}^{d}$ \\ 
$d \geq 2, d \in \text{even}$&  $\Z_4^{Tf} \times G_0$  & $\Omega_{Pin^+,\text{Tor}}^{d}(BG_o)$& $\Z_2^T\times\Z_2^f \times G_0$ & $\Omega_{Pin^-,\text{Tor}}^{d}(BG_o)$   \\
\hline
\end{tabular}
\caption{The classification of inversion c-fSPTs. 
Let $I$ be the generator of inversion symmetry.
In the columns under ``$I^2=+1$'' or ``$I^2=(-1)^F$'', 
we list down the corresponding internal symmetry (again, in Minkowski signature, where TR operator is anti-unitary). The relation is explained in the main text.
We also list down $G_c \times G_o$-fSPTs where $G_c$ is the inversion symmetry and  $G_o$ is an additional internal symmetry.
}
\label{table:crystalline-internal-2}
\end{table}

In Table \ref{table:crystalline-internal-2}, we list down the complete classification of inversion c-fSPTs.
Let $I$ denote the inversion symmetry (up to the fermion parity, it is a $\Z_2$ group).
Again, for fermionic system, there is a choice of $I^2=+1$  or $I^2=(-1)^F$ locked with the fermion parity symmetry, as already explained in Sec.~\ref{subsec:Reflection/Mirror}. 

\begin{enumerate}
\item
In an even-$d$ dimensional spacetime, both
the inversion ($x_j \to -x_j$, inverting all the odd-dimensional spatial coordinates, which gives rise to an overall sign $(-1)^{d-1}=(-1)$) and the time-reversal symmetry ($t \to -t$) 
are exactly the same as the reflection symmetry of Euclidean field theory.
We can again map the ``$I^2=+1$'' or ``$I^2=(-1)^F$'' to ${Pin^+}$ or ${Pin^-}$ spacetime symmetries of Euclidean field theory, the same as for reflection/mirror symmetry in Sec.~\ref{subsec:Reflection/Mirror}.

\item In an odd-$d$ dimensional spacetime, 
the inversion ($x_j \to -x_j$, inverting all gives rise to an overall sign $+1$) is rather \emph{distinct} from the reflection symmetry.
We can map the inversion instead to
${(Spin \times \Z_4)/\Z_2}$ or ${(Spin \times \Z_2)}$ spacetime symmetries of
Euclidean field theory.\footnote{For even $d$,  the  internal symmetry corresponds to extensions
$$ 1 \to \Z_2^f \to G \to  \Z_2^T \to 1,$$ where $G= \Z_4^{Tf}$ ($Pin^+$) for $I^2=+1$, while $G=\Z_2^T\times\Z_2^f$ ($Pin^-$) for $I^2=(-1)^F$.
For odd $d$, the internal symmetry corresponds to extension
$$ 1 \to \Z_2^f \to G \to  \Z_2 \to 1,$$ where $G= \Z_4^{f}$ (${(Spin \times \Z_4)/\Z_2}$-structure in the cobordism approach) for $I^2=+1$, while $G=\Z_2 \times\Z_2^f$ 
(${(Spin \times \Z_2)}$-structure in the cobordism approach) for $I^2=(-1)^F$.
}
\end{enumerate}
This leads to the result summarized in Table \ref{table:crystalline-internal-2}.

\subsection{Rotation SPTs}

\label{subsec:Rotation}

Here we only consider a rotation within a 2-plane, say with coordinates $x_1$ and $x_2$, such that rotation by a $\theta$-angle on this 2-plane with respect to the origin is given by
\bea
x_1 \to x_1 \cos(\theta) - x_2 \sin(\theta),  \quad x_2 \to x_1 \sin(\theta) + x_2  \cos(\theta).
\eea
When $\theta=\pi$, we have, $x_1 \to -x_1$ and $x_2 \to -x_2$.
\begin{figure}[!h]
\centering
(a) \begin{tikzpicture}
\draw[->, ultra thick] (-2.,0) -- (2.,0) node[right]{$x_1$};
\draw[->, ultra thick] (0,-2.) -- (0,2.) node[right]{$x_3$};  
\draw[->, ultra thick] (-1.4,-1.4) -- (1.4,1.4) node[right]{$x_2$}; 
 \node[circle, fill=black, inner sep=0pt, minimum size=3.mm] at (0,0){}; 
    \node[circle, minimum width=8pt, draw, inner sep=0pt] at (1,0){}; 
      \node[circle, minimum width=8pt, draw, inner sep=0pt] at (-1,0){}; 
         \node[circle, minimum width=8pt, draw, inner sep=0pt] at (.7,.7){}; 
            \node[circle, minimum width=8pt, draw, inner sep=0pt] at (-.7,-.7){}; 
               \node[circle, minimum width=8pt, draw, inner sep=0pt] at (0,1){}; 
      \node[circle, minimum width=8pt, draw, inner sep=0pt] at (0,-1){}; 
         \node[draw, single arrow,
              minimum height=20mm, minimum width=5mm,
              double arrow head extend=2mm,
              anchor=east, rotate=0] at (5,0) {};
\draw[->, ultra thick] (2.+8,0) -- (-2.+8,0)  node[left]{$x_1$}; 
\draw[->, ultra thick] (0+8,-2.) -- (0+8,2.) node[right]{$x_3$};  
\draw[->, ultra thick]  (1.4+8,1.4) -- (-1.4+8,-1.4)  node[left]{$x_2$}; 
 \node[circle, fill=black, inner sep=0pt, minimum size=3.mm] at (8,0){}; 
     \node[circle, minimum width=8pt, draw, inner sep=0pt] at (1+8,0){}; 
    \node[circle, minimum width=8pt, draw, inner sep=0pt] at (-1+8,0){}; 
         \node[circle, minimum width=8pt, draw, inner sep=0pt] at (.7+8,.7){}; 
            \node[circle, minimum width=8pt, draw, inner sep=0pt] at (-.7+8,-.7){}; 
               \node[circle, minimum width=8pt, draw, inner sep=0pt] at (0+8,1){}; 
      \node[circle, minimum width=8pt, draw, inner sep=0pt] at (0+8,-1){}; 
 \end{tikzpicture}\\
 (b) \quad \includegraphics[width=5cm]{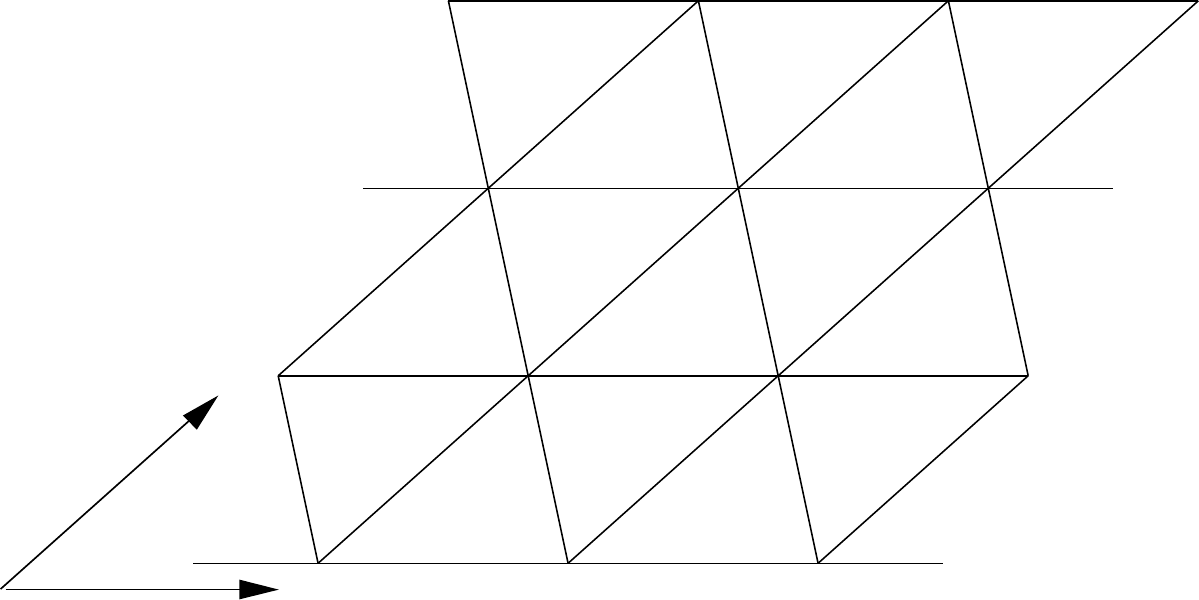}
    \caption{Rotation: (a) In particular the $\pi$-rotation from a cyclic group $C_2$ in the $x_1$-$x_2$-plane.
    See the caption in Fig.~\ref{fig:Reflection} for conventions. (b) In the bottom we show an example of 2D lattice (on the $x_1$-$x_2$-plane) with 
    $C_6$ rotational symmetry (the generator of $C_6$ is a rotation by angle $\frac{2\pi}{6}$ on any lattice site around the $x_3$ axis pointing out of the $x_1$-$x_2$-plane);
    in addition, this lattice also has two independent $\Z$-translational symmetries, reflection/mirror symmetry, inversion symmetry (which is the same as the $C_2$ as a subgroup of $C_6$ rotation symmetry), among others.
    }
        \label{fig:Rotation}
\end{figure}
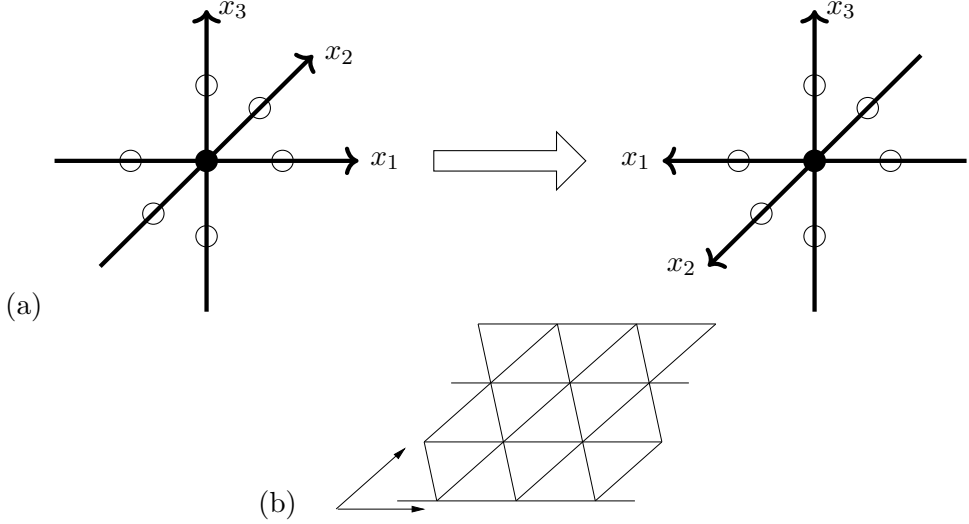
On a lattice or crystal, 
we have the rotation group $C_n$ ($C$ for cyclic) group, which is the finite $\Z_n$ group. The generator is the rotation by angle  $\theta=\frac{2\pi}{n}$, with $n=1,2,3,4,6$, due to the constraint of the lattice periodicity. See for example in Fig.~\ref{fig:Rotation}. 

\begin{table}[h!]
\centering
\begin{tabular}{c  | l l  |  l  l }
\hline
dim  $\backslash$ $G_c \times G_o$  &  $R^2=+1$  & c-fSPTs classes & $R^2=(-1)^F$ & c-fSPTs classes    \\
\hline
1+1D (2d)   & $\Z_4^{Tf}$ & $\Omega_{Pin^+,\text{Tor}}^{2}(pt)=\Z_2.$ & $\Z_2^T\times\Z_2^f$ & $\Omega_{Pin^-,\text{Tor}}^{2}(pt)=\Z_8.$   \\
 \hline
2+1D (3d) &$\Z_4^f$  &  $\Omega_{(Spin \times \Z_4)/\Z_2,\text{Tor}}^{3}(pt)=0.$ & $\Z_2\times\Z_2^f$ &  $\Omega_{Spin,\text{Tor}}^{3}(B(\Z_2  )) =\Z_8.$  \\
 \hline
3+1D (4d) &$\Z_4^f$  &  $\Omega_{(Spin \times \Z_4)/\Z_2,\text{Tor}}^{4}(pt)=0.$ & $\Z_2\times\Z_2^f$ &  $\Omega_{Spin,\text{Tor}}^{4}(B(\Z_2  )) =0.$  \\
\hline
$(d-1)+1$D ($d$d) &  $\Z_4^f$  &  $\Omega_{(Spin \times \Z_4)/\Z_2,\text{Tor}}^{d}(pt)$ & $\Z_2\times\Z_2^f$ &  $\Omega_{Spin,\text{Tor}}^{d}(B(\Z_2  ))$  \\
 $d > 2$ &  $\Z_4^f \times G_o$ &    $\Omega_{(Spin \times \Z_4)/\Z_2,\text{Tor}}^{d}(BG_o)$& $\Z_2\times\Z_2^f \times G_o$ & $\Omega_{Spin,\text{Tor}}^{d}(B(\Z_2 \times G_o))$ \\
 \hline
\end{tabular}
\caption{
The classification of rotation c-fSPTs. 
Let $R$ be the generator of rotation symmetry $C_2$.
In the columns under ``$R^2=+1$'' or ``$R^2=(-1)^F$'' crystalline symmetries, 
we list down the corresponding internal symmetry. The relation to the is explained in the main text.
We also list down $G_c \times G_o$-fSPTs where $G_c$ is the rotation symmetry and  $G_o$ is an additional internal symmetry.
}
\label{table:crystalline-internal-3}
\end{table}

\begin{enumerate}
\item
In an 1+1D (2d) spacetime, 
the $C_2$-rotation is the same as the inversion and the reflection.
So the first row of data in 1+1D (2d) of Table \ref{table:crystalline-internal-1},  \ref{table:crystalline-internal-2} and \ref{table:crystalline-internal-3} exactly coincide and give the same fSPTs.

\item
In an 2+1D (3d) spacetime, 
the $C_2$-rotation is the same as the spatial inversion, but not the same as the reflection.
So the first row of data in 2+1D (3d) of Table \ref{table:crystalline-internal-2} and \ref{table:crystalline-internal-3} exactly coincide and give the same fSPTs.

\item In any dimension larger than 1+1D ($d$d where $d>2$), 
the rotation symmetry acts on the 2-dimensional spatial subspace. 
We can map the $C_m$-rotation instead to
${(Spin \times \Z_{2m})/\Z_2}$ or ${(Spin \times \Z_m)}$ spacetime symmetry of 
Euclidean field theory.\footnote{For all $d$-dimensions where $d>2$, 
the internal symmetry corresponds to the extension
$$ 1 \to \Z_2^f \to G \to  \Z_{2m} \to 1,$$ where $G=\Z_{2m}^f$ (in terms of bordism of ${(Spin \times \Z_{2m})/\Z_2}$-structure) for $R^m=+1$, 
while $G=\Z_{m} \times\Z_2^f$ 
(in terms of bordism of ${(Spin \times \Z_m)}$-structure) for $R^m=(-1)^F$.
This is a natural generalization from $m=2$ of $C_2$-symmetry in 2+1D to generic $m$ in other dimensions. Note that $R$ is the generator of crystalline rotation symmetry group, not the internal symmetry $G$. The difference between two (factor $(-1)^F$ in the $m$-th power of the generator) has been discussed e.g. in \cite{Cheng2018aaz1810.12308}.
}
\end{enumerate}
This leads to the result summarized in Table \ref{table:crystalline-internal-3} and \ref{table:crystalline-internal-4}.

\begin{table}[h!]
\centering
\begin{tabular}{c  | l l  |  l  l }
\hline
dim  $\backslash$ $G_c \times G_o$  &   $R^m=+1$ &  c-fSPTs classes   & $R^m=(-1)^F$ &  c-fSPTs classes      \\
\hline
2+1D (3d) &$\Z_{2m}^f$  &  $\Omega_{(Spin \times \Z_{2m})/\Z_2,\text{Tor}}^{3}(pt)$ & $\Z_m\times\Z_2^f$ &  $\Omega_{Spin,\text{Tor}}^{3}(B(\Z_m  ))$  \\
  & $m=2$ & 0 & $m=2$ &  $\Z_8$ \\
  & $m=3$ &  {$\Z_3$}& $m=3$ &  {$\Z_3$}\\
 & $m=4$ & $\Z_2$& $m=4$ & $\Z_8 \times \Z_2$  \\
  & $m=6$ &  {$\Z_3$}& $m=6$ & {$\Z_{24}=\Z_{8} \times \Z_3$} \\
   & $\Z_{2m}^f \times G_o$ &    $\Omega_{(Spin \times \Z_{2m})/\Z_2,\text{Tor}}^{3}(BG_o)$&  $\Z_m\times\Z_2^f \times G_o$ & $\Omega_{Spin,\text{Tor}}^{3}(B(\Z_m \times G_o))$ \\
 \hline
3+1D (4d) &$\Z_{2m}^f$  &  $\Omega_{(Spin \times \Z_{2m})/\Z_2,\text{Tor}}^{4}(pt)$ & $\Z_m\times\Z_2^f$ &  $\Omega_{Spin,\text{Tor}}^{4}(B(\Z_m  )) $  \\
 & $m=2$ &  0& $m=2$ & 0 \\
  & $m=3$ & 0 & $m=3$ & 0 \\
 & $m=4$ & 0 & $m=4$ &0  \\
  & $m=6$ & 0 & $m=6$ & 0 \\
   & $\Z_{2m}^f \times G_o$ &    $\Omega_{(Spin \times \Z_{2m})/\Z_2,\text{Tor}}^{4}(BG_o)$&$\Z_m\times\Z_2^f \times G_o$   & $\Omega_{Spin,\text{Tor}}^{4}(B(\Z_m \times G_o))$ \\
\hline
$(d-1)+1$D ($d$d) &  $\Z_{2m}^f$  &  $\Omega_{(Spin \times \Z_{2m})/\Z_2,\text{Tor}}^{d}(pt)$ & $\Z_m\times\Z_2^f$ &  $\Omega_{Spin,\text{Tor}}^{d}(B(\Z_m  ))$  \\
 $d > 2$ &  $\Z_{2m}^f \times G_o$ &    $\Omega_{(Spin \times \Z_{2m})/\Z_2,\text{Tor}}^{d}(BG_o)$& $\Z_m\times\Z_2^f \times G_o$  & $\Omega_{Spin,\text{Tor}}^{d}(B(\Z_m \times G_o))$ \\
 \hline

\end{tabular}
\caption{Rotation symmetry generated by the $\frac{2\pi}{m}$-rotation (from a cyclic group $C_{m}$) in the $x_1$-$x_2$-plane, where $m=2,3,4,6$ due to the lattice packing constraint on a 2-plane.
    Our result can be compared with a recent Ref.~\cite{Cheng2018aaz1810.12308} on interacting rotation fSPTs which used a different method based on physical lattice models. It produces
    the same group classification output as far as we are concerned.
    Importantly, the last two rows show that our bordism group data calculated in Sec.~\ref{sec:computation} and \ref{sec:spin-twisted-bordism},
     coincide with $\Omega_{Spin,\text{Tor}}^{d}(B(\Z_m \times G_o))$ and $\Omega^{d}_{(Spin \times \Z_{2m})/\Z_2,\text{Tor}}(BG_o)$.
    This means that our previous fSPTs have the exact correspondence to the new rotation c-fSPTs found here (and in Ref.~\cite{Cheng2018aaz1810.12308}).}
\label{table:crystalline-internal-4}
\end{table}

\newpage
\subsection{Fermionic SETs (Symmetry Enriched Topologically ordered states)}

\label{subsec:fSET}

We can partially gauge the symmetry group of $G$-fSPT of an internal symmetry $G$. This becomes the so-called Symmetry Enriched Topologically ordered states (SETs): 
\begin{itemize}
\item Fermionic SETs (fSETs), if we only dynamically gauge a subgroup $G' \subset G$ and leave $\Z_2^f$ fermion parity ungauged.
\item Bosonic SETs (bSETs), if we dynamically gauge a subgroup $G' \subset G$ and also dynamically gauge $\Z_2^f$ fermion parity. The gauging 
$\Z_2^f$ process is known as a higher dimensional bosonization \cite{Kapustin2017jrc1701.08264}.
\end{itemize}
We can define the fSET partition function via generalizing the gauging discussion in Sec.~\ref{sec:gauging} .
Gauging subgroup $G' \subset G$ produces the fSET with symmetry $G_{\text{sym}}$ (i.e.\ $G_{\text{sym}}$-equivariant spin-TQFT), 
where $G_{\text{sym}}$ is the \emph{centralizer} of $G'$ in $G$, i.e.\ $G_{\text{sym}}=C_{G}(G')$. Then, for a given $g_{\text{sym}}: M^{n-1} \to BG_{\text{sym}}$, the Hilbert space of the resulting $n$-dimensional fSET on $M^{n-1}$ is given by a generalization of eqn.~(\ref{gauged-hilbert-space}) as:
\bea \label{eq:SET-gauged-hilbert-space}
 Z_{\text{gauged}}(M^{n-1}, g_{\text{sym}}) = \bigoplus_{[f']\in C_{M^{n-1},g_{\text{sym}}}} Z(M^{n-1},\mu\circ (g_{\text{sym}}\times f')),
\eea
where $\mu: BG_{\text{sym}} \times BG' \to BG$ 
is the map induced by the multiplication map $G_{\text{sym}} \times G' \to G$ 
(note that here, to define $\mu$, we use the fact that $G_{\text{sym}}$ and $G'$ are commuting subgroups in $G$ so that the map  
$G_{\text{sym}} \times G' \to G$ is a group homomorphism) and $C_{M^{n-1},g_{\text{sym}}}$ 
is the following subset of the set of homotopy classes of maps $M^{n-1}\rightarrow BG'$:
\bea
  C_{M^{n-1},g_{\text{sym}}}:=\left\{
   [f']\;\Big|
   \begin{array}{c}
\;Z(M^{n-1}\times S^1,\mu\circ(\text{pr}^*g_{\text{sym}}\times g') )\in \C \text{ is the same} \\
\text{for all } g':M^{n-1}\times S^1\rightarrow BG', \text{ s.t. } g'|_{M^{n-1}\times \text{pt}} \sim f'
   \end{array}
  \right\} \subset [M^{n-1},BG'].
\eea

Similarly, we can define bSETs via combining the formalism in Sec.~\ref{sec:relation-to-DW} to obtain bosonic theory
and the partial gauging in eqn.~(\ref{eq:SET-gauged-hilbert-space}). 
A related gauging procedure from 4d fSPTs to 4d bSETs and to 4d fSETs
was recently given in \cite{2018arXiv180108530Lan}.
A worthwhile remark is that Ref.~\cite{2018arXiv180108530Lan} gives a strong conjecture, stating that
gauging 4d fSPTs with finite group unitary internal symmetry, may give a large subclass of
all 4d bSETs of finite gauge group, and likely also a complete classification of
all 4d fSETs of finite gauge group. If so, our formulation provides a systematic study of all 
4d bSETs and fSETs of finite gauge group.


\subsection{Other aspects}

In this subsection, we make some remarks on how our work can be related to the previously existed literature.

\subsubsection{Fermionic higher global symmetries vs. Higher-form global symmetries}

Ref.~\cite{Gaiotto2014kfa1412.5148} proposes the generalized global symmetry for QFTs. For
TQFTs, the generalized global symmetry is intrinsically related to the link invariants between two set of extended operators;
one is regarded as a ``charged'' object (being measured), 
the other is regarded as a ``charge'' operator (measuring the ``charged'' object).

First let us note that the usual fermionic parity in many aspects behaves as a $\Z_2$ 0-form global symmetry. In particular, one can insert a Wilson loop that measures both holonomy of a gauge field for an ordinary 0-form symmetry and the action of invertible spin-TQFT $\eta$ coupled to the spin-structure induced from the ambient space via framing on the normal bundle.

Similarly, in many aspects, one can treat Arf-TQFT as a connection for fermionic $\Z_2$ 1-form symmetry. For our fermionic spin-TQFTs,
there are fermionic loops that are not only charged under 1-form global symmetry, but also ``charged under $\Z_2$ 1-form symmetry
generated by Arf invertible spin-TQFT.'' So when we parallel transport
them along some surface $\Sigma$ connecting $L_1$ loop with loop $L_2$ (i.e.
$\partial\Sigma=L_1 \sqcup L_2$), they obtain a phase given by
\bea
\exp( (2\pi i k \int_\Sigma B )+ \pi i \text{Arf}(\Sigma) )
\eea
where $B$ is a background 2-form field for 1-form symmetry ($\Z_N$ or
$U(1)$, with charge $k \in \Z_N$ or $\Z$ respectively) and the spin-structure on $\Sigma$ is induced from the spin structure of the ambient space using framing of the normal bundle to $\Sigma$.

In the context of generalized global symmetry, we not only have \cite{Gaiotto2014kfa1412.5148}'s higher-form global symmetries,
but also additional fermionic higher global symmetries which may \emph{not} be written as differential forms.

\subsubsection{Adams spectral sequence vs. Atiyah-Hirzebruch spectral sequence}
\label{sec:Adams-AH}

In this section we compare between our classifications and \cite{Wang2017moj1703.10937, Wang2018pdc1811.00536}'s classification scheme for fermionic SPTs (fSPTs).
Our classification relies on Adams spectral sequence.
To our best knowledge, \cite{Wang2017moj1703.10937, Wang2018pdc1811.00536}'s structure is similar to Atiyah-Hirzebruch spectral sequence (see also \cite{2016arXiv161202860BMorgan,2018arXiv180308147BMorgan}).

Based on a fermionic lattice model for fSPTs, Ref.~\cite{Wang2017moj1703.10937, Wang2018pdc1811.00536} derives a generalized cohomology group theory, such that there
are two layers of short exact sequences and other constraints upon their set of data.

For 3d (2+1D)  fSPTs  with a total symmetry $G_f=G\times \mathbb Z_2^f$, with a bosonic internal symmetry $G$ 
and  fermion parity $\mathbb Z_2^f$ symmetry, 
Ref.~\cite{2017PhRvB1610.08478,Wang2017moj1703.10937, Wang2018pdc1811.00536}  (and references therein) summarize three sets of group cohomology data of the symmetry group $G$, namely
\begin{equation}
	(H^3(G, U_T(1)), BH^2(G,\mathbb Z_2), H^1 (G, \mathbb Z_2 )).
\end{equation}
$BH^2(G,\mathbb Z_2)$ is the obstruction-free subgroup of $H^2(G,\mathbb Z_2)$, generated by $n_{2} \in H^2(G,\mathbb Z_2)$ that satisfy $Sq^2(n_{2})=0$ in
$H^{4}(G,U_T(1))$, where $Sq^2$ is the Steenrod square.\footnote{$U_T(1)$ is the notation indicating that the coefficient $U(1)$ of the cohomology group is non-trivially acted by the antiunitary time-reversal symmetry if exists. Here we do not pay attention to the antiunitary symmetry, and assume that all the (considered) symmetries are unitary.}
$H^3(G,U_T(1))$ is the classification of bosonic SPTs. 
Physically, the $H^1 (G,\mathbb Z_2 )$ layer can be constructed by decorating a 1+1D Kitaev fermionic chain \cite{2001KitaevWire}, 
which is a 2d invertible spin TQFT onto the $G$-symmetry domain walls. 
The $BH^2(G,\mathbb Z_2)$ layer is constructed by decorating complex fermions, 
which are 1d invertible spin TQFT onto the $G$-symmetry domain walls.

For 4d (3+1D) fSPTs with $G_f=G\times \mathbb Z_2^f$, with a bosonic symmetry $G$, 
Ref.~\cite{Wang2017moj1703.10937, Wang2018pdc1811.00536} propose three sets of group cohomology data of the symmetry group $G$, namely
\begin{equation}
	( H^4_{\rm rigid}(G, U_T(1)),  BH^3(G,\mathbb Z_2), \tilde{B}H^2 (G,\mathbb Z_2 ))
\end{equation}
to classify these 4d fSPT. For the meaning of the first two entries $H^4_\text{rigid}(G,U(1)_T)$ and $BH^3(G,\mathbb Z_2)$, see the original reference \cite{Wang2017moj1703.10937, Wang2018pdc1811.00536}.
The third entry $\tilde{B}H^2 (G,\mathbb Z_2 )$ is an obstruction-free subgroup of $H^2 (G,\mathbb Z_2 )$, generated by $\tilde n_{2} \in H^2(G, \mathbb Z_2)$ that obey 
$Sq^2(\tilde n_{2})=0$ in $H^{4}(G,\mathbb Z_2)$ and  $\mathcal O(\tilde n_{2})=0$ in $H^{5}(G,U_T(1))$. Here $\mathcal{O}$ is a certain cohomology operation
that maps $\tilde n_{2}$ satisfying $Sq^2(\tilde n_{2})=0$ in $H^{2}(G,\mathbb Z_2)$ into an element in $H^{5}(G,\mathbb Z_8) \subseteq H^{5}(G,U_T(1))$.
As far as we are concerned, 
the ``mysterious'' $\mathcal O$ discussed in \cite{Wang2017moj1703.10937} is actually, in Atiyah-Hirzebruch spectral sequence,
the dual of the secondary cohomology operation $\Theta: H^2(X,\Z) \to H^5(X,\Z_2)$ based on the relation $Sq^2Sq^2\rho=0$, where $\rho$ is the induced coefficient mod 2 reduction from 
$0\to \Z \to \Z \to \Z_2 \to 0$. In general, there is a bijection between stable homology operations $E_* \to F_{*-k}$ and stable cohomology operations $E^*\to F^{*+k}$.
So the dual of $\Theta$ is $H_5(X,\Z) \to H_2(X,\Z_2)$. Apply the functor $\Hom(-,U(1))$, we get $H^2(X,\Z_2)\to H^5(X,U(1))$, here $X=BG_b$ is the classifying space of $G_b$.
Other details of cohomology group notations are explained in \cite{Wang2017moj1703.10937, Wang2018pdc1811.00536}.

Below we fill in some background knowledge of Atiyah-Hirzebruch spectral sequence for comparison.
Here
\begin{equation}
	H^p(X,\Omega^q_{Spin})\Rightarrow\Omega^{p+q}_{Spin}(X)
\end{equation}
where $\Omega^n_{Spin}(X)=\Hom(\Omega^{Spin}_n(X),U(1))$.

\begin{figure}[!h]
\center
\begin{sseq}[grid=none,labelstep=1,entrysize=1.5cm]{0...4}{0...4}
\ssdrop{U(1)}
\ssmoveto 1 0
\ssdrop{\star}
\ssmoveto 2 0
\ssdrop{\star}
\ssmoveto 3 0
\ssdrop{\star}
\ssmoveto 4 0
\ssdrop{\star}
\ssmoveto 0 1
\ssdrop{\Z_2}
\ssmoveto 1 1
\ssdrop{\star}
\ssmoveto 2 1
\ssdrop{\star}
\ssmoveto 3 1
\ssdrop{\star}
\ssmoveto 0 2
\ssdrop{\Z_2}
\ssmoveto 1 2
\ssdrop{\star}
\ssmoveto 2 2
\ssdrop{\star}
\ssmoveto 0 3
\ssdrop{0}
\ssmoveto 1 3
\ssdrop{0}
\ssmoveto 0 4
\ssdrop{U(1)}

\end{sseq}
\center
\caption{Atiyah-Hirzebruch spectral sequence}
\label{fig:AHSS}
\end{figure}

There is a filtration 
\begin{equation}
	0=F^{n+1,-1}\subset F^{n,0}\subset F^{n-1,1}\subset\cdots\subset F^{p,n-p}\subset\cdots\subset \Omega^n_{Spin}(X)
\end{equation}
and an isomorphism
\begin{equation}
	F^{p,q}/F^{p+1,q-1}\cong E_{\infty}^{p,q}.
\end{equation}
The first layer is the extension 
\begin{equation}
	E^{n,0}_{\infty}\to F^{n-1,1}\to E^{n-1,1}_{\infty}.
\end{equation}
Their second layer is the extension
\begin{equation}
	F^{n-1,1}\to F^{n-2,2}\to E^{n-2,2}_{\infty}.
\end{equation}

Following Ref.~\cite{Wang2017moj1703.10937, Wang2018pdc1811.00536}'s notation, 
we find  the following correspondence for
$X=BG_b$ (where $G_b$, denoted as $G$ previously, is the bosonic internal symmetry group). If $n=3$:
\bea
E_{\infty}^{3,0}=H^3(G_b,U_T(1)), \quad E_{\infty}^{2,1}=BH^2(G_b,\Z_2), \quad E_{\infty}^{1,2}=H^1(G_b,\Z_2).
\eea
If $n=4$:
\bea
E_{\infty}^{4,0}=H_{\text{rigid}}^4(G_b,U_T(1)), \quad E_{\infty}^{3,1}=BH^3(G_b,\Z_2), \quad E_{\infty}^{2,2}=\tilde{B}H^2(G_b,\Z_2).
\eea
On the other hand, our work uses Adams spectral sequence. Our ``layer'' structure has physical and mathematical interpretations
(To recall, the topological terms underlined with a single or double lines are our notations introduced in Sec.~\ref{sec:computation}, such as eq.~(\ref{eq:topZ4Z2})): 
\begin{itemize}
\item Non-underlined topological terms are bosonic (i.e.\ belong to $H^n(BG,U(1))$ subgroup), they corresponds to the elements in $E_{\infty}^{n,0}$ of Atiyah-Hirzebruch spectral sequence. 
\item Topological terms underlined with a single line are fermionic that provide refinement of the bosonic elements from $H^n(BG,U(1))$. 
\item Topological terms underlined with double lines are fermionic and do not refine any elements of $H^n(BG,U(1))$. Here ``refine'' means the fermionic topological term is a nontrivial extension of the elements of $H^n(BG,U(1))$.
\end{itemize}
Examples by examples, our fSPT classification results show general agreements with Ref.~\cite{Wang2017moj1703.10937, Wang2018pdc1811.00536}.
It is worth noticing that similar explicit layer structures of the spin bordism group have been explored in \cite{2016arXiv161202860BMorgan, 2018arXiv180308147BMorgan}.
However, we remain direct comparison of our construction to \cite{2016arXiv161202860BMorgan, 2018arXiv180308147BMorgan} to be studied.

\subsubsection{Relations to other recent works}

\label{sec:recent}

Here are some final remarks related to the literature and recent works. We also highlight potential future directions.
\begin{enumerate}
\item We have discussed the topological invariants and link invariants of fermionic spin $G$-TQFTs.
In condensed matter physics,
the topological invariants corresponds to SPTs partition function;
the link invariants corresponds the braiding statistical Berry phases of time-evolution trajectory of world-line/world-volume of anyonic particles of anyonic strings
(either weakly gauged as probed field defect, or dynamically gauged as topological orders).
For the physical meanings of link invariants, one can refer to systematic work \cite{Wang2014oya1404.7854, Wang:2014wka, Wang2016jdt1602.05569Thesis, Wang2016qhf1602.05951, Tiwari2016zru1603.08429, Putrov:2016qdo, Chan2017eov, Cheng2017ftw1705.08911, Wang2018iwz1810.13428}.  
It will be interesting to explore the link invariants in terms of geometric-topology aspects like the surgery theory on submanifolds, along the ideas of \cite{Witten:1988hf, Wang2016qhf1602.05951}
in 3 and 4 dimensions.

\item The SPTs protected by crystalline lattice symmetry has generated broad interests recently, ranging from
 the earlier work on crystalline topological insulator \cite{2011PhRvLFu} to the recent work on intrinsic interacting 
crystalline insulator/superconductor \cite{2018arXiv181012317RYMLu}, and recent reviews \cite{Ando:2015sia} and references therein.

Apart from our discussion and Ref.~\cite{Thorngren2018wwt1612.00846}'s 
Crystalline Equivalence Principle, 
other development of general theory of cSPTs include 
general real-space recipe construction of  topological crystalline states 
\cite{2018arXiv1810.11013YangQi} and field theory approaches  \cite{Han2018xsvPYe1807.10844} (and references therein).

It is worthwhile to mention that recent
Ref.~\cite{Shiozaki2018yyjKen1810.00801} applies generalized homology and
Atiyah-Hirzebruch spectral sequence (AHSS) in crystalline SPTs. Therefore, it will be interesting to 
compare our understanding in Sec.~\ref{sec:Adams-AH} based on
our work (Adams spectral sequence) not only to \cite{Wang2017moj1703.10937, Wang2018pdc1811.00536}'s work (AHSS) on fSPT with internal symmetry, but also to \cite{Shiozaki2018yyjKen1810.00801}'s work (AHSS) on
crystalline symmetry. 
This may guide us to construct analogous lattice models, from our models, for internal \cite{Wang2017moj1703.10937, Wang2018pdc1811.00536} and crystalline symmetry \cite{Shiozaki2018yyjKen1810.00801}. Another approach to construct our models on a lattice or condensed mater systems can be 
fermion decoration construction \cite{2018arXiv180901112Lan}, similar to our discussion in Sec.~\ref{sec:Z2Z4-234}.

\end{enumerate}

\section{Acknowledgments}

The authorship is listed in the alphabetical order.
PP would like to thank Anna Beliakova, Ryan Thorngren for useful discussions and the hospitality of SCGP, Weizmann Institute and KITP where parts of the work were done. 
JW warmly thanks the participants and organizers of the PCTS workshop on Fracton Phases of Matter and Topological Crystalline Order 
(December 3-5, 2018)
for many inspiring conversations near the completion of this work. 
JW also thanks Meng Cheng, Dominic Else, Sheng-Jie Huang,
and Hao Song, for clarifying their own works.
MG thanks the support from U.S.-Israel Binational Science Foundation. 
KO gratefully acknowledges the support from NSF Grant PHY-1606531 and Paul Dirac fund.
PP gratefully acknowledges the support from Marvin L. Goldberger Fellowship and the DOE Grant
51 DE-SC0009988 during his appointment at IAS. 
ZW gratefully acknowledges support from NSFC grants 11431010, 11571329. 
JW gratefully acknowledges the Corning Glass Works Foundation Fellowship and NSF Grant PHY-1606531.
This work was also supported in part by NSF Grant DMS-1607871 ``Analysis, Geometry and Mathematical
Physics'', Center for Mathematical Sciences and Applications at Harvard University, and the National Science Foundation under Grant No. NSF PHY-1748958.

\bibliographystyle{plain}
\bibliography{spin-tqft-bib-3,spin-tqft-bib-3-JW-add.bib}

\providecommand{\href}[2]{#2}\begingroup\raggedright\begin{thebibliography}{10}

\bibitem{atiyah1988topological}
M.~Atiyah, \emph{Topological quantum field theories}, {\emph{Publications
  Math{\'e}matiques de l'Institut des Hautes {\'E}tudes Scientifiques} {\bf 68}
  175--186 (1988)}.

\bibitem{kirby1990pin}
R.~C. Kirby and L.~R. Taylor, \emph{Pin structures on low-dimensional
  manifolds}, {\emph{Geometry of low-dimensional manifolds} {\bf 2} 177--242
  (1990)}.

\bibitem{blanchet1992invariants}
C.~Blanchet, \emph{Invariants on three-manifolds with spin structure},
  {\emph{Commentarii Mathematici Helvetici} {\bf 67} 406--427 (1992)}.

\bibitem{blanchet1996topological}
C.~Blanchet, G.~Masbaum et~al., \emph{Topological quantum field theories for
  surfaces with spin structure}, {\emph{Duke Mathematical Journal} {\bf 82}
  229--268 (1996)}.

\bibitem{beliakova2014spin}
A.~Beliakova, C.~Blanchet and E.~Contreras, \emph{Spin modular categories},
  {\emph{arXiv preprint arXiv:1411.4232} (2014)}.

\bibitem{FH}
D.~S. Freed and M.~J. Hopkins, \emph{Reflection positivity and invertible
  topological phases}, {\emph{arXiv preprint arXiv:1604.06527} (2016)}.

\bibitem{Dijkgraaf:1989pz}
R.~Dijkgraaf and E.~Witten, \emph{{Topological Gauge Theories and Group
  Cohomology}}, \href{http://dx.doi.org/10.1007/BF02096988}{\emph{Commun. Math.
  Phys.} {\bf 129} 393 (1990)}.

\bibitem{Freed:1991bn}
D.~S. Freed and F.~Quinn, \emph{{Chern-Simons theory with finite gauge group}},
  \href{http://dx.doi.org/10.1007/BF02096860}{\emph{Commun. Math. Phys.} {\bf
  156} 435--472 (1993)}, [\href{https://arxiv.org/abs/hep-th/9111004}{{\tt
  arXiv:hep-th/9111004}}].

\bibitem{LurieLectures}
J.~Lurie, ``Finiteness and ambidexterity in k(n)-local stable homotopy
  theory.'' Lectures at Notre Dame Graduate Summer School on Topology and Field
  Theories, 2012.

\bibitem{Kapustin:2014tfa}
A.~Kapustin, \emph{{Symmetry Protected Topological Phases, Anomalies, and
  Cobordisms: Beyond Group Cohomology}},
  \href{https://arxiv.org/abs/1403.1467}{{\tt arXiv:1403.1467}}.

\bibitem{Kapustin:2014dxa}
A.~Kapustin, R.~Thorngren, A.~Turzillo and Z.~Wang, \emph{{Fermionic Symmetry
  Protected Topological Phases and Cobordisms}},
  \href{http://dx.doi.org/10.1007/JHEP12(2015)052}{\emph{JHEP} {\bf 12} 052
  (2015)}, [\href{https://arxiv.org/abs/1406.7329}{{\tt arXiv:1406.7329}}].

\bibitem{Yonekura:2018ufj}
K.~Yonekura, \emph{{On the cobordism classification of symmetry protected
  topological phases}},  \href{https://arxiv.org/abs/1803.10796}{{\tt
  arXiv:1803.10796}}.

\bibitem{Wang2014oya1404.7854}
J.~Wang and X.-G. Wen, \emph{{Non-Abelian string and particle braiding in
  topological order: Modular SL(3,$\mathbb{Z}$) representation and (3+1)
  -dimensional twisted gauge theory}},
  \href{http://dx.doi.org/10.1103/PhysRevB.91.035134}{\emph{Phys. Rev.} {\bf
  B91} 035134 (2015)}, [\href{https://arxiv.org/abs/1404.7854}{{\tt
  arXiv:1404.7854}}].

\bibitem{2017RMPWen1610.03911}
X.-G. {Wen}, \emph{{Colloquium: Zoo of quantum-topological phases of matter}},
  \href{http://dx.doi.org/10.1103/RevModPhys.89.041004}{\emph{Reviews of Modern
  Physics} {\bf 89} 041004 (2017 Oct.)},
  [\href{https://arxiv.org/abs/1610.03911}{{\tt arXiv:1610.03911}}].

\bibitem{Prakash2018ugo1804.11236}
A.~Prakash, J.~Wang and T.-C. Wei, \emph{{Unwinding Short-Range Entanglement}},
  \href{http://dx.doi.org/10.1103/PhysRevB.98.125108}{\emph{Phys. Rev.} {\bf
  B98} 125108 (2018)}, [\href{https://arxiv.org/abs/1804.11236}{{\tt
  arXiv:1804.11236}}].

\bibitem{2001KitaevWire}
A.~Y. {Kitaev}, \emph{{6. QUANTUM COMPUTING: Unpaired Majorana fermions in
  quantum wires}},
  \href{http://dx.doi.org/10.1070/1063-7869/44/10S/S29}{\emph{Physics Uspekhi}
  {\bf 44} 131 (2001 Oct.)},
  [\href{https://arxiv.org/abs/cond-mat/0010440}{{\tt
  arXiv:cond-mat/0010440}}].

\bibitem{Belov:2005ze}
D.~Belov and G.~W. Moore, \emph{{Classification of Abelian spin Chern-Simons
  theories}},  \href{https://arxiv.org/abs/hep-th/0505235}{{\tt
  arXiv:hep-th/0505235}}.

\bibitem{kirby19913}
R.~Kirby and P.~Melvin, \emph{{The 3-manifold invariants of Witten and
  Reshetikhin-Turaev for sl(2,C)}}, {\emph{Inventiones mathematicae} {\bf 105}
  473--545 (1991)}.

\bibitem{reshetikhin1991invariants}
N.~Reshetikhin and V.~G. Turaev, \emph{Invariants of 3-manifolds via link
  polynomials and quantum groups}, {\emph{Inventiones mathematicae} {\bf 103}
  547--597 (1991)}.

\bibitem{abp}
D.~W. Anderson, E.~Brown and F.~P. Peterson, \emph{The structure of the spin
  cobordism ring}, {\emph{Annals of Mathematics} 271--298 (1967)}.

\bibitem{beaudry2018guide}
A.~Beaudry and J.~A. Campbell, \emph{A guide for computing stable homotopy
  groups}, {\emph{arXiv preprint arXiv:1801.07530} (2018)}.

\bibitem{yu1995connective}
C.-Y. Yu, \emph{The connective real K-theory of elementary abelian 2-groups}.
\newblock PhD thesis, University of Notre Dame, 1995.

\bibitem{bruner2010connective}
R.~R. Bruner and J.~P.~C. Greenlees, \emph{Connective real K-theory of finite
  groups}.
\newblock No.~169. American Mathematical Soc., 2010.

\bibitem{Kitaev2009mg0901.2686}
A.~Kitaev, \emph{{Periodic table for topological insulators and
  superconductors}}, \href{http://dx.doi.org/10.1063/1.3149495}{\emph{AIP Conf.
  Proc.} {\bf 1134} 22--30 (2009)},
  [\href{https://arxiv.org/abs/0901.2686}{{\tt arXiv:0901.2686}}].

\bibitem{Putrov:2016qdo}
P.~Putrov, J.~Wang and S.-T. Yau, \emph{{Braiding Statistics and Link
  Invariants of Bosonic/Fermionic Topological Quantum Matter in 2+1 and 3+1
  dimensions}}, \href{http://dx.doi.org/10.1016/j.aop.2017.06.019}{\emph{Annals
  Phys.} {\bf 384} 254--287 (2017)},
  [\href{https://arxiv.org/abs/1612.09298}{{\tt arXiv:1612.09298}}].

\bibitem{milnor2016characteristic}
J.~Milnor and J.~D. Stasheff, \emph{Characteristic Classes.(AM-76)}, vol.~76.
\newblock Princeton university press, 2016.

\bibitem{sato1984cobordisms}
N.~Sato, \emph{Cobordisms of semi-boundary links}, {\emph{Topology and its
  Applications} {\bf 18} 225--234 (1984)}.

\bibitem{Wang:2018edf}
J.~{Wang}, K.~{Ohmori}, P.~{Putrov}, Y.~{Zheng}, Z.~{Wan}, M.~{Guo} et~al.,
  \emph{{Tunneling Topological Vacua via Extended Operators: (Spin-)TQFT
  Spectra and Boundary Deconfinement in Various Dimensions}},
  \href{http://dx.doi.org/10.1093/ptep/pty051}{\emph{PTEP} {\bf 2018} 053A01
  (2018)}, [\href{https://arxiv.org/abs/1801.05416}{{\tt arXiv:1801.05416}}].

\bibitem{kirby2004local}
R.~Kirby and P.~Melvin, \emph{Local surgery formulas for quantum invariants and
  the arf invariant}, {\emph{arXiv preprint math/0410358} (2004)}.

\bibitem{saito1993unoriented}
M.~Saito, \emph{On the unoriented sato-levine invariant}, {\emph{Journal of
  Knot Theory and Its Ramifications} {\bf 2} 335--358 (1993)}.

\bibitem{cochran1984invariant}
T.~D. Cochran, \emph{On an invariant of link cobordism in dimension four},
  {\emph{Topology and its Applications} {\bf 18} 97--108 (1984)}.

\bibitem{Debray:2018wfz}
A.~Debray and S.~Gunningham, \emph{{The Arf-Brown TQFT of Pin$^-$ Surfaces}},
  \href{https://arxiv.org/abs/1803.11183}{{\tt arXiv:1803.11183}}.

\bibitem{gunningham2016spin}
S.~Gunningham, \emph{Spin hurwitz numbers and topological quantum field
  theory}, {\emph{Geometry \& Topology} {\bf 20} 1859--1907 (2016)}.

\bibitem{khovanov1999categorification}
M.~Khovanov, \emph{A categorification of the jones polynomial}, {\emph{arXiv
  preprint math/9908171} (1999)}.

\bibitem{campbell}
J.~A. Campbell, \emph{Homotopy theoretic classification of symmetry protected
  phases}, {\emph{arXiv preprint arXiv:1708.04264} (2017)}.

\bibitem{hsieh}
C.-T. Hsieh, \emph{Discrete gauge anomalies revisited}, {\emph{arXiv preprint
  arXiv:1808.02881} (2018)}.

\bibitem{Gaiotto:2017zba}
D.~Gaiotto and T.~Johnson-Freyd, \emph{{Symmetry Protected Topological phases
  and Generalized Cohomology}},  \href{https://arxiv.org/abs/1712.07950}{{\tt
  arXiv:1712.07950}}.

\bibitem{2018arXiv180408628F}
L.~{Fidkowski}, A.~{Vishwanath} and M.~A. {Metlitski}, \emph{{Surface
  Topological Order and a new 't Hooft Anomaly of Interaction Enabled 3+1D
  Fermion SPTs}}, {\emph{ArXiv e-prints} (2018 Apr.)},
  [\href{https://arxiv.org/abs/1804.08628}{{\tt arXiv:1804.08628}}].

\bibitem{Garcia-Etxebarria:2017crf}
I.~García-Etxebarria, H.~Hayashi, K.~Ohmori, Y.~Tachikawa and K.~Yonekura,
  \emph{{8d gauge anomalies and the topological Green-Schwarz mechanism}},
  \href{http://dx.doi.org/10.1007/JHEP11(2017)177}{\emph{JHEP} {\bf 11} 177
  (2017)}, [\href{https://arxiv.org/abs/1710.04218}{{\tt arXiv:1710.04218}}].

\bibitem{Debray2018wfz1803.11183}
A.~Debray and S.~Gunningham, \emph{{The Arf-Brown TQFT of Pin$^-$ Surfaces}},
  \href{https://arxiv.org/abs/1803.11183}{{\tt arXiv:1803.11183}}.

\bibitem{0904.2197FK}
L.~{Fidkowski} and A.~{Kitaev}, \emph{{Effects of interactions on the
  topological classification of free fermion systems}},
  \href{http://dx.doi.org/10.1103/PhysRevB.81.134509}{\emph{Phys. Rev. B} {\bf
  81} 134509 (2010 Apr.)}, [\href{https://arxiv.org/abs/0904.2197}{{\tt
  arXiv:0904.2197}}].

\bibitem{ChenLu1303.4301}
X.~{Chen}, Y.-M. {Lu} and A.~{Vishwanath}, \emph{{Symmetry-protected
  topological phases from decorated domain walls}},
  \href{http://dx.doi.org/10.1038/ncomms4507}{\emph{Nature Communications} {\bf
  5} 3507 (2014 Mar.)}, [\href{https://arxiv.org/abs/1303.4301}{{\tt
  arXiv:1303.4301}}].

\bibitem{GuWW2015lfa1503.01768}
Z.-C. Gu, J.~C. Wang and X.-G. Wen, \emph{{Multi-kink topological terms and
  charge-binding domain-wall condensation induced symmetry-protected
  topological states: Beyond Chern-Simons/BF theory}},
  \href{http://dx.doi.org/10.1103/PhysRevB.93.115136}{\emph{Phys. Rev.} {\bf
  B93} 115136 (2016)}, [\href{https://arxiv.org/abs/1503.01768}{{\tt
  arXiv:1503.01768}}].

\bibitem{Wang2017loc1705.06728}
J.~Wang, X.-G. Wen and E.~Witten, \emph{{Symmetric Gapped Interfaces of SPT and
  SET States: Systematic Constructions}},
  \href{http://dx.doi.org/10.1103/PhysRevX.8.031048}{\emph{Phys. Rev.} {\bf X8}
  031048 (2018)}, [\href{https://arxiv.org/abs/1705.06728}{{\tt
  arXiv:1705.06728}}].

\bibitem{2017arXiv1712.09542Yuji}
Y.~{Tachikawa}, \emph{{On gauging finite subgroups}}, {\emph{ArXiv e-prints}
  (2017 Dec.)}, [\href{https://arxiv.org/abs/1712.09542}{{\tt
  arXiv:1712.09542}}].

\bibitem{2011PhRvLFu}
L.~{Fu}, \emph{{Topological Crystalline Insulators}},
  \href{http://dx.doi.org/10.1103/PhysRevLett.106.106802}{\emph{Phys.Rev.Lett}
  {\bf 106} 106802 (2011 Mar.)}, [\href{https://arxiv.org/abs/1010.1802}{{\tt
  arXiv:1010.1802}}].

\bibitem{Thorngren2018wwt1612.00846}
R.~Thorngren and D.~V. Else, \emph{{Gauging Spatial Symmetries and the
  Classification of Topological Crystalline Phases}},
  \href{http://dx.doi.org/10.1103/PhysRevX.8.011040}{\emph{Phys. Rev.} {\bf X8}
  011040 (2018)}, [\href{https://arxiv.org/abs/1612.00846}{{\tt
  arXiv:1612.00846}}].

\bibitem{2017Po1703.00911}
H.~C. {Po}, A.~{Vishwanath} and H.~{Watanabe}, \emph{{Complete theory of
  symmetry-based indicators of band topology}},
  \href{http://dx.doi.org/10.1038/s41467-017-00133-2}{\emph{Nature
  Communications} {\bf 8} 50 (2017 June)},
  [\href{https://arxiv.org/abs/1703.00911}{{\tt arXiv:1703.00911}}].

\bibitem{PRX1604.08151}
H.~{Song}, S.-J. {Huang}, L.~{Fu} and M.~{Hermele}, \emph{{Topological Phases
  Protected by Point Group Symmetry}},
  \href{http://dx.doi.org/10.1103/PhysRevX.7.011020}{\emph{Physical Review X}
  {\bf 7} 011020 (2017 Jan.)}, [\href{https://arxiv.org/abs/1604.08151}{{\tt
  arXiv:1604.08151}}].

\bibitem{2017PhRvB1705.09243}
S.-J. {Huang}, H.~{Song}, Y.-P. {Huang} and M.~{Hermele}, \emph{{Building
  crystalline topological phases from lower-dimensional states}},
  \href{http://dx.doi.org/10.1103/PhysRevB.96.205106}{\emph{Physical Review B}
  {\bf 96} 205106 (2017 Nov.)}, [\href{https://arxiv.org/abs/1705.09243}{{\tt
  arXiv:1705.09243}}].

\bibitem{Cheng2018aaz1810.12308}
M.~Cheng and C.~Wang, \emph{{Rotation Symmetry-Protected Topological Phases of
  Fermions}},  \href{https://arxiv.org/abs/1810.12308}{{\tt arXiv:1810.12308}}.

\bibitem{Kapustin2017jrc1701.08264}
A.~Kapustin and R.~Thorngren, \emph{{Fermionic SPT phases in higher dimensions
  and bosonization}},
  \href{http://dx.doi.org/10.1007/JHEP10(2017)080}{\emph{JHEP} {\bf 10} 080
  (2017)}, [\href{https://arxiv.org/abs/1701.08264}{{\tt arXiv:1701.08264}}].

\bibitem{2018arXiv180108530Lan}
T.~{Lan} and X.-G. {Wen}, \emph{{A classification of 3+1D bosonic topological
  orders (II): the case when some point-like excitations are fermions}},
  {\emph{ArXiv e-prints} arXiv:1801.08530 (2018 Jan.)},
  [\href{https://arxiv.org/abs/1801.08530}{{\tt arXiv:1801.08530}}].

\bibitem{Gaiotto2014kfa1412.5148}
D.~Gaiotto, A.~Kapustin, N.~Seiberg and B.~Willett, \emph{{Generalized Global
  Symmetries}}, \href{http://dx.doi.org/10.1007/JHEP02(2015)172}{\emph{JHEP}
  {\bf 02} 172 (2015)}, [\href{https://arxiv.org/abs/1412.5148}{{\tt
  arXiv:1412.5148}}].

\bibitem{Wang2017moj1703.10937}
Q.-R. Wang and Z.-C. Gu, \emph{{Towards a Complete Classification of
  Symmetry-Protected Topological Phases for Interacting Fermions in Three
  Dimensions and a General Group Supercohomology Theory}},
  \href{http://dx.doi.org/10.1103/PhysRevX.8.011055}{\emph{Phys. Rev.} {\bf X8}
  011055 (2018)}, [\href{https://arxiv.org/abs/1703.10937}{{\tt
  arXiv:1703.10937}}].

\bibitem{Wang2018pdc1811.00536}
Q.-R. Wang and Z.-C. Gu, \emph{{Construction and classification of symmetry
  protected topological phases in interacting fermion systems}},
  \href{https://arxiv.org/abs/1811.00536}{{\tt arXiv:1811.00536}}.

\bibitem{2016arXiv161202860BMorgan}
G.~{Brumfiel} and J.~{Morgan}, \emph{{The Pontrjagin Dual of 3-Dimensional Spin
  Bordism}}, {\emph{ArXiv e-prints} (2016 Dec.)},
  [\href{https://arxiv.org/abs/1612.02860}{{\tt arXiv:1612.02860}}].

\bibitem{2018arXiv180308147BMorgan}
G.~{Brumfiel} and J.~{Morgan}, \emph{{The Pontrjagin Dual of 4-Dimensional Spin
  Bordism}}, {\emph{ArXiv e-prints} (2018 Mar.)},
  [\href{https://arxiv.org/abs/1803.08147}{{\tt arXiv:1803.08147}}].

\bibitem{2017PhRvB1610.08478}
C.~{Wang}, C.-H. {Lin} and Z.-C. {Gu}, \emph{{Interacting fermionic
  symmetry-protected topological phases in two dimensions}},
  \href{http://dx.doi.org/10.1103/PhysRevB.95.195147}{\emph{Phys. Rev. B} {\bf
  95} 195147 (2017 May)}, [\href{https://arxiv.org/abs/1610.08478}{{\tt
  arXiv:1610.08478}}].

\bibitem{Wang:2014wka}
C.~Wang and M.~Levin, \emph{{Topological invariants for gauge theories and
  symmetry-protected topological phases}},
  \href{http://dx.doi.org/10.1103/PhysRevB.91.165119}{\emph{Phys. Rev.} {\bf
  B91} 165119 (2015)}, [\href{https://arxiv.org/abs/1412.1781}{{\tt
  arXiv:1412.1781}}].

\bibitem{Wang2016jdt1602.05569Thesis}
J.~C.-F. Wang, \emph{{Aspects of Symmetry, Topology and Anomalies in Quantum
  Matter}}.
\newblock PhD thesis, MIT, Cambridge, Dept. Phys., 2015.
\newblock \href{https://arxiv.org/abs/1602.05569}{{\tt arXiv:1602.05569}}.

\bibitem{Wang2016qhf1602.05951}
J.~Wang, X.-G. Wen and S.-T. Yau, \emph{{Quantum Statistics and Spacetime
  Surgery}},  \href{https://arxiv.org/abs/1602.05951}{{\tt arXiv:1602.05951}}.

\bibitem{Tiwari2016zru1603.08429}
A.~Tiwari, X.~Chen and S.~Ryu, \emph{{Wilson operator algebras and ground
  states of coupled BF theories}},
  \href{http://dx.doi.org/10.1103/PhysRevB.95.245124}{\emph{Phys. Rev.} {\bf
  B95} 245124 (2017)}, [\href{https://arxiv.org/abs/1603.08429}{{\tt
  arXiv:1603.08429}}].

\bibitem{Chan2017eov}
A.~P.~O. Chan, P.~Ye and S.~Ryu, \emph{{Braiding with Borromean Rings in
  (3+1)-Dimensional Spacetime}},
  \href{http://dx.doi.org/10.1103/PhysRevLett.121.061601}{\emph{Phys. Rev.
  Lett.} {\bf 121} 061601 (2018)},
  [\href{https://arxiv.org/abs/1703.01926}{{\tt arXiv:1703.01926}}].

\bibitem{Cheng2017ftw1705.08911}
M.~Cheng, N.~Tantivasadakarn and C.~Wang, \emph{{Loop Braiding Statistics and
  Interacting Fermionic Symmetry-Protected Topological Phases in Three
  Dimensions}}, \href{http://dx.doi.org/10.1103/PhysRevX.8.011054}{\emph{Phys.
  Rev.} {\bf X8} 011054 (2018)}, [\href{https://arxiv.org/abs/1705.08911}{{\tt
  arXiv:1705.08911}}].

\bibitem{Wang2018iwz1810.13428}
Q.-R. Wang, M.~Cheng, C.~Wang and Z.-C. Gu, \emph{{Topological Quantum Field
  Theory for Abelian Topological Phases and Loop Braiding Statistics in
  $(3+1)$-Dimensions}},  \href{https://arxiv.org/abs/1810.13428}{{\tt
  arXiv:1810.13428}}.

\bibitem{Witten:1988hf}
E.~Witten, \emph{{Quantum Field Theory and the Jones Polynomial}},
  \href{http://dx.doi.org/10.1007/BF01217730}{\emph{Commun. Math. Phys.} {\bf
  121} 351--399 (1989)}.

\bibitem{2018arXiv181012317RYMLu}
A.~{Rasmussen} and Y.-M. {Lu}, \emph{{Intrinsically interacting topological
  crystalline insulators and superconductors}}, {\emph{ArXiv e-prints} (2018
  Oct.)}, [\href{https://arxiv.org/abs/1810.12317}{{\tt arXiv:1810.12317}}].

\bibitem{Ando:2015sia}
Y.~Ando and L.~Fu, \emph{{Topological Crystalline Insulators and Topological
  Superconductors: From Concepts to Materials}},
  \href{http://dx.doi.org/10.1146/annurev-conmatphys-031214-014501}{\emph{Ann.
  Rev. Condensed Matter Phys.} {\bf 6} 361 (2015)},
  [\href{https://arxiv.org/abs/1501.00531}{{\tt arXiv:1501.00531}}].

\bibitem{2018arXiv1810.11013YangQi}
Z.~{Song}, C.~{Fang} and Y.~{Qi}, \emph{{Real-space recipes for general
  topological crystalline states}}, {\emph{ArXiv e-prints} (2018 Oct.)},
  [\href{https://arxiv.org/abs/1810.11013}{{\tt arXiv:1810.11013}}].

\bibitem{Han2018xsvPYe1807.10844}
B.~Han, H.~Wang and P.~Ye, \emph{{Symmetry-protected topological phases with
  both spatial and internal symmetries}},
  \href{https://arxiv.org/abs/1807.10844}{{\tt arXiv:1807.10844}}.

\bibitem{Shiozaki2018yyjKen1810.00801}
K.~Shiozaki, C.~Z. Xiong and K.~Gomi, \emph{{Generalized homology and
  Atiyah-Hirzebruch spectral sequence in crystalline symmetry protected
  topological phenomena}},  \href{https://arxiv.org/abs/1810.00801}{{\tt
  arXiv:1810.00801}}.

\bibitem{2018arXiv180901112Lan}
T.~{Lan}, C.~{Zhu} and X.-G. {Wen}, \emph{{Fermion decoration construction of
  symmetry protected trivial orders for fermion systems with any symmetries
  $G\_f$ and in any dimensions}}, {\emph{ArXiv e-prints} (2018 Sept.)},
  [\href{https://arxiv.org/abs/1809.01112}{{\tt arXiv:1809.01112}}].

\end{thebibliography}\endgroup

\end{document}